\documentclass[11pt]{article}
\usepackage{old-arrows}
\usepackage[scaled=.9]{helvet}
\usepackage{charter}
\usepackage{XoohmH}
 \newmdenv[linecolor=blue!60!teal,
 linewidth=1,roundcorner=2pt,backgroundcolor=yellow!3!red!1,
 innerleftmargin=5pt,innerrightmargin=5pt,leftmargin=0pc,rightmargin=0pt,
 fontcolor=blue!60!black,
 ]{bBox}
\usepackage{cite}

\def\SI#1{\setbox9\hbox{$\SSS#1$}\5[-1pt]{\copy9\kern-.125\wd9}{\mathscr{I}}}
\newcommand{\hD}{\ensuremath{\widehat{D}}}
\newcommand{\tD}{\ensuremath{\tilde{D}}}

\newcommand{\tT}[1][-4mu]{\mkern1mu^\intercal\mkern#1}
\newcommand{\tW}[1][-4mu]{\mkern1mu^{\sss\wtd}\mkern#1}
\newcommand{\tX}{{\mkern2mu^\wtd\mkern-5muX}}
\newcommand{\FF}[2][n]{F^{\sss(#1)}_{#2}}
\newcommand{\MF}[2][n]{\tW F^{\sss(#1)}_{#2}}
\newcommand{\EE}[2][n]{E^{\sss(#1)}_{#2}}
\newcommand{\ME}[2][n]{\tW E^{\sss(#1)}_{#2}}
\newcommand{\XX}[2][n-1]{X^{\sss(#1)}_{#2}}

\newcommand{\pDN}[2][]{{\pD^{\mkern-3mu\raisemath{1pt}{#1}}_{\mkern-1mu\smash{#2}}}}
\newcommand{\pDs}[2][]{\pDN[\star#1]{#2}}
\newcommand{\pFn}[2][]{{\pS^{\raisemath{1pt}{#1}}_{\mkern-2mu\smash{#2}}}}
\newcommand{\pFN}[2][]{\pFn[\wtd#1]{#2}}
\newcommand{\cKs}[1]{{\cK^*_{\!\smash{#1}}}}

\def\zZ#1#2{\big(\ZZ_{#1}{:}\,{#2}\big)}
\long\def\oMit#1{}
\let\vtl=\vartriangleleft

 \SfTitles
 \allowdisplaybreaks
 \numberwithin{equation}{section}

 \omitBackreferences

\begin{document}
\displayBAstretch[.667]
\thispagestyle{empty}
\setcounter{page}{0}
\vglue5mm
\begin{center}
{\LARGE\sf\bfseries\boldmath
  Chern Characteristics and Todd--Hirzebruch Identities\\[1mm]
  for Transpolar Pairs of Toric Spaces
}
\vspace{2mm}

\begin{tabular}{rcl}
\makebox[70mm][r]{\sf\bfseries Per Berglund$^{*\dag}$} &and&
\makebox[70mm][l]{\sf\bfseries Tristan H\"{u}bsch$^\ddag$}\\*[1mm]
\MC3c{\small\it
  $^*$%
      Department of Physics, University of New Hampshire, Durham, NH 03824, USA}\\[-1mm]
\MC3c{\small\it
  $^\dag$%
      Theoretical Physics Department, CERN, CH-1211 Geneve, Switzerland}\\[-1mm]
\MC3c{\small\it
  $^\ddag$%
      Department of Physics \&\ Astronomy,
      Howard University, Washington, DC 20059, USA}\\[-1mm]
 {\tt per.berglund@unh.edu} &and& {\tt  thubsch@howard.edu}
\end{tabular}
\vspace{2mm}

{\sf\bfseries ABSTRACT}\\[3mm]
\parbox{156mm}{\addtolength{\baselineskip}{-3pt}\parindent=2pc\noindent
Standard algebraic toric geometry used to construct 
Calabi--Yau varieties extends to hypersurfaces in non-Fano varieties by star-triangulating non-convex polytopes.
Generalizing the original construction of Batyrev and Borisov,
transposition mirror symmetry then naturally includes via a generalized {\em\/transpolar\/} duality certain flip-folded, multi-layered multihedral so-called {\em\/VEX multitopes.} Long since known in presymplectic geometry, these VEX multitopes are found to correspond to (non-algebraic) unitary toric spaces with equivariant Chern classes that satisfy the standard Todd--Hirzebruch identities. The computation of diffeomorphism invariants, including characteristic submanifold intersection numbers, corroborates the recent inclusion of such non-Fano embeddings in the connected web of Calabi--Yau spaces, together with standard (semi-Fano/reflexive polytope) constructions, within explicit deformation families of generalized complete intersections in products of projective spaces. In turn, VEX multitopes suggest how to construct such unitary toric spaces from toric varieties by surgery.
}

\vfill
\begin{minipage}{.75\hsize}\small
  \baselineskip=10pt plus1pt minus 1pt
  \renewcommand{\cftbeforetoctitleskip}{\smallskipamount}
  \renewcommand{\cftaftertoctitle}{\vskip-5pt}
  \tocloftpagestyle{empty}
  \setcounter{tocdepth}{3} 
  \tableofcontents
\end{minipage}
\end{center}
\vfill
\clearpage


\section{Introduction, Rationale and Summary}
\label{s:IRS}
Calabi--Yau (complex, compact, Ricci-flat) spaces have been studied intensely in physics as factors of spacetime through which superstrings propagate consistently\cite{rCHSW}. String dynamics easily permits various types of singularization\cite{rDHVW1}, some of which serve to connect topologically distinct models\cite{rReidK0, rCYCI1, rGHC, rGHPT, rCGH1, rCGH2, rBeast2, rSingS, Avram:1995pu, Avram:1997rs, McNamara:2019rup}.
 Within a decade, constructions and analysis refocused on the (abelian) {\em\/gauged linear sigma models\/} (GLSMs,\cite{rPhases, rMP0} and\cite{Fan:2015vca} for a more formal but primarily algebraic description), where the supersymmetry-complexified (abelian) gauge group and other symmetries govern the corresponding ({\em\/complex-algebraic\/}) toric geometry\cite{rKKMS-TE1, rD-TV, rO-TV, rF-TV, rGE-CCAG, rCLS-TV, rCox, rCox95, Cox:1997Rec}.\footnote{\label{fn:StrSuSy}The worldsheet-supersymmetric gauge-group complexification, $\U(1)\<\into\U(1;\IC)\<=\IC^*$,
admits some degeneration and boundary sources (target-space 'branes), loosening these underlying assumptions.
String theory with less supersymmetry couples to target space structures more general than its (co)tangent bundle\cite{Distler:1993mk, Schafer-Nameki:2016cfr, Sharpe:2024dcd};
 worldsheet $(0,1)$-supersymmetry is minimal as it suffices to guarantee stability (absence of tachyons), though not also a complex structure in the target spacetime\cite{rUDSS08,rUDSS09}.}
 In turn, the worldsheet quantum field theory underlying superstrings in general has a manifestly $T$-dual and mirror-symmetric formulation, whence our fundamental reliance on it\cite{rJPS, Freidel:2015pka, Berglund:2021hbo, rBeast2}.
 
 This framework produced a database of almost half a billion {\em\/reflexive\/} polytopes\cite{Kreuzer:2000xy}, each encoding at least one (semi) Fano {\em\/complex-algebraic\/} toric 4-fold, $X$, with a deformation family of anticanonical (Calabi--Yau) hypersurfaces.
 This vast pool of constructions exhibits {\em\/mirror symmetry\/}:
The (quantum) cohomology rings of pairs of Calabi--Yau hypersurfaces, $(Z_f,Z_g)$, satisfy $H^{p,q}(Z_f)\approx H^{n-p,q}(Z_g)$\cite{rGP1},
 are defined by anticanonical sections, $f\<\in\G(\cKs{X})$ and $g\<\in\G(\cKs{Y})$, and encoded by a polar pair of convex, reflexive polytopes,
 $(\pDN{X},\pDN{Y}\<=\pDN[\,\circ]{X})$\cite{Aspinwall:1993nu, rBaty01, rCOK, rBatyBor2, rBatyBor3}, 
\begin{equation}
  Z_f ~\lhook\joinrel\too{~f(x)=0~}~
  X\fif{\text{1--1}}\pDN{X} ~~\fif{~\circ~}~~
  \pDN{Y}\fif{\text{1--1}} Y 
  ~\fro{~g(y)=0~}\joinrel\rhook~ Z_g.
 \label{e:MM}
\end{equation}
This overlaps with early (transposition, $g\<=\tT f$) mirrors\cite{rBH, rBH-LGO+EG, rFJR-07b, Krawitz:2009aa, rMK-PhD, Ebeling:2012Mir} (also\cite{rLB-MirrBH, rACG-BHK, rA+P-MM, rF+K-BHK, Parkhomenko:2022kju, Belavin:2023ldz, Parkhomenko:2024mxq, Cho:2024Ber}), their judiciously chosen default finite quotients and desingularizations being encoded automatically when {\em\/smooth\/} star-triangulations\footnote{\label{fn:smooth}A lattice cone, $\s$, is {\em\/smooth\/} if its minimal generators form a lattice $\ZZ$-basis, so it encodes a smooth affine chart, $U_\s\<=\Spec(\IC[\check\s])\<=\IC^n$\cite{rF-TV, rGE-CCAG, rCLS-TV}. The minimal generators of a smooth $k$-dimensional cone form a {\em\/unit simplex,} of minimal Euclidean volume, $1/k!$. A (subdivision/triangulation of a) fan is smooth if all its cones are smooth. A triangulation of a polytope is smooth if it consists of unit simplices only; a star-triangulation consists of simplices that all share a common vertex, the {\em\/center.}} of $\pDN{X},\pDN{Y}$,
 encode the ``MPCP-desingularization'' of $X$ and $Y$, as well as their Calabi--Yau hypersurfaces, $Z_f\<\subset X$ and $Z_g\<\subset Y$\cite{Aspinwall:1993nu, rBaty01}.
 Linear combinations of anticanonical sections generate explicit continuous deformation families of models, while various polytope retriangulations form a web of discrete topology changes\cite{Aspinwall:1993nu}.
 The vast collection of polar pairs of {\em\/reflexive\/} (and so convex) polytopes\cite{Kreuzer:2000xy} thereby encodes a connected web\cite{rReidK0} of string vacua corresponding to Calabi--Yau 3-fold mirror pairs\cite{Avram:1995pu, Avram:1997rs} embedded as hypersurfaces in (semi) Fano complex-algebraic toric varieties ($X$ and $Y$).

However, Calabi--Yau hypersurfaces (and corresponding GLSMs) are also found in {\em\/non-Fano\/} varieties (e.g., Hirzebruch $n$-fold scrolls, $\FF{m}$ for $m\<>2$) encoded by {\em\/non-convex,} smooth, star-triangulated polytopes, $\pDs{X}$, with facets at unit distance from the origin\cite{rBH-gB, Berglund:2022dgb}.
 The same hypersurfaces also occur as special generalized complete intersections (gCICY) in products of projective spaces\cite{rgCICY1, rBH-Fm, rGG-gCI}, and {\em\/within explicit deformation families\/} that contain also their (semi) Fano ``cousin'' embeddings\cite{Berglund:2022dgb} --- generalizing Gross' and Ruan's {\em\/non-algebraic deformation equivalence\/} example, 
$\FF[4]{\!\sss(2,1,1,0)}\<{\approx_\IR}\FF[4]0$\cite{Gross:1994The, Ruan:1996top}.
The Newton polytope, $\pDN{X}$ (encoding anticanonical sections
in terms of Cox coordinates specified by the spanning polytope,
$\pDs{X}$),
was shown to agree computationally (Euler and Hodge numbers, select cohomology) also with the gCICY and 
$\FF{m}\<=\IP(\cO\<\oplus\cO(m)^{\oplus(n-1)})$ realizations\cite{rBH-gB, Berglund:2022dgb} {\em\/precisely if\/}:
({\small\bf1})~the standard polar operation ($\mkern2mu^\circ\mkern1mu$) is refined to its face-wise iterative {\em\/transpolar\/} operation 
($\mkern2mu^\wtd\mkern1mu$),\footnote{We thank Hal Schenck for letting us know that many practitioners of toric geometry in fact routinely use precisely this face-wise iterative procedure (Definition~\ref{D:tP}, below), the origins of which are ``lost in the myst of time.''}
so $\pDN{X}{=}(\pDs{X})^\wtd$, which
({\small\bf2})~perforce generalizes $\pDN{X}$ to include also certain (``VEX'') flip-folded, oriented and multilayered {\em\/multitopes.}\footnote{Corresponding to ``multifans''\cite{rM-MFans, Masuda:2000aa, rHM-MFs}, we use the ``multitope'' contraction of ``multi-polytope''\cite{rHM-MFs, rAM-MFs}, meant as a synonym of ``virtual polytopes''\cite{rK+P-vrtPlytp0, rK+P-vrtPlytp}, ``twisted polytope''\cite{rK+T-pSympTM}, and ``polyhedral complex''\cite{Hibi:1995aa}.}
 
Swapping the roles $\pDN{X}\<\iff\pDs{X}$ in a transpolar pair perfectly reproduces the {\em\/transpose\/}\cite{rBH}, $\tT f(y)$, of each anticanonical section, $f(x)\<\in\G(\cKs{X})$,
if one includes the {\em\/extension,} $(\pDs{X})^\wtd\<\ssm(\pDs{X})^\circ$.
The corresponding rational monomials
desingularize the Calabi--Yau hypersurface, $Z_f$, in such non-Fano toric spaces and enable all combinatorial computations\cite{rBH-gB, Berglund:2022dgb}; see below.
There is a continuum of {\em\/unitary (weakly almost complex) torus manifolds\/} corresponding to 
$(\pDs{\tX},\pDN{\tX})\<\coeq(\pDN{X},\pDs{X})$\cite{rM-MFans, Masuda:2000aa, rHM-MFs, Civan:2003aa, Masuda:2006aa, rHM-EG+MF, rH-EG+MFs2, Nishimura:2006vs, Ishida:2013aa, Ishida:2013ab, buchstaber2014toric}. 
Among these, we expect at least one, 
$\tX$, to admit a ring of complex $\pDs{\tX}\<\coeq\pDN{X}$-vertex specified $y$-coordinates,
so $y$-monomials corresponding to lattice points in 
$\pDN{\tX}\<\coeq\pDs{X}$
constitute $\tT f(y)$, the transpose of $f(x)$, as a section of its known ``anticanonical'' line-bundle\cite{rM-MFans, Masuda:2000aa, rHM-MFs}; see Remark~\ref{r:UTM}, below.

\begin{remk}[GLSM]
\label{r:GLSM}
A worldsheet-supersymmetric GLSM (underlying string theory models; see footnote~\ref{fn:StrSuSy}) is well specified by $n{+}r{+}1$ {\em\/complex\/} superfields, $\{x_i,P\}$, where
anomaly-free $\U(1)^r$-gauge invariance of 
the superpotential $P{\cdot}f(x)$ 
is insured by $\deg(P)\<={-}\big(\sum_{i=1}^{n+r}\deg(x_i)\<=\deg(f)\big)$ 
if $f(x)$ is a transverse deformation of the ``fundamental monomial'',
$\Pi x$\cite{rHY-SL2, Hubsch:2025sph, Hubsch:2025teh}, reducing a complementary $\U(1)^n$-action
to the discrete ``geometric symmetry''\cite{rBH, rMP0}.
Transposition\footnote{Originally framed for Landau--Ginzburg orbifolds (LGO), {\em\/transposition\/}\cite{rBH} generalizes the original mirror construction\cite{rGP1}, overlaps with the toric construction\cite{rBaty01}, and extends to GLSMs via their LGO phase\cite{rPhases}; see also\cite{rBH-gB, Berglund:2022dgb} and\cite{Maxfield:2019czc}.} then {\em\/defines\/} an anomaly-free (dual) $\U(1)^{\Tw{r}}$-gauge invariant $\tT f(y)$ in terms of {\em\/complex\/} superfields, $y_I$, so $\tT f(y)$ is a deformation of $\Pi y$ and specifies the mirror Calabi--Yau GLSM model, Laurent polynomials (and L'Hopital-like ``intrinsic limit''-defined zero-loci) included --- all combinatorially encoded by a transpolar pair of VEX multitopes,
$\big(\pDs{X},\pDN{X})$\cite{rBH-gB, Berglund:2022dgb, Hubsch:2025sph, Hubsch:2025teh}.
Here, we seek to identify the geometry of the ground states in such transposition-paired GLSMs:
the embedding (unitary) torus spaces, $(X,\tX)$
(specified in the ``geometric phases'' by omitting the superpotential),
which have ``anticanonical'' hypersurfaces, $(Z_f,Z_{\tT f})$
(the zero locus of the now included superpotential, possibly including rational monomials\footnote{$f(x)$ includes rational monomials from the extension, 
$(\pDs{X})^\wtd\<\ssm(\pDs{X})^\circ$, in terms of $\pDs{X}$-specified (Cox) variables, $x_i$. Defining the zero-locus, $Z_f$, requires a L'Hopital-like ``intrinsic limit'' --- a priori {\em\/not algebraic\/}\cite{rBH-gB} --- but which nevertheless admits at least two purely algebro-geometric approaches of analysis\cite{Berglund:2022dgb}, which remain to be explored.}),
generalizing the transposition mirror framework~\eqref{e:MM} to that in Figure~\ref{f:gMM} and described in more detail below; see Conjecture~\ref{C:gCYh} in particular.
\end{remk}
\begin{figure}[htb]
\setbox9\hbox{$\eqco$}
\centering
 \begin{tikzpicture}[scale=1.5, 
                         every node/.style={inner sep=0,outer sep=3pt}]
      \path[use as bounding box](-.6,-0.8)--(9.85,1.4);
      \node(K)  at(1.4,1.1)   {$\cKs{X}$\,\rlap{$\iff\sfa\pDN{X}$}};
      \node(G)  at(0,1.1)     {$f\<\in\G(\cKs{X})$};
      \node(P)  at(1.4,.3)  {$X$\,\rlap{$\iff\pFn{X}\<\smt\pDs{X}$}};
      \node(X)  at(1.4,-.6) {\llap{$X[c_1]\<\ni$}$Z_f$};
      \node(*K) at(7.5,1.1)   {\llap{$\sfa(\pDs{X}\<{\copy9}\pDN{\tX})\iff$}%
                                \,$\cKs{\tX}$};
      \node(*G) at(9.0,1.1)   {$\G(\cKs{\tX})\<\ni\tT f$};
      \node(*P) at(7.5,.3)  {\llap{$(\pDN{X}\<{\copy9}\pDs{\tX})\<\lat\pFn{\tX}\iff$}\,$\tX$};
      \node(*X) at(7.5,-.6) {$Z_{\,\tT f}$\rlap{$\in\tX[c_1]$}};
      \draw[semithick, ->>](K)--(P);
      \draw[semithick, <-{Hooks[left]}](P)--node[right=2pt]{\scriptsize$f{=}0$}(X);
      \draw[semithick, ->>](*K)--(*P);
      \draw[semithick, <-{Hooks[right]}](*P)--node[left=2pt]{\scriptsize$\tT f{=}0$}(*X);
      \draw[blue, thick, densely dotted, -stealth](G)
            --node[below, rotate=-50]{\scriptsize def.\ section\quad~~}(X);
      \draw[blue, thick, stealth-stealth](K)--(G);
      \draw[blue, thick, densely dotted, -stealth](*G)
            --node[below, rotate=50]{\scriptsize\quad def.\ section}(*X);
      \draw[blue, thick, stealth-stealth](*K)--(*G);
      \draw[purple, thick, Stealth-Stealth](2.55,1.1) to
             node[above]{\scriptsize(trans)polar\cite{rBH-gB}}
             node[below=-1pt]{\scriptsize$(\pDN{X})^\wtd\<=\pDs{X}$} (5.3,1.1);
      \draw[purple, thick, Stealth-Stealth](2.95,.3) to
             node[above]{\scriptsize(trans)polar\cite{rBH-gB}}
             node[below=-1pt]{\scriptsize$(\pDs{X})^\wtd\<=\pDN{X}$} (4.8,.3);
      \draw[purple, thick, Stealth-Stealth](X)to
             node[above]{\scriptsize transpose-mirrors\cite{rBH, rBH-LGO+EG} \& \cite{rBH-gB, Berglund:2022dgb}} (*X);
   \end{tikzpicture}
 \caption{The transposition mirror framework includes the well-studied cases\cite{rBaty01} when restricted to the standard polar pairs of (convex, plain\,=\,single-layer) reflexive polytopes, $(\pDs{X})^\circ\<=\pDN{X}\<\eqco\pDs{\tX}$.
 }
 \label{f:gMM}
\end{figure}

In lieu of formal proofs, our purpose is to motivate a more rigorous study of Calabi--Yau hypersurfaces in such ($\IC$-ringed) unitary torus spaces,
$(X,\tX)$, in Figure~\ref{f:gMM}:
To this end, we show that standard results about $\rT$-equivariant diffeomorphism invariants of toric varieties are readily computable and continue to hold also for the much broader collection including non-convex and flip-folded VEX multitopes --- with but minor modifications encoded only by a {\em\/continuous orientation\/} of VEX multitopes and multifans.
This being discrete and finite additional information, such toric embedding spaces in Figure~\ref{f:gMM} are expected to still be sparse within the {\em\/continuum\/} of all toric spaces.\footnote{\label{fn:TTMs}
We thank Prof.~Masuda for clarifying the non-injective (not 1--1) correspondence from torus manifolds to {\em\/multifans,} which complicates
the identification of a (transpolar) pair of unitary toric spaces, 
$(X,\tX)$, suitable for Figure~\ref{f:gMM}.
For example, {\em\/topological torus manifolds}\cite{Ishida:2013aa} (TTMs) correspond to {\em\/pairs\/} of (multi)fans in a $(\IC\<\times\ZZ)^r$-like {\em\/smooth\/} generalization of the algebraic lattice $N\<=\Hom_{\text{alg}}(\IC^*,(\IC^*)^r)\<\approx\ZZ^r$ of complex-algebraic toric varieties\cite{rKKMS-TE1, rD-TV, rO-TV, rF-TV, rGE-CCAG, rCLS-TV, rCox, rCox95, Cox:1997Rec}. TTMs also admit toric actions with a {\em\/continuum\/} of $\IC^r$-parametrized ``radial'' actions {\em\/in addition\/} to a multifan-encoded, ``circle'' $((S^1)^r\<\approx\ZZ^r)$-action\cite{Ishida:2013ab}.}
The mutually consistent combinatorial data reported herein then aims to help in this ``needle in a sea ($\IC$)'' search for such toric embedding spaces, 
$(X,\tX)$, and thereby the transposition mirror pairs of Calabi--Yau hypersurfaces, $Z_f\<\in X[c_1]$ and $Z_{\,\tT f}\<\in \tX[c_1]$. 
Whereas the existence of such transposition mirror embedding pairs 
is unclear from the mathematics of (unitary) torus manifolds, the corresponding framework in GLSMs is rather straightforward; see Remark~\ref{r:GLSM}.

The transposition mirror framework in Figure~\ref{f:gMM} agrees with the rigorous and general combinatorial results in 2 dimensions\cite{rP+RV-12, Higashitani:2017aa}, and the general GLSM computational results about the structure of the secondary fan reported earlier\cite{rBH-gB}. Additional corroboration by cumulative world-sheet instanton effects on the K{\"a}hler  class discriminant loci being mirrored in the complex structure moduli space will be reported separately.
The wealth of these computational results we hope provides ample and convincing motivation for further and more systematic study, aiming for rigorous proofs, clarified qualifications, and other refinements and corrections to the presented claims and conjectures.

\paragraph{Organization:}
In the remainder of this introduction, we provide more detail to the overarching transposition mirror-symmetric framework of Figure~\ref{f:gMM}, the implied key Claims (about VEX multitopes and multifans) and Conjectures (about corresponding unitary toric spaces), and for convenience list several key toric geometry results that extend throughout this multifan generalization.
Section~\ref{s:CCTH} verifies the multifan evaluation of ($\rT$-equivariant) Chern classes, invariants and Todd--Hirzebruch identities,
while the multifan-encoded intersection number results are examined in Section~\ref{s:iN}.
Section~\ref{s:VEX} shows how combinatorial features of VEX multitopes imply (with indicated qualifications) the Poincar\'e series and $\rT$-equivariant cohomology of corresponding toric spaces. Consistent with the foregoing, Section~\ref{s:surg} then proposes the ``surgical'' relation (see Conjecture~\ref{c:preC}) between complex-algebraic toric varieties and their almost complex and pre-complex torus manifold generalizations, respectively corresponding to (plain and convex) reflexive polytopes, and to multi-layered VEX multitopes, uniform and flip-folded.
We summarize our conclusions in Section~\ref{s:Coda}, deferring
several technical details and examples to Appendix~\ref{s:MM}.

\subsection{The Transposition Mirror-Symmetric Framework}
\label{s:gMM}
The transposition mirror framework generalizing~\eqref{e:MM} to include all VEX multitopes\footnote{Most triangulations of
the reflexive polytopes in the Kreuzer--Skarke database\cite{Kreuzer:2000xy} (already over 72\% by $h^{1,1}\<\leqslant6$) are not stellar but {\em\/vex\/}: they define fans spanned by (non-reflexive but VEX) sub-polytopes\cite{MacFadden:2025ssx}; see a 3d example in Figure~\ref{f:3pairs}, bottom right. We thank Elijah Sheridan and Nate MacFadden for sharing this impressive result before its publication.} as in Figure~\ref{f:gMM}, uses the ``twin definitions''\cite{rBH-gB, Berglund:2022dgb}, adapted here to conform to the literature\cite{rBaty01, rM-MFans, Masuda:2000aa, rHM-MFs, Civan:2003aa, Masuda:2006aa, rHM-EG+MF, rH-EG+MFs2, Nishimura:2006vs, Ishida:2013aa, Ishida:2013ab, buchstaber2014toric}:
\begin{defn}[VEX multitopes and multifans]\label{D:VEX}
An $L$-lattice {\em\/VEX multitope,}
$\pD\<\subset L_\IR\<\coeq\big((L\<\approx\ZZ^r)\<{\otimes_\ZZ}R\big)$,
is a continuously oriented, multihedral and possibly multi-layered in
$L_\IR$, real $n$-dimensional body with every facet at unit lattice distance\cite[Def.\;4.1.4]{rBaty01} from $0\<\in L$.
It is star-subdivided/triangulated by a $0$-centered complete {\em\/multifan}\cite{rM-MFans, Masuda:2000aa, rHM-MFs, Civan:2003aa, Masuda:2006aa, rHM-EG+MF, rH-EG+MFs2, Nishimura:2006vs, Ishida:2013aa, Ishida:2013ab, buchstaber2014toric}, $\pS\<\smt\pD$, providing a primitive (lattice coprime) central cone over every face:
$\pD\<\lat\pS\<=(\{\s_\a\};\pRec)$ is a facet-ordered poset of its $0$-centered cones,
$\s_\a\<=\sfa(\q_\a\<\in\vd\pD)$, were
``$\mkern2mu\vs\<\pRec\s$'' denotes $\vs\<\subset\vd\s$ and 
$\codim(\vs,\s)\<=1$;~
$\vd\pD\<=(\{\q_\a\};\pRec)$ has the same poset structure.
\end{defn}
\begin{defn}[transpolar]\label{D:tP}
\addtolength{\baselineskip}{-2pt}
 Given an $L$-lattice multitope,
 $\pD\<\subset L_\IR\<\coeq\big((L\<\approx\ZZ^r)\<{\otimes_\ZZ}R\big)$,
 with a \,$0$-centered $L$-lattice star-triangulation:
 \begin{enumerate}[itemsep=-3pt, topsep=-2pt]

 \item Recursively subdivide $\pD$ into a facet-ordered poset of convex faces, $\pD\<=(\{\q_\a\};\pRec)$.

 \item Compute the polar image of each convex face:
\begin{equation}
   \q^\wtd_\a=\{v\!: \vev{u,v}\<={-}1,~~u\<\in\q_\a\<\in\pD\,\}
    \subset (L^\vee\<{\otimes_\ZZ}\IR).
 \label{e:StdPq}
\end{equation}

 \item Dually to the poset $\pD\<=(\{\q_\a\};\pRec)$, assemble the transpolar poset $\pDN[\wtd]{}\<=(\{\q^\wtd_\a\};\pRec)$ satisfying the universal inclusion-reversing relations:
\begin{equation}
     \vd\pD\ni\big(\vq \subseteq \q_1\cap\cdots\cap\q_k\big)
      \quad\Longleftrightarrow\quad
     \big(\vd\vq^\wtd \supseteq \q_1^\wtd\cup\cdots\cup\q_k^\wtd\big) \in \vd\pDN[\wtd]{},
 \label{e:dualRules}
\end{equation}
which {\bfseries delimits}/{\bfseries bounds} the polar of each $k$-face by the polars of its adjacent $(k{+}1)$-faces.
\end{enumerate}\backUp
\end{defn}
A subset of the poset $\vd\pD\<=(\{\q_\a\};\pRec)$ often suffices to complete the procedure of Definition~\ref{D:tP}, and it simplifies for any {\em\/convex\/} $L$-lattice polytope, $\pD$:
\begin{equation}
  \pD\<=\Conv(\pD) \quad\To\quad
  \pD^\wtd =
  \pD^\circ \coeq \{\, v\!: \vev{u,v}\<\geqslant-1,~~u\<\in\pD\,\}
  ~\subset L^\vee_\IR\<\coeq L^\vee{\otimes_\ZZ}\,\IR,
 \label{e:StdP}
\end{equation}
to the standard {\em\/polar\/} ($\mkern1mu^\circ$) operation\cite{rF-TV, rGE-CCAG, rCLS-TV}.
Convex VEX polytopes are {\em\/reflexive,} and each spans a {\em\/complete} and {\em\/plain\/} (single-layer) central ($0\<\in L$) fan, $\pS\<\smt\pD$, encoding a {\em\/compact\/} (semi) Fano complex-algebraic toric variety by the fan-specified {\em\/local\/} chart-gluing construction\cite{rD-TV, rO-TV, rF-TV, rGE-CCAG, rCLS-TV}.
A polar pair of reflexive polytopes, $(\pDs{X},\pDN{X}{=}\pDs{\tX})$, thus encodes polar a pair of toric {\em\/varieties,}\footnote{The same pair, $(X,\tX)$, is also encoded by the inherently {\em\/global\/} quotient construction\cite{rCox, rCox95, Cox:1997Rec, rCLS-TV}, after swapping the polar pair of polytopes, $(\pDN{X},\pDs{X}{=}\pDN{\tX})$; how this might extend to non-convex VEX multitopes is not clear, and we eschew this method.} $(X,\tX)$, with deformation families of transpose-mirror Calabi--Yau hypersurfaces~\eqref{e:MM}.
The framework of Figure~\ref{f:gMM} offers to generalize this:
The {\em\/transpolar\/} operation ($\mkern1mu^\wtd$) maps involutively between all {\em\/VEX multitopes,} and
the chart-gluing construction extends to VEX multitopes stemming from unitary torus spaces; see below.
Extending completeness to {\em\/multifans\/} is technically more involved\cite{rM-MFans, Masuda:2000aa, rHM-MFs}, but insures that
each facet of each $k$-cone is a facet of another $k$-cone, and the multifan
covers all of $L_\IR\<\approx\IR^n$, albeit perhaps with a self-intersecting image.

\begin{remk}[unitary torus manifolds]
\label{r:UTM}%
For a telegraphic summary from Refs.\cite{rM-MFans, Masuda:2000aa, rHM-MFs}:
A compact, {\em\/unitary\/} (i.e., {\em\/weakly almost complex\/}) {\em\/torus manifold,} $M$, corresponds to a complete $n$-dimensional {\em\/multifan,} $\pFn{M}$,
and is a real $2n$-dimensional manifold that admits a (``half-dimensonal'') {\em\/torus\/} $\rT\<=(S^1)^n$-action with nonempty isolated $\rT$-fixed points, compatible with a complex structure on its ($\IR^{2k}$-extended) {\em\/stable tangent bundle,} $T_M\<\oplus\2{\IR}^{2k}$ for some $0\<\leq k\<\in\ZZ$.
The $\rT$-equivariant Chern class, $c^\rT(M)\<\coeq\prod_i(1{+}\x_i)$, is generated by the elements, $\x_i\<\in H_\rT^2(M)$,
that correspond to the 1-cone generators, $\n_i\<\smt\pFn{M}$, and are
Poincar\'e-dual to $\rT$-characteristic closed connected real codimension-2 submanifolds, analogous to toric divisors.
The ``anticanonical'' complex line-bundle, $\cKs{M}\<\coeq\wedge^{n+k}(T_M\<\oplus\2{\IR}^{2k})$, has
$c_1^\rT(\cKs{M})\<=\sum_i\x_i$\cite{rM-MFans, rHM-MFs}.
Also, unless $\Td(M)\<=0$ (see examples~\eqref{e:w=1,0} and~\eqref{e:*w=3,2,1,0}, below), $M$ is {\em\/cohomologically symplectic\/}: there is an element $\x\<\in H^2(M)$ such that
$\x^n\<\neq0$\cite[Cor.\;4.3]{rM-MFans}.
The {\em\/orientation\/} (winding or wrapping index) of an $n$-cone, 
$w(\s)\<={+}1$ (${-}1$), indicates that the orientation induced from the complex structure of $T_M\,\oplus\2{\IR}^{2k}$ in the affine chart $U_\s\<\subset M$ does (does not) agree with that of $M$ itself.
Throughout, we use $w(\s)$ as the sign of the degree of both a cone and its facet, $\s\<=\sfa(\q)$,\footnote{\label{fn:deg}The degree of a $k$-face, $d(\q^{\sss(k)}):=(k{+}1)!{\cdot}\!\Vol_{k+1}(\q^{\sss(k)})$, is the $(k+1)!$-fold volume of the {\em star-pyramid\/} over $\q^{\sss(k)}$\cite{rBaty01}. Faces with $|d(\q)|\<>1$ and the cones they span encode $(\IC^n/\ZZ_{d(\q)})$-charts in the toric space; their desingularization is encoded by a subdivision of the cone.
For simplicity, we tacitly identify the torus manifold-specific $\e$-function with the multifan-specific $w$-function\cite{rM-MFans, Masuda:2000aa, rHM-MFs}. Subtleties in ``reading'' a multifan (affecting its fit for our application) are exemplified by the case of $S^4$\cite{rMP-TO+MFs}.}
which determines a Duistermaat--Heckman measure over the multifan $\pFn{M}$\cite{Masuda:2000aa,rHM-MFs} (see also\cite{Berglund:2022zto}), and an {\em\/omniorientation\/} on $M$\cite{Buchstaber:2001aa, Masuda:2006aa, Ishida:2013aa, buchstaber2014toric}.
\end{remk}
Motivated by consistent correspondence with GLSMs, and since
1-cone generators of a multifan, $\n_i\<\smt\pS$, are so prominent in describing the geometry of corresponding unitary torus manifolds\cite{rM-MFans, Masuda:2000aa, rHM-MFs},
we assign a (Cox-like) complex variable,
$x_i$ to each $\n_i$, 
so complex sections of $\cKs{M}$ (with $c_1^\rT(\cKs{M})\<=\sum_i\x_i$) are deformations of $\Pi x$ --- precisely as needed in the above introduction of the GLSM-motivated transposition mirror framework in Figure~\ref{f:gMM} and Remark~\ref{r:GLSM}. With this notation and motivation:
\begin{conj}
\label{C:gCYh}
Corresponding to each smooth VEX multifan, $\pS$,
there is a unitary torus manifold, $M_\pS$, on which the complex ``anticanonical'' sections, $f(x)$, are (Laurent) polynomial deformations of\/ $\Pi x$ 
in terms of $\,\pS(1)$-assigned complex variables, $x_i$, and corresponding to lattice points in the transpolar multitope, $\pDN{M}\<=(\pDs{M}\<\lat\pFn{M})^\wtd$.
Zero loci, $Z_g\<\coeq\{f(x)\<=0\}$, are Calabi--Yau hypersurfaces in $M$.
\end{conj}

\paragraph{Two Example Sequences:}
As an illustrative guide and to set notation, consider the two infinite sequences of transpolar pairs of VEX multigons in Figure~\ref{f:Fm+Em}.
\begin{figure}[htb]
 \begin{center}
  \begin{picture}(160,62)(0,-7)
   \put(5,-5){\TikZ{[scale=.85]\path[use as bounding box](-1,-4)--(2,2);
              \foreach\y in{-1,...,2}
               \foreach\x in{-1,...,2} \fill[gray](\x,\y)circle(.4mm);
               \foreach\x in{-1,...,1} \path[gray!50!cyan]
                   (\x,-1.75)node{\Large$\vdots$};
               \fill[gray](-1,-4)circle(.4mm);
              \fill[Turque!25,opacity=.8](2,-1)--(-1,2)--(-1,-4)--(0,-1);
              \draw[blue,thick,midarrow=stealth](2,-1)
                    to node[above=1mm]{\footnotesize$\n_4^\wtd$}(-1,2);
              \draw[blue,thick,midarrow=stealth](-1,2)
                    to node[left]{\footnotesize$\n_1^\wtd$}(-1,-4);
              \draw[blue,thick,midarrow=stealth](-1,-4)
                    to node[right]{\footnotesize$\n_2^\wtd$}(0,-1);
              \draw[blue,thick,midarrow=stealth](0,-1)
                    to node[below]{\footnotesize$\n_3^\wtd$}(2,-1);
              \draw[thick,-stealth](0,0)--(-1,-4);
               \draw(-1,-4)node[right]
                    {\footnotesize$\n_{12}^\wtd=(-1,-m{-}1)$};
              \draw[thick,-stealth](0,0)--(0,-1);
               \path[red](0,-1)node{$\star$};
               \path[red](.25,-1.25)node{\footnotesize$\n_{23}^\wtd$};
              \draw[thick,-stealth](0,0)--(2,-1);
               \draw(2,-1)node[right]{\footnotesize$\n_{34}^\wtd$};
              \draw[thick,-stealth](0,0)--(-1,2);
               \draw(-1,2)node[left]{\footnotesize$\n_{41}^\wtd$};
              \foreach\y in{-3,...,2} \draw[densely dotted](0,0)--(-1,\y);
              \foreach\y in{0,...,1} \draw[densely dotted](0,0)--(\y,1-\y);
               \draw[densely dotted](0,0)--(1,-1);
              \filldraw[fill=white,thick](0,0)circle(.4mm);
              \path(-.75,-3.2)node[right]
                    {\large$\pDN{E_m^{\sss(2)}}\color{blue}
                             \<=\pDs{\ME[2]m}\<\lat\pFn{\ME[2]m}$};
              \draw[ultra thick, densely dotted,<->](1.4,0)
                    --node[below=-1pt, rotate=8]
                    {transpolar}++(4,.6);
             }}
   \put(40,25){\TikZ{\path[use as bounding box](-3,-1)--(1,1);
              \foreach\y in{-1,...,1}
               \foreach\x in{-1,...,1}
                \fill[gray](\x,\y)circle(.4mm);
               \foreach\y in{0,...,1}
                \path[gray!50!red](-2,\y)node{\Large$\cdots$};
               \fill[gray](-3,1)circle(.4mm);
              \fill[yellow,opacity=.8](0,0)--(-3,1)--(1,0)--(-1,-1)--(0,0);
              \draw[blue,thick,midarrow=stealth](1,0)--(-3,1);
               \path[blue](.2,.2)node[above right,rotate=-10]
                    {\footnotesize$\n_{12}$};
              \draw[thick,-stealth](0,0)--(-3,1);
               \draw(-3,1)node[below]{\footnotesize$\n_2$};
              \fill[red, opacity=.5](0,0)--(-3,1)--(0,1);
              \fill[yellow,opacity=.8](0,0)--(-1,-1)--(0,1)--(0,0);
              \draw[Red, very thick,midarrow=stealth](-3,1)
                    to node[above]{\footnotesize$\n_{23}$ \tiny(CW)}(0,1);
              \draw[blue,thick,midarrow=stealth](0,1)--(-1,-1);
               \path[blue](-.9,-.33)node[rotate=75]{\footnotesize$\n_{34}$};
              \draw[blue,thick,midarrow=stealth](-1,-1)
                    to node[below right,rotate=35]
                    {\footnotesize$\n_{41}$}(1,0);
              \draw[thick,-stealth](0,0)--(1,0);
               \draw(1,0)node[right]{\footnotesize$\n_1$};
              \draw[thick,-stealth](0,0)--(0,1);
               \draw(0,1)node[right]{\footnotesize$\n_3$};
              \draw[thick,-stealth](0,0)--(-1,-1);
               \draw(-1,-1)node[left]{\footnotesize$\n_4$};
               \draw(-3,.2)node[above]{\scriptsize$(-m,1)$};
              \foreach\x in{-2,...,-1} \draw[densely dotted](0,0)--(\x,1);
              \filldraw[fill=white,thick](0,0)circle(.4mm);
              \path(-1.7,-2)node[right, rotate=atan(1/2)]
                    {\large$\C3{\pDN{\ME[2]m}\<=}\pDs{\EE[2]m}
                            \<\lat\pFn{\EE[2]m}$};
             }}
   \put(70,0){\TikZ{\path[use as bounding box](-4,-1)--(1,1);
              \foreach\y in{-1,...,1}
               \foreach\x in{-1,...,1} \fill[gray](\x,\y)circle(.4mm);
               \foreach\y in{-1,...,0}
                \path[gray!50!cyan](-2.5,\y)node{\Large$\cdots$};
              \fill[yellow,opacity=.8](1,0)--(0,1)--(-1,0)--(-4,-1);
               \fill[gray](-4,-1)circle(.4mm);
              \draw[blue,thick,midarrow=stealth](1,0)
                    to node[above,rotate=-45]{\footnotesize$\n_{12}$}(0,1);
              \draw[blue,thick,midarrow=stealth](0,1)
                    to node[above,rotate=45]{\footnotesize$\n_{23}$}(-1,0);
              \draw[blue,thick,midarrow=stealth](-1,0)
                    to node[above,rotate=20]{\footnotesize$\n_{34}$}(-4,-1);
              \draw[blue,thick,midarrow=stealth](-4,-1)
                    to node[below,rotate=10]{\footnotesize$\n_{41}$}(1,0);
              \draw[Rouge, thick,-stealth](0,0)--(-1,0);
               \path[Rouge](-1,0)node{$\star$};
               \draw[red](-1.2,.2)node{\footnotesize$\n_3$};
              \draw[thick,-stealth](0,0)--(1,0);
               \draw(1,0)node[right]{\footnotesize$\n_1$};
              \draw[thick,-stealth](0,0)--(0,1);
               \draw(0,1)node[left]{\footnotesize$\n_2$};
              \draw[thick,-stealth](0,0)--(-4,-1);
               \draw(-4,-1)node[above]{\footnotesize$\n_4$};
               \draw(-4,-.7)node[above]{\scriptsize$(-m,-1)$};
               \path(-2.9,-1.4)node[right, rotate=atan(1/6)]
                    {\large$\C3{\pDN{\MF[2]m}\<=}\pDs{\FF[2]m}
                            \<\lat\pFn{\FF[2]m}$};
              \filldraw[fill=white,thick](0,0)circle(.4mm);
              \draw[ultra thick, densely dotted,<->](0,1.2)
                    to[out=60,in=165]
                    node[above, rotate=20]{transpolar}++(1.3,.5);
             }}
   \put(130,-10){\TikZ{[scale=.75]\path[use as bounding box](-1,-3)--(1,5);
              \foreach\y in{-1,...,3}
               \foreach\x in{-1,...,1}
                \fill[gray](\x,\y)circle(.4mm);
               \foreach\x in{0,...,1} 
                \path[gray!50!red](\x,-1.8)node{\Large$\vdots$};
               \foreach\x in{-1,...,0} 
                \path[gray!50!cyan](\x,4.2)node{\Large$\vdots$};
               \fill[gray](-1,5)circle(.4mm); \fill[gray](1,-3)circle(.4mm);
              \fill[Turque!25,opacity=.8](0,0)--(1,-1)--(-1,-1)--(-1,5);
              \draw[blue,thick,midarrow=stealth](-1,-1)
                    to node[below left]{\footnotesize$\n_2^\wtd$}(1,-1);
              \fill[red, opacity=.50](0,0)--(1,-1)--(1,-3);
              \draw[thick,-stealth](0,0)--(1,-1);
               \draw(1,-1)node[right]{\footnotesize$\n_{23}^\wtd$};
              \fill[Turque!25,opacity=.8](0,0)--(1,-3)--(-1,5);
              \draw[Red,very thick,midarrow=stealth](1,-1) to node[right]
                    {\footnotesize$\n_3^\wtd$ \tiny(CW)}(1,-3);
              \draw[blue,thick,midarrow=stealth](1,-3)
                    to node[above right]{\footnotesize$\n_4^\wtd$}(-1,5);
              \draw[blue,thick,midarrow=stealth](-1,5)
                    to node[left]{\footnotesize$\n_1^\wtd$}(-1,-1);
              \draw[thick,-stealth](0,0)--(-1,-1);
               \draw(-1,-1)node[left]{\footnotesize$\n_{12}^\wtd$};
              \draw[thick,-stealth](0,0)--(1,-3);
               \draw(1,-2.8)node[right]{\footnotesize$\n_{34}^\wtd$};
               \draw(1,-2.8)node[left]{\scriptsize$(1,1{-}m)$};
              \draw[thick,-stealth](0,0)--(-1,5);
               \draw(-1,5)node[left]{\footnotesize$\n_{41}^\wtd$};
               \draw(-1,4.5)node[left]{\scriptsize$(-1,1{+}m)$};
              \foreach\y in{0,...,3} \draw[densely dotted](0,0)--(-1,\y);
               \draw[densely dotted](0,-1)--(0,1);
               \path(-.3,3.35)node[right]{\large$\pDN{\FF[2]m}\color{blue}
                                          \<=\pDs{\MF[2]m}$};
               \path(1.33,2.6)node[right]{\large$\color{blue}
                                          \<\lat\pFn{\MF[2]m}$};
              \filldraw[fill=white,thick](0,0)circle(.4mm);
             }}
  \end{picture}
 \end{center}
 \caption{Two infinite sequences of transpolar (Definition~\ref{D:tP}) pairs of VEX multigons and multifans they span; the reversely, CW-oriented flip-folded facets $\n_{23}\<\subset\pDs{E_m^{\sss(2)}}$ and 
$\n_3^\wtd\<\in\pDN{\FF[2]m}$ are transpolar to a non-convex vertex}
 \label{f:Fm+Em}
\end{figure}
Vertices of the spanning multigons ($\pDs{X}$) are labeled $\n_i$, so
$\n_{ij}$ is the edge connecting $\ora{\n_i\n_j}$; their transpolar images in the Newton multigons ($\pDN{X}$) then are $\n_i^\wtd$ and $\n_{ij}^\wtd$, respectively.
In particular, $\pDs{\FF[2]m}$ with $m\<>2$ is a non-convex, star-triangulated polytope with facets at unit distance from the origin;
its Newton multitope, $\pDN{\FF[2]m}\<\coeq(\pDs{\FF[2]m})^\wtd$, is flip-folded as depicted in Figure~\ref{f:Fm+Em}, far right.
Its {\em\/extension,} $\pDN{\FF[2]m}\<\ssm(\pDs{\FF[2]m})^\circ$, shaded red and labeled ``CW,''
is the transpolar image of the non-convex vertex, $\n_3\in\pDs{\FF[2]m}$.
It encodes the additional, {\em\/rational\/} monomials that render the anticanonical sections transverse\cite{rBH-gB}, smooth the Tyurin-degenerate Calabi--Yau hypersurfaces\cite{Berglund:2022dgb}, are key
to the transpolar operation being an involution, and
enable the computational agreements mentioned above and shown explicitly in subsequent sections.

Continuous VEX multitope/multifan orientation means that the number of (possibly partially overlapping) cones in all directions is $(w^+{-}w^-)\<=\textit{const.}$ However, this property does not suffice to select among the various types of torus manifolds studied in the literature\cite{rM-MFans, Masuda:2000aa, rHM-MFs, Civan:2003aa, Masuda:2006aa, rHM-EG+MF, rH-EG+MFs2, Nishimura:2006vs, Ishida:2013aa, Ishida:2013ab, buchstaber2014toric}, esp.\cite[Figure~7.5]{buchstaber2014toric}. To wit, in Ref.\cite{Masuda:2000aa} only their Figure~2, Example~2.2 and cases specified {\em\/between\/} Examples~2.3 and~2.4 are VEX.\footnote{Modifying \cite[Example~2.4]{Masuda:2000aa} to
$\pS\<=(\{0,\,\n_1,\n_2,\n_3,\n_4,\n_5,\,\n_{12},\n_{23},\n_{34},\n_{45},\n_{51}\};\;\n_j\<\pRec\n_{ij},\n_{jk})$ would make it VEX.}
 Also, $\rT$-equivariant bundles on torus manifolds and their sections are currently actively studied, with (topological, smooth, holomorphic) structures on such bundles over the class of {\em\/topological torus manifolds\/} having been classified only recently\cite{Cui:2025Kly, Cui:2025Equ}.
These ``ingredients'' are simply not yet known well enough to determine which specific type of unitary torus manifolds admits
the $\pFn{\tX}(1)\<\iff y_I$-generated complex coordinate rings and
the ``anticanonical'' section $\tT f(y)$ specified by transposition.

Finally, non-convex, flip-folded and other multi-layered multitopes turn up routinely in the study of {\em\/presymp\-lec\-tic\/} toric spaces, where the (putative) symplectic structure degenerates at isolated locations\cite{rK+P-vrtPlytp0, rK+P-vrtPlytp, rK+T-pSympTM, Silva:2010aa, rMP-TO+MFs, godinho201612}.
 Since mirror symmetry relates symplectic and complex structures
\cite{Kontsevich:1995wk, rCK, rSYZ-Mirr, Berglund:2021hbo}, the mirror-symmetric framework in Figure~\ref{f:gMM} most likely does include also {\em\/precomplex\/} spaces, where a (putative) complex structure degenerates at isolated locations.
 After all, string dynamics in general
 expects complex and symplectic structure in target-space\,\footnote{\label{fn:CtC}The mirror-pairing of these two target-space structures stems from the fact that the underlying worldsheet quantum field theory likewise relates rings of ``chiral'' and ``twisted-chiral'' complex superfields\cite{rGHR, rSSQM3, rW-TSM, rW-MMTFT, rJPS}.}
 and requires its Ricci-flatness,
 but is found to accommodate a growing class of singularizations (some requiring the inclusion of suitable 'brane-like sources)\cite{rCHSW, rDHVW1, rReidK0, rCYCI1, rGHC, rGHPT, rCGH1, rCGH2, rBeast2, rSingS, Avram:1995pu, Avram:1997rs, McNamara:2019rup}.\footnote{\label{fn:preStr}As {\em\/hybrid\/} GLSM phases typically involve {\em\/stratified pseudo-manifolds\/}\cite{Aspinwall:1993nu, Hubsch:2002st, rAR01, rB-IS+ST}, at least some spaces with the complex and symplectic structures degenerating at isolated locations clearly cannot be avoided. Also, string dynamics is well known to admit the non-Riemannian (and so non-K{\"a}hler) ``Hull--Strominger system,'' with (geometric) torsion\cite{Hull:1985zy, Strominger:1986uh, Hull:1986kz, Becker:2009df, Dasgupta:1999ss}; for very recent results on almost complex Calabi--Yau manifolds with torsion and brief reviews, see Ref.\cite{Ivanov:2025Alm}, and\cite{McOrist:2025zwf} for $\a'{}^2$-corrections.}
 The focus on {\em\/complex-algebraic\/} toric geometry then appears to stem primarily from its impressive effectiveness. The mirror-symmetric framework in Figure~\ref{f:gMM} and known non-Fano embeddings highlight this as a technical convenience, and posit including also ``anticanonical'' hypersurfaces in unitary toric spaces, where (pre)complex or (pre)symplectic structures may degenerate at isolated locations, but which may be minimally (or not at all) intersected by the hypersurfaces; see footnote~\ref{fn:StrSuSy}, and Conjecture~\ref{C:obstruct} below.

\paragraph{Poset Structure of Multifans and Multitopes:}
Each fan specifies the full poset hierarchy of common facets, 
$\vs\<\pRec\s,\s'$, i.e., $\vs\<\coeq\s\<\cap\s'$ and
$\rlnt(\s)\<\cap\rlnt(\s')\<=\varnothing$.
For this to hold in a {\em\/multifan\/} when a $\s'$ {\em\,folds over\/} (a part of) a $\s$ as in Figure~\ref{f:Fm+Em}
($\pDs{\EE[2]m}$ and $\pDN{\FF[2]m}$), we regard $\s$ and $\s'$ as lying in distinct layers, connected by folding along the common facet, $\vs\<\pRec\s,\s'$, satisfying in this sense the ``Separation Lemma'' condition~\cite[Lemma~V.1.13, p.\,147]{rGE-CCAG} and allowing to generalize the chart-gluing construction; see, e.g.,\cite{Ishida:2013aa, Cui:2025Kly, Cui:2025Equ}.
The transpolar images of $\q,\q'\in\vd\pD$ are distinguished,
 $(\q^\wtd\<\neq\q'{}^\wtd)\<\in\vd\pD^\wtd$, {\em\/unless\/}
 $\q,\q'$ are subdividing parts in a single face in $\vd\pD$.\footnote{A non-convex multitope can have distinct but coplanar faces; their transpolar images are in the same location in the dual lattice, but are distinguished by being in different layers of the transpolar multitope.}
\begin{clam}
\label{CC:D**=D}
For an $L$-lattice VEX multitope, $\pD$:
({\small\bfseries1})~$\pD^\wtd\<\subset L^\vee_\IR$ is a VEX multitope, and
({\small\bfseries2})~$(\pD^\wtd)^\wtd\<=\pD$: the transpolar duality~\ref{D:tP} closes on VEX multitopes as an involution.
({\small\bfseries3})~The star-triangulation of $\pD$ is {\em\/$\boldsymbol{L}$-primitive}: lattice points of each star-triangulating simplex are only 
$\,0\<\in L$ and in $\vd\pD$.
If $\pD$ has a {\em\/smooth\/} star-triangulation,
({\small\bfseries4})~$\pD^\wtd$ has a corresponding smooth ($L^\vee$-)lattice star-triangulation, so there exist unitary torus manifolds corresponding to each $\pD$ and $\pD^\wtd$.
\end{clam}
In more than two dimensions, the degree of a lattice-primitive facet can be arbitrarily large\cite{rReeve1, Belyaev:2019aa}, so the transpolar~\eqref{e:StdPq} of its facets in $\vd\pD$ need not be integral, i.e., need not be in $L^\vee$. The unit-distance\cite[Def.\;4.1.4]{rBaty01} requirement in Definition~\ref{D:VEX} seems to exclude such facets from VEX multitopes and so imply part~(1), but we invite a rigorous verification of this and all other parts of Claim~\ref{CC:D**=D}:
These statements seem so inherent in the transposition mirror framework of Figure~\ref{f:gMM} and Corollary~\ref{CC:MMM} that, if any were found to fail, Definition~\ref{D:VEX} should be refined so they all do hold.

Each reflexive polytope, $\pDs{X}$, is star-subdivided/triangulated by a complete fan, with a unit winding (wrapping) number, $w(\pFn{X})\<=1$, equal to $\Td(X)\<=\c^h(X)$.
The star-subdividing/triangulating multifan of a general VEX multitope, $\pDs{X}$, is also complete, but may {\em\/flip-fold\/} or be otherwise multi-layered so its winding number and Todd genus, $w(\pFn{X})\<=\Td(X)$, may be any integer\cite{rM-MFans, Masuda:2000aa, rHM-MFs}. All VEX multitopes (reflexive polytopes included) being continuously orientable, the orientation of adjacent facets differs only when they flip-fold\cite{rBH-gB}.
There do exist VEX multitopes that flip-fold {\em\/around\/} $0\<\in L$ with no relative interior, so star-subdivisions refer to a ``center,'' avoiding references to an interior; see~\eqref{e:w=1,0} and~\eqref{e:*w=3,2,1,0} for examples.

\paragraph{Chart-Gluing and Obstructions:}
Each multifan defines an atlas $\{U_{\s_i}\<=\Spec(\IC[\check\s_i])\<\approx\IC^n/\ZZ_{d(\s_i)},\, \s_i\<\in\pFn[\sss(n)]M\}$.
Smooth multifans (footnote~\ref{fn:smooth}) stem from unitary torus manifolds covered by $\IC^n$-charts, since $d(\s)\<=1$ for a smooth cone.
Unlike with fans and (complex-algebraic) toric varieties, 
a general (multi-layered) multifan, $\pS$, does not uniquely specify a unitary torus manifold\cite{rM-MFans, Masuda:2000aa, rHM-MFs, rAM-MFs} (footnote~\ref{fn:TTMs}), nor imply a complex (or symplectic) structure.
When all $n$-cones $\s_i,\s_j$ are on opposite sides of their common facet,
$\s_i\<\cap\s_j\<\eqco\vs_{ij}\<\in\pFn[\sss(n-1)]M$,
a VEX multifan $\pFn{M}$ is orientable {\em\/uniformly,} i.e.,
$(w^+{-}w^-)\<=\mathit{const.}$ and $w^+\,w^-\<=0$,
and $M$ is (at least) almost complex;
see\cite[Thm.\;5.1]{rM-MFans} for $\dim(\pFn{M})\<=2$, and\cite[Cor.\;5.1]{Civan:2003aa} in general.
\begin{conj}
\label{C:obstruct}
When two adjacent cones, $\s_i,\s_j$, flip-fold ($w(\s_i)\,w(\s_j)\<<0$) at the common facet $\vs_{ij}$, a chart-wise local complex structure cannot transfer holomorphically from $U_{\s_i}$ to $U_{\s_j}$, so that flip-folded parts of a multifan encode local obstructions to a complex structure;
see \SS\:\ref{s:pCpx}.
\end{conj}
\begin{remk}[optimal models]
\label{r:optimal}
Remarkably many numerical consequences of Conjectures~\ref{C:gCYh} and~\ref{C:gMM} and Corollary~\ref{CC:MMM} extend from reflexive polytopes to non-convex VEX multitopes, and merely by meticulous inclusion of the continuous multitope orientation $w$-data\cite{rBH-gB}.
This suggests that corresponding to $(\pDs{X},\pDN{X})$, there exists a pair of embedding/ambient unitary toric spaces, $(X,\tX)$, where the obstructions minimally affect the complex (and symplectic) structure and {\em\/relevant\/} cohomology; see, e.g., Refs.\cite{rBeast2, Sharpe:2024dcd}.%
\footnote{Calabi--Yau hypersurfaces that intersect such obstructions may end up with $h^{1,1}\<=0$ and no K{\"a}hler structure, but may still be connected by conifold transition (continuously in the Gromov--Hausdorff topology\cite{Friedman:2024zid}) to the more familiar models, and may well also be mirrors to rigid ($h^{2,1}\<=0$) models, where in turn one would expect the complex structure to be obstructed.}
\end{remk}
While the transposition mirror of an invertible hypersurface $f^{-1}(0)\<\in X[c_1]$ was originally found as a desingularization of a specific finite quotient of the transposed polynomial zero locus\cite{rBH, rBH-LGO+EG, Krawitz:2009aa, rMK-PhD, rFJR-07b, Ebeling:2012Mir, rLB-MirrBH, rACG-BHK, rA+P-MM, rF+K-BHK, Parkhomenko:2022kju, Belavin:2023ldz, Parkhomenko:2024mxq, Cho:2024Ber}, the transposition mirror of $f^{-1}(0)\<\subset X$ in Figure~\ref{f:gMM} is simply its transposed polynomial zero locus,
 $\tT f^{-1}(0)\<\subset\tX$\cite{rBH-gB, Berglund:2022dgb}:
The transpolar embedding space, $\tX$, with the
MPCP-desingularizing exceptional set for each lattice-point in $\pDN{X}{=}\pDs{\tX}$, {\em\/automatically\/} encodes both the finite group quotients and all required desingularizations --- even when these do not stem from any global finite group action.

Originally focused on transposition mirror pairs of Calabi--Yau hypersurfaces (ground ``floor'' in Figure~\ref{f:gMM}), the transpolar map between VEX multitopes, $\pDs{X}\<{\fif{\wtd}}\pDN{X}$, also implies a (transpolar) duality between the corresponding (optimal choice of) ambient toric spaces, $X\<\iff\tX$, (middle ``floor'' in Figure~\ref{f:gMM}), and their respective ``anticanonical'' bundles, $\cKs{X}\<{\fif{\wtd}}\cKs{\tX}$,
(top ``floor'' in Figure~\ref{f:gMM}), the total spaces of which are themselves {\em\/non-compact\/} Calabi--Yau spaces.
The Calabi--Yau {\em\/hypersurfaces\/} in Figure~\ref{f:gMM} are typically expected to admit complex, symplectic and (complexified by geometric torsion) K{\"a}hler  structures, with some degree of degeneration/singularization generally admissible in stringy applications\cite{rDHVW1,rCGH2,rSingS,rBeast2,Aspinwall:1993nu,McNamara:2019rup} (footnote~\ref{fn:preStr}).
This implies the leeway in Conjecture~\ref{C:obstruct} and Remark~\ref{r:optimal}, a precise determination of which continues to evolve as finer details of string theory dynamics are being uncovered.
Furthermore, $\big((X\<\ssm Z_f),(\tX\<\ssm Z_{\tT f})\big)$ is a transpolar pair of {\em\/non-compact\/} Calabi--Yau $n$-folds\cite{rTY1,rTY2}, as is $(\cKs{X},\cKs{\tX})$; the possible mirror-symmetry in these remaining two ``floors'' in Figure~\ref{f:gMM} is yet to be explored\cite{Hubsch:2025sph, Hubsch:2025teh}.

\paragraph{Multiple Mirrors:}
Most transpolar VEX multitope pairs,\footnote{Except for {\em\/some\/} $w(\pDs{X})\<=0$ cases such as $\pD_4^\wtd$ in~\eqref{e:*w=3,2,1,0}, and some other, ``too simple or too symmetric'' exceptions.}
$(\pDs{X},\pDN{X})$, generate a web of sub-simplex reductions, $\pD\<\leadsto\Red\pD$, each to a 0-enclosing
simplex hull of $n{+}1$ of lattice points in $\pD$: Each such pair, $(\Red\pDs{X},\Red\pDN{X})$, corresponds to a transpose pair of invertible polynomials, $(f(x),\tT f(y))$, and corresponding transpose-mirror models; see\cite{rBH-gB, Berglund:2022dgb} and Appendix~\ref{s:MM}.
By changing $(\Red\pDs{X},\Red\pDN{X})$ one point (in either multitope) at a time, such pairs link into a web\footnote{Changing $\Red\pDN{X}$ deforms the complex structure of $Z_f$, while changing $\Red\pDs{X}$ varies (including flops) $H^2(X,\ZZ)$, the K{\"a}hler structure and Cox coordinates; the ``VEXing'' operation (closing passage in \SS\:\ref{s:Coda}) may combine both types of changes.}:
\begin{corl}[multiple mirror master]
\label{CC:MMM}
Most transpolar pairs of VEX multitopes, $(\pDs{X},\pDN{X})$ with $w(\pDs{X})\<\neq0$, generate a web of multiple mirror pairs of Calabi--Yau spaces, $(Z_f,Z_{\tT f})$.
\end{corl}
\begin{remk}[multiple mirrors]
\label{r:MMM}
For each sub-simplex reduction, $(\Red\pDs{X},\Red\pDN{X})$,
the vertices of $\Red\pDN{X}$ encode the monomials of $f(x)$ in terms of
the (Cox-like) $x$-coordinates corresponding to vertices of $\Red\pDs{X}$; its transpose, $\tT f(y)$, is defined by swapping the respective roles,
$\Red\pDs{X}\<\iff\Red\pDN{X}$.
This embeds a pair of {\em\/invertible\/}\cite{rBH, rBH-LGO+EG, rFJR-07b, Krawitz:2009aa, rMK-PhD} (Laurent) hypersurfaces,
$(Z_f,\,Z_{\tT f})$, in weighted $\IP^n$'s and defined the corresponding simplified GLSMs\cite{rBH-gB} {\em\/precisely
if\/} both $\Red\pDN{X}$ and $\Red\pDs{X}$ are 0-enclosing 
top-dimensional simplices;\footnote{\label{fn:encl}For simplices with primitive vertices, enclosing the lattice origin seems necessary for the so-encoded polynomial to be transverse (to have its base-points in the Stanley--Reisner excluded set per spanning multitope) and to be usable for transpose-mirror construction\cite{rBH, rBH-LGO+EG, rFJR-07b, Krawitz:2009aa, rMK-PhD}, but we are not aware of a formal proof of this statement; see Appendices~\ref{s:0E} and~\ref{s:0nE}.} see\cite[\SS\:3.4]{Berglund:2022dgb} and Appendix~\ref{s:MM} for more details and examples where not enclosing the origin obstructs the transposition mirror construction.
\end{remk}
This $\big\{(\Red\pDs{X},\Red\pDN{X})\<\subset(\pDs{X},\pDN{X})\big\}$ web 
suggests the 
{\em\bfseries\/double conjecture}:
\begin{conj}
\label{C:gMM}
For any transpolar pair of VEX multitopes, $(\pDs{X},\pDN{X})$, of Corollary~\ref{CC:MMM}:
({\small\bfseries1})~All $0$-enclosing 
sub-simplex reductions of Remark~\ref{r:MMM} encode transpose-mirror Calabi--Yau model pairs\cite{rBH, rBH-gB, Berglund:2022dgb}, interspersed throughout the pair of deformation families, $(X[c_1],\tX[c_1])$. 
({\small\bfseries2})~Mirror pairs of Calabi--Yau spaces may be embedded in {\em\/suitable choices\/} of unitary toric spaces corresponding to the {\em\/unreduced\/} multitopes themselves:
$Z_{f(x)}\<\into(X\<\iff\pFn{X}\<\smt\pDs{X})$ and
$Z_{\tT f(y)}\<\into(\tX\<\iff\pFn{\tX}\<\smt\pDN{X})$; see Conjecture~\ref{C:gCYh}, and Remark~\ref{r:optimal}.
\end{conj}

\subsection{Main Results}\label{s:Main}
Verified by explicit computations for all $m\<\in\ZZ$ cases of transpolar pairs of VEX multitopes
$(\FF{m},\MF{m})$ for $n\<=2,3,4$, $(\EE{m},\ME{m})$ for $n\<=2,3$, and many other (mostly $n\<=2$) testing examples, the main claims motivated by the multifan/multitope computations herein may then be summarized as follows:
\begin{clam}
\label{CC:THCEP}
Each transpolar pair of $n$-dimensional multitopes, $(\pDs{X},\pDN{X})$, admits a continuous, global and mutually compatible orientation so the multitope/multifan-encoded computations of:
\begin{enumerate}[itemsep=-3pt, topsep=-4pt]

 \item\label{i:THC}
  Todd--Hirzebruch identities and all other ($\,\rT$-equivariant) Chern invariants,

 \item\label{i:intC}
  intersection numbers of toric divisors, i.e., $\rT$-characteristic real codimension-2 submanifolds (see Remark~\ref{r:UTM}), and their relation to Chern invariants,

 \item\label{i:P+Betti}
  the Poincar\'e polynomials,
  $P_\pD(t)\<=\F_\pD(t)/(1{-}t)^{n+1}$ with $\F_\pD(t)\<=\sum_{k=0}^r h_k\,t^k$
  (both for $\pD\<=\pDs{X}$ and for $\pD\<=\pDN{X}$),
  and (with exceptions noted below) even Betti numbers,
\end{enumerate}
  are all mutually consistent, and are consistent with independently computed results where available.%
\footnote{\label{fn:Hirz}For example, Hirzebruch scrolls are also realized, 
a l\`a\cite{rH-Fm, rGrHa}, both 
 as projective bundles, $\FF{\vec{m}}\<=\IP(\oplus_{i=1}^r\cO_{\IP^1}(m_i))$ and also
 as specific bi-projective hypersurfaces in the explicit deformation family
 $\FF{\ora{m}}\<\in\ssK[{r||c}{\IP^r&1\\\IP^1&m}]$, 
 with $m\<=\sum_im_i$\cite{rBH-Fm, rBH-gB, Berglund:2022dgb}.}
\begin{enumerate}[resume, itemsep=-3pt, topsep=-1pt]
 \item\label{i:pcsK}
  Optimal transpolar pairs of toric spaces, $X\<{\fif{\sss\wtd}}\tX$ (Remark~\ref{r:optimal}), corresponding to transpolar pairs of VEX multitopes (spanning multifans) admit requisite (pre)complex and (pre)symplectic structures and bundles/sheaves or suitable generalizations, as needed in~\ref{i:THC}--\ref{i:P+Betti}, above; see Remark~\ref{r:C-infty}, below.
\end{enumerate}

\noindent
Furthermore, regarding the explicit deformation families of transposition mirror pairs of Calabi--Yau hypersurfaces of Conjecture~\ref{C:gMM}:
\begin{enumerate}[resume, itemsep=-3pt, topsep=-1pt]

 \item\label{i:C22-C22}
  The transpolar relations, $\pDs{X}\<=\pDN{\tX}$ and  $\pDN{X}\<=\pDs{\tX}$, imply some novel relations among the Chern invariants. For example, for $\,n\<=\dim\pD\<=4$ (Ref.\cite{Berglund:2021ztg} found~\eqref{e:C3C1=C2C12*} and~\eqref{e:dd*ll*} differently):
\begin{alignat}9
    C_3C_1|_X &= \!\!\sum_{\n_{ijk}\in\pDs{X}(3)}\!\! d(\n_{ijk})\,d(\n_{ijk}^\wtd)
   \overset\wtd=
   \!\!\sum_{\m_{IJ}\in\pDN{X}(2)}\!\! d(\m_{IJ}^\wtd)\,d(\m_{IJ}) = C_2C_1\!^2|_{\tX};
 \label{e:C3C1=C2C12*} \\[2mm]
   C_3(Z_f) &= C_2^{~2}|_X - C_2^{~2}|_{\tX} = -C_3(Z_{\tT f}),
 \label{e:C2-C2}\\
   &= 2\big(d(\pDN{X}) \<-d(\pDs{X})\big)
             -12\big(\ell(\pDN{X}) - \ell(\pDs{X})\big).
 \label{e:dd*ll*}
\end{alignat}

 \item\label{i:DehnS}
  The {\em\/Dehn--Sommerville relations\/} (\/$h_i\<=h_{n-i}$\,) hold and imply, e.g., for $\,n\<=\dim\pD\<=4$:
\begin{gather}
  \ell(3\pD)=5\ell(2\pD){-}9\ell(\pD){+}5,\qquad
  \ell(4\pD)=15\ell(2\pD){-}35\ell(\pD){+}21, \label{e:DS34}\\
  d(\pD) = \ell(2\pD){-}3\ell(\pD){+}2, \label{e:DSd}
\iText[-2mm]{and for the Calabi--Yau hypersurfaces (also in Ref.\cite{Berglund:2021ztg}):}
  C_3(Z_f)
   = 2\big( \ell(2\pDN{X}) {-} \ell(2\pDs{X}) \big)
     \<-18\big( \ell(\pDN{X}) {-} \ell(\pDs{X}) \big). \label{e:l2D-lD}
\end{gather}

 \item\label{i:uniM}
  {\em Unimodality\/} inequalities
 (\/$h_{i-1}\<\leqslant h_i$ for $i\<=1,\dots,\lfloor n/2\rfloor$\,) hold and imply, e.g., for $\,n\<=\dim\pD\<=4$:
\begin{equation}
  d(\pD)\geqslant 3\ell(\pD)-13,\qquad
  \ell(2\pD)\geqslant6\ell(\pD)-15,\qquad \text{i.e.},\qquad
  d(\pD)\geqslant \frc12\ell(2\pD)-\frc{11}2. \label{e:uM}
\end{equation}
\end{enumerate}\backUp[.5]
The standard correspondence with Betti numbers\cite[\SS\:4.5]{rF-TV} relates~\#\ref{i:DehnS} to Poincar\'e duality, and~\#\ref{i:uniM} to the Hard Lefschetz Theorem and $\SL(2)$-action on $H^{q,q}$.
\end{clam}
All statements of Claim~\ref{CC:THCEP} hold within the convex (and so reflexive) subset of VEX multitopes, where $X$ and $\tX$ are both (semi) Fano complex-algebraic toric varieties.
The complementing multi-layered VEX multitopes and multifans do not fully specify the underlying unitary toric spaces (statements~\ref{i:P+Betti} and~\ref{i:uniM} of Claim~\ref{CC:THCEP}; see footnote~\ref{fn:TTMs}), nor their $\rT$-equivariant (co)homology and Betti numbers\cite{rM-MFans, Masuda:2000aa, rHM-MFs, rM-EqH*TV, Masuda:2006aa}; see Remark~\ref{r:optimal}, Conjecture~\ref{c:optimal} and other computations below for examples.
Nevertheless, all of Claim~\ref{CC:THCEP} has been verified to hold for numerous transpolar pairs of VEX multitopes such as the infinite sequence, 
$(\pDs{\FF{m}},\pDN{\FF{m}})$,
--- {\em\/except\/} for unimodality (Claim~\ref{CC:THCEP}.\ref{i:uniM}) and the standard correspondence with Betti numbers, which fail {\em\/in some cases,} such as for some $(n,m)$-ranges in the infinite sequence 
$(\pDs{\EE{m}},\pDN{\EE{m}})$.

\begin{remk}\label{r:C3"T}
The top-dimensional $\rT$-equivariant Chern characteristic number of the generic Calabi--Yau (``anticanonical'') hypersurface in the deformation family $X[c_1]$,
\begin{equation}
  C_{n-1}(Z_f\<\in X[c_1])
  =\int_X c_1(X) \Big(\sum\nolimits_{k=2}^{n-1} (-1)^{n-1-k}
                       \,c_k(X)\,c_1^{~n-1-k}(X)\Big),\qquad
  n\<=\dim(X),
 \label{e:Euler}
\end{equation}
equals the standard, topological Euler characteristic, $\c(Z_f)$, only for almost complex $X$. For $X$ only precomplex, the precise nature of the ($\rT$-equivariant\cite{Atiyah:1989ty, Chen:2000cy} and ``stringy''\cite{rBatyrevDais}) characteristic number~\eqref{e:Euler} remains to be determined, but should be ``corrected'' by $Z_f$ intersecting the complex structure obstructions; see Conjectures~\ref{C:gCYh} and~\ref{C:obstruct}. We then use $C_n(X)$ and $C_{n-1}(Z_f)$, imply $\rT$-equivariance without noting so explicitly, and for simplicity drop the quotation marks from ``anticanonical.''
\end{remk}

For a non-Fano variety, $X$, the entire system of holomorphic anticanonical sections may even factorize, forcing its Calabi--Yau hypersurfaces to {\em\/reduce,\/} $Z\<=Z_1\<\cup Z_2$, i.e., Tyurin-degenerate\cite{tyurin2003fano}, with $\Sing(Z)\<=Z_1\<\cap Z_2$ itself a (codimension-2 in $X$) Calabi--Yau subspace. Even such ``Constructive Calabi--Yau manifolds''\cite{Lee:2006wf} admit the transposition mirror model construction\cite{rBH, rBH-gB, Berglund:2022dgb}, and so provide a broader testbed for the so-called {\em\/DHT~conjecture\/}\cite{doran2016mirror,Kanazawa:2017wp,Doran:2020vo,Doran:2018um,Barrott:2021wx,doran2021mirror,Doran:2021ud}.
 In turn, many of these singular and degenerate models in fact are smoothed by certain {\em\/Laurent deformations\/}\cite{rBH-gB,Berglund:2022dgb}, consistent with the continued validity of various Todd--Hirzebruch identities and relevant Chern data calculations, to which we now turn.

\section{Todd--Hirzebruch Identities and Characteristic Classes}
\label{s:CCTH}
For easy reference, we list below five key standard results about toric (complex-algebraic) varieties\cite{rD-TV, rO-TV, rF-TV, rGE-CCAG, rCLS-TV}, where each
1-cone $\n_i\<\in\pFn{X}$ and $\pDs{X}$-vertex corresponds to a divisor class, $D_i$, and its defining (Cox) coordinate, $x_i$\cite{rCox}; see also\cite[Thm.\;12.4.4]{rCLS-TV}.
In unitary torus manifolds (see Remarks~\ref{r:UTM}, \ref{r:optimal} and~\ref{r:C3"T} and Conjectures~\ref{C:gCYh} and~\ref{C:obstruct}), $D_i$ are elements of $\rT$-equivariant homology, the key standard results involve
($\rT$-equivariant) Chern classes ($\pDs{X}$-data) and
the line bundle $\cKs{X}$ ($\pDN{X}$-data)\cite{rM-MFans, Masuda:2000aa, rHM-MFs, Masuda:2006aa, rHM-EG+MF, rH-EG+MFs2}.
\begin{bBox}
{\bsf Key standard results:}
\begin{enumerate}[itemsep=-3pt, topsep=-2pt, label=R.\arabic*]

 \item\label{i:C1}
  The Chern class\cite[Prop.\,11.4]{rD-TV}\cite[Prop.\,13.1.2\,(a)]{rCLS-TV} ({\em\/\&\cite[Thm.\;3.1]{rM-MFans} for torus manifolds\/}) of the toric space $X$ encoded by the fan $\pFn{X}$ with generator 1-cones $\n_i$ is $c(T_X)=\prod_i(1{+}D_i)$.

 \item\label{i:C2}
  Expanding $c(T_X)$ yields
$c_k(T_X)=\sum_{i_1<\cdots<i_k}D_{i_1}{\cdots}D_{i_k}|_{S.R.}$,
the summands subject to the multifan-encoded Stanley--Reisner ideal of relations; $c_1(X)\<=\sum_iD_i\<=[\cKs{X}]$, the anticanonical divisor.

 \item\label{i:C3}
  Corollary\,11.5~\cite{rD-TV}: $c_k(X)=\sum_{\s\in\pS(k)}[F_\s]$. That is, the $k^\text{th}$ Chern class is generated by the (cohomology classes of the) subvarieties $F_\s$ corresponding to the $k$-cones $\s\<\in\pFn{X}$.

 \item\label{i:C4}
  Corollary\,11.6~\cite{rD-TV}:
 $C_n(X)\<=\int_X\!c_n\<=|\pFn{X}(\dim(X))|\<=d(\pDs{X})$.
 For convex polytopes, $d(\pDs{X})\<=\c(X)$, but {\em\,for multi-layered multitopes, only if $X$ is almost complex\cite{rM-MFans}.\/}

 \item\label{i:C5}
  Statement\,11.12.2~\cite[\PP\,5.8]{rD-TV}: $[-\!K_X]^r\<=C_1^{~n}(X)\<=d(\pDN{X})$, where 
 $\pDN{X}\<=(\pDs{X})^\wtd$ is the Newton polytope, the (external\cite{rGE-CCAG}, i.e., internal\cite{rCLS-TV}) {\em\/normal fan\/} of which is
 $\pFn{X}\<\smt\pDs{X}$.\qedhere

\end{enumerate}
\end{bBox}
\begin{remk}
\label{r:C-infty}
In physics literature, Chern and other characteristic classes are typically used on compact complex manifolds, but the Whitney product formula (the $C^\infty$ ``splitting principle'')\cite{rGrHa} allows the classic extension to smooth manifolds\cite{Atiyah:1959aa}.
The Hirzebruch--Riemann--Roch (HRR) theorem for $\rT$-equivariant line bundles is also well understood on large classes of (presymplectic) toric spaces that correspond to non-convex and flip-folded multitopes\cite{rK+P-vrtPlytp0, rK+P-vrtPlytp, rK+T-pSympTM}.
It then should not surprise if related results about toric varieties in fact do extend, and rather straightforwardly, to the much larger class of {\em\/precomplex\/} torus manifolds corresponding to VEX multitopes, consistent with the computations below. In fact, fairly routine occurrence of even {\em\/stratified pseudo-manifolds\/} in GLSM phase diagrams\cite{Aspinwall:1993nu, Hubsch:2002st, rAR01, rB-IS+ST} and other ``defects''\cite{McNamara:2019rup} in string theory suggest much broader generalizations to be likely.
\end{remk}

\subsection{Todd--Hirzebruch Identities}
Todd--Hirzebruch combinatorial identities stem from a special-case application of the HRR theorem\cite{rHirz}:
\begin{equation}
 \Td(X)\<=\c^h(X)\<=\c(X,\cO_X)
  =\int_X\td(X)\Big(\ch(\cO_X)=e^{c_1(\cO_X)=0}=1\Big),
 \label{e:RRO}
\end{equation}
which is identifiable with $w(\pDs{X})=w(\pFn{X})$, the winding number of the spanning multitope and multifan of the toric space $X$\cite{rM-MFans, Masuda:2000aa, rHM-MFs}. 
 We will need the standard results\cite{rHirz,rT-ArrInv}:
\begin{equation}
 \label{e:td}
 \td_2=\frc1{12}\big(c_2\<+c_1\!^2\big),\qquad
 \td_3=\frc1{24}\big(c_2\,c_1\big),\qquad
 \td_4=\frc1{720}\big({-}c_4\<+c_3c_1\<+3c_2\!^2\<+4c_2c_1\!^2
                      \<-c_1\!^4\big).
\end{equation}

\subsubsection{The 12-Theorem}
Integrating the $\td_2$ expression in~\eqref{e:td} over complex-algebraic surfaces yields the {\em\/Noether formula\/}\cite{rGrHa},
\begin{equation}
  C_2(X)\<+C_1\!^2(X)\<=12\Td(X).
 \label{e:12N}
\end{equation}
For smooth, projective toric varieties ($\Td(X)\<=1\<=\c(\cO_X)$) and with the results~\ref{i:C4} and~\ref{i:C5}, this yields:
\begin{thrm}[``12-Theorem'']
\label{T:12}
 For all polar pairs of reflexive ($M$ and $N$) lattice polygons, $(\pD,\pD^\circ)$,
\begin{equation}
  |\vd\pD\cap M|+|\vd\pD^\circ\cap N|=12.
 \label{e:12T}
\end{equation}
\backUp
\end{thrm}
\noindent
Its evident $\pD\<\iff\pD^\circ$ symmetry relates it to Batyrev's mirror manifold construction\cite{rBaty01,rBatyBor3}, the toric rendition\cite{rCOK, rLB-MirrBH, rACG-BHK, rA+P-MM, rF+K-BHK} of the transposition construction\cite{rBH, rBH-LGO+EG, rFJR-07b, Krawitz:2009aa, rMK-PhD}, and the VEX generalization\cite{rBH-gB, Berglund:2022dgb}, as in Figure~\ref{f:gMM}.
Indeed, Ref.\cite{rP+RV-12} generalizes the ``12-Theorem''~\eqref{e:12T} to an infinite class of {\em\/generalized legal loops,} $\cL$,
each of which is easily shown to bound a (2-dimensional) VEX multigon, with their {\em\/duality\/} operation $\cL\<\to\cL^\vee$\cite{rP+RV-12} (itself with no higher-dimensional generalization\,\footnote{We thank B.~Poonen for confirming this.}) easily shown to always reproduce the transpolar operation (Definition~\ref{D:tP}).
Transpolar pairs of VEX multigons\cite{rBH-gB} are thus identical to dual pairs of generalized legal loops, $(\cL,\cL^\vee)$, so the multiple different proofs in Refs.\cite{rP+RV-12, Higashitani:2017aa} imply:
\begin{corl}
\label{c:12VEX}
For every transpolar pair of VEX multigons, $(\pD,\pD^\wtd)$, the 12-Theorem implies:
\begin{equation}
   d(\vd\pD)+d(\vd\pD^\wtd) = d(\pD)+d(\pD^\wtd)
   =12\,\big(w(\pD)\<=w(\pD^\wtd)\big).
 \label{e:12Td}
\end{equation}
\end{corl}
\backUp\noindent
Formula~\eqref{e:12T} is here recast in terms of degrees, being easier to asses for negatively oriented faces than counting lattice points therein\cite{rBH-gB}; see also \SS\:\ref{s:StarSh}.

Figure~\ref{f:Fm+Em} illustrations exemplify the necessary orientation dependence (for all $m\<\in\ZZ$):
\begin{subequations}
 \label{e:EF=12}
\begin{alignat}9
  \big[d(\pDs{E_m^{\sss(2)}}) &=1{+}(\C1{-m}){+}1{+}1\big] &~+~
  \big[d(\pDN{E_m^{\sss(2)}}) &=(3\C2{+m}){+}1{+}2{+}3\big]&&=12.\\
  \big[d(\pDs{\FF[2]m}) &=1{+}1{+}1{+}1\big] &~+~
  \big[d(\pDN{\FF[2]m}) &=(2\C2{+m}){+}2{+}(2\C1{-m}){+}2\big]&&=12.
\end{alignat}
\end{subequations}
For $-3\<\leqslant m\<\leqslant0$ for $E_m^{\sss(2)}$ and $|m|\<\leqslant2$ for $\FF[2]m$, these (now convex) polytope pairs reproduce 11 of the 16 reflexive polytopes;
 $\pDs{E_{-3}^{\sss(2)}}\<=\pDN{E_{-3}^{\sss(2)}}$, 
 $\FF[2]{-m}\<\approx\FF[2]m$ and
 $E_{-1}^{\sss(2)}\<\approx\FF[2]{\pm1}$. 
For all other $m$, one of the multigons in each transpolar pair has a flip-folded, ``CW'' edge, 
which contributes negatively to the balance~\eqref{e:EF=12}. However,
whereas $d(\pDs{E_m^{\sss(2)}})$ balances $d(\pDN{E_m^{\sss(2)}})$,
$d(\pDN{\FF[2]m})$ balances itself; this difference will reemerge below.

Each multigon in Figure~\ref{f:Fm+Em} spans a multifan with a smooth subdivision, winding number equal to 1, and stems from some compact unitary torus manifold with unit Todd genus\cite{rM-MFans, Masuda:2000aa, rHM-MFs, rAM-MFs}.
As plain but non-convex polytopes, $\pDs{\EE[2]m}$ for $m\<>0$ and $\pDs{\FF[2]m}$ for $m\<>2$ in fact encode non-Fano toric (complex-algebraic) varieties.
Their {\em\/flip-folded\/} transpolar{s}, $\pDs{E^{(2)}_m}$ and
$\pDs{\MF[2]m}$, stem from torus manifolds, $\EE[2]m$ and $\MF[2]m$ are assumed to be only {\em\/precomplex\/}; see Conjecture~\ref{c:preC} and footnote~\ref{fn:S4}~p.\,\pageref{fn:S4}.

\subsubsection{Higher-Dimensional Todd--Hirzebruch Identities}
The 3-dimensional case of~\eqref{e:td} provides
\begin{equation}
  24\Td(X) = \int_X c_1\, c_2 = C_1C_2(X).
 \label{e:T24}
\end{equation}
With $c(X)\<=1{+}c_1{+}c_2{+}c_3$, anticanonical hypersurfaces,
$Z_c\<\in X[c_1]$, have
$c(Z_f\<\in X[c_1])
      \<= 1{+}(0){+}(c_2),
$
identifying $c_2(Z_f)\<=c_2(X)$. With $c_1(X)$ pulling back $Z_f$-integrals to $X$, the identity~\eqref{e:T24} implies
\begin{equation}
  \dim(X)\<=3:\quad
  24\Td(X) =\int_X c_1\, c_2,~ = \int_{X[c_1]}c_2 = C_2(X[c_1]),
\end{equation}
consistent with $Z_f\<\in X[c_1]$ being K3 surfaces as long as
 $\Td(X)\<=1$. This is true for all semi-Fano toric (complex-algebraic) 3-folds $X$, but also holds for torus manifolds corresponding to flip-folded multifans with winding number $w(\pFn{X})\<=1$; see below.
 The identity~\eqref{e:T24} involves a Chern invariant, $C_1C_2(X)$, which is not readily identified among the key standard results~\ref{i:C1}--\ref{i:C5} listed above.

In four dimensions,~\eqref{e:td} yields
\begin{equation}
  \dim(X)\<=4:\quad
  720\Td(X) =\int_X\big({-}c_4\<+c_3c_1\<+3c_2\!^2\<+4c_2c_1\!^2
                        \<-c_1\!^4\big),
 \label{e:td4}
\end{equation}
which involves three additional Chern invariants not readily identified within the key standard results~\ref{i:C1}--\ref{i:C5} in terms of corresponding multitopes, i.e., the multifans, $\pDs{X}\<\lat\pFn{X}$.
 We do find such concrete identifications below, while also verifying the extension of the key standard results~\ref{i:C2}--\ref{i:C3} to VEX multitopes.
With $\cKs{X}$ defined on unitary torus manifolds (Remark~\ref{r:UTM}) and the Todd genus~\eqref{e:RRO} being a special case of ($\rT$-equivariant) plurigenera, the above identities have many further generalizations.

\subsection{Chern Classes and Invariants}
In fact, the separate Chern numbers are also computable consistently with the results~\ref{i:C1}--\ref{i:C5}.

\subsubsection{Two Dimensions}
\paragraph{Hirzebruch Scrolls:}
For the pair $(\pDs{\FF[2]m},\pDN{\FF[2]m})$ in the right-hand half of Figure~\ref{f:Fm+Em}, Table~\ref{t:2Fmlist}
\begin{table}[htb]
$$
 \begin{array}{r|r@{~}lcr@{~}l|lcr@{~}r@{~}l}
\dim& \MC2c{\pFn{\FF[2]m}\<\smt\pDs{\FF[2]m}} &\MC1c{\fif\wtd}
    & \MC2{c|}{\pDN{\FF[2]m}\<\lat\pFN{\FF[2]m}} &\dim &
    & \MC3c{\sum d(\n_*)d(\n_*^\wtd)}\\[-1pt] \toprule\nGlu{-2pt}
  0 & d(0)&=1           && d(0^\wtd)&=8            & 3 &\to &  8     \\[-1pt]
\midrule\nGlu{-2pt}
  1 & d(\n_1)&=1        && d(\n_1^\wtd)&=2\C2{+m}& 2 &\to &2\C2{+m}
     & \MR3*[2pt]{$\left.\rule{0pt}{8mm}\right\}\,8$} 
      & \MR3*[2pt]{$=C_1\!^2(\FF[2]m)$} \\[-2pt]
    &d(\n_2)\<=d(\n_4)&=1&&d(\n_2^\wtd)\<=d(\n_4^\wtd)&=2&&\to&2{\times}2     \\[-2pt]
    & d(\n_3)&=1        && d(\n_3^\wtd)&=2\C1{-m}&   &\to &2\C1{-m}\\[-1pt]
\midrule\nGlu{-2pt}
  2 & d(\n_{i,i+1})&=1     && d(\n_{i,i+1}^\wtd)&=1  & 1 &\to &4{\times}1      
     & 4 & =C_2(\FF[2]m)\\[-1pt]
\midrule\nGlu{-2pt}
  3 & d(\n_{1234})&=4   && d(\n_{1234}^\wtd)&=1    & 0 &\to &  4     \\[-1pt]
\bottomrule
 \end{array}
$$
\caption{The cones and their degrees, spanned by the transpolar pair $(\pDs{\FF[2]m},\pDN{\FF[2]m})$}
\label{t:2Fmlist}
\end{table}
lists the degrees of transpolar pairs of cones in the 3-dimensional (lattice height-1) pyramids over these multitopes, corresponding to $\cKs{\FF[2]m}$ and $\cKs{\MF[2]m}$.
The straightforward tally,
\begin{subequations}
 \label{e:C1C12}
\begin{alignat}9
       C_2(\FF[2]m)&=\ttt\sum_{\n_{ij}\in\pS(2)} d(\n_{ij})\,d(\n_{ij}^\wtd)
       &&=4&&=d(\pDs{\FF[2]m}),\\
  C_1\!^2(\FF[2]m)&=\ttt\sum_{\n_i\in\pS(1)} d(\n_i)\,d(\n_i^\wtd)
       &&=8&&=d(\pDN{\FF[2]m}),
\end{alignat}
\end{subequations}
is in agreement with key standard results~\ref{i:C4} and~\ref{i:C5}, as well as the independent and standard computation done, e.g., for the bi-projective embedding, $\FF[2]m\<=\{x_0y_0\!^m{+}x_1y_1\!^m\<=0\}$ in $\IP^2{\times}\IP^1$\cite{rH-Fm} --- precisely because of the orientation dependent sign of $d(\n)$ in Table~\ref{t:2Fmlist}.

More importantly, reversing the roles of the multitopes, $\pDs{\FF[2]m}\fif{\sss\wtd}(\pDN{\FF[2]m}\<=\pDs{\MF[2]m})$, informs us:
the multifan $\pFn{\MF[2]m}\<\smt\pDs{\MF[2]m}$ being smoothly subdivided, a unitary torus manifold, $\MF[2]m$, exists for each $m\<\in\ZZ$ (Conjecture~\ref{C:gCYh}), with 
$C_1\!^2(\MF[2]m)\<=4$ and $C_2(\MF[2]m)\<=8$.
For $|m|\<\leqslant2$, $\MF[2]m$, are known (complex-algebraic) toric varieties, but
for $|m|\<\geqslant3$ they must be precomplex, since $\pFn{\MF[2]m}$ is flip-folded; see Conjecture~\ref{C:obstruct}.
 
\paragraph{The $\EE[2]m$ Sequence:}
The analogous computations also hold for $\EE[2]m$ (left-hand half in Figure~\ref{f:Fm+Em}), for all\footnote{For simplicity of the discussion and computations, we omit the special cases $m\<=0,~{-}3$, when these quadrangular multitopes reduce to triangles, encoding the well known $\EE[2]0\<=\IP^2$ and $\EE[2]{-3}\<=\IP^2_{(1:2:3)}$. In turn, $\EE[2]1$ is discussed briefly in\cite{rHM-MFs}, while
 $\EE[2]{-1}\<=\FF[2]1$ is a 1-point blowup of $\IP^2$ and $\EE[2]{-2}$ its (secondary, infinitesimally near) blowup; see \SS\:\ref{s:pCpx}.}
 $m\<\in\ZZ$:
\begin{table}[htb]
$$
 \begin{array}{r|r@{~}lcr@{~}l|lcr@{~}r@{~}l}
\dim& \MC2c{\pFn{\EE[2]m}\<\smt\pDs{\EE[2]m}} &\MC1c{\fif\wtd}
    & \MC2{c|}{\pDN{\EE[2]m}\<\lat\pFN{\EE[2]m}} &\dim &
    & \MC3c{\sum d(\n_*)d(\n_*^\wtd)}\\[-1pt] \toprule \nGlu{-2pt}
  0 & d(0)&=1                  && d(0^\wtd)&=9\C2{+m}   
     & 3 &\to &9\C2{+m}\\[-1pt] \midrule \nGlu{-2pt}
  1 & d(\n_1)&=1               && d(\n_1^\wtd)&=3\C2{+m}
     & 2 &\to &3\C2{+m}
      &\MR4*[2pt]{$\left.\rule{0pt}{10mm}\right\}\,9\C2{+m}$}
       &\MR4*[2pt]{$=C_1^{~2}(\EE[2]m)$} \\[-1pt]
    & d(\n_2)&=1               && d(\n_2^\wtd)&=1      &   &\to &1     \\[-1pt]
    & d(\n_3)&=1               && d(\n_3^\wtd)&=2      &   &\to &2     \\[-1pt]
    & d(\n_4)&=1               && d(\n_4^\wtd)&=3      &   &\to &3     \\[-1pt] 
 \midrule \nGlu{-2pt}
  2 &d(\n_{i,i+1})&=1,~\rlap{\footnotesize$i\<\neq2$} &&d(\n_{i,i+1}^\wtd)&=1&1&\to&3{\times}1
      & \MR2*[1pt]{$\left.\rule{0pt}{5mm}\right\}\,3\C1{-m}$} 
       &\MR2*[1pt]{$=C_2(\EE[2]m)$}\\[-1pt]
    & d(\n_{23})&=\C1{-m}      && d(\n_{23}^\wtd)&=1   &   &\to &\C1{-m}\\[-1pt]
 \midrule \nGlu{-2pt}
  3 & d(\n_{1234})&=3\C1{-m} && d(\n_{1234}^\wtd)&=1 & 0 &\to &3\C1{-m}\\[-1pt]
\bottomrule
 \end{array}
$$
\caption{The cones and their degrees, spanned by the transpolar pair $(\pDs{\EE[2]m},\pDN{\EE[2]m})$}
 \label{t:2Emlist}
\end{table}
While we do not have an independent verification of
 $C_1^{~2}(\EE[2]m)\<=d(\pDN{\EE[2]m})$ and $C_2(\EE[2]m)\<=d(\pDs{\EE[2]m})$, the data in Table~\ref{t:2Emlist} (also in Table~\ref{t:2Fmlist}) does verify the consistency relations\cite{rBH-gB}:
\begin{equation}
  \sum_{\n_*\in\pS(0)} d(\n_*)\,d(\n_*^\wtd)
  =\sum_{\n_*\in\pS(1)} d(\n_*)\,d(\n_*^\wtd)
   \qquad\text{and}\qquad
  \sum_{\n_*\in\pS(2)} d(\n_*)\,d(\n_*^\wtd)
  =\sum_{\n_*\in\pS(3)} d(\n_*)\,d(\n_*^\wtd),
\end{equation}
This is guaranteed by both multifans,
$\pFn{\EE[2]m}\<\smt\pDs{\EE[2]m}$ and
$(\pFn{\ME[2]m})\<\smt(\pDs{\ME[2]m}\<=\pDN{\EE[2]m})$, 
having a smooth stellar subdivision and
so stem from smooth toric manifolds, $\EE[2]m$ and $\ME[2]m$.
With the ($w$-signed!) degree function, we thus find the key results~\ref{i:C4} and~\ref{i:C5} to hold more generally:
\begin{clam}
\label{CC:CnC1n}
A transpolar pair of $n$-dimensional VEX multitopes, $(\pDs{X},\pDN{X})$, and the multifans they span,
 $(\pFn{X},\pFN{X})$, correspond to a transpolar pair of toric spaces, $(X,\tX)$ by identifying
 $\pDs{\tX}\<\coeq\pDN{X}$, $\pDN{\tX}\<\coeq\pDs{X}$, i.e.,
 $\pFn{\tX}\<\coeq\pFN{X}$ and $\pFN{\tX}\<\coeq\pFn{X}$, such that:
\begin{equation}
  C_1^{~n}(X)=d(\pDN{X}\<=\pDs{\tX})=C_n(\tX)
   \qquad\text{and}\qquad
  C_n(X)=d(\pDs{X}\<=\pDN{\tX})=C_1^{~n}(\tX).
 \label{e:CnC1n}
 \backUp
\end{equation}
\end{clam}

\subsubsection{Three Dimensisons}
\paragraph{Hirzebruch Scrolls:}
Without further ado, the 3-dimensional analogue of Hirzebruch surfaces was analyzed in Ref.\cite{rBH-gB}, and we simply tabulate the cones and degrees of its transpolar pair of multifans in Table~\ref{t:3Fmlist}; see also Figure~\ref{f:3F3MM} in Appedix~\ref{s:2MM}.
\begin{table}[htb]
$$
  \begin{array}[t]{r|r@{~}l@{}c@{}r@{~}l|l@{}cr@{~}r@{~}l}
\dim&\MC2c{\pFn{\FF[3]m}\<\smt\pDs{\FF[3]m}} &\MC1c{\fif\wtd}
    &\MC2{c|}{\pDN{\FF[3]m}\<\lat\pFN{\FF[3]m}} &\dim &&\MC3c{\sum d(\n_*)d(\n_*^\wtd)}\\
\toprule\nGlu{-2pt}
  0 & d(0)&=1             && d(0^\wtd)&=54              & 4 &\to &  54      \\[-1pt]
\midrule\nGlu{-2pt}
  1 & d(\n_1)&=1          && d(\n_1^\wtd)&=12\C1{-6m}   & 3 &\to &12\C1{-6m}
   & \MR3*[2pt]{$\left.\rule{0pt}{8mm}\right\}\,54$}   &\MR3*[2pt]{$=C_1\!^3(\FF[3]m)$}\\[-2pt]
    &d(\n_2)\<=d(\n_3)&=1&&d(\n_2^\wtd)\<=d(\n_2^\wtd)&=12\C2{+3m}&&\to&2(12\C2{+3m})\\[-2pt]
    &d(\n_4)\<=d(\n_5)&=1&&d(\n_4^\wtd)\<=d(\n_5^\wtd)&=9&&\to&2{\times}9   \\[-1pt]
\midrule\nGlu{-2pt}
  2 & d(\n_{12})\<=d(\n_{13})&=1&&d(\n_{12}^\wtd)\<=d(\n_{13}^\wtd)&=2\C1{-m}&2&\to
                                                               &2(2\C1{-m})
   & \MR3*[2pt]{$\left.\rule{0pt}{8mm}\right\}\,24$}   &\MR3*[2pt]{$=C_2C_1(\FF[3]m)$}\\
    & d(\n_{ij})^\star&=1 && d(\n_{ij}^\wtd)^\star&=3   &   &\to &6{\times}3\\[-2pt]
    & d(\n_{23})&=1       && d(\n_{23}^\wtd)&=2\C2{+2m} &   &\to & 2\C2{+2m}\\[-1pt]
\midrule\nGlu{-2pt}
  3 & d(\n_{ijk})^\diamond&=1&& d(\n_{ijk}^\wtd)^\diamond&=1&1&\to&6{\times}1
     &6&=C_3(\FF[3]m)\\
\midrule\nGlu{-2pt}
  4 & d(\n_{1\cdots5})&=6 && d(\n_{1\cdots5}^\wtd)&=1       & 0 &\to &  6    \\[-2pt]
\bottomrule
 \MC{11}l{\text{\footnotesize$^{\ttt\star}$ for $(ij)\<=14,\,15,\,24,\,25,\,34,\,35$;\quad
  $^{\diamond}$ for $(ijk)\<=412,\,423,\,431,\,521,\,532,\,513$;\quad
  ordering encodes orientation}}
  \end{array}
$$
\caption{The cones and their degrees, spanned by the transpolar pair
         $(\pDs{\FF[3]m},\pDN{\FF[3]m})$ as specified in\cite{rBH-gB}}
 \label{t:3Fmlist}
\end{table}
 The degree of each non-simplex (but {\em\/flat\/}\footnote{A lattice $k$-dimensional cone is {\em\/flat\/} if its lattice-primitive generators (defining its base) lie in a $(k{-}1)$-plane.}) cone has been computed in several simplex subdivisions, verifying that they all produce the same result; see\cite{rBH-gB} for details.
 The tabulated degrees clearly satisfy the ``24-Theorem''~\eqref{e:T24}, and also reproduce the individual Chern characteristic results obtained for the bi-projective embedding, $\FF[3]m\<\in\ssK[{r||c}{\IP^3&1\\\IP^1&m}]$. Moreover, these results evidently hold {\em\/precisely if\/} the degrees of the cones
 $\n_1^\wtd,\n_{12}^\wtd,\n_{13}^\wtd\subset\pFN{\FF[3]m}$
are negative for $m\<\geqslant3$ --- exactly as argued in Ref.~\cite{rBH-gB}.
 Since $\MF[3]m$ is defined so $\pDs{\MF[3]m}\<=\pDN{\FF[3]m}$ and {\em\/vice versa,\/} $\pDN{\MF[3]m}\<=\pDs{\FF[3]m}$, we learn that:
\begin{subequations}
 \label{e:C3Fm=*C*3Fm}
\begin{alignat}9
  C_1^{~3}(\MF[3]m)=d(\pDN{\MF[3]m})&=~6&&=d(\pDs{\FF[3]m})=C_3(\FF[3]m),\\*
  C_3(\MF[3]m)=d(\pDs{\MF[3]m})&=54&&=d(\pDN{\FF[3]m})=C_1^{~3}(\FF[3]m).
\iText[-1mm]{Finally, swapping $\pDs{\FF[3]m}\iff\pDN{\FF[3]m}$:}
  C_2C_1(\MF[3]m)&=24&&=C_2C_1(\FF[3]m), \label{e:C2=EU}
\end{alignat}
\end{subequations}
corroborates that $\MF[3]m[c_1]$ is also a family of \KCY\ surfaces.
The $\dim\<=2$ summation in the middle rows of Table~\ref{t:3Fmlist} exactly reproduces Batyrev's Euler characteristic formula\cite{rBaty01, rBatyrevDais} for $X\<=\FF[3]m$:
\begin{equation}
  \chi(\hat{Z}_f) 
  =\sum_{k=1}^{\dim X-2}(-1)^{k-1}\sum_{\substack{\dim(\q)=k\\ \q\subset\pD_X}} d(\q)\,d(\q^*)
  ~=\mkern-12mu
    \sum_{\n\<\subset\pS(2)}\mkern-12mu d(\n)\,d(\n^\wtd)
 \label{e:EuXm}
\end{equation}
since $\dim\FF[3]m=3$ and $\dim(\n)\<=d(\q){+}1$ for $\n\<=\sfa\q$.

\begin{figure}[htb]
$$
 \vC{\begin{picture}(160,75)(5,0)
   \put(18,42){\rotatebox[origin=c]{6}
               {\includegraphics[height=30mm]{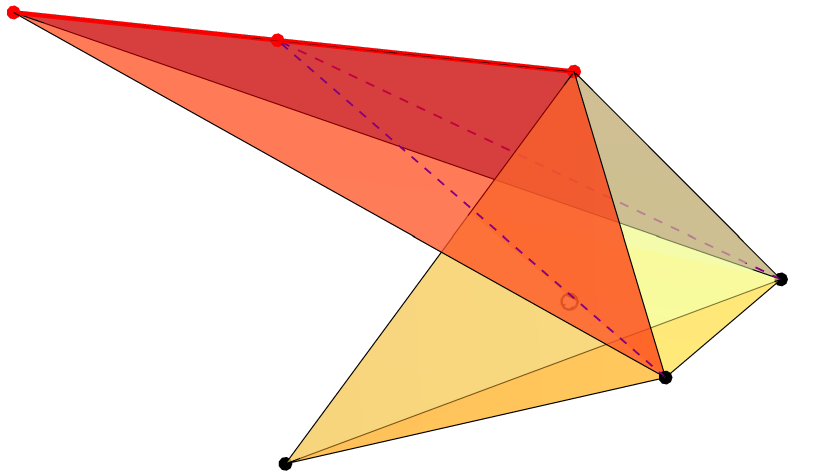}}}
   \put(52,43){$\pDs{\EE[3]m}$}
   \put(62,47){\scriptsize$\n_1$}
   \put(70,58){\scriptsize$\n_2$}
   \put(57,68){\scriptsize$\C1{\n_3}$}
   \put(40,40){\scriptsize$\n_4$}
   \put(18,66){\scriptsize$\C1{\n_5}$}
   \cput(19,63){\tiny$\C1{(-m,-m,1)}$}
   \put(34,70){\scriptsize$\C5{(\n_6)}$}
   \put(69,54){\footnotesize
    $\begin{array}
           {r|@{~}c@{~~}c@{~~}c@{~~}c@{~}|@{~}c@{~}c@{\quad}r@{~~}r@{~}l}
      \pFn{\EE[3]m} &\nu_1 & \nu_2 & \nu_3 & \nu_4 & \nu_5\\
     \cmidrule[1pt]{1-7}\nGlu{-2pt}
             & 1 & 0 & 0 &-1 &-m &&\To&D_1-D_4-mD_5&\sim0 \\[-2pt]
             & 0 & 1 & 0 &-1 &-m &&\To&D_2-D_4-mD_5&\sim0 \\[-2pt]
             & 0 & 0 & 1 &-1 &~~1&&\To&D_3-D_4+D_5 &\sim0 \\[-1pt]
     \cmidrule[.4pt]{1-6}\cmidrule[.4pt]{9-10}\nGlu{-2pt}
         Q^1 & 0 & 0 &1{+}m& m &-1 &&&
          \MC2l{D_2\sim D_1,~~D_1\sim D_4{+}mD_5} \\
         Q^2 & m & m &  -1 & 0 &~~1&&&\MC2l{D_3\sim D_4{-}D_5} \\[-1pt]
     \cmidrule[1pt]{1-6}
     \end{array}$}
   \put(0,0){\includegraphics[height=70mm]{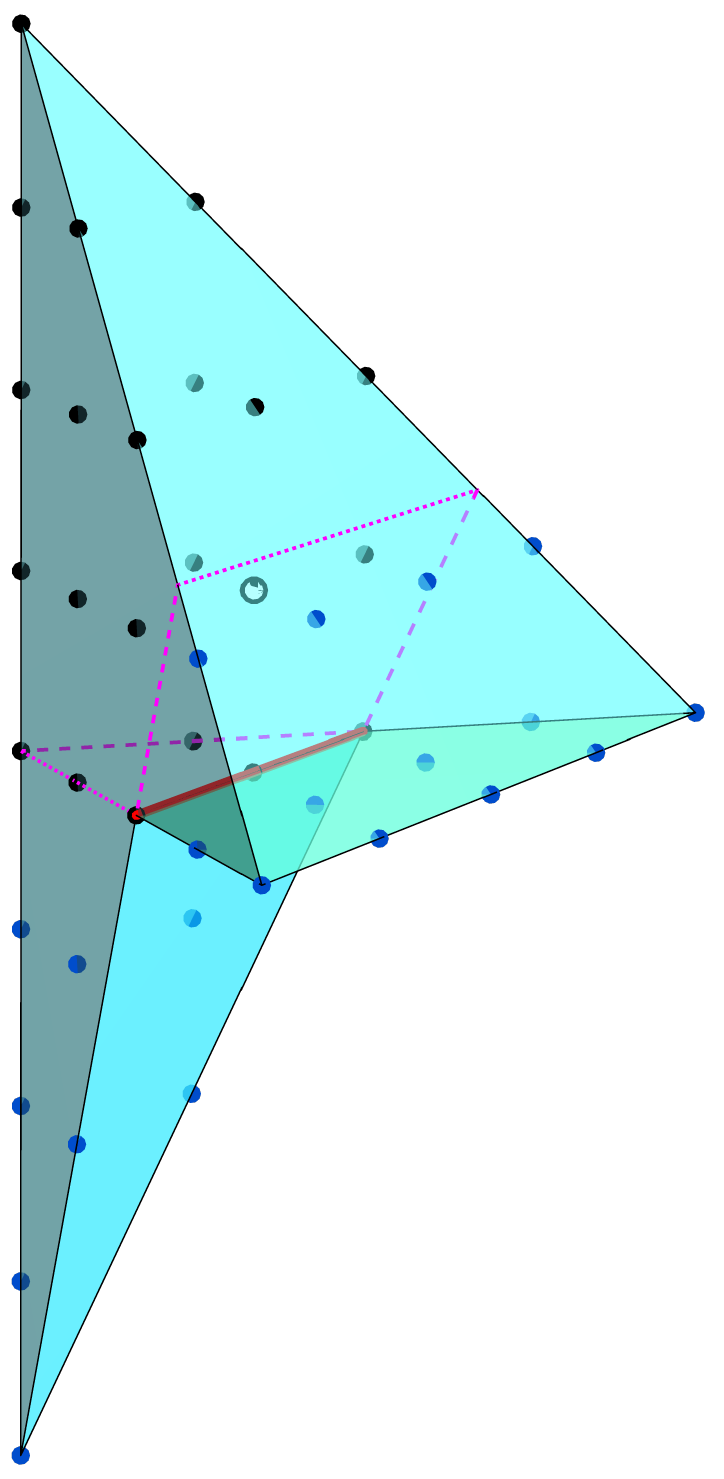}}
   \put(12,19){$\pDN{\EE[3]m}\<=\pDs{\ME[3]m}$}
   \put(2,69){\scriptsize$\n_{142}^\wtd$}
   \put(2,1){\scriptsize$\n_{125}^\wtd$, \tiny$(-1,{-}1,{-}1{-}2m)$}
   \put(31,33){\scriptsize$\n_{134}^\wtd$}
   \put(13,25){\scriptsize$\n_{243}^\wtd$}
   \put(22,38){\scriptsize$\C1{\n_{153}^\wtd}$}
   \put(5.5,23){\scriptsize$\C1{\n_{235}^\wtd}$}
   \put(0,0){\TikZ{\path[use as bounding box](0,0);
     \draw[red, thick, densely dashed, -stealth](.7,2.5)--++(95:.6);
     \draw[red, thick, densely dashed, -stealth](2.25,3.75)--++(202:.55);
     \draw[ultra thick, densely dotted, <->]
         (2.2,5)--node[above=-1mm, rotate=15]{\small transpolar}++(15:2.2);
              }}
   \put(30,13){\footnotesize
    $\begin{array}
           {r|c@{}c@{~}c@{~}c@{~}|@{~}c@{~}c@{~}c@{\quad}r@{~~}r@{~}l}
      \pFN{\EE[3]m} &\n_{142}^\wtd &\n_{125}^\wtd &\n_{134}^\wtd 
                    &\n_{243}^\wtd &\n_{153}^\wtd &\n_{235}^\wtd\\[-1pt]
     \cmidrule[1pt]{1-7}\nGlu{-2pt}
                    &-1 &-1     &-1 &~~3&-1 &~~1&&\To
                     &-\tD_1{-}\tD_2{-}\tD_3{+}3\tD_4{-}\tD_5{+}\tD_6
                      &\sim0\\[-2pt]
                    &-1 &-1     &~~3&-1 &~~1&-1 &&\To
                     &-\tD_1{-}\tD_2{+}3\tD_3{-}\tD_4{+}\tD_5{-}\tD_6
                      &\sim0\\[-2pt]
                    &~~3&-1{-}2m&-1 &-1 &-1 &-1 &&\To
                     &3\tD_1{-}(1{-}2m)\tD_2{-}\tD_3{-}3\tD_4{-}\tD_5
                            {-}\tD_6
                     &\sim0 \\[-1pt]
     \cmidrule[.4pt]{1-7}\cmidrule[.4pt]{10-11}\nGlu{-2pt}
      \Tw{q}\,^1& 0 & 1 & 1 & 0 &-2{-}m&-m &&&\MC2l{\tD_4\sim\tD_1
       {+}\tD_2{-}\tD_3,~~ \tD_5\sim2\tD_1{-}m\tD_3{-}2\tD_3}\\[-2pt]
      \Tw{q}\,^2& 1 & 0 & 1 & 0 & 0    & 2 &&&\MC2l{D_6\sim-(2{+}m)\tD_2
       {-}2\tD_3}\\[-2pt]
      \Tw{q}\,^3& 1 & 1 & 0 & 2 & 2{-}m&-2{-}m&&&\\[-1pt]
     \cmidrule[1pt]{1-7}
     \end{array}$}
   \put(45,34){e.g.,~\small
               $[\n_{243}^\wtd,\n_{134}^\wtd]=\n_{34}^\wtd$,~
               $[\n_{142}^\wtd,\n_{125}^\wtd]=\n_{12}^\wtd$,~
               $[\n_{142}^\wtd,\n_{243}^\wtd,\n_{134}^\wtd]=\n_4^\wtd$,~
               \etc}
 \end{picture}}
$$
 \caption{The non-convex VEX polytope $\pDN{\EE[3]m}$ (far left) and
 its transpolar multitope, $\pDs{\EE[3]m}$ (mid-left, top), their vertex systems, (proto) Mori vectors and the so-implied equivalences among the Cox divisors}
 \label{f:3EmSNP}
\end{figure}

\paragraph{The $\EE[3]m$ Sequence:}
Consider the $m\<>0$ infinite sequence of transpolar pairs of 3-dimensional multitope analogues of the multigons in the left-hand half of Figure~\ref{f:Fm+Em}, the $m\<=2$ case shown in Figure~\ref{f:3EmSNP}.
The (continuous) orientation of the top-dimensional cones is encoded by the order of their multi-index: e.g., $\n_{142}$ and $\n_{134}$ are oriented positively\footnote{\label{fn:d}Stacking the vertex-vectors in the given order associates to each simplicial cone a square matrix, the determinant of which is the {\em\/signed\/} degree of the cone:
 $d(\n_{142})\<=\det\bM{1&-1&0\\0&-1&1\\0&-1&0}\<=1$,
 $d(\n_{125})\<=\det\bM{1&0&-m\\0&1&-m\\0&0&~~1}\<=1$, but
 $d(\n_{153})\<=\det\bM{1&-m&0\\0&-m&0\\0&~~1&1}\<={-}m$, etc. Cones over non-simplex faces in a multitope are subdivided simplicially and the sub-cone degrees added; this sum does not depend on the choice of the subdivision\cite{rBH-gB}. In three dimensions, this sign agrees with the (quicker) ``right-hand rule.''} 
(away from the origin). Transporting this orientation continuously along the surface of $\pDs{\EE[3]m}$ orients $\n_{153}$ and $\n_{235}$ (shaded red) negatively (towards the origin) --- and then $\n_{125}$ again positively. With this signed multiplicity, $\pFn{\EE[3]m}$ covers every direction effectively once, so $w(\pFn{\EE[3]m})\<=1$\cite{rM-MFans, Masuda:2000aa, rHM-MFs}. Its transpolar, $\pDN{\EE[3]m}$, is a plain polytope, uniformly orientable away from the origin, but non-convex at the $[\n_{235}^\wtd,\n_{153}^\wtd]=\n_{35}^\wtd$ edge and the two adjacent facets.

The cones and their degrees, listed in Table~\ref{t:3Emlist},
\begin{table}[htb!]
$$
  \begin{array}{@{}r@{~}|@{~}r@{\,}l@{}c@{}r@{\,}l|l@{}c@{\,}r@{}r@{\,}l@{}}
\dim&\MC2c{\pFn{\EE[3]m}\<\smt\pDs{\EE[3]m}} &\MC1c{\!\!\!\fif\wtd\!\!\!}
    &\MC2{c|}{\pDN{\EE[3]m}\<\lat\pFN{\EE[3]m}} &\dim &&\MC3c{\sum d(\n_*)d(\n_*^\wtd)}\\
\toprule\nGlu{-2pt}
  0 & d(0)&=1             && d(0^\wtd)&=64\C2{+8m}      & 4 &\to &64\C2{+8m}\\[-1pt]
\midrule\nGlu{-2pt}
  1 &d(\n_1)\<=d(\n_2)&=1&&d(\n_1^\wtd)\<=d(\n_2^\wtd)&=16\C2{+4m}&3&\to&2(16\C2{+4m})
   & \MR4*[4pt]{$\left.\rule{0pt}{10mm}\right\}\,64\C2{+8m}$}
    &\MR4*[4pt]{$=C_1\!^3(\EE[3]m)$} \\
    & d(\n_3)&=1          && d(\n_3^\wtd)&=12           &   &\to &  12      \\[-2pt]
    & d(\n_4)&=1          && d(\n_4^\wtd)&=16           &   &\to &  16      \\[-2pt]
    & d(\n_5)&=1          && d(\n_5^\wtd)&=4            &   &\to &  4       \\[-1pt]
\midrule\nGlu{-2pt}
  2 & d(\n_{12})&=1       && d(\n_{12}^\wtd)&=4\C2{+2m} & 2 &\to &4\C2{+2m}
   & \MR4*[4pt]{$\left.\rule{0pt}{10mm}\right\}\,24$} &\MR4*[4pt]{$=C_2C_1(\EE[3]m)$}\\
    & d(\n_{ij})^\star&=1 && d(\n_{ij}^\wtd)^\star&=2   &   &\to &4{\times}2\\[-2pt]
    & d(\n_{k4})^\diamond&=1&&d(\n_{k4}^\wtd)^\diamond&=4&  &\to &3{\times}4\\[-2pt]
    & d(\n_{35})&=\C1{-m} && d(\n_{35}^\wtd)&=2         &   &\to &\C1{-2m}  \\[-1pt]
\midrule\nGlu{-2pt}
  3 & d(\n_{ijk})^{\natural}&=1&&d(\n_{ijk}^\wtd)^{\natural}&=1&1&\to&4{\times}1       
   & \MR2*[1pt]{$\left.\rule{0pt}{6mm}\right\}\,4\C1{-2m}$} &\MR2*[1pt]{$=C_3(\EE[3]m)$}\\
    &d(\n_{153})\<=d(\n_{235})&=\C1{-m}&&d(\n_{234}^\wtd)\<=d(\n_{253}^\wtd)&=1&&\to
     &2(\C1{-m})   \\[-1pt]
\midrule\nGlu{-2pt}
  4 & d(\n_{1\cdots5})&=4\C1{-2m}&& d(\n_{1\cdots5}^\wtd)&=1 & 0 &\to &4\C1{-2m}\\[-2pt]
\bottomrule
 \MC{11}l{\text{\footnotesize$^{\ttt\star}$\,for $(ij)\<=13,\,15,\,23,\,25$;\quad
  $^\diamond$\,for $k\<=1,\,2,\,3$;\quad
  $^{\natural}$\,for $(ijk)\<=142,\,125,\,134,\,243$;\quad ordering encodes orientation}}
  \end{array}
$$
\caption{The cones spanned by the transpolar pair
         $(\pDs{\EE[3]m},\pDN{\EE[3]m})$ and their degrees}
 \label{t:3Emlist}
\end{table}
not only satisfy the ``24-Theorem''~\eqref{e:T24} but also compute the two remaining Chern numbers:
\begin{alignat}9
  C_1^{~3}(\EE[3]m) &=d(\pDN{\EE[3]m})
  &&=64\C2{+8m}   &&=d(\pDs{\ME[3]m}) &&=C_3(\ME[3]m),\\
  C_3(\EE[3]m)      &=d(\pDs{\EE[3]m})
  &&=~\,4\C1{-2m} &&=d(\pDN{\ME[3]m}) &&=C_1^{~3}(\ME[3]m),
\end{alignat}
satisfying the key standard results~\ref{i:C4} and~\ref{i:C5} above, again precisely because of the orientation-depen\-dent signed degree values.
 Again, $C_2C_1(\EE[3]m)\<=24\<=C_2C_1(\ME[3]m)$ holds by simply swapping
 $\pDs{\EE[3]m}\<\iff\pDN{\EE[3]m}$, and is consistent with 
 $(\EE[3]m[c_1],\ME[3]m[c_1])$ being deformation families of K3 surface mirrors.
Tables~\ref{t:2Fmlist}---\ref{t:3Emlist} and alike, for many other examples, verify, with $n\<=\dim(X)$:
\begin{equation}
  \sum_{\n\in\pS(0)} d(\n)d(\n^\wtd)=\sum_{\n\in\pS(1)} d(\n)d(\n^\wtd)
   \qquad\text{and}\qquad
  \sum_{\n\in\pS(n)} d(\n)d(\n^\wtd)= d(\sfa\pS)\,d(\pFN[(0)]{}),
\end{equation}
where the degree of the ($n{+}1$-dimensional) height-1 cone over a multifan, $\sfa\pS$, equals the sum of the degrees of its top-dimensional cones, $d(\pS)=\sum_{\s\in\pFn[\sss(n)]{}}d(\s)$, and
$\pFN[(0)]{}\<=\n_{123\cdots}^\wtd\<=(\pS)^\wtd$ is the transpolar of the entire multifan, i.e., the center of the dual multifan, $\pFN{}$.

\subsubsection{Four Dimensions}
\paragraph{Hirzebruch Scrolls:}
The CY 3-folds $\XX[3]m\<\subset\FF[4]m$ have been analyzed in\cite{rBH-gB}, where Batyrev's formula~\eqref{e:EuXm} was found to confirm the direct Chern class computation. The relevant data is summarized here in Table~\ref{t:4Fmlist}, completed with $d(\n_*^\wtd)$, which we compute using a simplicial subdivision as needed.
\begin{table}[htb]
$$
  \begin{array}[t]{r|r@{~}l@{}c@{}r@{~}l|l@{}cr@{~}r@{~}l}
\dim&\MC2c{\pFn{\FF[4]m}\<\smt\pDs{\FF[4]m}} &\MC1c{\fif\wtd}
    &\MC2{c|}{\pDN{\FF[4]m}\<\lat\pFN{\FF[4]m}} &\dim &&\MC3c{\sum d(\n_*)d(\n_*^\wtd)}\\
\toprule\nGlu{-2pt}
  0 & d(0)&=1             && d(\sfa0^\wtd)&=512              & 5 &\to &512        \\[-1pt]
\midrule\nGlu{-2pt}
  1 & d(\n_1)&=1          && d(\n_1^\wtd)&=48(2\C1{-m})   & 4 &\to &(1)[48(2\C1{-m})]
   & \MR3*[2pt]{$\left.\rule{0pt}{8mm}\right\}\,512$}   &\MR3*[2pt]{$=C_1\!^4(\FF[4]m)$}\\[-2pt]
    &d(\n_i)&=1&&d(\n_i^\wtd)&=16(6\C2{+m})&&\to&3(1)[16(6\C2{+m})]\\[-2pt]
    &d(\n_\ell)&=1&&d(\n_\ell^\wtd)&=64&&\to&2{\times}(1)(64)   \\[-1pt]
\midrule\nGlu{-2pt}
  2 & d(\n_{1i})&=1&&d(\n_{1i}^\wtd)&=8(3\C1{-m})&3&\to     &3(1)[8(2\C1{-m})]
   & \MR3*[2pt]{$\left.\rule{0pt}{8mm}\right\}\,224$}   &\MR3*[2pt]{$=C_2C_1\!^2(\FF[4]m)$}\\
    & d(\n_{a\ell})&=1 && d(\n_{a\ell}^\wtd)&=16   &   &\to &4{\times}2(1)(16)\\[-2pt]
    & d(\n_{ij})&=1&&d(\n_{ij}^\wtd)&=8(3\C2{+m})&3&\to     &3(1)[8(2\C2{+m})]\\[-1pt]
\midrule\nGlu{-2pt}
  3 & d(\n_{1ij})&=1&& d(\n_{1ij}^\wtd)&=2\C1{-m}&2&\to&3(1)(2\C1{-m})
   & \MR3*[2pt]{$\left.\rule{0pt}{8mm}\right\}\,56$}   &\MR3*[2pt]{$=C_3C_1(\FF[4]m)$}\\
    & d(\n_{ab\ell})&=1&& d(\n_{ab\ell}^\wtd)&=4& &\to&6{\times}2(1)(4)\\[-2pt]
    & d(\n_{234})&=1&& d(\n_{234}^\wtd)&=2\C2{+3m}& &\to&1(1)(2\C2{+3m})\\[-1pt]
\midrule\nGlu{-2pt}
  4 & d(\n_{abc\ell})&=1 && d(\n_{abc\ell}^\wtd) &=1 & 1 &\to & 4{\times}2(1)(1)
     &8&=C_4(\FF[4]m)\\
\midrule\nGlu{-2pt}
  5 & d(\sfa\n_{1\cdots6})&=8 && d(\n_{1\cdots6}^\wtd)&=1       & 0 &\to &  8       \\[-2pt]
\bottomrule
 \MC{11}l{\text{\footnotesize Throughout: $a,b,c\<\in\{1,2,3,4\}$;\quad
  $i,j,k\<\in\{2,3,4\}$;\quad and $\ell\<\in\{5,6\}$;\quad
  ordering encodes orientation
  }}
  \end{array}
$$
\caption{The cones and their degrees, spanned by the transpolar pair
         $(\pDs{\FF[4]m},\pDN{\FF[4]m})$}
 \label{t:4Fmlist}
\end{table}
We compare the row-computations in Table~\ref{t:4Fmlist} with the explicit computation a Calabi--Yau hypersurface in any toric 4-fold:
\begin{subequations}
 \label{e:X4[5]cc}
\begin{alignat}9
 c(X^4[c_1])
 &= \Big\lfloor\frac{1{+}c_1{+}c_2{+}c_3{+}c_4}{1{+}c_1}\Big\rfloor_{\leqslant\deg3}
  =1 +c_2 +(c_3{-}c_2c_1);\\
 \c(X^4[c_1])
 &=\int_{X^4}\!\!c_1\cdot c(V^4[c_1])
  =[C_3C_1{-}C_2C_1\!^2]_X,
 \label{e:C3-C21=EU}
\end{alignat}
\end{subequations}
as well as the specific case of $X^4\<=\FF[4]m$. To this end, We note that for all
 $\FF{m}\in\ssK[{r||c}{\IP^r&1\\\IP^1&m}]$, the identity
 $\frac{(1+J_1)^2}{(1+J_1+mJ_2)}=1+J_1-mJ_2$ is implied by $J_2^{~2}\<=0$. Therefore,
\begin{equation}
  \frac{(1+J_1)^{n+1}(1+J_2)^2}{(1+J_1+mJ_2)}=(1+J_1)^{n-1}(1+J_2)^2(1+J_1-mJ_2),
 \label{e:cc4Fm}
\end{equation}
identifying the Mori (charge) vectors in the toric realization of $\FF{m}$\cite[Eq.\,(33)]{rBH-gB}. For $\FF[4]m$ in particular, we identify:
\begin{alignat}9
  c_1 &= 4J_1 +(2{-}m)J_2,&\qquad
  c_2 &= 6J_1^{~2} +(8{-}3m)J_1J_2,\\
  c_3 &= 4J_1^{~3} +(12{-}3m)J_1^{~2}J_2,&\qquad
  c_4 &= J_1^{~4} +(8{-}m)J_1^{~3}J_2,
\end{alignat}
and compute:
\begin{alignat}9
 C_4|_{\smash{\FF[4]m}}
  &=[(J_1{+}mJ_2)(J_1^{~4}{+}(8{-}m)J_1^{~3}J_2)]_{\smash{\FF[4]m}}
 &&=8;\\
 C_3C_1|_{\smash{\FF[4]m}}
  &=[(J_1{+}mJ_2)(16 J_1^{~4}{+}8(7{-}2 m)J_1^{~3}J_2)]_{\smash{\FF[4]m}}
 &&=56;\\
 C_2^{~2}|_{\smash{\FF[4]m}}
  &=[(J_1{+}mJ_2)(36 J_1^{~4}{+}12(8{-}3m)J_1^{~3}J_2)]_{\smash{\FF[4]m}}
 &&=96;\\
 C_2C_1^{~2}|_{\smash{\FF[4]m}}
  &=[(J_1{+}mJ_2)(96 J_1^{~4}{+}32(7{-}3m)J_1^{~3}J_2)]_{\smash{\FF[4]m}}
 &&=224;\\
 C_2^{~2}|_{\smash{\FF[4]m}}
  &=[(J_1{+}mJ_2)(256 J_1^{~4}{+}256(2{-}m)J_1^{~3}J_2)]_{\smash{\FF[4]m}}
 &&=512.
\end{alignat}
These clearly confirm both the 4-dimensional Todd--Hirzebruch identity~\eqref{e:td4},
\begin{equation}
  720 =-C_4+C_3C_1+3C_2\!^2+4C_2C_1\!^2-C_1\!^4,
\end{equation}
as well as the individual computations in Table~\ref{t:4Fmlist}, which however conspicuously omit $C_2^{~2}|_{\smash{\FF[4]m}}$.

Surveying the entries in Table~\ref{t:4Fmlist},
 we seek to represent $c_2^{~2}$ by a pair of 2-cones, $\n_{ij}{\otimes}\n_{kl}$,
 where no four indices $i,j,k,l$ are in the Stanley--Reisner ideal, so that all combinations contribute.
 On $\FF[4]m$, the Stanley--Reisner ideal excludes four or more of the $\{1,2,3,4\}$-valued indices and two or more of the $\{5,6\}$-valued ones, so that the 2-cone product contributing to $C_2^{~2}$ must be of the form
\begin{equation}
   \#[d(\n_{ab})d(\n_{c\ell})+d(\n_{a\ell})d(\n_{bc})]
   =(6{\cdot}1)(8{\cdot}1)+(8{\cdot}1)(6{\cdot}1) = 96,
\end{equation}
where $a,b,c\<\in\{1,2,3,4\}$ and $\ell\<\in\{5,6\}$, as in Table~\ref{t:4Fmlist}. This completes the multifan/multitope computation of Chern characteristic invariants for $\FF[4]m$, and offers the immediate transpolar consequences:
\begin{subequations}
 \label{e:C4Fm=*C*4Fm}
\begin{alignat}9
   C_4|_{\smash{\FF[4]m}}
  &=\ttt\sum_{\n\in\pS(4)} d(\n_{abc\ell})\,d(\n_{abc\ell}^\wtd)
  &&=C_1^{~4}|_{\smash{\MF[4]m}}; \label{e:4DC4}\\
   C_3C_1|_{\smash{\FF[4]m}}
  &=\ttt\sum_{\n\in\pS(3)} d(\n_{ijk})\,d(\n_{ijk}^\wtd)
  &&=C_2C_1^{~2}|_{\smash{\MF[4]m}}; \label{e:4DC3C1}\\
   C_2^{~2}|_{\smash{\FF[4]m}}
  &=\ttt\sum_{\n\in\pS(2)}
         \big(d(\n_{ab})\,d(\n_{c\ell})+d(\n_{a\ell})\,d(\n_{bc})\big)\,d(\n_{abc\ell}^\wtd),
  &&
   \qquad\text{\footnotesize$a,b,c\<\in\{1,2,3,4\},~~\ell\<\in\{5,6\}$}; \label{e:4DC22}\\
   C_2C_1^{~2}|_{\smash{\FF[4]m}}
  &=\ttt\sum_{\n\in\pS(2)} d(\n_{ij})\,d(\n_{ij}^\wtd)
  &&=C_3C_1|_{\smash{\MF[4]m}}; \label{e:4DC2C12}\\
   C_1^{~4}|_{\smash{\FF[4]m}}
  &=\ttt\sum_{\n\in\pS(1)} d(\n_i)\,d(\n_i^\wtd)
  &&=C_4|_{\smash{\MF[4]m}}. \label{e:4DC14}
\end{alignat}
\end{subequations}
The index selections are specified in more detail in Table~\ref{t:4Fmlist}.
 While we defer the analogous (and hopefully computer-aided) computations for other VEX multitopes of interest for another time, the transpolar swap, $\FF{m}\fif{\sss\wtd}\MF{m}$, in Table~\ref{t:4Fmlist} easily implies that
\begin{equation}
  C_4|_{\smash{\MF[4]m}}\<=512,\quad
 C_3C_1|_{\smash{\MF[4]m}}\<=224,\quad
 C_2^{~2}|_{\smash{\MF[4]m}}\<=96,\quad
 C_2C_1^{~2}|_{\smash{\MF[4]m}}\<=56,\quad
 C_1^{~4}|_{\smash{\MF[4]m}}\<=8.
 \label{e:C*4Fm}
\end{equation}

\paragraph{Detailed Chern Class Relations:}
Since $\FF[4]m\<\in\ssK[{r||c}{\IP^3&1\\ \IP^1&m}]$, we may use the explicit, coordinate-level identification with the toric rendition\cite{Berglund:2022dgb} to compare the Chern class computations:
\begin{alignat}9
 \int_{X_m}c(X_m)
 &= \bigg[(J_1{+}mJ_2)(4J_1{+}(2{-}m) J_2)\,
     \bigg(\!\frac{(1{+}J_1)^3(1{+}J_2)^2(1{+}J_1{-}m J_2)}
                  {(1{+}4J_1{+}(2{-}m)J_2)}\!\bigg)\bigg]_{J_1\!^4J_2},\nonumber\\
 &= \bigg[\underbrace{(J_1{+}mJ_2)}_{p(x,y)=0}
          \underbrace{(1{+}J_1{-}m J_2)}_{x_1\to\Fs(x,y)}
          \underbrace{(1{+}J_1)^3}_{c(X_{2,3,4})}
          \underbrace{(1{+}J_2)^2}_{c(X_{5,6})}
     \sum_{j=0}^3(-1)^j\big(4J_1{+}(2{-}m) J_2\big)^{j+1}
    \bigg]_{J_1\!^4J_2}, \nonumber\\
 &= \bigg[\underbrace{[(J_1{+}mJ_2)+(J_1^{~2})]}_{c(p^*\Fs)-1}\,
          \underbrace{(1{+}J_1)^3}_{c(X_{2,3,4})}\,
          \underbrace{(1{+}J_2)^2}_{c(X_{5,6})}\,
     \sum_{k=-1}^4(-1)^{k-1}\!
      \big(\underbrace{4J_1{+}(2{-}m) J_2}_{c_1(\sN)}\big)^k
    \bigg]_{J_1\!^4J_2}, \nonumber\\
 &=:\sum_{k=-1}^4(-1)^{k-1}\hat{c}_k\qquad\To\qquad
 \hat{c}_k=(8,8,56,224,512,512).
 \label{e:chatj}
\end{alignat}
This identifies, in order, the pullback of the embedding
 $\FF[4]m\lhook\joinrel\too{~p=0~}\IP^4\<\times\IP^1$, followed the Chern classes:
of the deg-$\pM{~~1\\-m}$ directrix,\footnote{This is the unique such holomorphic section on
 $\FF[4]m\<\subset\IP^4{\times}\IP^1$; it's irreducible zero-locus has self-intersection $[\Fs^{-1}(0)]^4\<={-}3m$.}
 $\Fs(x,y)$, that replaces
 $x_1\<\in\G\cO\pM{1\\0}$ on $\FF[4]m\<=(p^{-1}(0){\subset}\IP^4{\times}\IP^1)$, 
of the $x_{2,3,4}$-coordinate (toric $X'\<\coeq X_{2,3,4}$) line bundles
 $\cO\pM{1\\0}^{\oplus3}$, and 
of the $y_{0,1}$-coordinate (toric $X''\<\coeq X_{5,6}$) line bundles $\cO\pM{0\\1}^{\oplus2}$.
 We then combined the first two factors, and also extend the summation to $k\to{-}1$,
 defining $c_1^{~0}(\sN)\<\coeq1$ and assign $c_1^{-1}(\sN)\<\coeq(4J_1{+}(2{-}m) J_2)^{-1}$ the formal linear combination $(4J_1^{-1}{+}(2{-}m) J_2^{-1})$.

Not only do the numerical (and $m$-independent) values of the so-defined contributions $\hat{c}_1$ and $\hat{c}_2$ equal the total sums~\cite[Eqs.\,(69) and~(70)]{rBH-gB}, respectively, but they match {\em\/term-by-term\/} the contributions~\cite{rBH-gB} in Batyrev's formula for the Euler characteristic~\cite[Thm.\;4.5.3]{rBaty01} (denoting $X'{=}X_{2,3,4}$ and $X''{=}X_{5,6}$):
\begin{alignat}9
 \hat{c}_{-1}
    &=c_2(p^*\Fs)\,c_3(X')\,c_1(X'')\,c_1^{-1}(\sN)
     =(J_1^{~2})(J_1^{~2})(2J_2)\big(4J_1^{-1}{+}(2{-}m)J_2^{-1}\big)
     =1{\cdot}1{\cdot}2{\cdot}4 =8.\\
 \hat{c}_0
    &=\big[c_1(p^*\Fs)\,c_3(X')\,c_1(X'')
          +c_2(p^*\Fs)\,c_2(X')\,c_1(X'')\big](c_1^{~0}(\sN)\<=1)
     =[1{\cdot}1{\cdot}2 + 1{\cdot}3{\cdot}2]{\cdot}1 =8.\\
 \hat{c}_1
    &=\big[c_1(p^*\Fs)\,c_3(X')\,c_0(X'')
          {+}c_1(p^*\Fs)\,c_2(X')\,c_1(X'')
          {+}c_2(p^*\Fs)\,c_1(X')\,c_1(X'')
          {+}c_2(p^*\Fs)\,c_2(X')\,c_0(X'')\big]c_1(\sN),\nonumber\\
    &=\underbrace{[1{\cdot}1{\cdot}1{\cdot}(2{-}m){+}m{\cdot}1{\cdot}1{\cdot}4]}_{2\C2{+3m}}
     +\underbrace{[1{\cdot}3{\cdot}2{\cdot}4]
                 +[1{\cdot}3{\cdot}2{\cdot}4]}_{2{\times}6\<\cdot4}
     +\underbrace{[1{\cdot}3{\cdot}1{\cdot}(2{-}m)]}_{3(2\C1{-m})} =56.\\
 \hat{c}_2
    &=\big[c_1(p^*\Fs)\,c_2(X')\,c_0(X'')
          {+}c_1(p^*\Fs)\,c_1(X')\,c_1(X'')
          {+}c_2(p^*\Fs)\,c_0(X')\,c_1(X'')
          {+}c_2(p^*\Fs)\,c_1(X')\,c_0(X'')\big]c_1^{~2}(\sN),\nonumber\\
    &=\underbrace{[1{\cdot}3{\cdot}1{\cdot}8(2{-}m){+}m{\cdot}3{\cdot}1{\cdot}16]}
                 _{3{\cdot}8(2\C2{+m})}
     +\underbrace{[1{\cdot}3{\cdot}2{\cdot}16]
                 +[1{\cdot}1{\cdot}2{\cdot}16]}_{4{\cdot}2\<\cdot16}
     +\underbrace{[1{\cdot}3{\cdot}1{\cdot}8(2{-}m)]}_{3{\cdot}8(2\C1{-m})} =224.\\
 \hat{c}_3
    &=\big[c_1(p^*\Fs)\,c_1(X')\,c_0(X'')
          {+}c_1(p^*\Fs)\,c_0(X')\,c_1(X'')
          {+}c_2(p^*\Fs)\,c_0(X')\,c_0(X'')\big]c_1^{~3}(\sN),\nonumber\\
    &=\underbrace{[1{\cdot}3{\cdot}1{\cdot}48(2{-}m){+}m{\cdot}3{\cdot}1{\cdot}64]}
                 _{3{\cdot}16(6\C2{+m})}
     +\underbrace{[1{\cdot}1{\cdot}2{\cdot}64]}_{2\<\cdot64}
     +\underbrace{[1{\cdot}1{\cdot}1{\cdot}48(2{-}m)]}_{48(2\C1{-m})} =512.\\
 \hat{c}_4
    &=c_1(p^*\Fs)\,c_0(X')\,c_0(X'')\,c_1^{~4}(\sN)
     =[1{\cdot}1{\cdot}1{\cdot}256(2{-}m){+}m{\cdot}1{\cdot}1{\cdot}256] =512.
 \label{e:c4=512}
\end{alignat}
These match {\em\/each and every\/} entry in Table~\ref{t:4Fmlist} (compare under-braced terms), clearly identifying the contributions that turn negative for $m\<>2$ and providing them with a more geometric interpretation.

The same term-by-term identifications hold also in lower dimensions, and we expect this to be true generally. Following\cite{Berglund:2022dgb}, a more general ``family picture'' emerges: identifying each $\FF{m}\<\in\ssK[{r||c}{\IP^r&1\\\IP^1&m}]$ with an $m$-twisted $\IP^r$-bundle over $\IP^1$, explicit deformations within the family distribute the twisting, $m\<\to\vec{m}\<=(m_1,m_2,\cdots)$, with $\sum_{i=1}^rm_i\<=m$, each $\vec{m}$ specifying a distinct toric variety. By explicit deformation, all such Hirzebruch scrolls explicitly form a diffeomorphism class, as do their (Laurent-smoothed for $m\<>2$\cite{rBH-gB,Berglund:2022dgb}) Calabi--Yau hypersurfaces. Since the total number of so-constructed diffeomorphism classes is limited by Wall's Theorem\cite{rWall} to $[m\!\pmod n]$, the number of distinctly different (weakly/pre-)complex Calabi--Yau manifolds (dubbed {\em\/discrete deformations\/}\cite{rBH-Fm} and {\em\/non-algebraic deformation equivalence\/}\cite{Gross:1994The, Ruan:1996top}) within each diffeomorphism class grows unbounded with increasing $m$.

\paragraph{A Novel Euler Characteristic Formula:}
For the anticanonical hypersurface, $Z_f\<=\{f(x)\<=0\}\<\subset X$,
\begin{equation}
  C_3(Z_f)\<\coeq\int_{Z_f}\!\!c_3(Z_f) =\int_Xc_1\,(c_3{-}c_1c_2)=\big[C_3C_1{-}C_1\!^2C_2\big]_X.
\end{equation}
The key standard results~\ref{i:C4} and~\ref{i:C5},
  $C_4|_X\<=d(\pDs{X})$ and $C_1\!^4|_X=d(\pDN{X})$
yields for $X$:
\begin{alignat}9
  \Td(X)
   &=\frc1{720}\big[{-}C_4\<+C_3C_1\<+3C_2\!^2\<+4C_2C_1\!^2\<-C_1\!^4\big]_X,\\
 \text{i.e.,}\qquad
  720\Td(X)
   &={-}d(\pDs{X})\<+C_3(Z_f)\<+3C_2\!^2|_X\<+5C_2C_1\!^2|_X\<-d(\pDN{X}).
 \label{e:Td4X}
\iText[-1mm]{Swapping $\pDs{\tX}\<=\pDN{X}$ and $\pDN{\tX}\<=\pDs{X}$ produces the analogous result for the mirror $Z_{\tT f}\<\subset\tX$ (using that $C_3(Z_{\tT f})=-C_3(Z_f)$):}
  720\Td(\tX)
   &={-}d(\pDN{X})\<-C_3(Z_f)\<+3C_2\!^2|_{\tX}\<+5C_2C_1\!^2|_{\tX}\<-d(\pDs{X}).
 \label{e:Td4Y}
\iText[-2mm]{Since $\Td(X)\<=w(\pDs{X})\<=w(\pDN{X})\<=\Td(\tX)$, subtracting~\eqref{e:Td4Y} from~\eqref{e:Td4X} yields:}
 0&=2C_3(Z_f) +3\big(C_2\!^2|_X\<-C_2\!^2|_{\tX}\big)
          +5\big(C_2C_1\!^2|_X\<-C_2C_1\!^2|_{\tX}\big).
 \label{e:tmp1}
\end{alignat}
Comparing now the assignments~\eqref{e:4DC3C1} and~\eqref{e:4DC2C12}, we observe the identity (previously derived in\cite{Berglund:2021ztg} using the so-called Libgober--Wood formula\cite{Libgober:1990aa}):
\begin{equation}
   C_3C_1|_X = \ttt\sum_{\n_{ijk}\in\pFn{X}(3)} d(\n_{ijk})\,d(\n_{ijk}^\wtd)
   \overset\wtd=
   \ttt\sum_{\m_{IJ}\in\pFn{\tW \!X}(2)} d(\m_{\sss IJ}^\wtd)\,d(\m_{\sss IJ})
   = C_2C_1\!^2|_{\tX},
 \tag{\ref{e:C3C1=C2C12*}$'$}
\end{equation}
where the middle equality is implied by the transpolar relation.
 This identifies 1--1 not only the total sum of the contributions, but the individual factors in each contribution:
 $\n_{ijk}^\wtd\<=\m_{\sss IJ}\subset\pFn{\tX}{=}\pFN{X}$ and
 $\m_{\sss IJ}^\wtd\<=\n_{ijk}\subset\pFn{X}{=}\pFN{\tX}$
are the cones over individual transpolar pairs of faces.
 Using the identity~\eqref{e:C3C1=C2C12*} to substitute
 $C_2C_1\!^2|_{\tX}\to C_3C_1|_X$ in~\eqref{e:tmp1} yields:
\begin{gather}
 0=2C_3(Z_f) +3\big(C_2\!^2|_X\<-C_2\!^2|_{\tX}\big)
  +5\Big(\big[C_2C_1\!^2\<-C_3C_1\big]_X = -\c(Z_f)\Big),\\
 \text{and so}\qquad
 C_3(Z_f)=C_2\!^2|_X\<-C_2\!^2|_{\tX}=-C_3(Z_{\tT f}).
 \tag{\ref{e:C2-C2}$'$}
\end{gather}
We used that $\tX$ is a unitary torus manifold,
corresponding to the multifan spanned by (the MPCP-subdivided) $\pDN{X}$, wherein anticanonical hypersurfaces in the deformation family $Z_{\tT f}\<\subset \tX[c_1]$ (Conjecture~\ref{C:gCYh}) are transposition mirrors of the anticanonical hypersurfaces $Z_f\<\subset X[c_1]$; see Conjecture~\ref{C:gMM}.

\paragraph{HRR for $\cK^*$:}
The HRR theorem for the anticanonical bundle,\footnote{Standard for (semi) Fano $X$ encoded by convex polytopes\cite{rD-TV, rO-TV, rF-TV, rGE-CCAG, rCLS-TV}, the final equality extends to unitary torus manifolds with simplicial multifans\cite{rM-MFans, Masuda:2000aa, rHM-MFs}, which includes all VEX multifans, as they have a simplicial subdivision by Definition~\ref{D:VEX}.}
\begin{equation}
 \c(X,\cK^*)=\int_X\mathop{\mathrm{td}}(X)\mathop{\mathrm{ch}}(\cK^*)
  =\sum_{k=0}^r(-1)^k H^k(X,\cK^*)= \ell(\pDN{X})
 \label{e:RRK*}
\end{equation}
interprets the (negative-contributing) flip-folded extension(s) in $\pDN{X}$ as counting $H^1(X,\cK^*)$, and implies
\begin{alignat}9
 \c(X,\cK^*)
 &=\int_X\Big(1{+}\frac{c_1}2{+}\frac{c_2{+}c_1\!^2}{12}{+}\frac{c_1c_2}{24}
              {+}\frac{{-}c_4{+}c_3c_1{+}3c_2\!^2{+}4c_2c_1\!^2{-}c_1\!^4}{720}\Big)
   \Big(1{+}c_1{+}\frac{c_1\!^2}{2!}{+}\frac{c_1\!^3}{3!}{+}\frac{c_1\!^4}{4!}\Big)\Big\rfloor^4\nn\\
 &=\int_X\Big(\frac{c_1\!^4}{4!} \<+\frac{c_1}2\,\frac{c_1\!^3}{3!}
              \<+\frac{c_2\<+c_1\!^2}{12}\,\frac{c_1\!^2}{2!} \<+\frac{c_1c_2}{24}c_1
              \<+\frac{{-}c_4\<+c_3c_1\<+3c_2\!^2\<+4c_2c_1\!^2\<-c_1\!^4}{720}\Big),\\
 \ell(\pDN{X})
 &=\frc1{720}\big(119\,C_1\!^4|_X +64\,C_1\!^2C_2|_X +3\,C_2\!^2|_X +C_1C_3|_X -C_4|_X\big).
 \label{e:tmp2}
\iText[-1mm]{The analogous result for the transpolar uses that $\pDN{\tX}\<=\pDs{X}$:}
 \ell(\pDs{X})
 &=\frc1{720}\big(119\,C_1\!^4|_{\tX} +64\,C_1\!^2C_2|_{\tX}
     +3\,C_2\!^2|_{\tX} +C_1C_3|_{\tX} -C_4|_{\tX}\big). \label{e:tmp2*}
\iText[-1mm]{The relation~\eqref{e:C3C1=C2C12*} allows replacing
 $C_1\!^2C_2|_{\tX}\<\to C_1C_3|_X$~ and
 $C_1C_3|_{\tX}\<\to C_1\!^2C_2|_X$,
 and~\eqref{e:CnC1n} allows replacing
 $C_1\!^4|_{\tX}\<\to C_4|_X\<=d(\pDs{X})$~ and~
 $C_4|_{\tX}\<\to C_1\!^4|_X\<=d(\pDN{X})$:}
 \ell(\pDs{X})
 &=\frc1{720}\big(119\,C_4|_X +64\,C_1C_3|_X +3\,C_2\!^2|_{\tX}
     +C_1\!^2C_2|_X -C_1\!^4|_X\big).
\iText[-1mm]{Subtracting from~\eqref{e:tmp2} yields:}
 \ell(\pDN{X}) {-} \ell(\pDs{X})
 &=\frc1{720}\big(120\,C_1\!^4|_X \<+64[C_1\!^2C_2{-}C_1C_3]_X
                   \<+3(C_2\!^2|_X{-}C_2\!^2|_{\tX}) \<+[C_1C_3{-}C_1\!^2C_2]_X
                    \<-120\,C_4|_X\big),\nn\\
 &=\frc1{720}\big(120\,d(\pDN{X}) \<-64\,C_3(Z_f) \<+3\,C_3(Z_f) 
                  \<+C_3(Z_f) \<-120\,d(\pDs{X})\big),\\
 &=\frc16\big(d(\pDN{X}) \<-d(\pDs{X})\big) \<-\frc1{12}\,C_3(Z_f),
\end{alignat}\vglue-3mm\noindent
so that finally:
\begin{equation}
  C_3(Z_f) = 2\big(d(\pDN{X}) \<-d(\pDs{X})\big)
             -12\big(\ell(\pDN{X}) - \ell(\pDs{X})\big).
 \tag{\ref{e:dd*ll*}$'$}
\end{equation}
This relation was derived (slightly differently) and was already used in a machine-learning exploration\cite{Berglund:2021ztg} of the Kreutzer--Skarke database of (convex) reflexive polytopes and associate (complex-algebraic) toric varieties\cite{Kreuzer:2000xy}. The above derivations would imply~\eqref{e:C2-C2} and~\eqref{e:dd*ll*} to hold also for the unitary toric manifolds corresponding to VEX polytopes and anticanonical hypersurfaces therein (Conjecture~\ref{C:gCYh}).

By considering also the sums, \eqref{e:Td4X}+\eqref{e:Td4Y} and~\eqref{e:tmp2}+\eqref{e:tmp2*}, we obtain
\begin{subequations}
 \label{e:otherCiii}
\begin{alignat}9
  C_2\!^2|_X &=
  3 d(\pDN{X}) +d(\pDs{X}) -16 \ell(\pDN{X}) -4\ell(\pDs{X}) +260\Td(X),\\
  C_2\!^2|_\tX &=
  3 d(\pDs{X}) +d(\pDN{X}) -16 \ell(\pDs{X}) -4\ell(\pDN{X}) +260\Td(X),
\iText{in addition to~\eqref{e:C2-C2}. Similar manipulations can extract analogous, multitope-counting expressions for other Chern characteristics; for example,}
  C_3C_1|_X &=
  2\big( 6\ell(\pDs{X}) -d(\pDs{X}) -6 \Td(X) \big),\\
  C_3C_1|_\tX &=
  2\big( 6\ell(\pDN{X}) -d(\pDN{X}) -6 \Td(X) \big).
\end{alignat}
\end{subequations}

\section{Intersection Numbers}
\label{s:iN}
Adapting from the standard literature (esp.~\cite[Thm.\;10.4.4 and Lemma~12.5.2]{rCLS-TV}) and the algorithm given explicitly in\cite{rBKK-tvMirr}, we use:
\begin{proc}[Intersection numbers, Mori cone]\label{P:InMc}
\addtolength{\baselineskip}{-2pt}
In a simplicialy subdivided multifan,\footnote{All non-simplicial cones in
$\pFn{X}$ are subdivided into simplicial (sub-)cones, $\s_\a$; a subdivision is {\em\/smooth\/} if $|d(\s_\a)|\<=1$ for all top-dimensional cones, and {\em\/coarse\/} otherwise; ``$\s\uplus\s'$'' is the disjoint (indexed) union of $\s$ and $\s'$ that retains $\t\<=\s\<\cap\s'$ in the facet ($\pRec$)-generated poset structure of the multifan\cite{rHM-MFs}.}
$\pFn{X}\<=\{\uplus_\a\s_\a,\pRec\}$,
let $\n_i\<\smt\pFn{X}$ denote the 1-cone generators.
\begin{enumerate}[itemsep=-1pt, topsep=-1pt]
  \item For each codimension-1 simplicial cone, $\t\<\in\pFn{X}(n{-}1)$, its two adjacent
   top-dimensional simplicial cones, $\s\<\cap\s'\<=\t$, and generators,
   $\n_i\<\smt\,\s,\s'$, find $\ell_\t^i\in\IQ$ such that
\begin{subequations}
 \label{e:ells}
\begin{gather}
  \sum_{\n_i\smt\,\s,\s'}\ell^i_\t\,\n_i=0,\qquad\text{and}\qquad
  \sum_{\n_i\smt\,\s,\s'}\ell^i_\t = \frac{d(\s|\s')}{d(\s)\,d(\s')}
  =d(\t^\wtd),
 \label{e:ells1}
\iText[-3mm]{where}
  d(\s)=\det[\n_i{\smt}\s],\quad
  d(\s')=\det[\n_i{\smt}\s'],\quad
  d(\s|\s')=\det[(\n_i{\smt}\s)\<-\n_*]_
        {\shortstack[l]{$\sss \n_i\neq\n_*$\\[-4pt]
                        $\sss\n_*\in(\s\<\uplus\s')$}}.
 \label{e:ells2}
\end{gather}
The degree, $d(\s)$, of a simplicial top-dimensional cone is the volume of its central (primitive) simplex, the determinant of the row-stacked matrix of generating vectors, $\n_i\<\smt\s$. The (relative) degree $d(\s|\s')$ is the volume of the simplex hull of the adjoined bases of $\s,\s'$, and equals the degree of their intersection's transpolar:
\begin{equation}
 \vC{\TikZ{\path[use as bounding box](0,.2)--(5,3.2);
             \fill[teal!30, opacity=.75](2.2,.5)--(0,2)--(2,3);
             \draw[blue, thick, -stealth](2.2,.5)--(0,2);
              \path(0,2)node[left]{\footnotesize$\n_k$};
             \draw[blue, thick, -stealth](2.2,.5)--(2,3);
              \path(2,3)node[right]{\footnotesize$\n_{i_{n-1}}$};
             \draw[blue, thick, -stealth](2.2,.5)--(1.5,2.2);
              \path(1.35,2.2)node[below]{\footnotesize$\n_{i_1}$};
             \draw[thick, blue, densely dotted, -stealth](2.2,.5)--(4,1);
             \filldraw[very thick, fill=yellow, opacity=.85,
                       line join=round](2,3)--(0,2)--(1.5,2.2);
             \filldraw[very thick, fill=yellow, opacity=.85,
                       line join=round](2,3)--(4,1)--(1.5,2.2);
             \draw[line width=2pt, purple](2,3)--(1.5,2.2);
              \path(4,1)node[right]{\footnotesize$\n_{k'}$};
              \path(.8,2.7)node{$\s$};
              \path(1.3,1.5)node{$d(\s)$};
              \path[purple](1.9,2.45)node{$\t$};
              \path(3.1,2.4)node{$\s'$};
             \fill(0,2)circle(.5mm);
             \fill(2,3)circle(.7mm);
             \fill(1.5,2.2)circle(.7mm);
             \fill(4,1)circle(.5mm);
             \filldraw[thick, fill=white](2.2,.5)circle(.5mm);
              \path(2.2,.45)node[left]{\footnotesize$0$};
            }}
 \vC{\TikZ{\path[use as bounding box](0,.2)--(5,3.2);
             \draw[blue, thick, -stealth](2.2,.5)--(2,3);
              \path(2,3)node[right]{\footnotesize$\n_{i_{n-1}}$};
             \fill[teal!30, opacity=.75](2.2,.5)--(1.5,2.2)--(4,1);
             \draw[thick, blue, densely dotted, -stealth](2.2,.5)--(0,2);
              \path(0,2)node[left]{\footnotesize$\n_k$};
             \draw[blue, thick, -stealth](2.2,.5)--(1.5,2.2);
              \path(1.35,2.2)node[below]{\footnotesize$\n_{i_1}$};
             \draw[blue, thick, -stealth](2.2,.5)--(4,1);
             \filldraw[very thick, fill=yellow, opacity=.85,
                       line join=round](2,3)--(0,2)--(1.5,2.2);
             \filldraw[very thick, fill=yellow, opacity=.85,
                       line join=round](2,3)--(4,1)--(1.5,2.2);
             \draw[line width=2pt, purple](2,3)--(1.5,2.2);
              \path(4,1)node[right]{\footnotesize$\n_{k'}$};
              \path(.8,2.7)node{$\s$};
              \path[purple](1.9,2.45)node{$\t$};
              \path(3.1,2.4)node{$\s'$};
              \path(2.7,1.1)node{$d(\s')$};
             \fill(0,2)circle(.5mm);
             \fill(2,3)circle(.7mm);
             \fill(1.5,2.2)circle(.7mm);
             \fill(4,1)circle(.5mm);
             \filldraw[thick, fill=white](2.2,.5)circle(.5mm);
              \path(2.2,.45)node[left]{\footnotesize$0$};
            }}
 \vC{\TikZ{\path[use as bounding box](0,.2)--(4,3.2);
             \draw[thick, blue, densely dotted, -stealth](2.2,.5)--(2,3);
             \fill[teal!30, opacity=.75](0,2)--(1.5,2.2)--(4,1);
              \path(0,2)node[left]{\footnotesize$\n_k$};
              \path(2,3)node[right]{\footnotesize$\n_{i_{n-1}}$};
              \path(1.35,2.2)node[below]{\footnotesize$\n_{i_1}$};
             \draw[thick, blue, densely dotted, -stealth](2.2,.5)--(0,2);
             \draw[thick, blue, densely dotted, -stealth]
                 (2.2,.5)--(1.5,2.2);
             \draw[thick, blue, densely dotted, -stealth](2.2,.5)--(4,1);
             \draw[blue, thick](0,2)--(4,1);
             \filldraw[very thick, fill=yellow, opacity=.85,
                       line join=round](2,3)--(0,2)--(1.5,2.2);
             \filldraw[very thick, fill=yellow, opacity=.85,
                       line join=round](2,3)--(4,1)--(1.5,2.2);
             \draw[line width=2pt, purple](2,3)--(1.5,2.2);
              \path(4,1)node[right]{\footnotesize$\n_{k'}$};
              \path(.8,2.7)node{$\s$};
              \path[purple](1.9,2.45)node{$\t$};
              \path(3.1,2.4)node{$\s'$};
              \path(1.8,1.3)node[rotate=-13]{$d(\s|\s')$};
             \fill(0,2)circle(.5mm);
             \fill(2,3)circle(.7mm);
             \fill(1.5,2.2)circle(.7mm);
             \fill(4,1)circle(.5mm);
             \filldraw[thick, fill=white](2.2,.5)circle(.5mm);
              \path(2.2,.45)node[left]{\footnotesize$0$};
            }}
 \label{e:ss'}
\end{equation}
\end{subequations}
This $d(\s|\s')$ may be computed as the determinant of the row-stacked matrix of the generator vector differences $(\n_i{-}\n_*)_{i\neq*}$ and is independent of the choice of
 $\n_*\<\smt(\s\uplus\s')$. The row-stacked matrices are ordered compatibly with a continuous orientation of the (yellow) facets, $\s,\s'$, in~\eqref{e:ss'}.

  \item Solutions to~\eqref{e:ells} provide the \`a\:priori nonzero (divisor class) intersection numbers:
\begin{equation}
  \k_{i_1,\cdots,\,i_{n-1},\,k} \coeq\rule[-2mm]{0mm}{1mm}
  \smash{\underbrace{D_{i_1}\<\cdots D_{i_{n-1}}}_{\t}}
  \cdot D_k = \ell_\t^k,\quad
  \n_{i_1},{\cdots},\n_{i_{n-1}}\smt \t,\quad \n_k\smt (\s\<\cup\s'\ssm\t),
 \label{e:0ks}
\end{equation}
and $\ell_\t^j\<=0$ for $\n_j\not\smt(\s\<\cup\s'\ssm\t)$.

  \item The so-obtained array of intersection numbers, $\k_{i_1,\cdots,i_n}$, is totally symmetric.

  \item The $n$-vectors computed in~\eqref{e:ells} in terms of which all others are non-negative linear combinations are the Mori vectors.
\end{enumerate}\backUp
\end{proc}
\begin{remk}[normalization]
\label{r:nDij}
The results~\eqref{e:ells} are {\em\/local\/} to $\s,\s'\in\pFn{X}$, i.e., independent of $\pFn{X}\ssm(\s\<\uplus\s')$. The whole array of intersection numbers, $\k_{i_1,\cdots i_n}$, in~\eqref{e:0ks} is however {\em\/global\/} over $\pFn{X}$, and its total symmetry under index-permutations constrains the results.
 The relative degree $d(\s|\s')$ in~\eqref{e:ss'} measures the dual of the ``inner'' dihedral angle between $\s$ and $\s'$ at $\t\<=\s\<\cap\s'$.
It vanishes when $\s$ and $\s'$ are co-planar, so the second equation~\eqref{e:ells1} fails to normalize the $\ell^k_\t$.
The overabundant conditions of total symmetry in $\k_{i_1,\cdots,i_n}$ however sufficed to normalize the $\ell^k_\t$ in all examples we checked. Most importantly, this determines the sign of $\ell^k_\t$ corresponding to $w(\s)$ and $w(\s')$; see examples below.
\end{remk}
\ping

Owing to the key standard results~\ref{i:C1}--\ref{i:C2}, the intersection numbers $\k_{i_1,\cdots,i_n}$ are related to Chern characteristics. They also provide essential information (certain Yukawa couplings) for superstring applications, and we now verify their computability and consistency also for VEX multitopes.

\subsection{Two Dimensions}
\begin{figure}[htb]
$$
 \vC{\TikZ{\path[use as bounding box](-1.4,-3.1)--(1.2,5.4);
      \path(.24,1)node
       {\includegraphics[width=27mm]{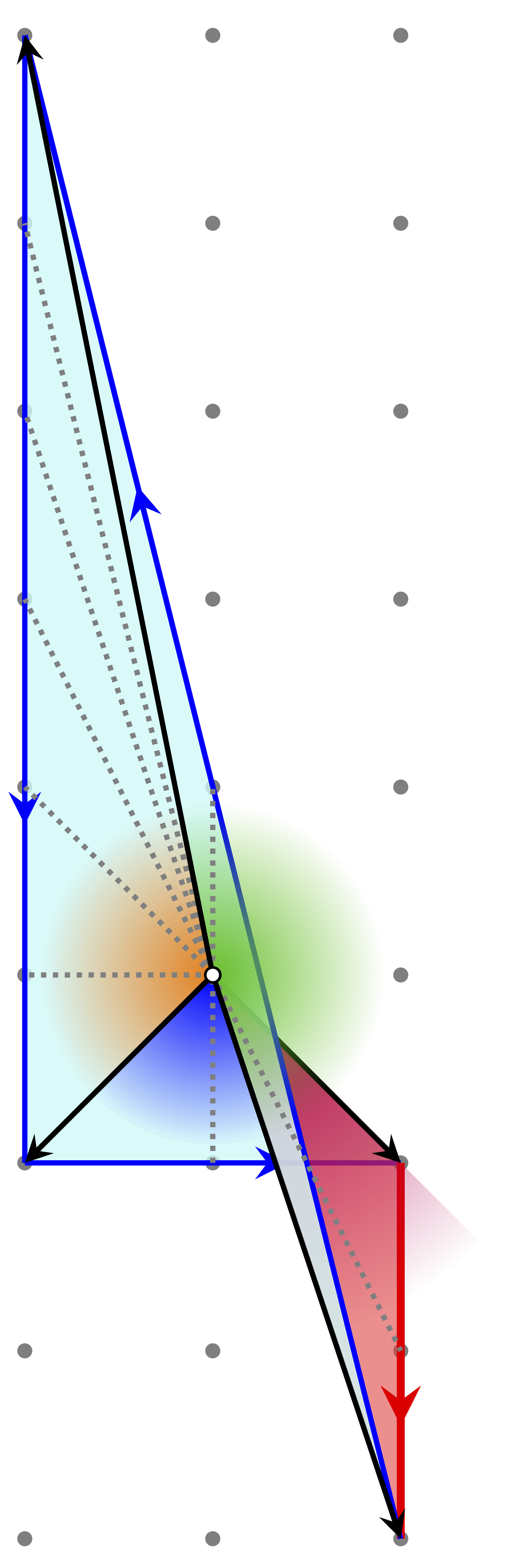}};
       \path(-1,5)node[left=0mm]{\scriptsize$\n_{41}^\wtd$};
        \path(-1,5)node[right=0mm]{\scriptsize$\tD_1$};
        \path[gray](-1,4)node[left=0mm]{\scriptsize$\tD_5$};
        \path[gray](-1,3)node[left=0mm]{\scriptsize$\tD_6$};
        \path[gray](-1,2)node[left=0mm]{\scriptsize$\tD_7$};
        \path[gray](-1,1)node[left=0mm]{\scriptsize$\tD_8$};
        \path[gray](-1,0)node[left=0mm]{\scriptsize$\tD_9$};
       \path(-1,-1)node[left=0mm]{\scriptsize$\n_{12}^\wtd$};
        \path(-1,-1)node[below=0mm]{\scriptsize$\tD_2$};
        \path[gray](0,-1)node[below=0mm]{\scriptsize$\tD_{10}$};
       \path(1,-1.1)node[right=0mm]{\scriptsize$\n_{2\C13}^\wtd$};
        \path(1.05,-.95)node[above=0mm]{\scriptsize$\C1{\tD_3}$};
        \path[gray][red](1,-2)node[right=0mm]{\scriptsize$\tD_{11}$};
       \path(1,-3)node[right=0mm]{\scriptsize$\n_{\C134}^\wtd$};
        \path(1,-3)node[left=0mm]{\scriptsize$\C1{\tD_4}$};
        \path[gray](.3,1.2)node{\scriptsize$\tD_{12}$};
       \path(.5,4.5)node{\large$\pDN{\FF[2]m}\lat\pFN{\FF[2]m}$};
      \filldraw[fill=white](0,0)circle(.4mm);
       \path(.6,1.9)node[rotate=-atan(4)]{$m\<=4$ case shown};
            }}
 \qquad
 \vC{\begin{picture}(130,85)(-5,-2)
  \put(-5,76){$
     \begin{array}[t]{r|c@{~}c@{~}c@{~}c}
      \pFN{\FF[2]m}\!\!
             & \n_{41}^\wtd & \n_{12}^\wtd & \n_{23}^\wtd & \n_{34}^\wtd \\
              \toprule\nGlu{-3pt}
             &-1    &-1 &\3-1 & 1    \\[-2pt]
             &1{+}m &-1 &  -1 &1{-}m \\[-2pt] \midrule\nGlu{-2pt}
      \tw{Q}^1 & 0 &m{-}2& m &-2 \\
      \tw{Q}^2 &m{-}2& 0 &-2 & m \\[-2pt] \bottomrule\nGlu{3pt}
      \MC5r{\text{\footnotesize
             ${-}\tD_1 {-}\tD_2 {+}\tD_3 {+}\tD_4 \<\sim0$}}\\
      \MC5r{\text{\footnotesize
             $(1{+}m)\tD_1 {-}\tD_2 {-}\tD_3 {+}(1{-}m)\tD_4 \<\sim0$}}\\
      \MC5r{\text{\footnotesize S.-R.: $\tD_1{\cdot}\tD_3\<\sim0$ and
                                       $\tD_2{\cdot}\tD_4\sim0$}}
     \end{array}$}
  \put(55,18){\TikZ{[every node/.style={thick, inner sep=.5, outer sep=0}]
      \path[use as bounding box](-1,-3)--(1,3);
      \node[rounded corners=5pt, inner sep=2, draw=black, fill=white](1) 
            at(-1,3)   {\scriptsize$\frac{m}{2(m+2)}$};
      \node[rounded corners=5pt, inner sep=2, draw=black, fill=white](2) 
            at(-1,-3)  {\scriptsize$\frac{m}{2(m+2)}$};
      \node[Purple, rounded corners=5pt, inner sep=2, draw=Purple, 
            fill=white](3)
            at(1,-1)  {\scriptsize$\frac{m}{2(m-2)}$};
      \node[Purple, rounded corners=5pt, inner sep=2, draw=Purple, 
            fill=white](4) 
            at(1,-3) {\scriptsize$\frac{m}{2(m-2)}$};
      \foreach\n in{1,...,4} 
       \draw[gray!50, thick, -stealth](0,-1)--(\n);
      \draw[blue, thick, midarrow=stealth](1)--node[below right=1pt]
            {\scriptsize$\frac1{m+2}$}(2);
      \draw[blue, thick, midarrow=stealth](2)--node[above left=-1pt]
            {\scriptsize$\fRc12$}(3);
      \draw[red, thick, midarrow=stealth](3)--node[right=2pt]
            {\scriptsize$-\frac1{m-2}$}(4);
      \draw[blue, thick, midarrow=stealth](4)--node[above left=3pt]
            {\scriptsize$\fRc12$}(1);
      \path(-.25,2.9)node[right]{$[\hD_i{\cdot}\hD_j]_{\MF[2,\sharp]m}$};
      \path(-.6,2.3)node[right]{\footnotesize shown upon};
      \path(-.45,1.9)node[right]{\footnotesize 
                                 $\bM{x\\y}{\to}\bM{1&0\\1&1}\bM{x\\y}$};
      \path(-.3,1.5)node[right]{\footnotesize $\GL(2;\ZZ)$ transf.};
      \filldraw[fill=white, thick](0,-1)circle(.4mm);
            }}
  \put(92,18){\TikZ{[every node/.style={thick, inner sep=.5, outer sep=0}]
      \path[use as bounding box](-1,-3)--(1,3);
      \node[rounded corners=5pt, inner sep=2, draw=black, fill=white](1) 
            at(-1,3)   {\scriptsize$-1$};
      \node[rounded corners=5pt, inner sep=2, draw=black, fill=white](2) 
            at(-1,-3)  {\scriptsize$-1$};
      \node[Purple, rounded corners=5pt, inner sep=2, draw=Purple, 
            fill=white](3)
            at(1,-1)  {\scriptsize$+1$};
      \node[Purple, rounded corners=5pt, inner sep=2, draw=Purple, 
            fill=white](4) 
            at(1,-3) {\scriptsize$+1$};
      \node[rounded corners=5pt, inner sep=2, draw=black, fill=white](5) 
            at(-1,2)   {\scriptsize$-2$};
      \foreach\n in{1,...,4}  \draw[gray!50, thick, -stealth](0,-1)--(\n);
      \node[rounded corners=5pt, inner sep=2, draw=black, fill=white](6) 
            at(-1,1)   {\scriptsize$-2$};
      \node[rounded corners=5pt, inner sep=2, draw=black, fill=white](7) 
            at(-1,0)   {\scriptsize$-2$};
      \node[rounded corners=5pt, inner sep=2, draw=black, fill=white](8) 
            at(-1,-1)  {\scriptsize$-2$};
      \node[rounded corners=5pt, inner sep=2, draw=black, fill=white](9) 
            at(-1,-2)  {\scriptsize$-2$};
      \node[rounded corners=5pt, inner sep=2, draw=black, fill=white](10) 
            at(0,-2)  {\scriptsize$-2$};
      \node[red, rounded corners=5pt, inner sep=2, draw=red, fill=white]
            (11) 
            at(1,-2)  {\scriptsize$+2$};
      \node[rounded corners=5pt, inner sep=2, draw=black, fill=white](12) 
            at(0,0)  {\scriptsize$-2$};
      \foreach\n in{5,...,12} \draw[gray, densely dotted](0,-1)--(\n);
      \draw[blue, thick, midarrow=stealth](1)--node[below right=1pt]
            {\scriptsize$1$}(5);
      \draw[blue, thick, midarrow=stealth](5)--node[right=2pt]
            {\scriptsize$1$}(6);
      \draw[blue, thick, midarrow=stealth](6)--node[right=2pt]
            {\scriptsize$1$}(7);
      \draw[blue, thick, midarrow=stealth](7)--node[right=2pt]
            {\scriptsize$1$}(8);
      \draw[blue, thick, midarrow=stealth](8)--node[right=2pt]
            {\scriptsize$1$}(9);
      \draw[blue, thick, midarrow=stealth](9)--node[above right=1pt]
            {\scriptsize$1$}(2);
      \draw[blue, thick, midarrow=stealth](2)--node[above=7pt]
            {\scriptsize$1$}(10);
      \draw[blue, thick, midarrow=stealth](10)--node[left=6pt]
            {\scriptsize$1$}(3);
      \draw[red, thick, midarrow=stealth](3)--node[right=2pt]
            {\scriptsize$-1$}(11);
      \draw[red, thick, midarrow=stealth](11)--node[right=2pt]
            {\scriptsize$-1$}(4);
      \draw[blue, thick, midarrow=stealth](4)--node[above left=6pt]
            {\scriptsize$1$}(12);
      \draw[blue, thick, midarrow=stealth](12)--node[below=3pt]
            {\scriptsize$1$}(1);
      \path(-.25,2.9)node[right]{$[\tD_i{\cdot}\tD_j]_{\MF[2]m}$};
      \path(-.6,2.3)node[right]{\footnotesize shown upon};
      \path(-.45,1.9)node[right]{\footnotesize 
                                 $\bM{x\\y}{\to}\bM{1&0\\1&1}\bM{x\\y}$};
      \path(-.3,1.5)node[right]{\footnotesize $\GL(2;\ZZ)$ transf.};
      \filldraw[fill=white, thick](0,-1)circle(.4mm);
            }}
  \put(0,23){$\begin{array}{@{}r@{~=~}l}
              C_2(\MF[2,\sharp]m) 
               &\overbrace{\ttt\sum_{i<j}\hD_i{\cdot}\hD_j}^
                {\sss\text{only edges, once}}\\
               &\frac{m^2{-}8}{m^2{-}4}
              \end{array}$}
  \put(0,5){$\begin{array}{@{}r@{\,}l@{\quad}r@{\,}l}
              C_1^{~2}(\MF[2,\sharp]m)
              &=\sum_{i,j}\hD_i{\cdot}\hD_j\<=4\<=d(\pDN{\MF[2]m}),
             &C_2(\MF[2]m)&=\overbrace{\ttt\sum_{i<j}\tD_i{\cdot}\tD_j}^
                            {\sss\text{only edges, once}}\<=8,\\
              &=\sum_{i,j}\tD_i{\cdot}\tD_j\<=C_1^{~2}(\MF[2]m)
             &&=d(\pDs{\MF[2]m})
              \end{array}$}
 \end{picture}}
$$
 \caption{Intersection and Chern numbers of $\MF[2]m$, the transpolar of the Hirzebruch surface $\FF[2]m$}
 \label{f:*2FmDD}
\end{figure}

\paragraph{Hirzebruch Surfaces:}
As toric varieties, they are encoded by their fan and spanning polytope:
\begin{gather}
 \vC{\TikZ{\path[use as bounding box](-3.8,-1)--(1.1,1);
           \path(-1.47,.05)node
           {\includegraphics[width=52mm]{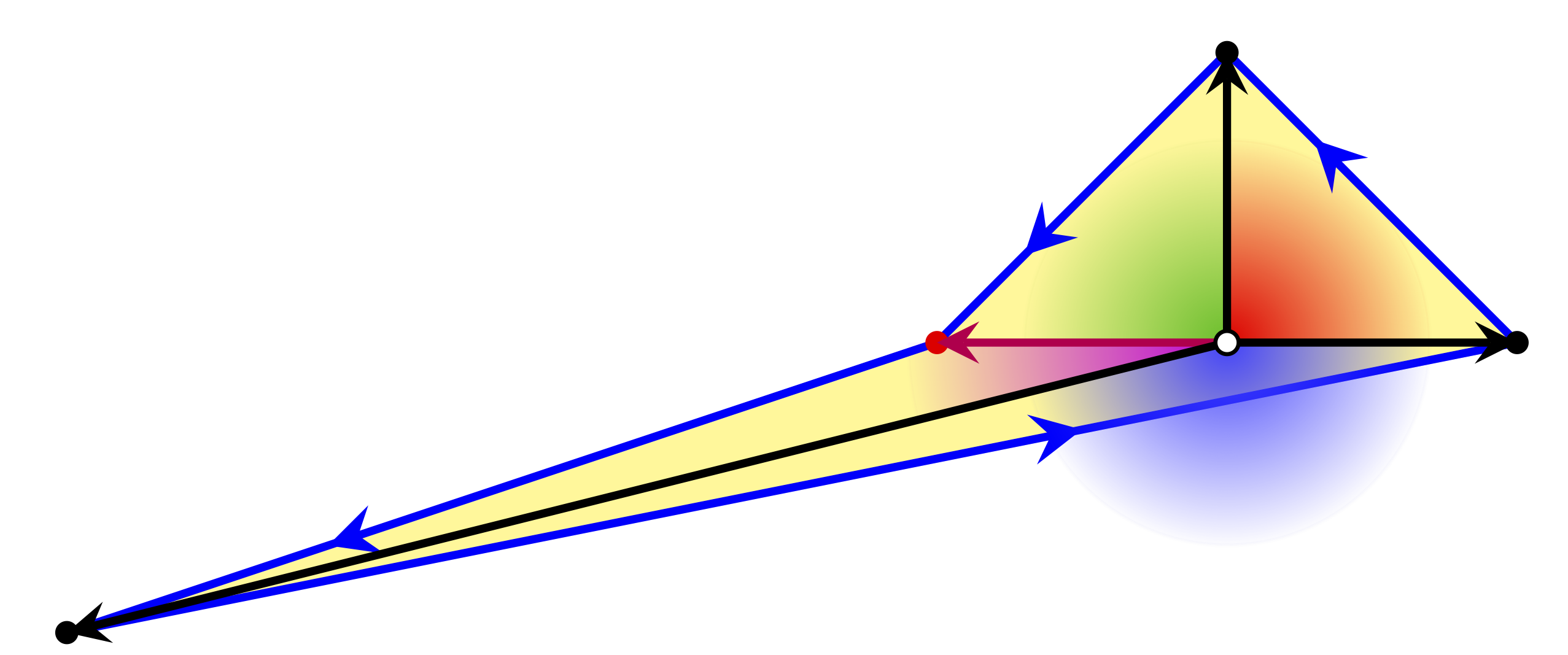}};
            \path(1.1,.2)node{\scriptsize$\n_1$};
            \path(0,1.2)node{\scriptsize$\n_2$};
            \path[red](-1.25,.2)node{\scriptsize$\n_3$};
            \path(-3.8,-.7)node{\scriptsize$\n_4$};
           \filldraw[fill=white](0,0)circle(.4mm);
           \path(-2,.67)node{$\pFn{\FF[2]m}\smt\pDs{\FF[2]m}$};
           \path(-1,-.75)node[rotate=10]{$m\<=4$ case shown};
            }}
 \quad
 \begin{array}{r|c@{~}c@{~}c@{~}c}
 \pFn{\FF[2]m}
     & \n_1 & \n_2 & \n_3 & \n_4 \\ \toprule\nGlu{-3pt}
     & 1 & 0 &-1 & -m \\[-2pt]
     & 0 & 1 & 0 & -1 \\ \hline
   Q^1 & 1 & 0 & 1 & 0 \\[-2pt]
   Q^2 & 0 & 1 &-m & 1 \\[-2pt] \bottomrule
 \end{array}
\quad
\vC{\TikZ{[every node/.style={thick, inner sep=.5, outer sep=0}]
         \path[use as bounding box](-4.33,-1)--(1,1);
      \node[circle, draw=black, fill=white](1) at(1,0)   {\scriptsize$m$};
      \node[circle, draw=black, fill=white](2) at(0,1)   {\scriptsize$0$};
      \node[red, rounded corners=5pt, inner sep=2, draw=red, fill=white](3)
                                               at(-1,0)  {\scriptsize$-m$};
      \node[circle, draw=black, fill=white](4) at(-4,-1) {\scriptsize$0$};
      \draw[gray!50, thick, -stealth](0,0)--(1);
      \draw[gray!50, thick, -stealth](0,0)--(2);
      \draw[red!30, thick, -stealth](0,0)--(3);
      \draw[gray!50, thick, -stealth](0,0)--(4);
      \draw[blue, thick, midarrow=stealth](1)--node[above right=1pt]
          {\scriptsize$1$}(2);
      \draw[blue, thick, midarrow=stealth](2)--node[above left=1pt]
          {\scriptsize$1$}(3);
      \draw[blue, thick, midarrow=stealth](3)--node[above=2pt]
          {\scriptsize$1$}(4);
      \draw[blue, thick, midarrow=stealth](4)--node[below right=1pt]
          {\scriptsize$1$}(1);
      \path(-1,.67)node[left]{$[\tD_i{\cdot}\tD_j]_{\FF[2]m}$};
      \filldraw[fill=white, thick](0,0)circle(.4mm);
            }}
 \label{e:2FmDDs}\\[2mm]
\mkern-12mu
 \begin{array}{r@{\,}l}
  \sum_i \n_i\,D_i\<\sim0\,\To~ D_1{-}D_3{-}mD_4\<\sim0,~~D_i&\<\sim D_j;\\
   \text{S.-R.}: D_1{\cdot}D_3\<\sim0,~~D_2{\cdot}D_4&\sim0.\\
 \end{array}
 \begin{array}{r@{\,}l@{\,}l@{\,}l}
   C_2(\FF[2]m) &=\sum_{i>j}D_i{\cdot}D_j
                &= 4{\times}(1)=4&=d(\pDs{\FF[2]m});\\
   C_1\!^2(\FF[2]m) &=\sum_{i,j}D_i{\cdot}D_j 
                &= 8{\times}(1)=8&=d(\pDN{\FF[2]m}).\\
  \end{array}
 \label{e:2FmDDC}
\end{gather}
The spanning multigon at far right in~\eqref{e:2FmDDs} is used to display the mutual and self-intersection numbers of the vertex-divisors. The analogous results for the novel transpolar multitope and multifan\cite{rBH-gB,Berglund:2022dgb} are shown in Figure~\ref{f:*2FmDD}.
Associating vertices to $\rT$-characteristic submanifolds,
$\n_{ij}^\wtd\<\mapsto\hD_j$ for $j\<=1,{\cdots},4$ results in the (mid-right) diagram of {\em\/fractional\/} intersection numbers for a {\em\/singular\/} orbifold, $\MF[2,\sharp]m$.
Interspersing the additional (exceptional) $\rT$-characteristic submanifolds $\tD_5,\cdots,\tD_{12}$ (see Figure~\ref{f:*2FmDD}, far-left) renders all intersection numbers integral for a {\em\/smooth\/} (but precomplex)
$\MF[2]m$, as shown in Figure~\ref{f:*2FmDD} far-right; see  Conjectures~\ref{C:gCYh} and~\ref{C:obstruct}, and Remarks~\ref{r:UTM} and~\ref{r:optimal}.

The indicated sums for a {\em\/singular\/} $\MF[2,\sharp]m$ fail to satisfy the Noether formula~\eqref{e:12N},
while those for a {\em\/smooth\/} $\MF[2]m$ corresponding to the MPCP-subdivided multifan do satisfy it.
Since all $\vec{Q}_{\sss(i)}\<=\tD_i{\cdot}\tD_j$ are non-negative linear combinations of
 $\tw{Q}^1\<\coeq\tD_3{\cdot}\tD_j$ and $\tw{Q}^2\<\coeq\tD_4{\cdot}\tD_j$, these two 4-vectors serve as Mori vectors, after ``clearing denominators.''
Note that $\sum_j\tD_*{\cdot}\tD_j\<=0$ for each subdivision $\tD_*$ of the smooth $\MF[2]m$.

The intersection numbers for $\MF[2]m$, shown at the right-hand side of Figure~\ref{f:*2FmDD} also exhibit a few characteristic and novel {\em\/orientation-dependent\/} features:
\begin{enumerate}[itemsep=-1pt, topsep=-1pt]
 \item The self-intersections $\tD_i^{~2}\<={-}2$ for $i\<=5,\cdots,10,12$, pertain to subdividing lattice-points within CCW-oriented facets, and encode the $\cO_{\IP^1}(-2)$ normal bundle of the exceptional $\IP^1$s.
In turn, the subdividing lattice-point within a flip-folded (CW-oriented) facet corresponds to $\tD_{11}^{~2}\<={+}2$ instead.
 \item The self-intersection of the submanifolds corresponding to the standard vertices in $\pDs{\MF[2]m}\<=\pDN{\FF[2]m}$ are negative, $\tD_1^{~2}\<={-}1\<=\tD_2^{~2}$, whereas those delimiting the ``extension'' are positive, $\tD_3^{~2}\<={+}1\<=\tD_4^{~2}$.
 \item Mutual intersections:
$\tD_i{\cdot}\tD_j\<=
  \big\{\begin{array}{@{}r@{~~}l@{}}
         \SSS +1, &\SSS\text{if $i,j$ are adjacent}\\[-6pt]
         \SSS  0, &\SSS\text{otherwise}
        \end{array}\big\}$,
as expected\cite{rF-TV, rGE-CCAG, rCLS-TV}, --- except
$\C1{\tD_{11}}{\cdot}\C1{\tD_3\<={-}1}$ and 
$\C1{\tD_{11}}{\cdot}\C1{\tD_4\<={-}1}$, because of the reversed, CW-orientation of the extension facet, 
$w[\n_{2\C13}^\wtd,\n_{\C134}^\wtd]\<=(-)1$.\qedhere
\end{enumerate}

The {\em\bfseries\/standard\/} Newton polytope, 
 $\pDN[\!^{\text{(std)}}]{\FF[2]m}\<=(\pDs{\FF[2]m})^\circ\<=(\Conv(\pDs{\FF[2]m}))^\circ$%
\cite{rD-TV, rO-TV, rF-TV, rGE-CCAG, rCLS-TV}, is the (plain) left-hand half of the multitope in Figure~\ref{f:*2FmDD}, completed on the right-hand side with the vertex, $\big(\frac2m,-1\big)$.
 For $m\<\geqslant3$, this polytope has no lattice point in the right-hand half-plane, its integral support spans an incomplete fan and so encodes a non-compact toric space. It contains $6{+}|m|$ lattice points, which cannot satisfy the 12-Theorem~\eqref{e:12T}; correspondingly,
  $d(\pDN[\!^{\text{(std)}}]{\FF[2]m})\<=4{+}|m|$ cannot satisfy~\eqref{e:12Td}.

\paragraph{The $\EE[2]m$ Sequence:}
\label{s:2Em}
The $\EE[2]m$ sequence (left-hand half od Figure~\ref{f:Fm+Em}) exhibits similar features\footnote{Here we focus on $m\<>0$; the $m\<\leqslant0$ cases are straightforward to analyze in the same manner, and in fact encode convex polytopes and fans for $m\<\in[0,3]$; see the discussion after~\eqref{e:EF=12}.}:
\begin{equation}
 \vC{\TikZ{\path[use as bounding box](-2,-1)--(1.5,1);
              \path(-.46,0)node
              {\includegraphics[width=32mm]{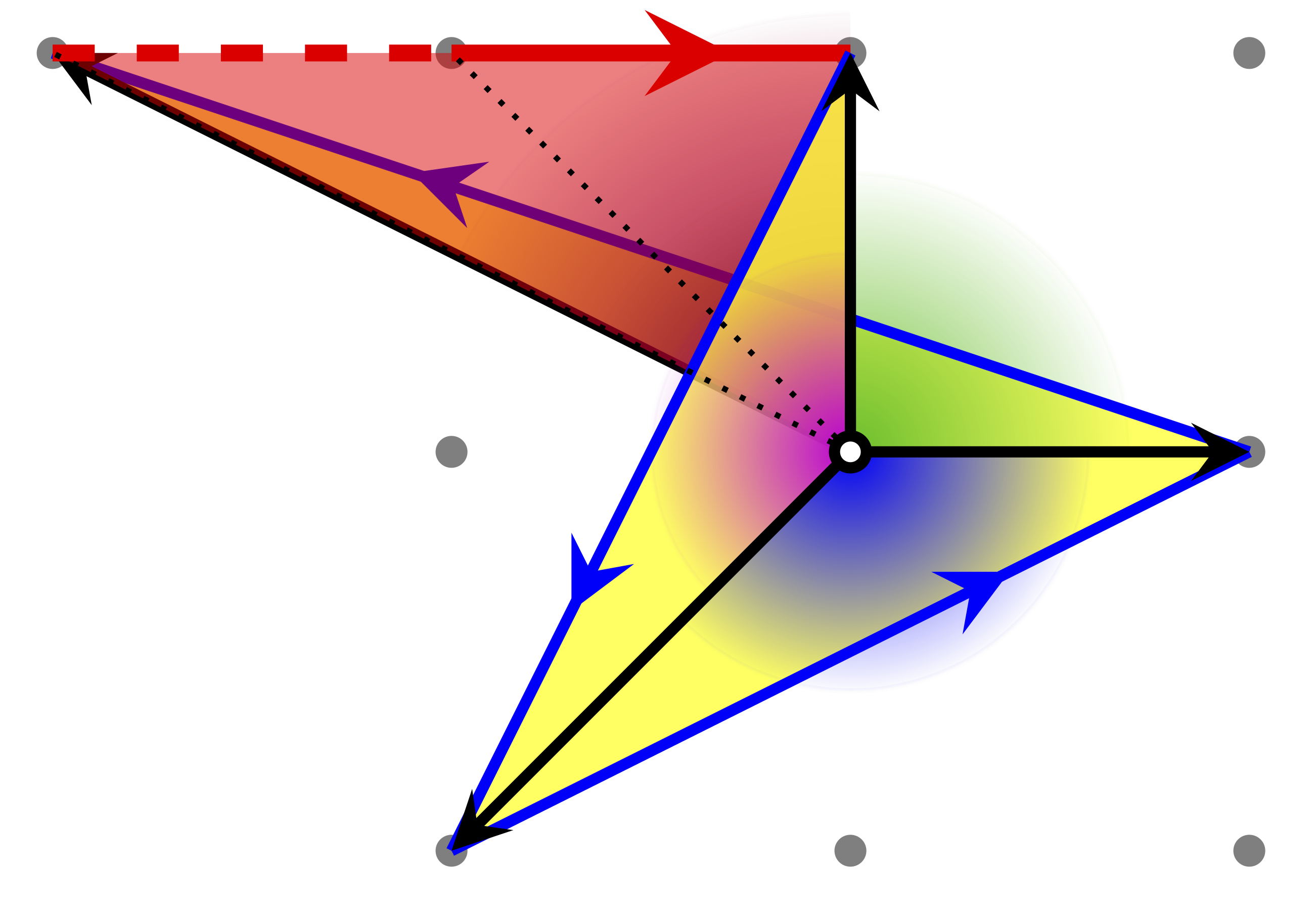}};
               \path[blue](.2,.2)node[above right,rotate=-10]
               {\footnotesize$\n_{12}$};
               \draw(-2,1)node[below]{\footnotesize$\n_2$};
              \path[Red](-.5,.95)node[above]
              {\footnotesize$\n_{23}$ \tiny(CW)};
               \path[blue](-.9,-.33)node[rotate=75]{\footnotesize$\n_{34}$};
              \path[blue,](0,-.5)node[below right,rotate=35]
              {\footnotesize$\n_{41}$}(1,0);
               \draw(1,0)node[right]{\footnotesize$\n_1$};
               \draw(0,1)node[right]{\footnotesize$\n_3$};
               \draw(-1,-1)node[left]{\footnotesize$\n_4$};
               \draw(-2,1)node[above=-2pt]{\scriptsize$(-m,1)$};
              \path(-2.7,.3)node[right]{\large$\pDs{E_m^{\sss(2)}}$};
              \path(-2.7,-.2)node[right]{\large$\<=\C3{\pDN{\EE[2]m}}$};
             }}
 \begin{array}{r|c@{~}c@{~}c@{~}c}
 \pFn{\EE[2]m}
     & \n_1 & \n_2 & \n_3 & \n_4 \\ \toprule\nGlu{-3pt}
     & 1 &  -m & 0 & -1 \\[-2pt]
     & 0 &\3-1 & 1 & -1 \\ \hline
   Q^1 & m &\3-1 &-1 & 0 \\[-2pt]
   Q^2 & 0 &  -1 &m{+}1 & m \\[-2pt] \bottomrule
 \end{array}
 \qquad
 \vC{\TikZ{[every node/.style={thick, inner sep=.5, outer sep=0}]
         \path[use as bounding box](-2,-1)--(1,1);
      \node[rounded corners=5pt, inner sep=2, draw=black, fill=white](1) 
                                               at(1,0)   {\scriptsize$1{+}m$};
      \node[Purple, rounded corners=6pt, inner sep=2, draw=Purple, fill=white](2) 
                                               at(-2,1)  {\scriptsize$\fRc1m$};
      \node[Purple, rounded corners=6pt, inner sep=2, draw=Purple, fill=white](3)
                                               at(0,1)  {\scriptsize$1{+}\fRc1m$};
      \node[circle, draw=black, fill=white](4) at(-1,-1) {\scriptsize$1$};
      \draw[gray!50, thick, -stealth](0,0)--(1);
      \draw[gray!50, thick, -stealth](0,0)--(2);
      \draw[gray!50, thick, -stealth](0,0)--(3);
      \draw[gray!50, thick, -stealth](0,0)--(4);
      \draw[blue, thick, midarrow=stealth](1)--node[above=2pt]{\scriptsize$1$}(2);
      \draw[red, very thick, dashed](2)--(-1,1);
      \draw[red, very thick, midarrow=stealth](-1,1)--
            node[above left=1pt]{\scriptsize$-\fRc1m$}(3);
      \draw[blue, thick, midarrow=stealth](3)--node[right=2pt]{\scriptsize$1$}(4);
      \draw[blue, thick, midarrow=stealth](4)--node[above=2pt]{\scriptsize$1$}(1);
      \path(-2,0)node[right]{$[\hD_i{\cdot}\hD_j]$};
      \path(0,-.8)node[right]{$\EE[2,\sharp]m$};
      \filldraw[fill=white, thick](0,0)circle(.4mm);
            }}
 \quad
 \vC{\TikZ{[every node/.style={thick, inner sep=.5, outer sep=0}]
         \path[use as bounding box](-2.5,-1)--(1,1);
      \node[rounded corners=5pt, inner sep=2, draw=black, fill=white](1) 
                                               at(1,0)   {\scriptsize$1{+}m$};
      \node[Purple, circle, draw=Purple, fill=white](2) at(-2.4,1) {\scriptsize$1$};
      \node[red, circle, draw=red, fill=white](5) at(-1.6,1) {\scriptsize$2$};
      \node[red, circle, draw=red, fill=white](6) at(-.8,1) {\scriptsize$2$};
      \node[Purple, circle, draw=Purple, fill=white](3) at(0,1)  {\scriptsize$2$};
      \node[circle, draw=black, fill=white](4) at(-1,-1) {\scriptsize$1$};
      \draw[Rouge!50, thick, densely dotted](0,0)--(5);
      \draw[Rouge!50, thick, densely dotted](0,0)--(6);
      \draw[gray!50, thick, -stealth](0,0)--(1);
      \draw[gray!50, thick, -stealth](0,0)--(2);
      \draw[gray!50, thick, -stealth](0,0)--(3);
      \draw[gray!50, thick, -stealth](0,0)--(4);
      \draw[blue, thick, midarrow=stealth](1)--node[above=2pt]{\scriptsize$1$}(2);
      \draw[red, very thick, midarrow=stealth](2)--node[above=1pt]{\scriptsize$-1$}(5);
      \draw[red, very thick, dotted](5)--(6);
      \draw[red, very thick, midarrow=stealth](6)--node[above=1pt]{\scriptsize$-1$}(3);
      \draw[blue, thick, midarrow=stealth](3)--node[right=2pt]{\scriptsize$1$}(4);
      \draw[blue, thick, midarrow=stealth](4)--node[above=2pt]{\scriptsize$1$}(1);
      \path(-2,0)node[right]{$[D_i{\cdot}D_j]$};
      \path(0,-.8)node[right]{$\EE[2]m$};
      \filldraw[fill=white, thick](0,0)circle(.4mm);
            }}
 \label{e:2EmDDs}
\end{equation}
The multifan $\pFn{\EE[2]m}$ encodes the divisor relations:
 $\hD_1\<-m\hD_2\<-\hD_4\<\sim0$ and $\hD_2\<+\hD_3\<-\hD_4\<\sim0$,
which simplify computations, but we continue with the unreduced intersection numbers in~\eqref{e:2EmDDs}, mid-right:
\begin{equation}
   C_1\!^2(\EE[2,\sharp]m) =\ttt\sum_{i,j}\hD_i{\cdot}\hD_j = 9{+}m
    \qquad\text{but}\qquad
   C_2(\EE[2,\sharp]m) =\ttt\sum_{i<j}\hD_i{\cdot}\hD_j = 3{-}\frc1m,
 \label{e:2EmDDC}
\end{equation}
which cannot possibly satisfy Noether's relation~\eqref{e:12N}. Again, smoothly subdividing the (CW-oriented) cone, $\n_{23}$, renders all intersection numbers integral (see~\eqref{e:2EmDDs} right), which satisfy Noether's formula~\eqref{e:12N} for all $m$. This desingularization induces
\begin{equation}
  C_2(\EE[2,\sharp]m) = 3{-}\frc1m  \quad\to\quad
  C_2(\EE[2]m)\<=3{-}m~~\text{(subdivision-desingularized)},
 \label{e:C2smoothing}
\end{equation}
as computed by adding up only the adjacent intersection numbers, once.
The Mori vectors are $\tw{Q}^1\!_j\<\coeq m\hD_2{\cdot}\hD_j$ and
$\tw{Q}^2\!_j\<\coeq m\hD_3{\cdot}\hD_j$, as the remaining vectors,
$\hD_1{\cdot}\vec{\hD}$ and $\hD_4{\cdot}\vec{\hD}$, are both their positive linear combinations.
 
The analogous results for the transpolar multitope and multifan\cite{rBH-gB,Berglund:2022dgb} are:
\begin{equation}
 \vC{\TikZ{[scale=.85]\path[use as bounding box](-1,-4.2)--(2,2.2);
      \path(.52,-.98)node
      {\includegraphics[width=28mm]{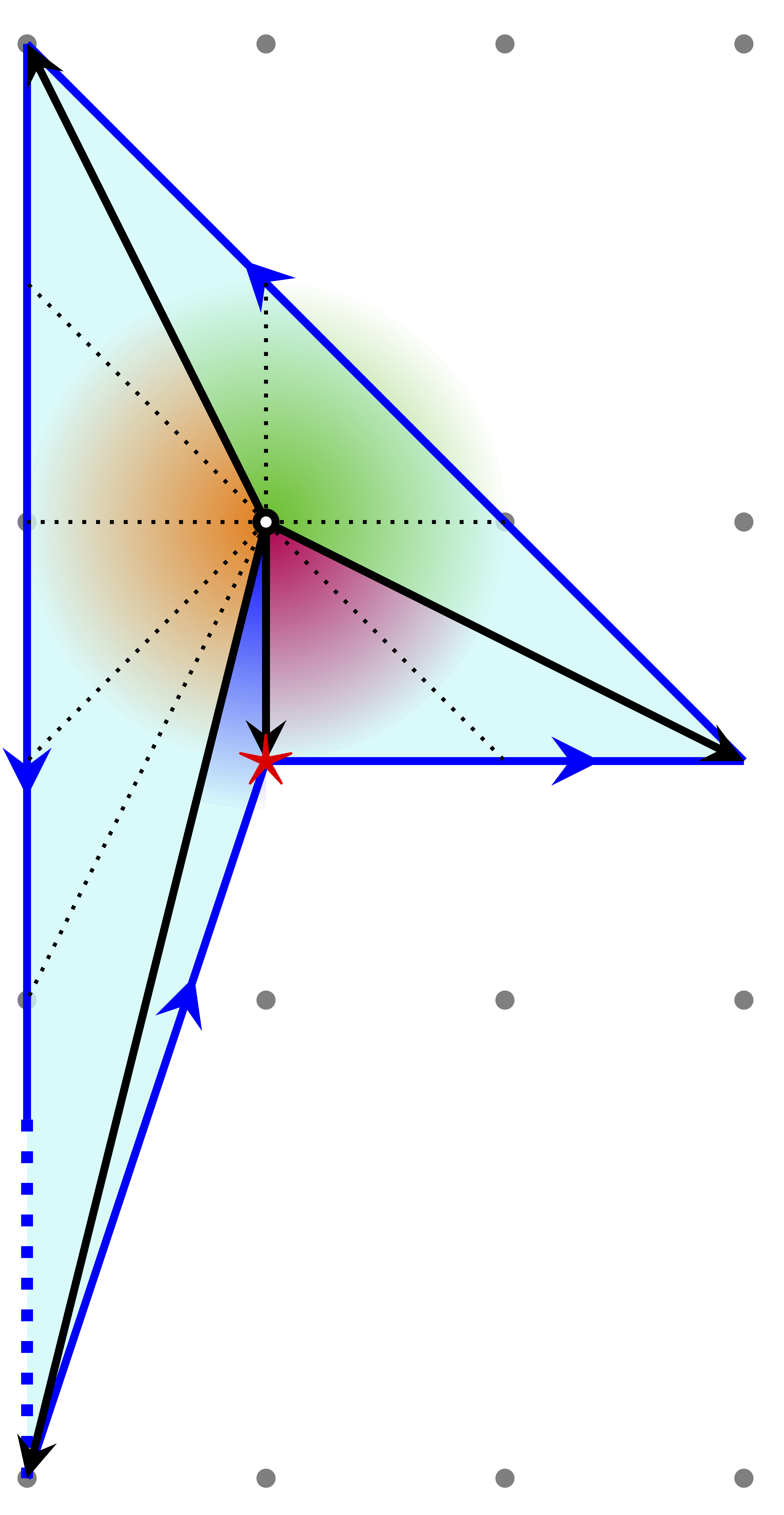}};
      \path[blue](.8,.8)node{\footnotesize$\n_4^\wtd$};
      \path[blue](-1,-.3)node[left]{\footnotesize$\n_1^\wtd$};
      \draw[blue, very thick, dotted](-1,-2.5)--(-1,-4);
      \path[blue](-.5,-2.5)node[right]{\footnotesize$\n_2^\wtd$};
      \path[blue](1,-1)node[below]{\footnotesize$\n_3^\wtd$};
       \path(-1,-3.8)node[right]{\footnotesize$\n_{12}^\wtd=(-1,-1{-}m)$};
       \path(-1,-3.75)node[left]{\footnotesize$\hD_2$};
       \path[red](.25,-1.25)node{\footnotesize$\n_{23}^\wtd$};
       \path[red](.25,-1.3)node[below]{\footnotesize$\hD_3$};
       \path(2.2,-.7)node{\footnotesize$\n_{34}^\wtd$};
       \path(2.1,-1.3)node{\footnotesize$\hD_4$};
       \path(-1,2)node[left]{\footnotesize$\n_{41}^\wtd$};
       \path(-1,2.1)node[right]{\footnotesize$\hD_1$};
       \draw[densely dotted](0,0)--(1,-1);
      \path(-.3,-2.9)node[right]{\large$\pDN{E_m^{\sss(2)}}\color{blue}
                \<=\pDs{\EE[2]m}$};
             }}
 \quad
 \begin{array}{r|c@{~}c@{~}c@{~}c}
 \MC5c{}\\
 \pFN{\EE[2]m}\!\!
         & \n_{41}^\wtd & \n_{12}^\wtd & \n_{23}^\wtd & \n_{34}^\wtd \\
 \toprule\nGlu{-2pt}
         &-1 &-1     &\3-0 &\3-2 \\[-1pt]
         & 2 &-1{-}m &  -1 &  -1 \\[-1pt] \midrule\nGlu{-2pt}
  \tw{Q}^1 & 1 & -1 & 3{+}m & 0 \\
  \tw{Q}^2 & 0 & 2 &-3{-}2m & 1 \\[-1pt] \bottomrule\nGlu{4pt}
  \MC5r{\text{\footnotesize$-\hD_1-\hD_2+2\hD_4\sim0$}}\\
  \MC5r{\text{\footnotesize$2\hD_1-(1{+}m)\hD_2-\hD_3-\hD_4\sim0$}}
 \end{array}
 \quad
 \vC{\TikZ{[scale=.85, 
            every node/.style={thick, inner sep=.5, outer sep=0}]
      \path[use as bounding box](-1.3,-4.2)--(2.3,2.2);
      \node[rounded corners=6pt, inner sep=2, draw=black, fill=white](1) 
            at(-1,2) {\scriptsize$\frac23{-}\frac{1}{m+3}$};
      \node[rounded corners=6pt, inner sep=2, draw=black, fill=white](2)
            at(-1,-4) {\scriptsize$-\frac{1}{m+3}$};
      \node[rounded corners=6pt, inner sep=2, rotate=30](3)
            at(0,-1) {\phantom{\scriptsize$-m{-}\fRc32$}};
      \node[circle, draw=black, fill=white](4)
            at(2,-1) {\scriptsize$\fRc16$};
      \draw[gray!50, thick, -stealth](0,0)--(1);
      \draw[gray!50, thick, -stealth](0,0)--(2);
      \draw[red!30, thick, -stealth](0,0)--(3);
      \draw[gray!50, thick, -stealth](0,0)--(4);
      \path(3)node[red, rounded corners=5pt, inner sep=2, draw=red, 
                   fill=white, rotate=30]
           {\scriptsize$-m{-}\fRc32$};
      \draw[blue, thick, midarrow=stealth](1)--node[right=2pt]
            {\scriptsize$\frac1{m+3}$}(-1,-2.5);
      \draw[blue, very thick, dotted](-1,-2.5)--(2);
      \draw[blue, thick, midarrow=stealth](2)--node[right=1pt]
          {\scriptsize$1$}(3);
      \draw[blue, thick, midarrow=stealth](3)--node[above=1pt]
          {\scriptsize$\fRc12$}(4);
      \draw[blue, thick, midarrow=stealth](4)--node[above=2pt]
          {\scriptsize$\fRc13$}(1);
      \path(-.1,-2.9)node[right]{$[\hD_i{\cdot}\hD_j]_{\EE[2,\sharp]m}$};
      \filldraw[fill=white, thick](0,0)circle(.4mm);
             }}
 \quad
 \vC{\TikZ{[scale=.85, every node/.style={thick, inner sep=.5,
            outer sep=0}]
      \path[use as bounding box](-1.3,-4.2)--(2.3,2.2);
      \node[rounded corners=4pt, inner sep=2, draw=black, fill=white](1) 
            at(-1,2){\scriptsize$-1$};
      \node[rounded corners=4pt, inner sep=2, draw=black, fill=white](2)
            at(-1,-4) {\scriptsize$-1$};
      \node[rounded corners=6pt, inner sep=2, rotate=45](3)
            at(0,-1){\phantom{\scriptsize$-2{-}m$}};
      \node[rounded corners=4pt, inner sep=2, draw=black, fill=white](4) 
            at(2,-1) {\scriptsize$-1$};
      \node[rounded corners=4pt, inner sep=2, draw=black, fill=white](8) 
            at(-1,1) {\scriptsize$-2$};
      \node[rounded corners=4pt, inner sep=2, draw=black, fill=white](9) 
            at(-1,0) {\scriptsize$-2$};
      \node[rounded corners=4pt, inner sep=2, draw=black, fill=white](10) 
            at(-1,-1) {\scriptsize$-2$};
      \node[rounded corners=4pt, inner sep=2, draw=black, fill=white](11) 
            at(-1,-2) {\scriptsize$-2$};
      \node[rounded corners=4pt, inner sep=2, draw=black, fill=white](5) 
            at(1,-1) {\scriptsize$-2$};
      \node[rounded corners=4pt, inner sep=2, draw=black, fill=white](6) 
            at(1,0) {\scriptsize$-2$};
      \node[rounded corners=4pt, inner sep=2, draw=black, fill=white](7) 
            at(0,1) {\scriptsize$-2$};
      \draw[gray!50, thick, -stealth](0,0)--(1);
      \draw[gray!50, thick, -stealth](0,0)--(2);
      \draw[red!30, thick, -stealth](0,0)--(3);
      \draw[gray!50, thick, -stealth](0,0)--(4);
      \path(3)node[red, rounded corners=5pt, inner sep=2, draw=red, 
                   fill=white, rotate=45]
          {\scriptsize$-2{-}m$};
      \foreach\n in{5,...,11}
       \draw[densely dotted](0,0)--(\n);
      \draw[blue, thick, midarrow=stealth](1)--node[right=2pt]
          {\scriptsize$1$}(8);
      \draw[blue, thick, midarrow=stealth](8)--node[right=2pt]
          {\scriptsize$1$}(9);
      \draw[blue, thick, midarrow=stealth](9)--node[right=2pt]
          {\scriptsize$1$}(10);
      \draw[blue, thick, midarrow=stealth](10)--node[right=2pt]
          {\scriptsize$1$}(11);
      \draw[blue, thick, midarrow=stealth](11)--node[right=2pt]
          {\scriptsize$1$}(-1,-3);
      \draw[blue, very thick, dotted](-1,-3)--(2);
      \draw[blue, thick, midarrow=stealth](2)--node[above=5pt]
          {\scriptsize$1$}(3);
      \draw[blue, thick, midarrow=stealth](3)--node[below=1pt]
          {\scriptsize$1$}(5);
      \draw[blue, thick, midarrow=stealth](5)--node[below=1pt]
          {\scriptsize$1$}(4);
      \draw[blue, thick, midarrow=stealth](4)--node[left=2pt]
          {\scriptsize$1$}(6);
      \draw[blue, thick, midarrow=stealth](6)--node[below left=1pt]
          {\scriptsize$1$}(7);
      \draw[blue, thick, midarrow=stealth](7)--node[below=3pt]
          {\scriptsize$1$}(1);
      \path(-.1,-2.9)node[right]{$[\tD_i{\cdot}\tD_j]_{\EE[2]m}$};
      \filldraw[fill=white, thick](0,0)circle(.4mm);
             }}
 \label{e:*2EmDD}
\end{equation}
where $\tD_i\<=\hD_i$ for $i,1\dots,4$ are the Cox divisors, and the exceptional MPCP divisors all have self-intersection $-2$, as indicated at far right in~\eqref{e:*2EmDD}.
 The Mori vectors are now $\tw{Q}^1\!_j\<\coeq(m{+}3)\Ht{D}_2{\cdot}\Ht{D}_j$ and
 $\tw{Q}^2\!_j\<\coeq2\,\Ht{D}_3{\cdot}\Ht{D}_j$, of which the remaining vectors,
 $\Ht{D}_1{\cdot}\vec{\Ht{D}}$ and $\Ht{D}_4{\cdot}\vec{\Ht{D}}$, are both positive linear combinations.

 As with $\EE[2]m$, reading from the intersection diagrams yields:
\begin{subequations}
 \label{e:*2EmDDC}
\begin{alignat}9
   C_2(\ME[2,\sharp]m) 
   &\ttt=\sum_{i<j}\hD_i{\cdot}\hD_j
   &&\ttt=\frac1{m{+}3}{+}1{+}\frac12{+}\frac13
   &&\ttt=\frac{11}6{+}\frac1{m+3};\\
   C_1^{~2}(\ME[2,\sharp]m)
   &\ttt=\sum\nolimits_{i,j}\tD_i{\cdot}\tD_j
   &&=\text{\footnotesize
           $(\frc23{-}\frc1{m+3}){+}({-}\frc1{m+3}){+}(-m{-}\frc32)
            {+}(\frc16){+}2(\frc1{m+3}{+}1{+}\frc12{+}\frc13)$}
   &&=3{-}m;\\
   C_2(\ME[2]m) &\ttt=\sum_{i<j}\tD_i{\cdot}\tD_j
   &&=(9{+}m)(1)
   &&=9{+}m;\\
   C_1^{~2}(\ME[2]m)
   &\ttt=\sum\nolimits_{i,j}\tD_i{\cdot}\tD_j
   &&=\text{\footnotesize$3(-1){+}(-2{-}m){+}(5{+}m)(-2){+}2(9{+}m)(1)$}
   &&=3{-}m.
\end{alignat}
\end{subequations}
The intersection numbers again {\em\/fail\/} to satisfy the Noether formula~\eqref{e:12N} for the {\em\/singular\/} $\ME[2,\sharp]m$ encoded by the un-subdivided fan~\eqref{e:*2EmDD}, but they {\em\/do satisfy\/} it for the {\em\/smooth\/} $\ME[2]m$ encoded by the MPCP-subdivision of
 $\pDs{\ME[2]m}$ and its fan.
 The evaluation $C_1^{~2}(\ME[2,\sharp]m)\<=C_1^{~2}(\ME[2]m)=3{-}m\<=C_2(\EE[2]m)$ is unaffected.
 
Throughout all 2-dimensional examples, the self-intersection of each subdivision-introduced $\rT$-charac\-te\-ris\-tic $\tD_*$ cancels the sum of its intersections with its adjacent divisors in the $\pm(+1{-}2{+}1)$-pattern, the overall sign chosen by the CCW/CW edge-orientation;
see Figure~\ref{f:*2FmDD}, and displays~\eqref{e:2EmDDs} and~\eqref{e:*2EmDD}.

\subsection{Three Dimensions}

\paragraph{Hirzebruch Scrolls:}
With five vertices, the unreduced array of (Cox) divisor intersection numbers, $D_i{\cdot}D_j{\cdot}D_k$, now form a $5{\times}5{\times}5$ {\em\/cube.} Their computation is as straightforward as for 2-folds, but decidedly less intuitive and harder to display.
 This simplifies on using the relations among the Cox divisors $D_i\<\mapsfrom\n_i$ implied by the fan and spanning polytope (far left in Figure~\ref{f:3F3MM}, Appendix~\ref{s:2MM}), 
 $\pFn{\FF[3]m}\<\smt\pDs{\FF[3]m}$:
\begin{equation}
\begin{array}{r|@{~}r|rrr@{~}|@{~}rr@{~}c@{\quad}r@{~~}r@{~}l}
 \pFn{\FF[3]m} & \nu_0 &\nu_1 & \nu_2 & \nu_3 & \nu_4 & \nu_5\\[-1pt]
\cmidrule[1pt]{1-7}\nGlu{-2pt}
        &0      &-1 & 1 & 0 & 0 &-m &&\To&-D_1+D_2-mD_5&\sim0  \\
        &0      &-1 & 0 & 1 & 0 &-m &&\To&-D_1+D_3-mD_5&\sim0  \\
\cline{2-7}
        &0      & 0 & 0 & 0 & 1 &-1 &&\To&D_4-D_5&\sim0  \\[-1pt]
\cmidrule[.4pt]{1-7}\cmidrule[.4pt]{9-11}\nGlu{-2pt}
    Q^1 & -3    & 1 & 1 & 1 & 0 & 0 &&&\MC2l{D_3\sim D_2,~~D_5\sim D_4} \\
    Q^2 & m{-}2 &-m & 0 & 0 & 1 & 1 &&&\MC2l{D_1\sim D_2-mD_4}         \\[-1pt]
\cmidrule[1pt]{1-7}
\end{array}
 \label{e:3Fm-SP}
\end{equation}
These relations permit using $\{D_2,D_4\}$ as basis.
Procedure~\ref{P:InMc} straightforwardly produces
(for $\t\<=\n_{23}$ and $\t\<=\n_{24}$) the intersection numbers (immediately using commutativity)
\begin{equation}
  D_2^{~3}\sim m,\qquad
  D_2^{~2}{\cdot}D_4\sim1,\qquad
  D_2{\cdot}D_4^{~2}\sim0,
 \label{e:Kijk}
\end{equation}
normalized by using $d(\n_{23}^\wtd)\<=2{+}2m$ and $d(\n_{24}^\wtd)\<=3$ from Table~\ref{t:3Fmlist}. For $\t\<=\n_{12}$ we have:
\begin{equation}
  D_1^{~2}{\cdot}D_2\sim-m,\quad
  D_1{\cdot}D_2^{~2}\sim0,\quad
  D_1{\cdot}D_2{\cdot}D_4\sim0,
\end{equation}
where the normalization by $d(\n_{12}^\wtd)\<=2{-}m$ crucially becomes negative for $m\<>2$, consistently with~\eqref{e:3Fm-SP}--\eqref{e:Kijk}.  The absence of the cones, $\n_{123},\n_{45}\<{\not\in}\pFn{\FF[3]m}$ implies the Stanley--Reisner equivalence relations:
\begin{equation}
  D_1D_2D_3\sim0\quad\text{and}\quad D_4D_5\sim0,
   \quad\too{\sss\eqref{e:3Fm-SP}}\quad
  D_2^{~3}\sim mD_2^{~2}D_4\quad\text{and}\quad D_4^{~2}\sim0.
 \label{e:SR3Fm}
\end{equation}
the latter of which then implies $D_4^{~2}{\cdot}D_i\<\sim0$, and $D_4^{~3}\<\sim0$ in particular.
Completing Procedure~\ref{P:InMc} for every $\t\<=\n_{ij}$ multiply verifies Table~\ref{t:3Fmlist}, the negative degrees included.

The Chern class is given by the expanding $\prod_{i=1}^5(1{+}D_i)$, subject to the above equivalence relations:
\begin{equation}
 c_1 \sim 3D_2\<+(2{-}m)D_4,\quad
 c_2 \sim 3D_2^{~2} +2(3{-}m)D_2D_4,\quad
 c_3 \sim6D_2^{~2}D_4.
\end{equation}
This allows computing the Chern numbers from the divisor intersection numbers:
\begin{alignat}9
 C_1\!^3 &\sim\big[3D_2{+}(2{-}m)D_4\big]^3\big|
     \sim[3D_2]^3{+}3[3D_2]^2[(2{-}m)D_4]\big|=27(m{+}2{-}m)=54.\\
 C_1C_2 &\sim\big[3D_2{+}(2{-}m)D_4\big]
              \big[3D_2^{~2}{+}(6{-}2m)D_2D_4\big]\big|
     \sim\big[9D_2^{~3}{+}[3(6{-}2m){+}3(2{-}m)]D_2^{~2}D_4\big]\big|\nn\\*
 &\quad~=9m+[(18{-}6m){+}(6{-}3m)]=24.\\
 C_3
 &\sim[6D_2^{~2}D_4]\big|=6,
\end{alignat}
in full agreement with Table~\ref{t:3Fmlist}.

The analogous intersection number are as straightforward to compute for the transpolar torus manifold $\MF[3]m$, corresponding to the flip-folded multifan,
$\pFn{\MF[3]m}\<\smt\pDs{\MF[3]m}\<=\pDN{\FF[3]m}$, shown (for $m\<=3$) at right in~\cite[Fig.~3]{rBH-gB}. As the smoothly subdivided multitope
$\pDs{\MF[3]m}$ contains $30{+}8\vq_3^m(m{-}2)$ lattice points\footnote{Here, $\vq_i^j=\{0\text{ if }j\<<i,~1\text{ if }j\<\geqslant i\}$ is the step-function.}, we forego computing here the
$\binom{32{+}8\vq_3^m(m{-}2)}3=O(m^3)$ (albeit mostly vanishing) triple intersection numbers.

\paragraph{The $\EE[3]m$ Sequence:}
Following Procedure~\ref{P:InMc} through all 2-cones ($\n_{45}\<{\not\in}\pFn{\EE[3]m}$) of the multifan spanned by $\pDs{\EE[3]m}$ (mid-left in Figure~\ref{f:3EmSNP}), we find the redundant listing
\begin{equation}
\begin{array}{r@{:~}c|ccccc|c@{~=~}c}
  \t&D_i{\cdot}D_j & D_1 & D_2 & D_3 & D_4 & D_5 &\!\text{sums}&d(\n_{ij}^\wtd)\\[1pt]
\toprule\nGlu{-2pt}
\n_{12} &D_1{\cdot}D_2 &m{+}1 &m{+}1 &0          &1 &1       &2(2{+}m) &d(\n_{12}^\wtd)\\[-2pt]
\n_{13} &D_1{\cdot}D_3 &0   &0   &\frac{m+1}m &1 &-\fRc1m &2      &d(\n_{13}^\wtd)\\[-2pt]
\n_{14} &D_1{\cdot}D_4 &1   &1   &1           &1 &0       &4      &d(\n_{14}^\wtd)\\[-2pt]
\n_{15} &D_1{\cdot}D_5 &1   &1   &-\fRc1m     &0 &\fRc1m  &2      &d(\n_{15}^\wtd)\\[-1pt]
 \midrule\nGlu{-2pt}
\n_{23} &D_2{\cdot}D_3 &0   &0   &\frac{m+1}m &1 &-\fRc1m &2      &d(\n_{23}^\wtd)\\[-2pt]
\n_{24} &D_2{\cdot}D_4 &1   &1   &1           &1 &0       &4      &d(\n_{24}^\wtd)\\[-2pt]
\n_{25} &D_2{\cdot}D_5 &1   &1   &-\fRc1m     &0 &\fRc1m  &2      &d(\n_{25}^\wtd)\\[-1pt]
 \midrule\nGlu{-2pt}
\n_{34} &D_3{\cdot}D_4 &1   &1   &1           &1 &0       &4      &d(\n_{34}^\wtd)\\[-2pt]
\n_{35} &D_3{\cdot}D_5 &1   &1   &-\fRc1m     &0 &\fRc1m  &2      &d(\n_{35}^\wtd)\\[-2pt]
\end{array}
 \label{e:3EmDijk}
\end{equation}
Each fractional entry (for $m\<>1$) stems from $d(\n_{153})\<=d(\n_{235})\<={-}m$, i.e., by $d(\n_{35})\<={-}m$. Each row in~\eqref{e:3EmDijk} is normalized by the right-most sum in~\eqref{e:ells1}, reading $d(\n_{ij}^\wtd)$ directly from the left-most (plain but non-convex) polytope in Figure~\ref{f:3EmSNP}; its only $m$-dependence stems from the downward growing ``icicle,'' at $\n_{125}^\wtd$.

The homology relations in Figure~\ref{f:3EmSNP} permit using $\{D_4,D_5\}$ as a basis, where
 $c_1(\EE[3]m)=4D_4{+}2mD_5$,
 $c_2(\EE[3]m)=6D_4^{~2}{+}(m^2{-}1)D_5^{~2}$, and
 $c_3(\EE[3]m)=4D_4^{~3}{-}2mD_5^{~3}$,
by the Stanley--Reisner relation, $D_4{\cdot}D_5\sim0$.
Then (for $m\<\neq0$),%
\footnote{Using that $D_2\<\sim D_1$, the intersection table~\eqref{e:3EmDijk} implies $D_1^{~3}\<\sim1{+}m$. With the Stanley--Reisner relation $D_1{\cdot}D_2{\cdot}D_3\<\sim0$ and translated via the divisor relations in Figure~\ref{f:3EmSNP} into the $\{D_4,D_5\}$ basis, these become $D_4^{~3}{+}m^3D_5^{~3}\<\sim1{+}m$ and $D_4^{~3}{-}m^2D_5^{~3}\<\sim0$, respectively, which imply~\eqref{e:3EmKijk}. When $m\<={-}1$, we find that $D_4^{~3}\<\sim D_5^{~3}$. We then also use $D_1{\cdot}D_3{\cdot}D_5\<\sim1$ to find $D_5^{~3}\<\sim1$.}
\begin{equation}
  D_4^{~3}\<\sim1 \quad\text{and}\quad D_5^{~3}\<\sim\fRc1{m^2},
 \label{e:3EmKijk}
\end{equation}
so:\backUp[.5]
\begin{subequations}
\label{e:3EmChN}
\begin{alignat}9
 c_1^{~3}
 &\sim 64 D_4^{~3}+8m^3D_5^{~3}&
 &\too{\sss\eqref{e:3EmKijk}} 64(1)+8m^3(\frc1{m^2})&
 &=64{+}8m; \label{e:3EmC111}\\
 c_1c_2
 &\sim 24D_4^{~3} +2(m^3{-}m)D_5^{~3}&
 &\too{\sss\eqref{e:3EmKijk}} 24(1) +2(m^3{-}m)(\frc1{m^2})&
 &=24{+}2m{-}\frc2m; \label{e:3EmC12}\\
 c_3
 &\sim 4D_4^{~3} -2mD_5^{~3}&
 &\too{\sss\eqref{e:3EmKijk}} 4(1) -2m\big(\frc1{m^2}\big)&
 &=4{-}\frc2m. \label{e:3EmC3}
\end{alignat}
\end{subequations}
Much as with the 2-dimensional $\EE[2,\sharp]m$, only the first of these,~\eqref{e:3EmC111},
 $C_1^{~3}(\EE[3,\sharp]m)\<=C_1^{~3}(\EE[2]m)\,=C_3(\ME[2]m)\,=64{+}8m$, agrees with Table~\ref{t:3Emlist}; the other two Chern numbers are fractional for $m\<>2$, disagree with Table~\ref{t:3Emlist} and fail to satisfy the Todd--Hirzebruch theorem~\eqref{e:T24}.
The smooth subdivision of $\n_{35}$, $\n_{153}$ and $\n_{235}$ identifies $m{-}1$ submanifolds, $D_{5+k}\mapsfrom(-k,-k,1)$ for $k\<=1,2,\dots,(m{-}1)$.
 While significantly increasing the array of intersection numbers, this effectively replaces%
 \footnote{The mechanics of this owes to the fact that when a cone is subdivided, $\s\<\to\vs_1\<\uplus\vs_2$, its sub-cones, $\vs_1,\vs_2$, {\em\/replace\/} $\s$ in the specification of the multifan, so that $\s$ enters the Stanley--Reisner ideal.
 Thus, when $\n_{35}$ in the top, mid-left ($m\<=2$) in Figure~\ref{f:3EmSNP} is subdivided by adding $\n_6$, the Stanley--Reisner ideal acquires contributions from both
 $\n_{35}$ and $\n_{46}$, and so also $\n_{35i}$ and $\n_{46i}$ for all $i=1,\dots,6$.
 The absence of the latter 3-cones from the (smoothly subdivided) multifan enforces the vanishing of the corresponding intersection numbers:
 $D_3{\cdot}D_5{\cdot}D_i\sim0$ and $D_4{\cdot}D_6{\cdot}D_i\sim0$.} 
 $\frc2m\<\to2m$ in~\eqref{e:3EmC12}--\eqref{e:3EmC3}, somewhat akin to~\eqref{e:C2smoothing}.
 This makes the intersection number computed Chern numbers~\eqref{e:3EmChN} perfectly reproduce the cone-degree results in Table~\ref{t:3Emlist}, and of course also satisfy the Todd--Hirzebruch identity~\eqref{e:T24}.

The analogous computations are as straightforward for the transpolar toric space $\ME[3]m$, encoded by the (plain) fan, $\pFn{\ME[3]m}\<\smt\pDs{\ME[3]m}\<=\pDN{\EE[3]m}$, shown (for $m\<=2$) in the left-most illustration in Figure~\ref{f:3EmSNP}. Since the MPCP-subdivided polytope 
$\pDs{\ME[3]m}$ has $34{+}4m$ lattice points, we forego here computing the $\binom{36{+}4m}3$ (albeit mostly vanishing) triple divisor intersection numbers.

\section{Combinatorial and Cohomological Features of VEX Multitopes}
\label{s:VEX}
\paragraph{Betti Numbers:}
Plain fans encode complex-algebraic toric varieties (\/``{\em toric crystals\/}''\cite[\SS\:0.6]{rD-TV}), for which the Betti numbers are completely determined by the fan:
\begin{thrm}[Danilov{\cite[p.~140]{rD-TV}}, Fulton{\cite[p.~92]{rF-TV}}]
\label{T:Betti}
For the complex $n$-dimensional toric variety, $X$, encoded by the fan, $\pFn{X}$,
 $\dim H^{p,q}(X)\<=0$ for $p\<\neq q$, and the Betti numbers, 
 $b_q(X)\<=\sum_{r=0}^q\dim H^{r,q-r}(X)$, are:
\begin{equation}
 b_{2k+1}(X)=0
   \qquad\text{and}\qquad
 b_{2k}(X)=\sum_{i=k}^r (-1)^{i-k} \binom{i}{k} ~e_{n-i},
   \qquad k\<=0,1,\ldots,n,
 \label{e:Betti}
\end{equation}
where $e_k$ is the number of $k$-dimensional cones in a smooth subdivision of
 $\,\pFn{X}\<\smt\pDs{X}$.
\end{thrm}
\begin{remk}
\label{r:qConv}
By definition, $e_0\<\coeq1$.
 Also, $e_n\<=d(\pDs{X})$ for all smooth (smoothly subdivisible) VEX multitopes. Rational fans of toric varieties for which $b_{2k+1}(X)\<=0$ are {\em\/quasi-convex,} with boundaries supported by {\em\/real homology manifolds\/}\cite[Thm.\;4.2--4]{Barthel:2002aa}.
\end{remk}
The related classic result\cite[Lemma\,1.5]{rD-TV}, expressing the Poincar\'e series of a complex $n$-dimensional algebraic variety encoded by the polytope $\pD$ as
 $P_\pD(t)=\F_\pD(t)(1{-}t)^{-n-1}$, has a suitable generalization:
\begin{lema}[T.\,Hibi\cite{Hibi:1995aa}]
\label{L:Hibi}
Let $\pD$ be an $(L\<\approx\ZZ^r)$-lattice polyhedral complex (a finite connected union of convex lattice polytopes) of dimension $n$ that contains the faces of all (component) polytopes as well as all their intersections, and which is star-shaped, i.e., homeomorphic to an $n$-ball. Then
\begin{equation}
 \F_\pD(t)\<=(1{-}t)^{n+1}\sum_{k=0}^\infty\ell_L(k\pD)\,t^k
                    \<=\sum_{k=0}^\infty h_k\,t^k,\qquad
 \Big\{\begin{array}{@{}r@{~}lr@{~}l}
        h_1&=\ell_L(\pD){-}(n{+}1),& h_0&=1,\\
        h_n&=\ell_L(\pD\<\ssm\vd\pD),& h_i&=0~~\text{for}~~i\<>n.
       \end{array}
 \label{e:DanilovF}
\end{equation}
 Also, $h_1\<\leqslant h_k$ for $k\<\in[2,n]$ and
\begin{equation}
  \sum_{i=0}^kh_i\<\leqslant\sum_{i=0}^kh_{n-i}~~\text{for}~~k\<\in[0,\lfloor n/2\rfloor],
 \quad\text{but}\quad
 \sum_{i=0}^{k+1}h_i\<\geqslant\sum_{i=0}^kh_{n-i}~~\text{for}~~k\<\in[0,\lfloor(n{-}1)/2\rfloor].
 \label{e:Hibi}
\end{equation}
 Finally, if the monoid $\oplus_k\,\vec{x}^{\,\vec\n_k}t^k$ for $\vec\n_k\<\in(k\pD\<\cap L)$ with $0\<\leqslant k\<\in\ZZ$ is Gorenstein (see\cite{Hibi:1995aa} for precise details),
 then $h_k\<=h_{n-k}$ for $0\<\leqslant k\<\leqslant n$.
\end{lema}
\begin{remk}
\label{r:PoiLef}
The identification $h_k(\pD)\<=b_{2k}(X_\pD)$\cite[\SS\:4.5]{rF-TV} and Poincar\'e duality then imply the Dehn--Sommerville relations, $h_k\<=h_{n-k}$. Also, the Hard Lefschetz Theorem and the K{\"a}hler  form generated $\SL(2)$-action on $H^{q,q}$ implies the so-called ``unimodality'':
 $h_{i-1}\leqslant h_i$ for $i\<=1,\dots,\lfloor n/2\rfloor$. 
\end{remk}

Multifan/multitope generalizations of $e_k$ and $h_k$ exist and define the $T_y$-genus%
\footnote{This extends the classic $\c_y$-characteristic\cite{rHirz}, interpolating between the Euler characteristic, $\c\<=T_{-1}$ and signature, $\t\<=T_1$},
\begin{equation}
   T_y[\pFn{}]=\sum_{k=0}^rh_k(\pFn{})(-y)^k=\sum_{k=0}^r e_{n-k}(\pFn{})(-1-y)^k,
   \qquad
   e_{n-k}(\pFn{})=\sum_{i=k}^r\binom{i}{k}\,h_i(\pFn{}),
 \label{e:Ty}
\end{equation}
but encode torus manifolds {\em\/incompletely\/}\cite{rHM-MFs}: Betti numbers satisfy~\eqref{e:Betti} only if the cohomology of $X$ is (independently known to be) generated in degree two\cite{Masuda:2006aa}.
 Multiple torus manifolds may therefore have the same multifan, but differing Betti numbers; see\cite{rM-MFans}, and\cite{Ishida:2013aa} for a (qualified) resolution. Akin to {\em\/minimal models\/} of birational geometry, we then propose (cf.\ Conjecture~\ref{C:gCYh} and Remark~\ref{r:optimal}):
\begin{conj}
\label{c:optimal}
From among the real $2n$-dimensional torus manifolds\cite{rM-MFans, Masuda:2000aa, rHM-MFs, Civan:2003aa, Masuda:2006aa, rHM-EG+MF, rH-EG+MFs2, Nishimura:2006vs, rMP-TO+MFs, Ishida:2013aa, buchstaber2014toric}, $X$, corresponding to a multifan, $\pFn{X}$, there exists an {\em\/optimal torus manifold\/} such that its ($\,\rT$-equivariant) Betti numbers are optimally non-negative: with minimal values
\begin{equation}
  b_{2k}(X)\geqslant h_k(\pFn{X})=\sum_{i=k}^r (-1)^{i-k} \binom{i}{k} ~e_{n-i}(\pFn{X}),
   \qquad k\<=0,1,\ldots,n,
 \label{e:BettiMF}
\end{equation}
that satisfy
 ({\small\em1})~the Poincar\'e duality, $b_{2n-r}\<=b_r$,
 ({\small\em2})~the Hard Lefschetz theorem, and
 ({\small\em3})~are consistent with all ($\,\rT$-equivariant) characteristic classes of $X$,
  as encoded by $\pFn{X}$.
\end{conj}
\begin{remk}
\label{r:ExtStd}
While VEX multitopes are star-shaped\cite{rBH-gB}, they may be flip-folded and otherwise multi-layered. Lemma~\ref{L:Hibi} was found to hold for many examples such as the entire $(\FF{m},\MF{m})$-sequence, but non-negativity~\eqref{e:Hibi} fails, e.g., in the $(\EE{m},\ME{m})$-sequence for some $m$; see~\eqref{e:2Emh1} and~\eqref{e:2*Emh1} as well as~\eqref{e:3Emh1} and~\eqref{e:3*Emh1}.
This agrees with Conjecture~\ref{C:obstruct}, that flip-folded (``extension''\cite{rBH-gB}) portions of a multifan/multitope encode locations in the corresponding torus manifold where the complex structure degenerates; see also footnote~\ref{fn:preStr}, p.\,\pageref{fn:preStr}.
Also, unimodality (Claim~\ref{CC:THCEP}, part~\ref{i:uniM}) and the corresponding non-negativity~\eqref{e:Hibi} stem from the $\SL(2)$-action in the Hard Lefschetz Theorem, which in turn is generated by multiplication with the K{\"a}hler form. Thus, it may well be that some unitary torus manifolds (such as $\EE{m}$ and $\ME{m}$, for some ranges of $m$) fail the non-negativity~\eqref{e:Hibi} ``merely'' owing to a degeneration of their K{\"a}hler metric, including by possible appearance of geometric torsion.
Finally, in examples such as $\pDs{\EE{m}}$, the relative sizes of the ``extension'' and the ``standard part'' depend on $m$, and may reverse roles.
\end{remk}

$\rT$-equivariant (co)homology is key in identifying the most general GLSM geometry (Remark~\ref{r:GLSM}) since $\rT\<=\U(1)^r$ is its gauge group, and is known to distinguish various {\em\/quasitoric manifolds\/}\cite{rM-EqH*TV}. A class of further generalizations is obtained by extending the
$\IC^*\<\ni\l\<=|\l|e^{i\vq}$-action to {\em\/separate\/} $|\l|$- and $e^{i\vq}$-actions\cite{Ishida:2013aa}.
 In turn, intersection homology mentioned in Remark~\ref{r:qConv} is known to be modified in ``middle dimension'' by the (co)homology theory consistent with both mirror symmetry and geometric (conifold) transitions, as needed in string theory\cite{rSSC,Hubsch:2002st,rAR01,rB-IS+ST}.
 It is at present not clear which combination (if any) of such generalizations may provide the necessary framework for Corollary~\ref{CC:MMM}, Claim~\ref{CC:THCEP} and Conjectures~\ref{C:gCYh}, \ref{C:gMM} and~\ref{c:optimal} to hold, especially regarding the balance between the ``standard'' and ``extension'' parts and the corresponding degenerations in the complex or symplectic structures; see Remark~\ref{r:ExtStd} and also applications in string theory\cite{rBatyrevDais, Mustata:2005aa}, but also, e.g.,\cite{Franco:2015tna, Franco:2023flw, Kho:2023dcm}.

Regarding the $\F_\pD(t)$-function~\eqref{e:DanilovF}, we now test the validity and consequences of the Dehn--Sommerville and unimodality relations, and so also Hibi's Lemma~\ref{L:Hibi} and Theorem~\ref{T:Betti}; see Remark~\ref{r:PoiLef}.

\subsection{Dehn--Sommerville and Unimodality Relations in Two Dimensions}
\label{s:PF2}
The Dehn--Sommerville relations for a 2-dimensional $\pD$ imply that $\F_{\!\pD}(t)\<=1{+}h_1t{+}t^2$. Then,
\begin{subequations}
 \label{e:PDlkP}
\begin{alignat}9
  \F_{\!\pD}(1)\<=d(\pD) \quad\To\quad
  h_1          &=d(\pD){-}2;
    \label{e:h1=dD-2}\\
  P_{\!\pD}(t) &=1+\big(1{+}d(\pD)\big)t +\big(1{+}3d(\pD)\big)t^2,
    \label{e:PD(t)}\\
  \ell(\pD)    &=d(\pD){+}1\qquad\text{and}\qquad\ell(2\pD)\<=3d(\pD){+}1.
    \label{e:lkP}
\end{alignat}
\end{subequations}
The first of the results~\eqref{e:lkP} holds for every VEX multitope $\pDs{X}$: whether plain, flip-folded or otherwise multilayered --- even when its winding number is 0, and there is no ``interior'': All vertices of $\pDs{X}$ are primitive and so the first lattice point from the origin. All central triangles over facets contain lattice points only in the facets and the origin, so that each lattice point in $\vd\pDs{X}$ is primitive. Thereby,
 $\pDs{X}$ has a smooth subdivision, the number of facet unit-segments equals the number of lattice points in $\vd\pDs{X}$, so $d(\pDs{X})\<=\ell(\vd\pDs{X})$. Including by definition the origin in $\ell(\pDs{X})$ results in $\ell(\pDs{X})\<=d(\pDs{X}){+}1$.
 This also provides a sign-assignment to lattice points, i.e., 1-cones:
\begin{enumerate}[itemsep=-1pt, topsep=-1pt]
 \item $d(\n_i)\<={+}1$ if $\n_i\<=\n_{ji}\<\cap\n_{ik}$ and $d(\n_{ji}),\,d(\n_{ik})>0$, i.e.,
  if the orientation remains $+1$ (CCW) through $\n_i$.
 \item $d(\n_i)\<={-}1$ if $\n_i\<=\n_{ji}\<\cap\n_{ik}$ and $d(\n_{ji}),\,d(\n_{ik})<0$, i.e.,
  if the orientation remains $-1$ (CW) through $\n_i$.
 \item $d(\n_i)\<=0$ if $\n_i\<=\n_{ji}\<\cap\n_{ik}$ and $d(\n_{ji}){\cdot}d(\n_{ik})<0$, i.e.,
  if the orientation changes through $\n_i$.
\end{enumerate}
This is exactly the counting that was used to consistently determine the Euler and Betti (\& Hodge) numbers of the Calabi--Yau hypersurfaces, Laurent-smoothed as needed\cite{rBH-gB}.

\paragraph{Explicit Verification:}
It then remains to verify the second of the relations~\eqref{e:lkP}:
\begin{equation}
  \vC{\begin{picture}(40,55)
        \put(0,37){\TikZ{[scale=.75]\path[use as bounding box](-3.5,-1.2)--(2,1.2);
            \fill[yellow, opacity=.5](1,0)--(0,1)--(-1,0)--(-3,-1)--cycle;
            \foreach\x in{-3,...,1}\draw[gray](\x,-1)--(\x,1);
            \foreach\y in{-1,...,1}\draw[gray](-3,\y)--(1,\y);
            \draw[blue, thick, line join=round](1,0)--(0,1)--(-1,0)--(-3,-1)--cycle;
            \fill(0,0)circle(.6mm);
            \fill(1,0)circle(.6mm);
             \path(1,0)node[below]{\tiny$(1,0)$};
            \fill(0,1)circle(.6mm);
             \path(0,1)node[above=-2pt]{\tiny$(0,1)$};
            \fill(-1,0)circle(.6mm);
             \path(-1.6,.2)node{\tiny$(-1,0)$};
            \fill(-3,-1)circle(.6mm);
             \path(-2.75,-1)node[above, rotate=30]{\tiny$(-3,-1)$};
             \path(-2.33,.75)node{$\pDs{\FF[2]3}$};
            \path(-3,-1.4)node[right]{$\ell(\pDs{\FF[2]m})=5$};
            \path(-3,-2)node[right]{$d(\pDs{\FF[2]m})=4$};
            \path(-3,-3.2)node[right]{$\ell(2\pDs{\FF[2]m})=13$};
            }}
        \put(0,2){\TikZ{[scale=.75]\path[use as bounding box](-3.5,-1.2)--(2,1.2);
            \fill[yellow, opacity=.5](1,0)--(0,1)--(-1,0)--(-3,-1)--cycle;
            \foreach\x in{-6,...,2}\draw[gray](\x/2,-1)--(\x/2,1);
            \foreach\y in{-2,...,2}\draw[gray](-3,\y/2)--(1,\y/2);
            \draw[blue, thick, line join=round](1,0)--(0,1)--(-1,0)--(-3,-1)--cycle;
            \fill(0,1)circle(.6mm);
            \foreach\x in{-1,...,1} \fill(\x/2,.5)circle(.6mm);
            \foreach\x in{-2,...,2} \fill(\x/2,0)circle(.6mm);
            \foreach\x in{-1,...,1} \fill(\x/2-1.5,-.5)circle(.6mm);
            \fill(-3,-1)circle(.6mm);
             \path(-2.33,.75)node{\large$2\pDs{\FF[2]3}$};
            }}
      \end{picture}}
  \vC{\TikZ{[scale=.67]\path[use as bounding box](-1.2,-3.2)--(1.5,5.0);
            \foreach\x in{-1,...,1}\draw[gray!50](\x,-3)--(\x,5);
            \foreach\y in{-3,...,5}\draw[gray!50](-1,\y)--(1,\y);
            \fill[Turque!50, opacity=.9](0,0)--(1,-1)--(-1,-1)--(-1,5)--(0,0);
             \draw[blue,thick,line join=round](1,-1)--(-1,-1)--(-1,5);
            \fill[red, opacity=.67](0,0)--(1,-1)--(1,-3)--(0,0);
             \draw[red, thick](1,-1)--(1,-3);
             \filldraw[gray,fill=white](1,-1)circle(.6mm);
             \draw[gray, -stealth](0,0)--(1,-1);
            \fill[Turque!50, opacity=.9](0,0)--(-1,5)--(1,-3)--(0,0);
             \draw[blue, line join=round](-1,5)--(1,-3);
             \path(-1,-1)node[below]{\tiny$(-1,-1)$};
             \path(1,-1)node[above, rotate=-45]{\tiny$(1,-1)$};
             \filldraw[gray,fill=white](1,-3)circle(.6mm);
             \path(1,-3)node[left=-1pt]{\tiny$(1,-3)$};
            \draw[gray, -stealth](0,0)--(1,-3);
            \draw[stealth-stealth](-1,5)--(0,0)--(-1,-1);
            \fill[red](1,-2)circle(.6mm);
             \path(-1,5)node[right]{\tiny$(-1,5)$};
            \foreach\y in{-1,...,1} \foreach\x in{-1,0} \fill(\x,\y)circle(.6mm);
            \foreach\y in{2,...,5} \filldraw[fill=green!75!black](-1,\y)circle(.6mm);
             \path(-.25,-2.2)node{$\pDN{\FF[2]4}$};
            \path(-.55,4.95)node[right, rotate=-72]
              {\footnotesize$\ell(\pDN{\FF[2]m})=6\C2{{+}m}\C1{{-}(m{-}3)}$};
            \path(.25,4.75)node[right, rotate=-72]
              {\scriptsize$d(\pDN{\FF[2]m})=8$};
            }}
\vC{\TikZ{[scale=.67]\path[use as bounding box](-1.7,-3.2)--(1.5,5.2);
            \foreach\x in{-2,...,2}\draw[gray!50](\x/2,-3)--(\x/2,5);
            \foreach\y in{-6,...,10}\draw[gray!50](-1,\y/2)--(1,\y/2);
            \fill[Turque!50, opacity=.9](0,0)--(1,-1)--(-1,-1)--(-1,5)--(0,0);
             \draw[blue,thick,line join=round](1,-1)--(-1,-1)--(-1,5);
            \fill[red, opacity=.67](0,0)--(1,-1)--(1,-3)--(0,0);
             \draw[red, thick](1,-1)--(1,-3);
             \filldraw[ultra thin, fill=green!75!black](.5,-1)circle(1.5mm);
             \filldraw[ultra thin, fill=red](.5,-1)circle(1mm);
             \filldraw[gray,fill=white](1,-1)circle(.6mm);
             \draw[gray, -stealth](0,0)--(1,-1);
            \fill[Turque!50, opacity=.9](0,0)--(-1,5)--(1,-3)--(0,0);
             \draw[blue, line join=round](-1,5)--(1,-3);
             \filldraw[ultra thin, fill=green!75!black](.5,-1)circle(.5mm);
             \draw[gray, -stealth](0,0)--(1,-3);
             \filldraw[gray,fill=white](1,-3)circle(.6mm);
            \draw[stealth-stealth](-1,5)--(0,0)--(-1,-1);
            \filldraw[gray, fill=white](.5,-.5)circle(.6mm);
            \filldraw[gray, fill=white](.5,-3/2)circle(.6mm);
            \foreach\y in{-5,...,-3} \fill[red](1,\y/2)circle(.6mm);
            \foreach\y in{-2,...,2}\foreach\x in{-2,...,0} \fill(\x/2,\y/2)circle(.6mm);
            \foreach\y in{3,...,6} \filldraw[fill=green!75!black](-.5,\y/2)circle(.6mm);
            \foreach\y in{3,...,10} \filldraw[fill=green!75!black](-1,\y/2)circle(.6mm);
             \path(-.4,-2.2)node{$2\pDN{\FF[2]4}$};
            \path(-.75,5.5)node[right, rotate=-72]
              {\scriptsize$\ell(2\pDN{\FF[2]m})=15\C2{{+}(3m{+}2)}\C1{{-}(3m{-}8)}$};
            }}
  \vC{\begin{picture}(45,55)
        \put(10,37){\TikZ{[scale=.75]\path[use as bounding box](-2.2,-1.2)--(2,1.2);
            \foreach\x in{-2,...,1}\draw[gray](\x,-1)--(\x,1);
            \foreach\y in{-1,...,1}\draw[gray](-2,\y)--(1,\y);
             \fill[yellow, opacity=.8](0,0)--(0,1)--(-1,-1)--(1,0)--(0,0);
            \draw[blue,thick,line join=round](0,1)--(-1,-1)--(1,0);
             \fill[red, opacity=.67](0,0)--(0,1)--(-2,1)--(0,0);
            \draw[red, thick](-2,1)--(0,1);
             \fill[red](-1,1)circle(.6mm);
             \filldraw[gray, fill=white](0,1)circle(.5mm);
             \draw[gray, -stealth](0,0)--(0,1);
             \fill[yellow, opacity=.8](0,0)--(1,0)--(-2,1)--(0,0);
            \draw[blue,thick,line join=round](1,0)--(-2,1);
             \draw[stealth-stealth](1,0)--(0,0)--(-1,-1);
             \filldraw[gray, fill=white](-2,1)circle(.5mm);
             \draw[gray, -stealth](0,0)--(-2,1);
            \fill(0,0)circle(.5mm);
            \fill(1,0)circle(.5mm);
             \path(1,0)node[right]{\tiny$(1,0)$};
             \path(0,1)node[right]{\tiny$(0,1)$};
             \path(-2.2,1)node[below=-1pt]{\tiny$(-2,1)$};
            \fill(-1,-1)circle(.5mm);
             \path(-1,-1)node[above left=-1pt]{\tiny$(-1,-1)$};
             \path(-1.5,0)node{\large$\pDs{\EE[2]2}$};
            \path(-2.5,-1.4)node[right]{$\ell(\pDs{\EE[2]m})=4{-}m$};
            \path(-2.5,-2)node[right]{$d(\pDs{\EE[2]m})=3{-}m$};
            \path(-2.5,-3.2)node[right]{$\ell(2\pDs{\EE[2]m})=10{-}3m$};
            }}
        \put(10,2){\TikZ{[scale=.75]\path[use as bounding box](-2.2,-1.2)--(2,1.2);
            \foreach\x in{-4,...,2}\draw[gray](\x/2,-1)--(\x/2,1);
            \foreach\y in{-2,...,2}\draw[gray](-2,\y/2)--(1,\y/2);
             \fill[yellow, opacity=.8](0,0)--(0,1)--(-1,-1)--(1,0)--(0,0);
            \draw[blue, thick, line join=round](0,1)--(-1,-1)--(1,0);
             \fill[red, opacity=.67](0,0)--(0,1)--(-2,1)--(0,0);
            \draw[red, thick](-2,1)--(0,1);
             \filldraw[ultra thin, draw=red!70!black, fill=red](-.5,.5)circle(.9mm); 
             \filldraw[gray, fill=white](0,1)circle(.5mm);
             \draw[gray, -stealth](0,0)--(0,1);
             \filldraw[gray, fill=white](0,.5)circle(.5mm);
            \fill[yellow, opacity=.8](0,0)--(1,0)--(-2,1)--(0,0);
             \foreach\x in{-3,...,-1} \fill[fill=red](\x/2,1)circle(.6mm);
            \draw[blue, thick, line join=round](1,0)--(-2,1);
             \filldraw[gray, fill=white](-2,1)circle(.5mm);
             \draw[gray, -stealth](0,0)--(-2,1);
             \filldraw[gray, fill=white](-1,.5)circle(.5mm);
            \draw[stealth-stealth](1,0)--(0,0)--(-1,-1);
            \fill(-.5,.5)circle(.5mm);
            \foreach\x in{-1,...,2} \fill(\x/2,0)circle(.5mm);
            \foreach\x in{-1,0} \fill(\x/2,-.5)circle(.5mm);
            \fill(-1,-1)circle(.5mm);
             \path(-1.5,0)node{\large$2\pDs{\EE[2]2}$};
            }}
      \end{picture}}
  \vC{\begin{picture}(20,55)
        \put(0,30){
          \TikZ{[scale=.5]\path[use as bounding box](-1,-3)--(2,2);
            \foreach\x in{-1,...,2}\draw[gray!80](\x,-3)--(\x,2);
            \foreach\y in{-3,...,2}\draw[gray!80](-1,\y)--(2,\y);
            \fill[fill=Turque, opacity=.5](-1,2)--(-1,-3)--(0,-1)--(2,-1);
            \draw[blue, thick, line join=round](-1,2)--(-1,-3)--(0,-1)--(2,-1)--cycle;
            \foreach\x in{-1,...,2}
             {\pgfmathsetmacro{\yM}{1-\x}
               \foreach\y in{-1,...,\yM}\fill(\x,\y)circle(.9mm);
             }
             \fill(-1,-2)circle(.9mm); \fill(-1,-3)circle(.9mm);
            \path(0,1.3)node[right]{\large$\pDN{\ME[2]2}$};
            \path(-.5,-1.7)node[right]{\small$\ell(\pD)\<=10{+}m$};
            \path(-.5,-2.7)node[right]{\small$d(\pD)\<=9{+}m$};
            }}
        \put(0,0){
          \TikZ{[scale=.5]\path[use as bounding box](-1,-3)--(2,2);
            \foreach\x in{-2,...,4}\draw[gray!70](\x/2,-3)--(\x/2,2);
            \foreach\y in{-6,...,4}\draw[gray!70](-1,\y/2)--(2,\y/2);
            \fill[fill=Turque, opacity=.5](-1,2)--(-1,-3)--(0,-1)--(2,-1);
            \draw[blue, thick, line join=round](-1,2)--(-1,-3)--(0,-1)--(2,-1)--cycle;
            \foreach\x in{-2,...,4}
             {\pgfmathsetmacro{\yM}{2-\x}
               \foreach\y in{-2,...,\yM}\fill(\x/2,\y/2)circle(.9mm);
             }
            \foreach\x in{-2,...,-1}
             {\pgfmathsetmacro{\yM}{2*\x-2}
               \foreach\y in{\yM,...,-3}\fill(\x/2,\y/2)circle(.9mm);
             }
             \fill(1.5,-1)circle(.9mm); \fill(1.5,-.5)circle(.9mm); \fill(2,-1)circle(.9mm);
            \path(0,1.3)node[right]{\large$2\pDN{\ME[2]2}$};
            \path(-.5,-2)node[right]{\small$\ell(2\pD)\<=28{+}3m$};
            }}
      \end{picture}}
 \label{e:pD2pD}
\end{equation}
As itemized above, lattice points occurring in the interior of negatively (CW) oriented cones (red-shaded in~\eqref{e:pD2pD}) contribute negatively, while lattice points occurring in the 1-cone intersections of positively (CCW) and negatively (CW) oriented 2-cones are counted with both signs, i.e., effectively do not count\cite{rBH-gB}.
 For both infinite sequences in Figure~\ref{f:Fm+Em}, this indeed confirms the relations~\eqref{e:PDlkP}:
\begin{equation}
 \begin{array}{r|c|r@{~\6\checkmark=~}l|r@{~\6\checkmark=~}l|c}
  \bS{\pD} &\bS{d(\pD)} &\MC1{c@{\iS~}}{\bS{\ell(\pD)}} &\bS{d(\pD){+}1}
                        &\MC1{c@{\iS~}}{\bS{\ell(2\pD)}} &\bS{3d(\pD){+}1}
           &\MC1c{\bS{\F_{\!\pD}(t)}}\\ \toprule
  \pDs{\FF[2]m} &4 &5 &4{+}1 &13 &3(4){+}1 &1 +2t +t^2 \\ 
  \pDN{\FF[2]m} &8 &9 &8{+}1 &25 &3(8){+}1 &1 +6t +t^2 \\ \midrule
  \pDN{\EE[2]m} &9{+}m &10{+}m &9{+}m{+}1 &28{+}3m &3(9{+}m){+}1 & 1 +(7{+}m)t +t^2 \\ 
  \pDs{\EE[2]m} &3{-}m &4{-}m  &3{-}m{+}1 &10{-}3m &3(3{-}m){+}1 & 1 +(1{-}m)t +t^2 \\
  \bottomrule
 \end{array}
 \label{e:dll2F}
\end{equation}
Furthermore, the first row in~\eqref{e:dll2F} confirms the independently known values $\dim h^{q,q}(\FF[2]m)\<=(1,2,1)$. The transpolar (Newton) multitope produces the results in the 2nd row,
which correspondingly imply $\dim H^{2q}(\MF[2]m)\<=(1,6,1)$, consistently with the 12-Theorem: $\F_{\FF[2]m}(1){+}\F_{\MF[2]m}(1)\<=12$. This is consistent with Lemma~\ref{L:Hibi} holding for both the Hirzebruch surfaces, $\FF[2]m$, and their transpolar, $\MF[2]m$.

\paragraph{The $\EE[2]m$ Sequence:}
The sum of $\F$-polynomials in the bottom half of~\eqref{e:dll2F} is also consistent with the 12-Theorem, the $m$-dependent contributions canceling between the two polynomials: $\F_{\!\EE[2]m}(1){+}\F_{\ME[2]m}(1)\<=12$.
 On the other hand, $h_1(\EE[2]m)\<=(1{-}m)$ becomes negative for $m\<>1$, fails the second of conditions~\eqref{e:Hibi}, and clearly cannot equal a Betti number.
 Similarly for $m\<<0$, $\EE[2]m$ is the $|m|$-fold iterated blowup of $\IP^2$, its exceptional set consisting of a daisy-chain of $\IP^1$s, for which $\dim H^{1,1}(\EE[2]{-|m|},\ZZ)\<=1{+}|m|$. The transpolar toric space, $\ME[2]{-|m|}$, however has $\F(t)\<=1{+}(7{-}|m|)t{+}t^2$, where $h_1\<=7{-}|m|$ cannot possibly equal a Betti (or a Hodge) number when $|m|\<>7$.
 
A {\em\/partial solution\/} is presented by the {\em\/choice\/} of the overall $\pD$-orientation (using~\eqref{e:h1=dD-2}):
\begin{alignat}9
  \begin{array}{rc@{~\To~}c|c@{~\To~}c}
              & d(\pDs{\EE[2]m}) & h_1(\pDs{\EE[2]m})
               & d(\pDs{\ME[2]m}) & h_1(\pDs{\ME[2]m}) \\[2pt] \toprule
 \textbf{CCW}:& 3{-}m & 1{-}m &~~\,9{+}m &~~~~\,7{+}m \\ \midrule
  \textbf{CW}:& m{-}3 & m{-}5 &-9{-}m &-11{-}m \\ \bottomrule
  \end{array}
\end{alignat} 
This shows that switching the overall orientation of the flip-folded multitope $\pDs{\EE[2]m}$ does permit a standard cohomological interpretation of $h_1$:
\begin{subequations}
\label{e:2Emh1}
\begin{alignat}9
   \textbf{CCW}&:&\quad b_2(\EE[2]m)&\6\checkmark=h_1(\EE[2]m)&&=1{-}m&&\geq1,&\quad
    &\text{for}&\quad m&\leqslant0,\\
    \textbf{CW}&:&\quad b_2(\EE[2]m)&\6\checkmark=h_1(\EE[2]m)&&=m{-}5&&\geq1,&\quad
    &\text{for}&\quad m&\geqslant6,
\end{alignat}
\end{subequations}
excluding $m\<=1,\cdots,5$.
 In turn,
\begin{subequations}
\label{e:2*Emh1}
\begin{alignat}9
   \textbf{CCW}&:&\quad b_2(\ME[2]m)&\6\checkmark=h_1(\ME[2]m)&&=7{+}m&&\geq1,&\quad
    &\text{for}&\quad m&\geqslant{-}6,\\
    \textbf{CW}&:&\quad b_2(\ME[2]m)&\6\checkmark=h_1(\ME[2]m)&&={-}11{-}m&&\geq1,&\quad
    &\text{for}&\quad m&\leqslant{-}12,
\end{alignat}
\end{subequations}
excludes a standard cohomological interpretation of $h_1(\ME[2]m)$ for $m\<={-}11,\cdots,{-}7$.
 Notably, {\em\/both\/} $h_1(\EE[2]m)$ {\em\/and\/} $h_1(\ME[2]m)$ can {\em\/simultaneously\/} have a standard cohomological interpretation only with overall CCW orientation and for $m\<={-}6,\dots,0$, when $\EE[2]{-|m|}\<=\Bl^{|m|}(\IP^2)$.

Thus, while reversing the overall multigon orientation may be consistent with a standard cohomological interpretation of $\F_\pD(t)$ for some multigons, this is not true for VEX multigons in general, nor for transpolar pairs of VEX multigons {\em\/simultaneously\/}; see also \SS\:\ref{s:surg}.

\subsection{Dehn--Sommerville and Unimodality Relations in Three Dimensions}
\label{s:PF3}
Dehn--Sommerville relations for a 3-dimen\-sional $\pD$ imply
 $\F_{\!\pD}(t)\<=1{+}h_1t{+}h_1t^2{+}t^3$. Then,
\begin{subequations}
 \label{e:PDlkP3}
\begin{alignat}9
   \F_{\!\pD}(1)&=d(\pD) \quad\To\quad h_1\<=\frc12d(\pD){-}1, \label{e:2h1=dD-2}\\
   P_{\!\pD}(t) &=\F_{\!\pD}(t)\big/(1{-}t)^4\big\rfloor_{t^4=0}
   =1+(h_1{+}4)t +(5h_1{+}10)t^2 +(14h_1{+}21)t^3,
   \label{e:PD(t)3} \\
   \ell(\pD) &=\frc12d(\pD){+}3,\quad \ell(2\pD) =\frc52d(\pD){+}5,\quad
   \text{and}\quad  \ell(3\pD)\<=7d(\pD){+}7.
   \label{e:lkP3}
\end{alignat}
\end{subequations}
It remains to verify the relations~\eqref{e:lkP3}, which we do by direct counting:
\begin{equation}
 \begin{array}{r|c|c|c|c|c}
           &            &\bS{\ell(\pD)}\iS &\bS{\ell(2\pD)}\iS &\bS{\ell(3\pD)}\iS &\\[-3pt]
  \bS{\pD} &\bS{d(\pD)} &\bS{\frc12d(\pD){+}3} &\bS{\frc52d(\pD){+}5} &\bS{7d(\pD){+}7}
           &\MC1c{\bS{\F_{\!\pD}(t)}}\\ \toprule
  \pDs{\FF[3]m} &6  &6~\checkmark &20~\checkmark &49~\checkmark &1 +2t +2t^2 +t^3 \\ 
  \pDN{\FF[3]m} &54 &30~\checkmark &140~\checkmark &385~\checkmark
                &1 +26t +26t^2 +t^3 \\ \midrule
  \pDN{\EE[3]m} &64{+}8m &35{+}4m~\checkmark  &165{+}20m~\checkmark
                &455{+}56m~\checkmark &1 +(31{+}4m)t +(31{+}4m)t^2 +t^3 \\
  \pDs{\EE[3]m} &4{-}2m  &5{-}m~\checkmark    &15{-}5m~\checkmark
                &35{-}14m~\checkmark  &1 +(1{-}m)t +(1{-}m)t^2 +t^3 \\ 
  \bottomrule
 \end{array}
 \label{e:dll3F}
\end{equation}
The lattice points in the scaled flip-folded multitopes,
 $k\pDN{\FF[3]m}$ and $k\pDs{\EE[3]m}$,
are counted according to the orientation of the cones containing them in their relative interior, directly extending the counting that was used in\cite{rBH-gB} to compute the Euler characteristic and the Betti numbers of the Calabi--Yau hypersurface.

The so-computed $\F_\pD(t)$-polynomial correctly identifies the known values
 $h^{1,1}(\FF[3]m)\<=2\<=h^{2,2}(\FF[3]m)$. The corresponding identifications
 $b_2(\MF[3]m)\<=26\<=b_4(\MF[3]m)$ are consistent with the general expectation. (The signed total of 81 subdividing unit-degree edges\footnote{Each VEX multitope is a star-shaped domain with a signed multiplicity, and the multifan spanned by $\pDN{\FF[3]m}\<=\pDs{\MF[3]m}$ covers $M\<{\otimes_\ZZ}\IR\approx\IR^3$ effectively once: the winding number $w(\pDs{\MF[3]m})\<=1$ and so is the Todd genus, $\Td(\MF[3]m)\<=1$. Therefore, $\pDN{\FF[3]m}\<=\pDs{\MF[3]m}$ is homotopic (with signed multiplicity) to a 3-ball, and $\vd\pDs{\MF[3]m}$ to $S^2$, for which Euler's well-known triangulation result determines the number of unit-degree edges as
 $(\ell(\pDN{\FF[3]m}){-}1){+}d(\pDN{\FF[3]m}){-}2\<=81$ --- which is verified by direct counting.} connecting the signed total of 29 lattice points of $\vd\pDN{\FF[3]m}$ encode a wealth of MPCP-desingularizations, which contribute to the $2q$-cohomology.)

On the other hand, $h_1(\EE[3]m)\<=(1{-}m)$ in $\F_{\!\EE[3]m}(1)$ becomes negative for $m\<>1$ and so clearly cannot equal a Betti number, just as in the 2-dimensional case~\eqref{e:dll2F}.
 Again, for $m\<<0$, $\EE[2]m$ is an $|m|$-fold iterated blowup of $\IP^3$. Now $h_1$ for the transpolar toric space, $\ME[3]{m}$, becomes negative for $m\<<{-}8$ and so cannot possibly equal a Betti number.
 As in \SS\:\ref{s:PF2} for $(\EE[2]m,\ME[2]m)$, we use~\eqref{e:2h1=dD-2} and now check consistency:
\begin{subequations}
\label{e:3Emh1}
\begin{alignat}9
   \textbf{CCW}&:&\quad b_2(\EE[3]m)&\6\checkmark=h_1(\EE[3]m)&&=1{-}m&&\geq1,&\quad
    &\text{for}&\quad m&\leqslant0,\\
    \textbf{CW}&:&\quad b_2(\EE[3]m)&\6\checkmark=h_1(\EE[3]m)&&=m{-}3&&\geq1,&\quad
    &\text{for}&\quad m&\geqslant4,
\end{alignat}
\end{subequations}
excluding $m\<=1,2,3$ --- which is better than~\eqref{e:2Emh1}.
 In turn,
\begin{subequations}
\label{e:3*Emh1}
\begin{alignat}9
   \textbf{CCW}&:&\quad b_2(\ME[3]m)&\6\checkmark=h_1(\ME[3]m)&&=&31{+}4m&\geq1,&\quad
    &\text{for}&\quad m&\geqslant{-}7,\\
    \textbf{CW}&:&\quad b_2(\ME[3]m)&\6\checkmark=h_1(\ME[3]m)&&=&{-}33{-}4m&\geq1,&\quad
    &\text{for}&\quad m&\leqslant{-}9,
\end{alignat}
\end{subequations}
excludes a {\em\/standard\/} cohomological interpretation of $h_1(\ME[2]m)$ only for $m\<={-}8$ --- again better than~\eqref{e:2*Emh1}. Notably, {\em\/both\/} $h_1(\EE[3]m)$ {\em\/and\/} $h_1(\ME[3]m)$ can {\em\/simultaneously\/} have a {\em\/standard\/} cohomological interpretation only with overall CCW orientation and for $m\<={-}6,\dots,0$, when $\EE[3]{-|m|}\<=\Bl^{|m|}(\IP^3)$.

This last result being the {\em\/same\/} as in the 2-dimensional case reinforces the closing conclusion of \SS\:\ref{s:PF2}, that the {\em\/standard\/} (co)homological interpretation of $\F_\pD(t)$ cannot hold for VEX multitopes in general. The above computations suggest the
$h_i$-coefficients in $\F_\pD(t)$ to correspond to a (flip-folding signed)
$\rT$-twisted trace over $H^\rT_{2i}(X_\pD)$, reproducing the {\em\/standard\/} $b_{2i}\<\coeq\dim H_{2i}(X_\pD)$ for non-folded $\pD$.

\subsection{Dehn--Sommerville and Unimodality Relations in Four Dimensions}
\label{s:PF4}
By definition, VEX polytopes are star-shaped with respect to a preferred $L$-lattice point (the origin)\cite{rBH-gB}, so Lemma~\ref{L:Hibi} applies to all of them. 
It has also been conjectured\cite{Ohsugi:2006aa} that unimodality, $h_{k-1}\<\leqslant h_k$ for $k\<\in[1,\lfloor n/2\rfloor]$, requires an additional (``integer decomposition'') property\cite{Adiprasito:2022aa} --- which the above computations show {\em\/fails\/} for at least some VEX multitopes, such as $\pDs{\EE{m}}$. 

Focusing thus on the Dehn--Sommerville relations (Poincar\'e duality), for 4-dimensional $\pD$,
\begin{equation}
 \F_\pD(t) =\big(1{+}\sum_{k=1}^4\ell(k\pD)t^k\big)(1{-}t)^5\big\rfloor_{t^5=0}
 =1+h_1t+h_2t^2+h_1t^3+t^4
\end{equation}
which implies
\begin{equation}
  \begin{aligned}
 \ell(3\pD){-}5\ell(2\pD)\<+10\ell(\pD)-10&\Is\ell(\pD){-}5
  &\To\quad
  &\ell(3\pD)=5\ell(2\pD){-}9\ell(\pD){+}5,\\
 \ell(4\pD){-}5\ell(3\pD)\<+10\ell(2\pD)-10\ell(\pD)+5&\Is1
  &\To\quad
  &\ell(4\pD)=15\ell(2\pD){-}35\ell(\pD){+}21.
\end{aligned}
 \tag{\ref{e:DS34}$'$}
\end{equation}
Using also that
\begin{equation}
   d(\pD) = \F_\pD(1)
   = 1 \<+(\underbrace{\ell(\pD){-}5}_{h_1})1
       \<+\big(\underbrace{\ell(2\pD){-}5\ell(\pD)\<+10}_{h_2}\big)1^2
       \<+(\ell(\pD){-}5)1^3 \<+1^4 ,
\end{equation}
implies that
\begin{equation}
  d(\pD) = \ell(2\pD){-}3\ell(\pD){+}2,
   \qquad\text{i.e.},\qquad
  \ell(2\pD)\<=3\ell(\pD){+}d(\pD)-2,
 \tag{\ref{e:DSd}$'$}
\end{equation}
which allows rewriting the formula~\eqref{e:dd*ll*} for the Euler number of the Calabi--Yau hypersurface as
\begin{equation}
  C_3(Z_f) \isBy{\eqref{e:DSd}}
            2\Big[ \big(\ell(2\pDN{X}) - \ell(2\pDs{X})\big)
                  -9\big(\ell(\pDN{X}) \<-\ell(\pDs{X})\big)\Big].
 \tag{\ref{e:l2D-lD}$'$}
\end{equation}

The weaker condition of unimodality would imply that
\begin{subequations}
 \label{e:Lefschetz}
\begin{alignat}9
   h_1&\leqslant h_2
   ~~\To&\quad&
   \ell(\pD)\leqslant\frc{13}3+\frc13d(\pD),
    \quad\text{i.e.}\quad
   d(\pD)\geqslant3\ell(\pD)-13,\\
   1\<\leqslant h_1&\leqslant h_2
   ~~\To&\quad&
   d(\pD)\geqslant5~~\&~~6\leqslant\ell(\pD)\leqslant\frc{13}3+\frc13d(\pD).
\end{alignat}
\end{subequations}
Based on the above results for $\EE[2]m$ and $\EE[3]m$, these are expected to fail for some multitopes, but perhaps hold for torus manifolds with the ``additional (integer decomposition) property\cite{Adiprasito:2022aa}.'' A systematic testing of the above relations and the inequalities~\eqref{e:Lefschetz} is clearly best left to computer-aided counting.

\section{Surgical Relations}
\label{s:surg}
A VEX multitope and its star-subdividing multifan, $\pDs{X}\<\lat\pFn{X}$,
both correspond to a toric space, $X$, {\em\/directly\/}\footnote{In turn, the transpolar Newton multitope and its star-subdividing multifan, $\pDN{X}\<\lat\pFN{X}$, encode anticanonical sections, $f\<\in\G(\cKs{X})$, {\em\/over\/} a toric space, $X$, whereby $\pDN{X}$ corresponds to $X$ {\em\/dually.}}: Top-dimensional cones, $\s\<\in\pFn{X}$, correspond to open affine charts in $X$, two charts ``gluing'' over $U_\s\<\cap U_{\s'}$ that corresponds to the common facet, $\vs\<\coeq\s\<\cap\s'$, as provided by the multifan
$\pFn{X}$ (=\,a facet-ordered poset\cite{Masuda:2000aa, rHM-MFs}).
 This restricts the choices among toric spaces that correspond to a given multifan.

\subsection{Complex vs.\ Pre-Complex Structure}
\label{s:pCpx}
\begin{figure}[htb]
$$
\begin{array}{@{}c@{\qquad\quad}c@{\qquad\quad}c@{\qquad\quad}c@{\qquad\quad}c@{}}
\vC{\TikZ{[scale=.8]\path[use as bounding box](-2.2,-2)--(1.2,2);
           \path(-.47,.15)node
           {\includegraphics[width=27mm]{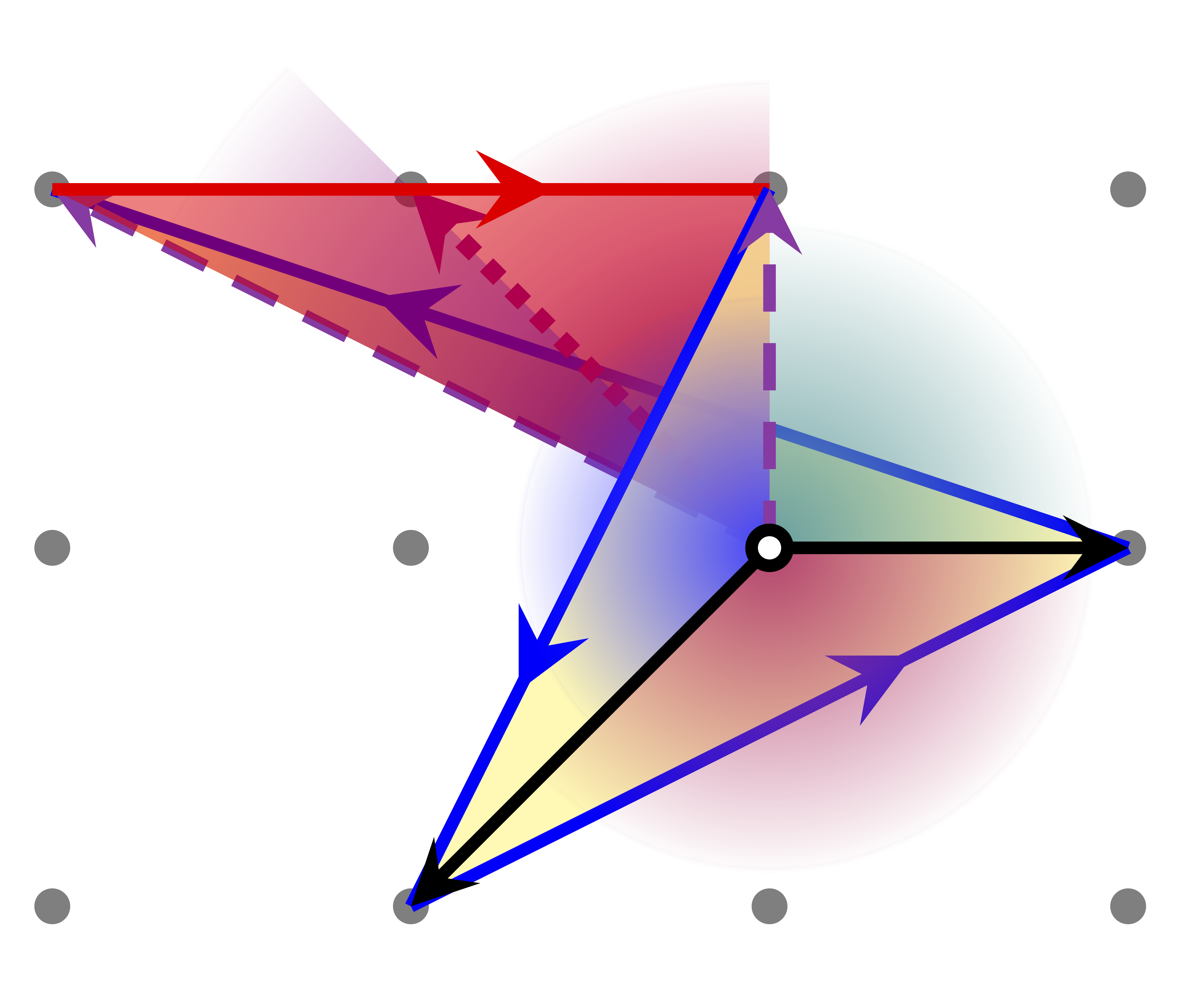}};
           \path(1,0)node[below]{$\sss1$};
           \path[Purple](-2,1)node[below]{$\sss2$};
           \path[Purple](0,1)node[right]{$\sss3$};
           \path(-1,-1)node[above left=-2pt]{$\sss4$};
           \path(-.5,-1.5)node{$\EE[2]2$};
            }}
 &
\vC{\TikZ{[scale=.8]\path[use as bounding box](-1.2,-2)--(1.2,2);
           \path(-.01,.1)node
           {\includegraphics[width=18mm]{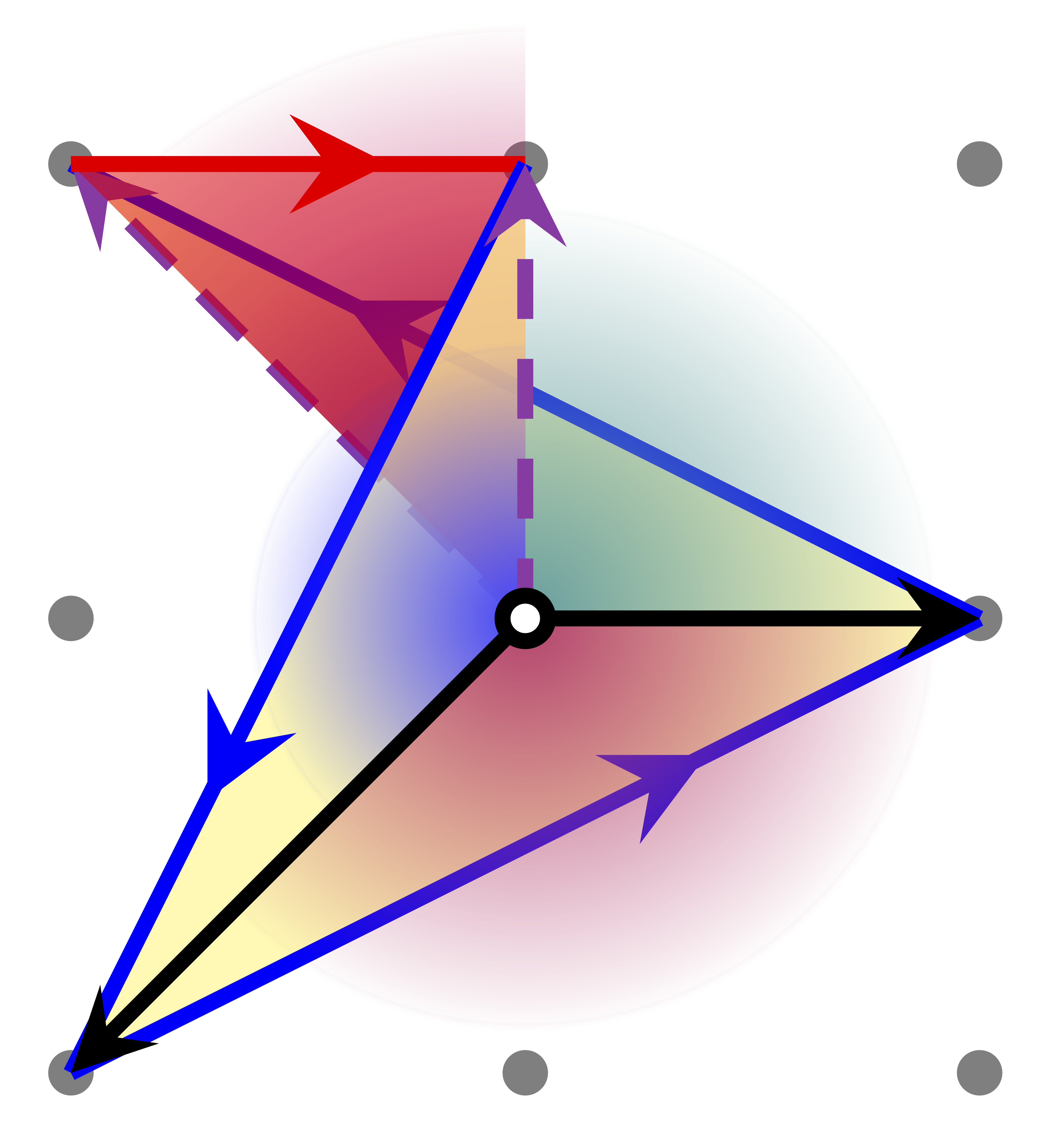}};
           \path(1,0)node[below]{$\sss1$};
           \path(-1,1)node[below]{$\sss2$};
           \path(0,1)node[right]{$\sss3$};
           \path(-1,-1)node[above left=-2pt]{$\sss4$};
           \path(0,-1.5)node{$\EE[2]1$};
            }}
 &
\vC{\TikZ{[scale=.8]\path[use as bounding box](-1.2,-2)--(1.2,2);
           \path(0,0)node
           {\includegraphics[width=18mm]{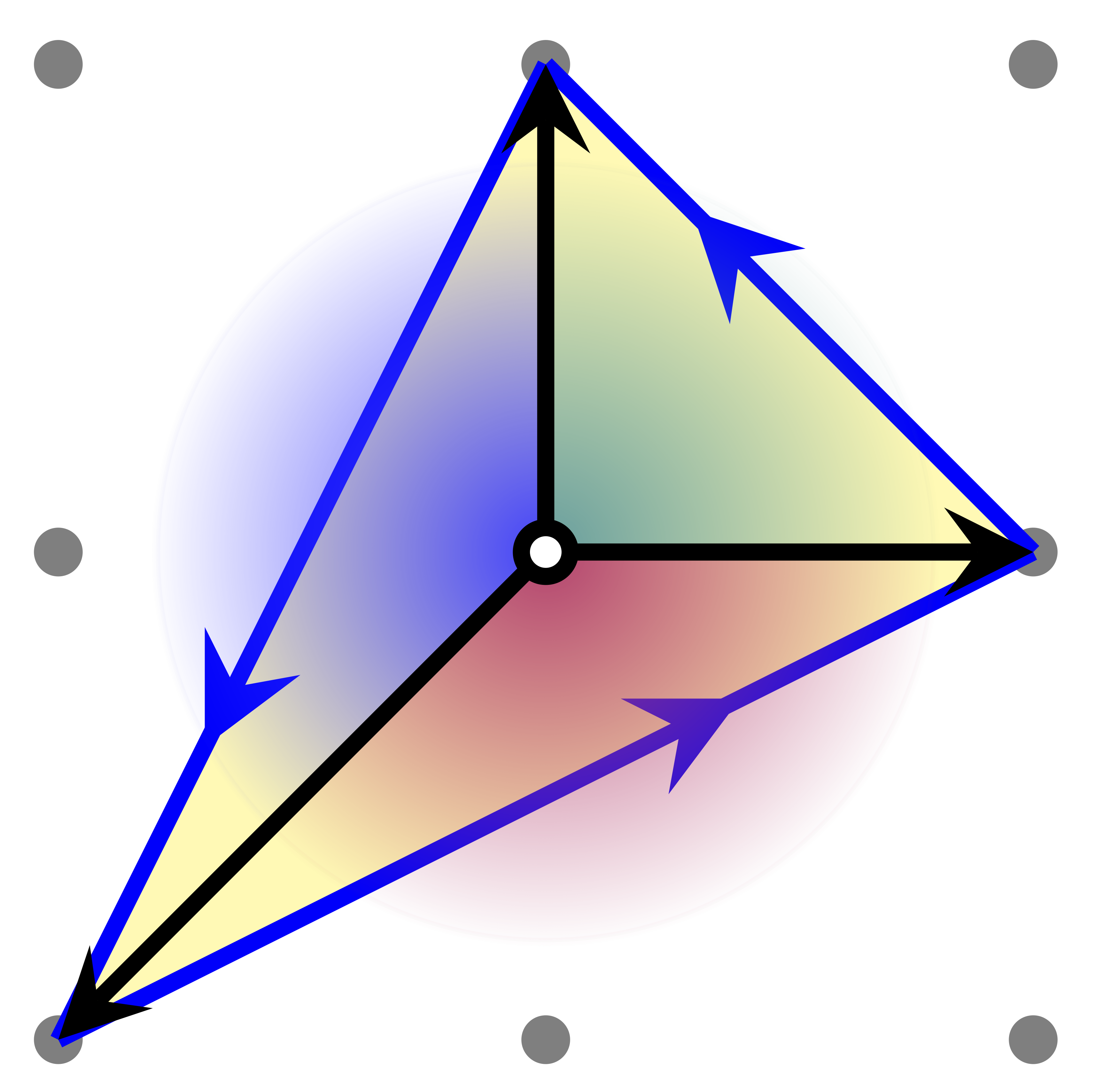}};
           \path(1,0)node[below]{$\sss1$};
           \path(0,1)node[left]{$\sss3$};
           \path(-1,-1)node[above left=-2pt]{$\sss4$};
           \path(0,-1.5)node{$\EE[2]0\<=\IP^2$};
            }}
 &
\vC{\TikZ{[scale=.8]\path[use as bounding box](-1.2,-2)--(1.2,2);
           \path(0,0)node
           {\includegraphics[width=19mm]{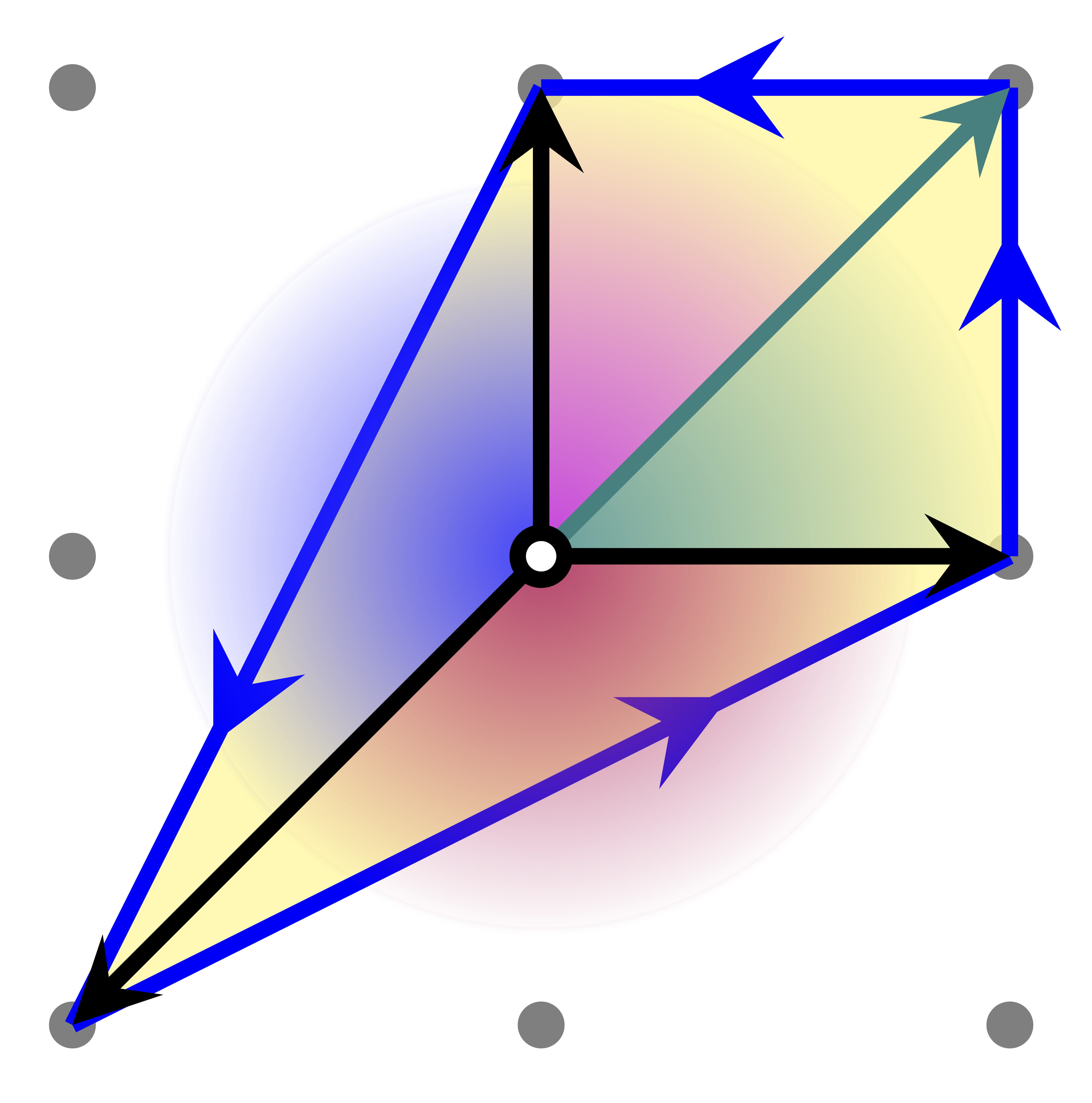}};
           \path(1,0)node[below]{$\sss1$};
           \path(1,1)node[right]{$\sss2$};
           \path(0,1)node[left]{$\sss3$};
           \path(-1,-1)node[above left=-2pt]{$\sss4$};
           \path(0,-1.5)node{$\EE[2]{-1}$\footnotesize
               $\<=\Bl_*(\IP^2)\<=\FF[2]1$};
            }}
 &
\vC{\TikZ{[scale=.8]\path[use as bounding box](-1.2,-2)--(2.2,2);
           \path(.55,0)node
           {\includegraphics[width=28mm]{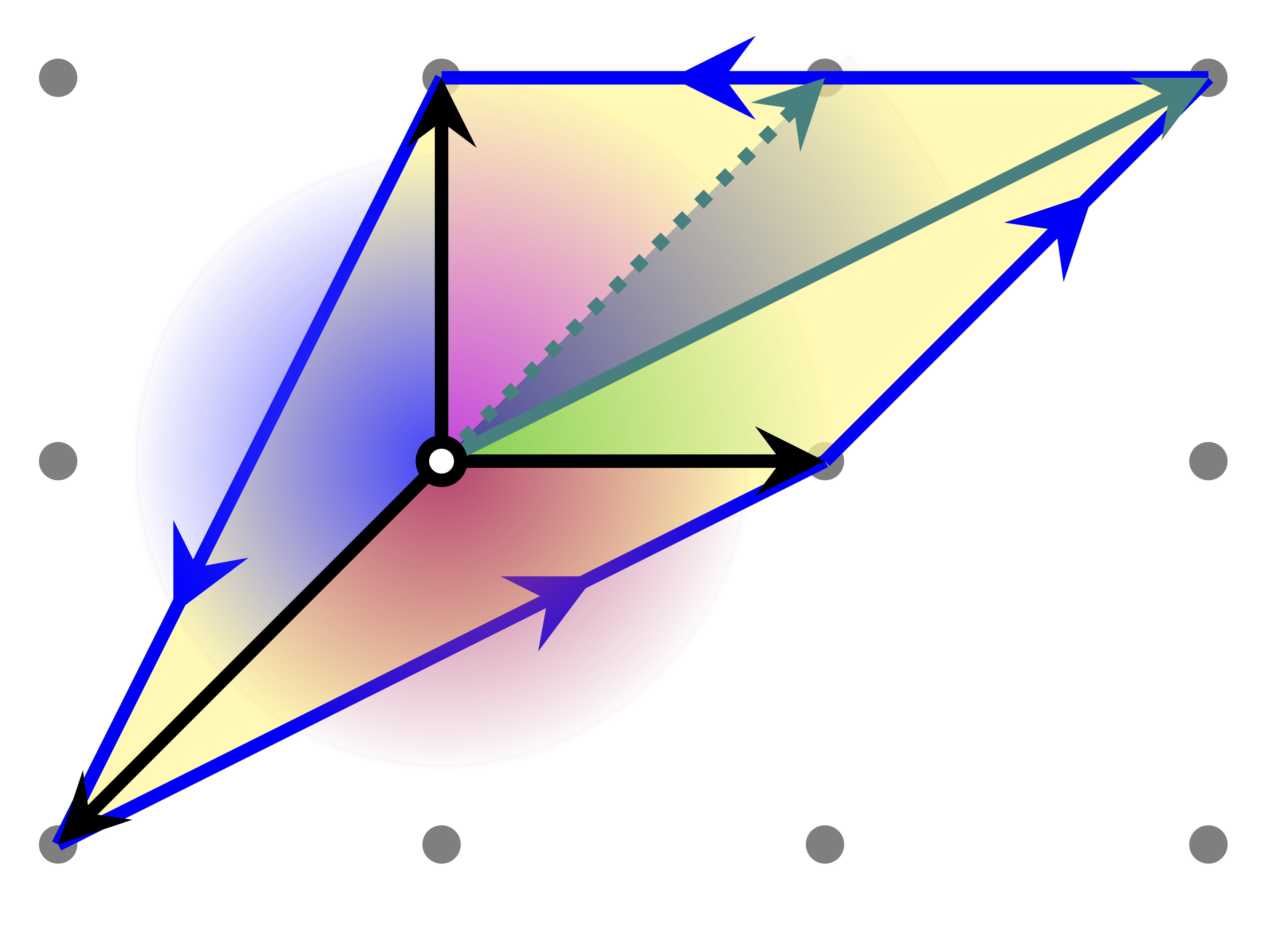}};
           \path(1,0)node[below]{$\sss1$};
           \path(2,1)node[below]{$\sss2$};
           \path(0,1)node[left]{$\sss3$};
           \path(-1,-1)node[above left=-2pt]{$\sss4$};
           \path(.5,-1.5)node{$\EE[2]{-2}$\footnotesize
               $\<=\Bl_*^2(\IP^2)$};
            }}
\\[5mm]
\vC{\TikZ{[scale=.8, every node/.style={thick, inner sep=.5, outer sep=0}]
         \path[use as bounding box](-2,-2)--(1,1.5);
      \node[circle, draw=blue, fill=white](1) at(1,0)
          {\scriptsize$3$};
      \node[Purple, circle, draw=Purple, fill=white](2) at(-2,1)
          {\scriptsize$1$};
      \node[red, circle, draw=red, fill=white](5) at(-1,1) 
          {\scriptsize$2$};
      \node[Purple, circle, draw=Purple, fill=white](3) at(0,1)
          {\scriptsize$2$};
      \node[circle, draw=blue, fill=white](4) at(-1,-1)
          {\scriptsize$1$};
      \draw[Rouge!50, very thick, densely dotted](0,0)--(5);
      \draw[gray!50, thick, -stealth](0,0)--(1);
      \draw[Purple!50, thick, -stealth](0,0)--(2);
      \draw[Purple!50, thick, -stealth](0,0)--(3);
      \draw[gray!50, thick, -stealth](0,0)--(4);
      \draw[blue, thick, midarrow=stealth](1)--node[above=2pt]
          {\scriptsize$1$}(2);
      \draw[red, thick, midarrow=stealth](2)--node[above=1pt]
          {\scriptsize$-1$}(5);
      \draw[red, thick, midarrow=stealth](5)--node[above=1pt]
          {\scriptsize$-1$}(3);
      \draw[blue, thick, midarrow=stealth](3)--node[right=2pt]
          {\scriptsize$1$}(4);
      \draw[blue, thick, midarrow=stealth](4)--node[above=2pt]
          {\scriptsize$1$}(1);
      \path(.25,-1.5)node{\footnotesize$[D_i{\cdot}D_j]_{\EE[2]2}$};
      \filldraw[fill=white, thick](0,0)circle(.5mm);
            }}
 &
 \vC{\TikZ{[scale=.8, every node/.style={thick, inner sep=.5, outer sep=0}]
         \path[use as bounding box](-1,-2)--(1,1.5);
      \node[circle, draw=blue, fill=white](1) at(1,0)
          {\scriptsize$2$};
      \node[Purple, circle, draw=Purple, fill=white](2) at(-1,1)
          {\scriptsize$1$};
      \node[Purple, circle, draw=Purple, fill=white](3) at(0,1)
          {\scriptsize$2$};
      \node[circle, draw=blue, fill=white](4) at(-1,-1)
          {\scriptsize$1$};
      \draw[gray!50, thick, -stealth](0,0)--(1);
      \draw[Purple!50, thick, -stealth](0,0)--(2);
      \draw[Purple!50, thick, -stealth](0,0)--(3);
      \draw[gray!50, thick, -stealth](0,0)--(4);
      \draw[blue, thick, midarrow=stealth](1)--node[above=-1pt, xshift=3mm]
          {\scriptsize$1$}(2);
      \draw[red, thick, midarrow=stealth](2)--node[above=1pt]
          {\scriptsize$-1$}(3);
      \draw[blue, thick, midarrow=stealth](3)--node[right=2pt]
          {\scriptsize$1$}(4);
      \draw[blue, thick, midarrow=stealth](4)--node[above=2pt]
          {\scriptsize$1$}(1);
      \path(.25,-1.5)node{\footnotesize$[D_i{\cdot}D_j]_{\EE[2]1}$};
      \filldraw[fill=white, thick](0,0)circle(.5mm);
            }}
 &
 \vC{\TikZ{[scale=.8, every node/.style={thick, inner sep=.5, outer sep=0}]
         \path[use as bounding box](-1,-2)--(1,1.5);
      \node[circle, draw=blue, fill=white](1) at(1,0)   {\scriptsize$1$};
      \node[circle, draw=blue, fill=white](3) at(0,1)   {\scriptsize$1$};
      \node[circle, draw=blue, fill=white](4) at(-1,-1) {\scriptsize$1$};
      \draw[gray!50, thick, -stealth](0,0)--(1);
      \draw[gray!50, thick, -stealth](0,0)--(3);
      \draw[gray!50, thick, -stealth](0,0)--(4);
      \draw[blue, thick, midarrow=stealth](1)--node[above right=1pt]
          {\scriptsize$1$}(3);
      \draw[blue, thick, midarrow=stealth](3)--node[right=2pt]
          {\scriptsize$1$}(4);
      \draw[blue, thick, midarrow=stealth](4)--node[above=2pt]
          {\scriptsize$1$}(1);
      \path(.25,-1.5)node{\footnotesize$[D_i{\cdot}D_j]_{\IP^2}$};
      \filldraw[fill=white, thick](0,0)circle(.5mm);
            }}
 &
 \vC{\TikZ{[scale=.8, every node/.style={thick, inner sep=.5, outer sep=0}]
         \path[use as bounding box](-1,-2)--(1,1.5);
      \node[circle, draw=blue, fill=white](1) at(1,0)   {\scriptsize$0$};
      \node[rounded corners=5pt, inner sep=2, draw=blue, fill=white](2)
                                              at(1,1)   {\scriptsize$-1$};
      \node[circle, draw=blue, fill=white](3) at(0,1)   {\scriptsize$0$};
      \node[circle, draw=blue, fill=white](4) at(-1,-1) {\scriptsize$1$};
      \draw[gray!50, thick, -stealth](0,0)--(1);
      \draw[gray!50, thick, -stealth](0,0)--(2);
      \draw[gray!50, thick, -stealth](0,0)--(3);
      \draw[gray!50, thick, -stealth](0,0)--(4);
      \draw[blue, thick, midarrow=stealth](1)--node[right=2pt]
          {\scriptsize$1$}(2);
      \draw[blue, thick, midarrow=stealth](2)--node[above=2pt]
          {\scriptsize$1$}(3);
      \draw[blue, thick, midarrow=stealth](3)--node[right=2pt]
          {\scriptsize$1$}(4);
      \draw[blue, thick, midarrow=stealth](4)--node[above=2pt]
          {\scriptsize$1$}(1);
      \path(.25,-1.5)node{\footnotesize$[D_i{\cdot}D_j]_{\EE[2]{-1}}$};
      \filldraw[fill=white, thick](0,0)circle(.5mm);
            }}
 &
 \vC{\TikZ{[scale=.8, every node/.style={thick, inner sep=.5, outer sep=0}]
         \path[use as bounding box](-1,-2)--(2,1.5);
      \node[rounded corners=5pt, inner sep=2, draw=blue, fill=white](1) 
                                              at(1,0)   {\scriptsize$-1$};
      \node[rounded corners=5pt, inner sep=2, draw=blue, fill=white](2)
                                              at(2,1)   {\scriptsize$-1$};
      \node[rounded corners=5pt, inner sep=2, draw=blue, fill=white](5)
                                              at(1,1)   {\scriptsize$-2$};
      \node[circle, draw=blue, fill=white](3) at(0,1)   {\scriptsize$0$};
      \node[circle, draw=blue, fill=white](4) at(-1,-1) {\scriptsize$1$};
      \draw[gray!50, thick, -stealth](0,0)--(1);
      \draw[gray!50, thick, -stealth](0,0)--(2);
      \draw[gray!50, thick, -stealth](0,0)--(3);
      \draw[gray!50, thick, -stealth](0,0)--(4);
      \draw[gray!50, very thick, densely dotted](0,0)--(5);
      \draw[blue, thick, midarrow=stealth](1)--node[right=2pt, yshift=-1mm]
          {\scriptsize$1$}(2);
      \draw[blue, thick, midarrow=stealth](2)--node[above=2pt]
          {\scriptsize$1$}(5);
      \draw[blue, thick, midarrow=stealth](5)--node[above=2pt]
          {\scriptsize$1$}(3);
      \draw[blue, thick, midarrow=stealth](3)--node[right=2pt]
          {\scriptsize$1$}(4);
      \draw[blue, thick, midarrow=stealth](4)--node[above=2pt]
          {\scriptsize$1$}(1);
      \path(.25,-1.5)node{\footnotesize$[D_i{\cdot}D_j]_{\EE[2]{-2}}$};
      \filldraw[fill=white, thick](0,0)circle(.5mm);
            }}
\\
  \IP^2\<\#\IP^2\<\#\2{\IP^2}
 &\IP^2\<\#\2{\IP^2}
 &\2{\IP^2}
 &\2{\IP^2}\<\#\7{\IP^2}
 &\2{\IP^2}\<\#\7{\IP^2}\<\#\7{\IP^2}
\end{array}
$$
 \caption{The sequence of multigons, $\EE[2]m$ for $m\<=2,\cdots,{-}2$, with the spanned multifans, and the corresponding intersection numbers of
 $\rT$-characteristic submanifolds underneath, identified as blowups (right-hand half) and blowup-like alterations (left-hand half) 
of an ``initial'' (underlined) $\IP^2$}
 \label{f:E2-E-2}
\end{figure}

Consider the sequence $\EE[2]m$ with $m\<=2,\cdots,{-}2$ depicted in Figure~\ref{f:E2-E-2}.%
With $\IP^2$ in the middle, the right-hand side depicts its standard blowup at a smooth point (center-right), followed by the (iterated) double blowup at far right. It is standard\cite[Prop.\,2.5.8, p.~102]{Huybrechts:2005aa} that
 $\EE[2]{-1}\<=\Bl_*(\IP^2)\<=\FF[2]1$ is diffeomorphic (as an oriented differentiable manifold) to the connected sum, $\IP^2\#\7{\IP^2}$, where 
$\7{\IP^2}$ is a complex projective space with its orientation opposite to the one naturally induced by the complex structure.\footnote{The orientation reversal is necessary for the complex structure of one component to extend through the other component compatibly with the orientation of the underlying real space: At the crossover/gluing region, the ``radially'' inward directions in one component must match the ``radially'' outward directions in the other component.}
Correspondingly, the 1-cone $(1,1)\<\in\pFn{\EE[2]{-1}}$
encodes the {\em\/exceptional set\/} of
self-intersection $[\sE_{\sss(1,1)}]^2\<={-}1$,
which is the base $\IP^1$ of a {\em\/local\/} $\cO_{\IP^1}(-1)$-like neighborhood\cite{Huybrechts:2005aa}. This exceptional set,
$\sE_{\sss(1,1)}\<\approx\IP^1$, may be identified with the hyperplane\footnote{It is also convenient to identify the hyperplane 
$\IP^1\<\subset\IP^2$ with the ``$\IP^1$ at infinity'' in the cell-decomposition, $\IP^2\simeq\IC^2\<\cup\IP^1$.} in the 2nd (orientation-reversed) summand in the connected sum, $\2{\IP^2}\#\7{\IP^2}$, with ``the rest'' of the fan $\pFn{\EE[2]{-1}}$ corresponding to the 1st (underlined) summand.
The iterates, $\EE[2]{-2}\<=\Bl_*^2(\IP^2)$ in Figure~\ref{f:E2-E-2} (far right) and $\EE[2]{-m}$ for $m\<\geqslant1$ in general, are blowups of
$\IP^2$ at ``infinitely near points,'' easily identified as a daisy-chain connected sum of $\IP^2$'s, with orientations opposite to the ``initial'' (underlined) one. 

On the other hand, the 1-cone $\n_2\<=(-1,1)\<\in\pFn{\EE[2]1}$ (Figure~\ref{f:E2-E-2}, left) encodes an exceptional set of self-intersection $[\sE_{\sss(-1,1)}]^2\<={+}1$, 
suggesting that $\EE[2]1$ is {\em\/akin to a blowup,} diffeomorphic to
$\EE[2]1\<{\simeq_\IR}\IP^2\<\#\2{\IP^2}$, however without the orientation-reversal in either summand,\footnote{We thank M.~Masuda for suggesting this interpretation; for a not too different but additionally specified example, see\cite[\SS\:5]{Ishida:2013aa}. The example of $\IP^2\<\#\IP^2$ as a torus manifold that is not even almost complex was given already in\cite{Davis:1991uz}.} and so not admitting a global complex structure.
Extending the chart-wise complex structure
from $U_{\n_{12}}$
across $(U_{\n_{12}}\<\cap U_{\n_{23}})\<\supset\sE_{(-1,1)}$
into $U_{\n_{23}}$
is obstructed by the flipping of the relative orientations of the stable tangent bundle and the underlying torus manifold:
from $w(\n_{12})\<={+}1$ to $w(\n_{23})\<={-}1$.
The non-compact space $\EE[2]1\ssm U_{\n_{23}}$ is complex,
but the compact $\EE[2]1$ is only precomplex;
the affine chart $U_{\n_{23}}$ being contractible, this obstruction to the complex structure can be made arbitrarily small, and so is localized.
This is somewhat akin to the fact that
$(S^4\<\ssm\{\text{pt}\})\<\approx\IC^2$ is complex, while $S^4$ is not.\footnote{\label{fn:S4}The precomplex structure on $S^4$ nevertheless suffices to define a Dolbeault (anti)holomorphic exterior derivative and cohomology\cite{rS-S4notCplx}, which may well suffice for most string theory applications.}
 Then, the $\EE[2]m$ for $m\<>1$ are daisy-chain connected sums of
 $(m{+}1)$ $\IP^2$'s without any orientation reversal, and so without a global complex structure and only precomplex.
\begin{conj}[cont.\ Conj.~\ref{C:obstruct}]
\label{c:preC}
All flip-folded multitopes, such as $\pDs{\MF{m}}$ for $\,m\<\geqslant3$ and $\pDs{\EE{m}}$ for $\,m\<\geqslant1$, correspond to only
{\em\/precomplex\/} torus manifolds, where the flip-folded ``extensions'' encode the obstruction to extending a chart-wise complex structure throughout the atlas, such as in
 $U_{\n^\wtd_3}\<\subset\MF{m}$ and $U_{\n_{23}}\<\subset\EE{m}$.
Since flip-folded portions of a multitope, $\pDs{X}$, are transpolar to non-convex regions of the transpolar multitope, $\pDN{X}\<=\pDs{\tX}$, non-convexity in $\pDs{\tX}$ encodes (by mirror symmetry) obstructions to extending a {\em\/presymplectic\/} structure of $\,\tX$ to a global symplectic structure, such as in $\FF{m}$ and $\ME{m}$.
\end{conj}
The ``optimal model'' of Remark~\ref{r:optimal} and Conjecture~\ref{c:optimal} is thereby seen as a torus manifold wherein the obstructions from Conjectures~\ref{C:obstruct} and~\ref{c:preC} are minimized in a suitable $\rT$-equivariant (co)homology, 
and especially within the Calabi--Yau hypersurfaces of interest; see footnote~\ref{fn:preStr} and Remark~\ref{r:C-infty}.

\subsection{Being Star-Triangulable}
\label{s:StarSh}
The definition of VEX multitopes as being star-triangulable (and with $w$-signed multiplicity) relates the numbers of their faces (i.e., cones in the multifans they span) by Euler's relation. Even/odd-dimensional multitopes $\pD$ have their boundary, $\vd\pD$, homotopic to an odd/even-dimensional sphere with $\c(\vd\pD)\<=0$, i.e., $\c(\vd\pD)\<=2$, respectively, while $\c(\pD)\<=1$ in both cases. Using the ($w$-signed) face-counting function, $\sharp(\n)$, shows this to hold regardless of the winding number, $w$:
\begin{gather}
  \vC{\TikZ{\path[use as bounding box](-1,-1.2)--(1,1.2);
       \foreach\x in{-1,...,1}
        \foreach\y in{-1,...,1}
         \fill[gray](\x,\y)circle(.5mm);
       \fill[cyan!33, opacity=.9](0,0)--(1,0)--(-1,1)--(0,-1)--(0,0);
        \draw[blue, thick, midarrow=stealth](1,0)--(-1,1);
        \draw[blue, thick, midArrow={pos=.35, end={Stealth[scale=.9]}}]
            (-1,1)--(0,-1);
        \draw[-Stealth](0,0)--(-1,1);
       \fill[cyan!33, opacity=.9](0,0)--(0,-1)--(1,1)--(-1,0)--(0,0);
        \draw[-Stealth](0,0)--(0,-1);
        \draw[blue, thick, midarrow=stealth](0,-1)--(1,1);
        \draw[blue, thick, midArrow={pos=.35, end={Stealth[scale=.9]}}]
            (1,1)--(-1,0);
        \draw[-Stealth](0,0)--(1,1);
       \fill[cyan!33, opacity=.9](0,0)--(-1,0)--(1,-1)--(0,1)--(0,0);
        \draw[-Stealth](0,0)--(-1,0);
        \draw[blue, thick, midarrow=stealth](-1,0)--(1,-1);
        \draw[blue, thick, midarrow=stealth](1,-1)--(0,1);
        \draw[-Stealth](0,0)--(1,-1);
       \fill[cyan!33, opacity=.9](0,0)--(0,1)--(-1,-1)--(1,0)--(0,0);
        \draw[blue, thick, midarrow=stealth](0,1)--(-1,-1);
        \draw[blue, thick, midarrow=stealth](-1,-1)--(1,0);
        \draw[-Stealth](0,0)--(0,1);
        \draw[-Stealth](0,0)--(1,0);
        \draw[-Stealth](0,0)--(-1,-1);
       \path(1,0)  node[above]{$\sss1$};
       \path(-1,1) node[below]{$\sss2$};
       \path(0,-1) node[right]{$\sss3$};
       \path(1,1)  node[left]{$\sss4$};
       \path(-1,0) node[above]{$\sss5$};
       \path(1,-1)node[left]{$\sss6$};
       \path(0,1)node[left]{$\sss7$};
       \path(-1,-1)node[right]{$\sss8$};
       \filldraw[fill=white, thick](0,0)circle(.5mm);
       \path(1.25,0)node[right]{$w\<=3$};
       \path(-.8,-.5)node[left]{$\pD_1$};
            }}
 \quad
  \vC{\small$\begin{array}{@{}r@{~=~}l}
   \sharp(\n_{i,i+1}),~ \sharp(\n_i) &1\\ \toprule
   \sum_i \sharp(\n_{i,i+1}) &8\\
   \sum_i \sharp(\n_i) &8\\ \bottomrule
   \sum_i \sharp(\n_{i,i+1}){-}\sum_i \sharp(\n_i) &0
  \end{array}$}
 \qquad\qquad
  \vC{\TikZ{\path[use as bounding box](-1,-1.2)--(1,1.2);
       \foreach\x in{-1,...,1}
        \foreach\y in{-1,...,1}
         \fill[gray](\x,\y)circle(.5mm);
       \fill[cyan!33, opacity=.9](0,0)--(1,0)--(-1,1)--(0,-1)--(0,0);
        \draw[blue, thick, midarrow=stealth](1,0)--(-1,1);
        \draw[blue, thick, midArrow={pos=.35, end={Stealth[scale=.9]}}]
            (-1,1)--(0,-1);
        \draw[-Stealth](0,0)--(-1,1);
       \fill[cyan!33, opacity=.9](0,0)--(0,-1)--(1,1)--(-1,0)--(0,0);
        \draw[-Stealth](0,0)--(0,-1);
        \draw[blue, thick, midarrow=stealth](0,-1)--(1,1);
        \draw[blue, thick, midarrow=stealth](1,1)--(-1,0);
        \draw[-Stealth](0,0)--(1,1);
       \fill[cyan!33, opacity=.9](0,0)--(-1,0)--(-1,-1)--(1,0)--(0,0);
        \draw[-Stealth](0,0)--(-1,0);
        \draw[blue, thick, midarrow=stealth](-1,0)--(-1,-1);
        \draw[blue, thick, midarrow=stealth](-1,-1)--(1,0);
        \draw[-Stealth](0,0)--(1,0);
        \draw[-Stealth](0,0)--(-1,-1);
       \path(1,0)  node[above]{$\sss1$};
       \path(-1,1) node[below]{$\sss2$};
       \path(0,-1) node[right]{$\sss3$};
       \path(1,1)  node[left]{$\sss4$};
       \path(-1,0) node[above]{$\sss5$};
       \path(-1,-1)node[right]{$\sss6$};
       \filldraw[fill=white, thick](0,0)circle(.5mm);
       \path(1.25,0)node[right]{$w\<=2$};
       \path(.4,-.6)node[right]{$\pD_2$};
        }}
 \quad
  \vC{\small$\begin{array}{@{}r@{~=~}l}
   \sharp(\n_{i,i+1}),~ \sharp(\n_i) &1\\ \toprule
   \sum_i \sharp(\n_{i,i+1}) &6\\
   \sum_i \sharp(\n_i) &6\\ \bottomrule
   \sum_i \sharp(\n_{i,i+1}){-}\sum_i \sharp(\n_i) &0
  \end{array}$} \label{e:w=3,2}
 \\
  \vC{\TikZ{\path[use as bounding box](-1,-1.2)--(1,1.2);
       \foreach\x in{-1,...,1}
        \foreach\y in{-1,...,1}
         \fill[gray](\x,\y)circle(.5mm);
       \filldraw[fill=white, draw=gray](1,0)circle(.5mm);
       \filldraw[fill=white, draw=gray](-1,0)circle(.5mm);
       \fill[red](0,1)circle(.5mm);
       \fill[blue](1,1)circle(.5mm);
       \fill[blue](-1,1)circle(.5mm);
       \fill[blue](0,-1)circle(.5mm);
       \fill[cyan!33, opacity=.9](0,0)--(1,0)--(-1,1)--(0,-1)--(0,0);
        \draw[blue, thick, midArrow={pos=.85, end={Stealth[scale=.9]}}]
            (1,0)--(-1,1);
        \draw[blue, thick, midarrow=stealth](-1,1)--(0,-1);
        \draw[-Stealth](0,0)--(-1,1);
       \fill[cyan!33, opacity=.9](0,0)--(0,-1)--(1,1)--(-1,0)--(0,0);
        \draw[-Stealth](0,0)--(0,-1);
        \draw[blue, thick, midArrow={pos=.45, end={Stealth[scale=.9]}}]
            (0,-1)--(1,1);
        \draw[blue, thick, midArrow={pos=.3, end={Stealth[scale=.9]}}]
            (1,1)--(-1,0);
        \draw[-Stealth](0,0)--(1,1);
       \fill[red!75, opacity=.75](0,0)--(-1,0)--(0,1)--(1,0)--(0,0);
        \draw[red, thick, midarrow=stealth](-1,0)--(0,1);
        \draw[red, thick, midarrow=stealth](0,1)--(1,0);
        \draw[Purple, densely dashed, thick, -stealth](0,0)--(-1,0);
        \draw[Purple, densely dashed, thick, -stealth](0,0)--(1,0);
        \draw[red!90!black, -Stealth](0,0)--(0,1);
       \path(1,0)  node[above]{$\sss1$};
       \path(-1,1) node[below]{$\sss2$};
       \path(0,-1) node[right]{$\sss3$};
       \path(1,1)  node[left]{$\sss4$};
       \path(-1,0) node[above]{$\sss5$};
       \path(0,1)node[left]{$\sss6$};
       \filldraw[fill=white, thick](0,0)circle(.5mm);
       \path(1.25,0)node[right]{$w\<=1$};
       \path(.4,-.6)node[right]{$\pD_3$};
            }}
 \quad
  \vC{\small$\begin{array}{@{}r@{~=~}l}
   \MC2c{\sharp(\n_{i,i+1}){=}\pm1,~ \sharp(\n_i){=}\pm1,0}\\ \toprule
   \sum_i \sharp(\n_{i,i+1}) &(4{-}2)\\
   \sum_i \sharp(\n_i) &(3{-}1)\\ \bottomrule
   \sum_i \sharp(\n_{i,i+1}){-}\sum_i \sharp(\n_i) &0
  \end{array}$}
 \qquad \qquad
  \vC{\TikZ{\path[use as bounding box](-1,-1.2)--(1,1.2);
       \foreach\x in{-1,...,1}
        \foreach\y in{-1,...,1}
         \fill[gray](\x,\y)circle(.5mm);
       \filldraw[fill=white, draw=gray](1,0)circle(.5mm);
       \filldraw[fill=white, draw=gray](-1,-1)circle(.5mm);
       \fill[red](0,1)circle(.5mm);
       \fill[blue](-1,1)circle(.5mm);
       \fill[cyan!33, opacity=.9](0,0)--(1,0)--(-1,1)--(-1,-1)--(0,0);
        \draw[blue, thick, midarrow=stealth](1,0)--(-1,1);
        \draw[blue, thick, midarrow=stealth](-1,1)--(-1,-1);
        \draw[-Stealth](0,0)--(-1,1);
       \fill[red!75, opacity=.75](0,0)--(-1,-1)--(0,1)--(1,0)--(0,0);
        \draw[red, thick, midarrow=stealth](-1,-1)--(0,1);
        \draw[red, thick, midarrow=stealth](0,1)--(1,0);
        \draw[Purple, densely dashed, thick, -stealth](0,0)--(-1,-1);
        \draw[Purple, densely dashed, thick, -stealth](0,0)--(1,0);
        \draw[red!90!black, -Stealth](0,0)--(0,1);
       \path(1,0)   node[above]{$\sss1$};
       \path(-1,1)  node[right]{$\sss2$};
       \path(-1,0)  node[opacity=.5, left]
           {$\stR[-.5pt]{4}{blue!75}\sss(3)$};
       \path(-1,-1) node[right]{$\sss4$};
       \path(0,1)   node[right]{$\sss5$};
       \filldraw[fill=white, thick](0,0)circle(.5mm);
       \path(1.25,0)node[right]{$w\<=0$};
       \path(-.2,-.6)node[right]{$\pD_4$};
            }}
 \quad
  \vC{\small$\begin{array}{@{}r@{~=~}l}
   \MC2c{\sharp(\n_{i,i+1}){=}\pm1,~ \sharp(\n_i){=}\pm1,0}\\ \toprule
   \sum_i \sharp(\n_{i,i+1}) &(2{-}2)\\
   \sum_i \sharp(\n_i) &(1{-}1)\\ \bottomrule
   \sum_i \sharp(\n_{i,i+1}){-}\sum_i \sharp(\n_i) &0
  \end{array}$}
 \label{e:w=1,0}
\end{gather}
In the left-most example of~\eqref{e:w=1,0},
 $\sharp(\n_{12}),\,\sharp(\n_{23}),\,\sharp(\n_{34}),\,\sharp(\n_{45})\<={+}1$ and
 $\sharp(\n_{56}),\,\sharp(\n_{61})\<={-}1$, which implies that
 $\sharp(\n_2),\,\sharp(\n_3),\,\sharp(\n_4)\<={+}1$, while $\sharp(\n_6)\<={-}1$ so
 $\sharp(\n_1),\,\sharp(\n_5)\<=0$.
In the right-most example of~\eqref{e:w=1,0}, which has no relative interior but does have lattice-primitive vertices and is centered at the origin, we have that
 $\sharp(\n_{12}),\,\sharp(\n_{23}),\,\sharp(\n_{34})\<={+}1$ and
 $\sharp(\n_{45}),\,\sharp(\n_{51})\<={-}1$, which implies that
 $\sharp(\n_2),\,\sharp(\n_3)\<={+}1$, while $\sharp(\n_5)\<={-}1$ so
 $\sharp(\n_1),\,\sharp(\n_4)\<=0$.
This face-counting function is related to~\eqref{e:Ty} and~\eqref{e:BettiMF} by $e_i\<=\sum_{\n\in\pS(i)}\sharp(\n)$.
There exist almost complex torus manifolds corresponding to the uniformly (CCW) oriented multifans~\eqref{e:w=3,2}, but only precomplex toric spaces should correspond to the flip-folded ones~\eqref{e:w=1,0}; see Conjecture~\ref{C:obstruct}.

For completeness, we include the VEX multitopes transpolar to those in~\eqref{e:w=3,2} and~\eqref{e:w=1,0}:
\begin{equation}
 \vC{\TikZ{[scale=.95]\path[use as bounding box](-2,-2)--(2,2);
            \fill[blue](2,1)circle(.5mm);
            \fill[blue](1,2)circle(.5mm);
            \fill[blue](-1,2)circle(.5mm);
            \fill[blue](-2,1)circle(.5mm);
            \fill[blue](-2,-1)circle(.5mm);
            \fill[blue](-1,-2)circle(.5mm);
            \fill[blue](1,-2)circle(.5mm);
            \fill[blue](2,-1)circle(.5mm);
            \fill[yellow, opacity=.85](1,0)--(-1,2)--(-1,-2)--(1,0);
             \draw[blue, thick, midarrow=stealth](1,0)--(-1,2);
             \draw[blue, thick](-1,2)--(-1,0);
             \draw[blue, thick, midarrow=stealth](-1,0)--(-1,-2);
             \draw[blue, thick, midarrow=stealth](-1,-2)--(1,0);
              \draw[thick, -stealth](0,0)--(-1,2);
              \draw[thick, -stealth](0,0)--(-1,-2);
            \fill[yellow, opacity=.85](1,0)--(2,1)--(-2,1)--(1,-2)--(1,0);
             \draw[blue, thick, midarrow=stealth](1,0)--(2,1);
             \draw[blue, thick, midarrow=stealth](2,1)--(-2,1);
             \draw[blue, thick](-2,1)--(0,-1);
             \draw[blue, thick, midarrow=stealth](0,-1)--(1,-2);
             \draw[blue, thick](1,-2)--(1,0);
              \draw[thick, -stealth](0,0)--(2,1);
              \draw[thick, densely dotted, -stealth](0,0)--(-1,1);
              \draw[thick, -stealth](0,0)--(-2,1);
              \draw[thick, -stealth](0,0)--(1,-2);
            \fill[yellow, opacity=.85](1,0)--(1,2)--(-2,-1)--(2,-1)--(1,0);
             \draw[blue, thick](2,-1)--(1,0);
             \draw[blue, thick, midarrow=stealth](-2,-1)--(2,-1);
             \draw[blue, thick, midarrow=stealth](1,2)--(-2,-1);
              \filldraw[thick, gray, double, fill=white](1,0)circle(.7mm);
             \draw[blue, thick, midarrow=stealth](1,0)--(1,2);
              \draw[thick, densely dotted, double, -stealth](0,0)--(1,0);
              \draw[thick, densely dotted, -stealth](0,0)--(1,1);
              \draw[thick, -stealth](0,0)--(1,2);
              \draw[thick, densely dotted, -stealth](0,0)--(0,1);
              \draw[thick, densely dotted, -stealth](0,0)--(-1,0);
              \draw[thick, -stealth](0,0)--(-2,-1);
              \draw[thick, densely dotted, -stealth](0,0)--(-1,-1);
              \draw[thick, densely dotted, -stealth](0,0)--(0,-1);
              \draw[thick, densely dotted, -stealth](0,0)--(1,-1);
              \draw[thick, -stealth](0,0)--(2,-1);
            \filldraw[thick, fill=white](0,0)circle(.5mm);
            \path(0,-1.7)node{$\pD_1^\wtd$};
            \path(0,1.7)node{$w\<=3$};
           }}
\quad
 \vC{\TikZ{[scale=.95]\path[use as bounding box](-2,-2)--(2,2);
            \fill[blue](2,1)circle(.5mm);
            \fill[blue](-1,2)circle(.5mm);
            \fill[blue](-2,1)circle(.5mm);
            \fill[blue](-1,-2)circle(.5mm);
            \fill[blue](1,-2)circle(.5mm);
            \fill[yellow, opacity=.85](1,0)--(-1,2)--(-1,-2)--(1,0);
             \draw[blue, thick, midarrow=stealth](1,0)--(-1,2);
             \draw[blue, thick, midarrow=stealth](-1,2)--(-1,-2);
             \draw[blue, thick](-1,-2)--(1,0);
              \draw[thick, -stealth](0,0)--(-1,2);
              \draw[thick, -stealth](0,0)--(-1,-2);
              \draw[thick, densely dotted, -stealth](0,0)--(-1,-1);
            \fill[yellow, opacity=.85](1,0)--(2,1)--(-2,1)--(1,-2)--(1,0);
             \draw[blue, thick, midarrow=stealth](1,-2)--(1,0);
             \draw[blue, thick, midarrow=stealth](2,1)--(-2,1);
             \draw[blue, thick, midarrow=stealth](-2,1)--(1,-2);
              \filldraw[thick, gray, double, fill=white](1,0)circle(.7mm);
             \draw[blue, thick, midarrow=stealth](1,0)--(2,1);
              \draw[thick, densely dotted, double, -stealth](0,0)--(1,0);
              \draw[thick, -stealth](0,0)--(2,1);
              \draw[thick, densely dotted, -stealth](0,0)--(1,1);
              \draw[thick, densely dotted, -stealth](0,0)--(0,1);
              \draw[thick, densely dotted, -stealth](0,0)--(-1,1);
              \draw[thick, -stealth](0,0)--(-2,1);
              \draw[thick, densely dotted, -stealth](0,0)--(-1,0);
              \draw[thick, densely dotted, -stealth](0,0)--(0,-1);
              \draw[thick, -stealth](0,0)--(1,-2);
              \draw[thick, densely dotted, -stealth](0,0)--(1,-1);
            \filldraw[thick, fill=white](0,0)circle(.5mm);
            \path(0,-1.7)node{$\pD_2^\wtd$};
            \path(0,1.7)node{$w\<=2$};
           }}
\quad
 \vC{\TikZ{[scale=.95]\path[use as bounding box](-2,-2)--(2,2);
            \fill[yellow, opacity=.85](0,0)--(1,0)--(2,1)--(-2,1)--(0,-1);
              \fill[blue](2,1)circle(.5mm);
              \fill[blue](-2,1)circle(.5mm);
             \draw[blue, thick, midarrow=stealth](2,1)--(-2,1);
             \draw[blue, thick, midarrow=stealth](-2,1)--(-1,0);
             \draw[blue, thick](-1,0)--(0,-1);
              \draw[thick, -stealth](0,0)--(2,1);
              \draw[thick, densely dotted, -stealth](0,0)--(1,1);
              \draw[thick, densely dotted, -stealth](0,0)--(0,1);
              \draw[thick, densely dotted, -stealth](0,0)--(-1,1);
              \draw[thick, -stealth](0,0)--(-2,1);
              \draw[thick, densely dotted, -stealth](0,0)--(-1,0);
            \fill[yellow, opacity=.85](0,-1)--(1,-2)--(1,-1)--(0,0);
              \fill[blue](1,-2)circle(.5mm);
             \draw[blue, thick, midarrow=stealth](0,-1)--(1,-2);
             \draw[blue, thick, midarrow=stealth](1,-2)--(1,-1);
              \draw[thick, -stealth](0,0)--(1,-2);
            \fill[red!60, opacity=.9](0,0)--(1,-1)--(-1,-1)--(0,0);
             \filldraw[thick, fill=white, draw=gray](1,-1)circle(.5mm);
             \draw[Rouge, thick, midarrow=stealth](1,-1)--(0,-1);
             \draw[Rouge, thick](0,-1)--(-1,-1);
              \draw[blue, thick, densely dashed, -stealth](0,0)--(1,-1);
            \fill[yellow, opacity=.85](0,0)--(-1,-1)--(-1,-2)--(1,0)--(0,0);
             \filldraw[thick, fill=white, draw=gray](-1,-1)circle(.5mm);
             \draw[blue, thick, midarrow=stealth](-1,-1)--(-1,-2);
              \draw[blue, thick, densely dashed, -stealth](0,0)--(-1,-1);
              \fill[blue](-1,-2)circle(.5mm);
             \draw[blue, thick, midarrow=stealth](-1,-2)--(2,1);
              \draw[thick, densely dotted, -stealth](0,0)--(1,0);
              \draw[thick, -stealth](0,0)--(-1,-2);
              \draw[thick, densely dotted, -stealth](0,0)--(0,-1);
            \filldraw[thick, fill=white](0,0)circle(.5mm);
            \path(0,-1.7)node{$\pD_3^\wtd$};
            \path(0,1.7)node{$w\<=1$};
           }}
\quad
 \vC{\TikZ{[scale=.95]\path[use as bounding box](-1,-2)--(2,2);
            \fill[red!60, opacity=.9](0,0)--(1,0)--(2,-1)--(-1,-1)--(0,0);
              \fill[Rouge](2,-1)circle(.5mm);
             \draw[Rouge, thick, midarrow=stealth](1,0)--(2,-1);
             \draw[Rouge, thick, midarrow=stealth](2,-1)--(0,-1);
             \draw[Rouge, thick](0,-1)--(-1,-1);
              \draw[densely dotted, -stealth](0,0)--(1,-1);
              \draw[thick, -stealth](0,0)--(2,-1);
            \fill[yellow, opacity=.85](0,0)--(-1,-1)--(-1,-2)--(1,0)--(0,0);
              \fill[blue](-1,-2)circle(.5mm);
              \filldraw[thick, fill=white, draw=gray](1,0)circle(.5mm);
              \filldraw[thick, fill=white, draw=gray](-1,-1)circle(.5mm);
             \draw[blue, thick, midarrow=stealth](-1,-1)--(-1,-2);
             \draw[blue, thick, midarrow=stealth](-1,-2)--(1,0);
              \draw[thick, -stealth](0,0)--(-1,-2);
              \draw[thick, densely dotted, -stealth](0,0)--(0,-1);
              \draw[blue, thick, densely dashed, -stealth](0,0)--(1,0);
              \draw[blue, thick, densely dashed, -stealth](0,0)--(-1,-1);
            \filldraw[thick, fill=white](0,0)circle(.5mm);
            \path(0,-1.7)node{$\pD_4^\wtd$};
            \path(0,1.7)node{$w\<=0$};
           }}
 \label{e:*w=3,2,1,0}
\end{equation}
and invite the Reader to verify that $\c(\vd\pD_i^\wtd)\<=0$.
The double-drawn lattice-vectors and double-circled lattice points in $\pD_1^\wtd$ and $\pD_2^\wtd$ indicate a choice of layer-crossing branch cuts/points; dashed lattice vectors indicate (CCW/CW) orientation-changing flip-folds and dotted lattice vectors are MPCP-subdivisions.
Again, there exist almost complex torus manifolds corresponding to the uniformly (CCW) oriented multifans with $w\<=3,2$~\eqref{e:*w=3,2,1,0}, but the flip-folded ones with $w\<=1,0$ should stem from precomplex toric spaces.

Note that $\sharp(\n)\<\neq d(\n)$ in general:
For example, in the right-hand side example~\eqref{e:w=1,0}, $\sharp(\n_{24})\<=1$ even for the un-subdivided cone $\n_{24}\<\in\pD_4$ (where $\n_3$ is omitted from the multifan), while $d(\n_{24})\<=2$. For lower-dimensional cones in turn, $d(\n_i)\<=1$ for all vertices in all examples, while $\sharp(\n_i)$ varies, depending on the orientation of adjacent higher-dimensional cones. For example, the 1-cone $\n_1\<\in\pD_3,\,\pD_4$ (depicted by a dashed arrow) counts $\sharp(\n_1)\<=0$, as it forms the interfacing boundary $(\n_1\<=\n_{61}\<\cap\n_{12})\<\in\pD_3$ and $(\n_1\<=\n_{51}\<\cap\n_{12})\<\in\pD_4$ between a positively and a negatively-oriented 2-cone\cite{rBH-gB}.

Suffice it here to also showcase the more intricate 3-dimensional case:
\begin{equation}
 \vC{\begin{picture}(160,30)(0,-2)
   \put(0,0){\rotatebox[origin=c]{6}{\includegraphics[height=30mm]{Pix/3E2SP.pdf}}}
    \put(5,10){$\pDs{\EE[3]m}$}
    \put(44,5){\scriptsize$\n_1$}
    \put(50,18){\scriptsize$\n_2$}
    \put(39,26){\scriptsize$\C1{\n_3}$}
    \put(18,1){\scriptsize$\n_4$}
    \put(0,24){\scriptsize$\C1{\n_5}$}
   \put(58,24){$\begin{array}[t]{@{}c@{~}|@{~}c@{~}|@{~}c@{~}|@{~}c@{~}?@{~}c@{}}
      \dim & \sharp(\n_*)\<={+}1 & \sharp(\n_*)\<=0 & \sharp(\n_*)\<={-}1 & \pS \\
       \toprule\nGlu{-2pt}
         3 & \n_{142},\, \n_{125},\, \n_{134},\, \n_{243}
           & \text{---}
           & \n_{153},\, \n_{235} & 4{-}2\\[-1pt] \midrule\nGlu{-2pt}
         2 & \n_{12},\, \n_{14},\, \n_{24},\, \n_{34} 
           & \n_{13},\, \n_{23},\, \n_{15},\, \n_{25}
           & \n_{35} & 4{-}1 \\[-1pt] \midrule\nGlu{-2pt}
         1 & \n_1,\, \n_2,\, \n_4
           & \n_3,\, \n_5
           & \text{---} & 3\\[-1pt] \bottomrule
      \MC5{l@{}}{\text{\footnotesize$\sum_{\n\in\pS(3)}\sharp(\n)-\sum_{\n\in\pS(2)}\sharp(\n)
            +\sum_{\n\in\pS(1)}\sharp(\n)=2=\c(S^2)=\c(\vd\pDs{\EE[3]m})$}}
                \end{array}$}
 \end{picture}}
 \label{e:3Em=S2}
\end{equation}
The facets $\q_{ijk}$ and the central cones over them, $\n_{ijk}\<=\sfa\q_{ijk}$, are oriented away (or toward) the center by their multi-index order, see footnote~\ref{fn:d}, p.\,\pageref{fn:d}. Each edge (and the 2-cone over it) is then oriented by continuity, as the interface between its two adjacent 2-faces. For example:
\begin{itemize}[itemsep=-1pt, topsep=-1pt]
 \item $\sharp(\n_{12})\<={+}1$ since $\n_{12}\<=\n_{142}\<\cap\n_{125}$, and
 $\sharp(\n_{142})\<=\sharp(\n_{125})\<={+}1$.
 \item $\sharp(\n_{13})\<=0$ since $\n_{13}\<=\n_{134}\<\cap\n_{153}$ is folding over between
 $\sharp(\n_{134})\<{+}1$ and $\sharp(\n_{153})\<={-}1$.
 \item $\sharp(\n_{35})\<={-}1$ since $\n_{35}\<=\n_{153}\<\cap\n_{235}$, and
 $\sharp(\n_{153})\<=\sharp(\n_{235})\<={-}1$.
\end{itemize}
Finally, vertices (and 1-cones over them) are also oriented by continuity:
\begin{enumerate}[itemsep=-1pt, topsep=-1pt]
 \item $\sharp(\n_4\<=\n_{142}\<\cap\n_{134}\<\cap\n_{243})\<={+}1$, since
  $\sharp(\n_{142})\<=\sharp(\n_{134})\<=\sharp(\n_{243})\<={+}1$. 
 \item $\sharp(\n_3\<=\n_{13}\<\cap\n_{32})\<=0$, since $\sharp(\n_{13})\<=\sharp(\n_{32})\<=0$,
  being folds between positive and negative facets.\\
  The same holds for $\n_5\<=\n_{15}\<\cap\n_{52}$ and implies $\sharp(\n_5)\<=0$.
 \item Approached from any direction outside the $\n_{153}$ cone, $\n_1$ inherits the outward
 ($+1$) orientation of the adjacent facets.
 Approached from {\em\/within\/} the $\n_{153}$ cone, the orientation is inherited from the 
 overlaying cones:
 $\sharp(\n_1\<=\n_{134}^+\<\cap\n_{125}^+\<\cap\n_{153}^-)\<={+}1{+}1{-}1={+}1$, verifying that
 $\sharp(\n_1)\<={+}1$ is constant, i.e., independent of the (limit) approach.
    The same holds for $\n_2$ and implies $\sharp(\n_2)\<={+}1$.
\end{enumerate}
As show in the bottom row in~\eqref{e:3Em=S2}, the Euler number of 
$\vd\pDs{\EE[3]m}$ is that of a 2-sphere, supporting the description of VEX multitopes as (possibly flip-folded or otherwise multi-layered) star-shaped, complete collections of star-pyramids akin to the multifans they span, and as required for Hibi's Lemma~\ref{L:Hibi}.

This combinatorial function, $\sharp(\n)$, defined here for VEX multitopes and the multifans they span, seems to be very closely related to the function $\ve(p)$ originally defined\cite{rM-MFans} over the tangent bundles of general torus manifolds corresponding to multifans. 
 While the fact that $n$-dimensional VEX multitopes are star-shaped domains (albeit multi-layered, with $w$-signed multiplicity) and homotopic to $n$-dimensional balls seems to provide a key restriction, we are not aware of a general, complete algorithm for the multifan function $\sharp(\n)$, nor its relation to the torus manifold function $\ve(p)$. It is clear however that $d(\n)$ and $\sharp(\n)$ should play distinct roles in (presumably $\rT$-equivariant) (co)homological, topological and other related computations, especially within the context of various string theory and related applications.

\subsection{Collapsing Exceptional Sets}
The above examples showcase the utility of multifans and the multitopes that span them in characterizing (if only partially; see footnote~\ref{fn:TTMs}, p.\,\pageref{fn:TTMs}) not only toric spaces and a host of their diffeomorphism invariants, but also toric surgeries generalizing the standard blowup MPCP desingularizations\cite{rD-TV, rO-TV, rF-TV, rGE-CCAG, rCLS-TV} and\cite{rBaty01}. In particular, consider the sequence of (multi)fans depicted in Figure~\ref{f:E2-E-2}: 
the surgery $\EE[2]1\<\onto\IP^2$ implies the collapse of the exceptional $\rT$-characteristic submanifold
$(\sE_{(-1,1)}\<\subset\EE[2]1)\<\onto(\{\text{pt.}\}\<\subset\IP^2)$ encoded by removing the ``2''-labeled 1-cone, $\n_2\<=(-1,1)$, from the multifan $\pFn{\EE[2]1}$.
This is very much akin to the familiar blowdown, 
$\EE[2]{-1}\<\supset\sE_{(1,1)}\<\onto\{\text{pt.}\}\<\subset\IP^2$,
which is encoded by removing the ``2''-labeled 1-cone from the fan
$\pFn{\EE[2]{-1}}$, and both encode ``vanishing cycles'' in a suitable (co)homology. 
Whereas the latter operation keeps within the class of complex-algebraic varieties, the former extends this to include also precomplex toric spaces such as $\EE[2]1$ (and then also $\EE[2]m$ for $m\<>1$), and so points to a substantial generalization of the standard toric blowup/blowdown\cite{rF-TV, rGE-CCAG, rCLS-TV, Jang:2023aa}.

In this vein, we may consider iteratively collapsing exceptional sets, such as in:
\begin{equation}
 \vC{\TikZ{[scale=.75]\path[use as bounding box](-1,-1.5)--(2,1.2);
            \foreach\x in{-1,...,1} \foreach\y in{-1,...,1} \fill[gray](\x,\y)circle(.5mm);
            \fill[yellow, opacity=.9](-1,-1)--(1,-1)--(0,1);
            \draw[blue, thick, -stealth](1,-1)--(0,1);
            \draw[blue, thick, -stealth](0,1)--(-1,-1);
            \draw[blue, thick, -stealth](-1,-1)--(1,-1);
            \draw[-stealth](0,0)--(1,-1);
            \draw[-stealth](0,0)--(0,1);
            \draw[-stealth](0,0)--(-1,-1);
            \draw[densely dotted, -stealth](0,0)--(0,-1);
            \path(-.75,.5)node{$\FF[2]2$};
            \path  (0,1)node[right]{\tiny$1$};
            \path(-1,-1)node [left]{\tiny$2$};
            \path (1,-1)node[right]{\tiny$5$};
            \filldraw[thick, fill=white](0,0)circle(.5mm);
            \draw[thick, ->>](2.5,0)--node[above]{~~$\SSS\widehat3$}++(-1,0);
            }}
 \vC{\TikZ{[scale=.75]\path[use as bounding box](-2,-1.5)--(2,1.2);
            \foreach\x in{-1,...,1}
             \foreach\y in{-1,...,1}
              \fill[gray](\x,\y)circle(.5mm);
            \fill[yellow, opacity=.9](0,0)--(1,-1)--(0,1);
            \draw[blue, thick, -stealth](1,-1)--(0,1);
            \fill[red, opacity=.67](0,0)--(1,0)--(1,-1);
            \fill[yellow, opacity=.9](0,0)--(0,1)--(-1,-1);
            \draw[blue, thick, -stealth](0,1)--(-1,-1);
            \draw[Magenta, densely dotted, thick, -stealth](0,0)--(1,0);
            \draw[red!90!black, thick, midarrow=stealth](1,0)--(1,-1);
            \draw[Magenta, thick, -stealth](0,0)--(0,1);
            \fill[yellow, opacity=.9](0,0)--(-1,-1)--(1,0);
            \draw[blue, thick, -stealth](-1,-1)--(1,0);
            \draw[-stealth](0,0)--(-1,-1);
            \draw[Magenta, densely dotted, thick, -stealth](0,0)--(1,-1);
            \path(-.75,.6)node{$\EE[2]1$};
            \path  (0,1)node[right]{\tiny$1$};
            \path(-1,-1)node [left]{\tiny$2$};
            \path  (1,0)node[above]{\tiny$3$};
            \path (1,-1)node[right]{\tiny$5$};
            \filldraw[thick, fill=white](0,0)circle(.5mm);
            \draw[thick, ->>](2.5,0)--node[above]{~~$\SSS\widehat4$}++(-1,0);
            \draw[thick, ->>](1.5,-.67)to[out=-30,in=180]
                node[above=10pt, left=10pt]{~~$\SSS\widehat5$}++(3,-.75)
                --++(5,0)to[out=0,in=-150]++(3,.75);
            }}
 \vC{\TikZ{[scale=.75]\path[use as bounding box](-3,-1.5)--(1,1.2);
            \foreach\x in{-2,...,1}
             \foreach\y in{-1,...,1}
              \fill[gray](\x,\y)circle(.5mm);
            \fill[yellow, opacity=.9](0,0)--(1,0)--(-2,1);
            \draw[blue, thick, midarrow=stealth](1,0)--(-2,1);
            \fill[yellow, opacity=.9](0,0)--(-2,1)--(1,-1);
            \draw[Magenta, thick, -stealth](0,0)--(-2,1);
            \draw[blue, thick, midarrow=stealth](-2,1)--(1,-1);
            \fill[yellow, opacity=.9](0,0)--(1,-1)--(0,1);
            \draw[blue, thick, -stealth](1,-1)--(0,1);
            \draw[-stealth](0,0)--(1,-1);
            \fill[yellow, opacity=.9](0,0)--(0,1)--(-1,-1);
            \draw[blue, thick, -stealth](0,1)--(-1,-1);
            \draw[Magenta, thick, -stealth](0,0)--(0,1);
            \fill[yellow, opacity=.9](0,0)--(-1,-1)--(1,0);
            \draw[blue, thick, -stealth](-1,-1)--(1,0);
            \draw[-stealth](0,0)--(-1,-1);
            \draw[-stealth](0,0)--(1,0);
            \path(-1.5,-.4)node{$\Tw{E}_2^{\sss(2)}$};
            \path  (0,1)node[right]{\tiny$1$};
            \path(-1,-1)node [left]{\tiny$2$};
            \path  (1,0)node[above]{\tiny$3$};
            \path (-2,1)node[below]{\tiny$4$};
            \path (1,-1)node[right]{\tiny$5$};
            \filldraw[thick, fill=white](0,0)circle(.5mm);
            \draw[thick, ->>](1.5,0)--node[above]{$\SSS\widehat5$~~~}++(1,0);
            }}
 \vC{\TikZ{[scale=.75]\path[use as bounding box](-4,-1.5)--(1,1.2);
            \foreach\x in{-2,...,1}
             \foreach\y in{-1,...,1}
              \fill[gray](\x,\y)circle(.5mm);
            \fill[yellow!85](0,0)--(1,0)--(-2,1);
            \draw[blue, thick, midarrow=stealth](1,0)--(-2,1);
            \fill[red, opacity=.67](0,0)--(-2,1)--(0,1);
            \draw[red!90!black, thick, densely dotted, -stealth]
                (0,0)--(-1,1);
            \draw[red!90!black, thick, midarrow=stealth](-2,1)--(0,1);
            \draw[Magenta, densely dotted, thick, -stealth](0,0)--(-2,1);
            \fill[yellow, opacity=.85](0,0)--(0,1)--(-1,-1)--(1,0);
            \draw[-stealth](0,0)--(-1,-1);
            \draw[-stealth](0,0)--(1,0);
            \draw[Magenta, densely dotted, thick, -stealth](0,0)--(0,1);
            \draw[blue, thick, midarrow=stealth](0,1)--(-1,-1);
            \draw[blue, thick, midarrow=stealth](-1,-1)--(1,0);
            \path(-1.5,-.4)node{$\EE[2]2$};
            \path  (0,1)node[right]{\tiny$1$};
            \path(-1,-1)node [left]{\tiny$2$};
            \path  (1,0)node[above]{\tiny$3$};
            \path (-2,1)node[below]{\tiny$4$};
            \filldraw[thick, fill=white](0,0)circle(.5mm);
            \draw[thick, ->>](1.5,0)--node[above]{$\SSS\widehat4$~~~}++(1,0);
            }}
 \vC{\TikZ{[scale=.75]\path[use as bounding box](-3,-1.5)--(1,1.2);
            \foreach\x in{-1,...,1}
             \foreach\y in{-1,...,1}
              \fill[gray](\x,\y)circle(.5mm);
            \fill[yellow!85](1,0)--(0,1)--(-1,-1);
            \draw[-stealth](0,0)--(1,0);
            \draw[-stealth](0,0)--(0,1);
            \draw[-stealth](0,0)--(-1,-1);
            \draw[blue, thick, midarrow=stealth](1,0)--(0,1);
            \draw[blue, thick, midarrow=stealth](0,1)--(-1,-1);
            \draw[blue, thick, midarrow=stealth](-1,-1)--(1,0);
            \path(-.75,.5)node{$\IP^2$};
            \path  (0,1)node[right]{\tiny$1$};
            \path(-1,-1)node [left]{\tiny$2$};
            \path  (1,0)node[above]{\tiny$3$};
            \filldraw[thick, fill=white](0,0)circle(.5mm);
            }}
\end{equation}
somewhat akin to a sequence of blowdown contractions. For example,
``$\6[2pt]{~\widehat{3}~}{\relbar\joinrel\onto}$'' denotes the collapsing of the exceptional set ($\rT$-characteristic submanifold) corresponding to the 1-cone labeled ``$3$'' to a point.
Noticing that $\FF[2]2$ and $\IP^2$ are complex-algebraic toric varieties, while $\Tw{E}_2^{\sss(2)}$ is an almost complex torus manifold\cite{rM-MFans, rHM-MFs}, such local surgery operations suggest:
\begin{clam}
\label{CC:flipRE}
Every {\em\/flip-folded\/} VEX multitope (multifan), wherein some faces (cones) have opposite orientation, may be surgically reverse-engineered in terms of only {\em\/uniformly\/} oriented multitopes (multifans), all of which correspond to almost complex torus manifolds, moreover to complex-algebraic toric varieties when the VEX multitopes (multifans) are plain (so VEX\,$\to$\,reflexive) polytopes (fans).
\end{clam}

\subsection{Flip-Folded Extension Replacement}
\begin{figure}[htb]
 \begin{center}
  \begin{picture}(160,107)(0,3)
   \put(30,60){\TikZ{[scale=.95]
        \path[use as bounding box](-3,-3.1)--(2,2.1);
		\foreach\x in{1,...,2} \fill[gray](\x,-3)circle(.4mm);
		\foreach\x in{0,...,2} \fill[gray](\x,-2)circle(.4mm);
		\foreach\x in{-1,...,2} \fill[gray](\x,-1)circle(.4mm);
		\foreach\x in{-2,...,2} \fill[gray](\x,0)circle(.4mm);
		\foreach\x in{-3,...,2}{\fill[gray](\x,1)circle(.4mm);
                                 \fill[gray](\x,2)circle(.4mm);};
		\fill[Turque!25, opacity=.9](0,0)--(0,1)--(2,-3)--(-3,2);
		 \draw[blue, thick, midarrow=stealth](2,-3)
			to node[above right=-1mm]{\footnotesize$\n_2^\wtd$}(0,1);
		\fill[red, opacity=.67](0,0)--(0,1)--(1,0);
		\draw[Rouge, ultra thick, midarrow=stealth](0,1)
			to node[above right=-3pt]{\footnotesize$\n_*^\wtd$}(1,0);
		 \draw[thick, -stealth](0,0)--(0,1);
		 \draw[thick,-stealth](0,0)--(2,-3);
		 \draw(0,1)node[above right=-1mm]{\footnotesize$\n_{2*}^\wtd$};
		 \draw[blue, thick, midarrow=stealth](-3,2)--(2,-3);
		 \path[blue](-.5,-.5)node[below left=-1mm]
				{\footnotesize$\n_1^\wtd$};
		 \draw(2,-3)node[right]{\footnotesize$\n_{12}^\wtd$};
		 \draw(1,0)node[above right=-1mm]{\footnotesize$\n_{*5}^\wtd$};
		\foreach\z in{0,...,3} \draw[densely dotted](0,0)--(1-\z,\z-2);
		\fill[Turque!25,opacity=.9](0,0)--(-3,2)--(1,0);
		 \draw[densely dotted](-1,1)--(1,-1);
		 \draw[thick,-stealth](0,0)--(1,0);
		 \draw[blue, thick, midarrow=stealth](1,0)
			to node[above]{\footnotesize$\n_5^\wtd$}(-3,2);
		 \draw[thick,-stealth](0,0)--(-3,2);
		 \draw(-3,2)node[below=1mm]{\footnotesize$\n_{51}^\wtd$};
		\filldraw[fill=white,thick](0,0)circle(.4mm);
		 \path(-.8,-1.5)node{$\pDN{\FF[2]3}\<\lat\pFn{\MF[2]3}$};
		 \draw[ultra thick, densely dotted,<->](1.2,-.2)
			to node[below=-3pt]{transpolar}++(2.5,0);
		 \draw[ultra thick, dashed, -Stealth](-.15,-2.1)
			to[out=-110,in=110] node[left]{surgery}++(0,-3);
		 \draw[ultra thick, <<-](.15,-1.9)
			to[out=-70,in=70] node[right]
			{$\Ht{\SSS{\n_{34}^\wtd}}$}++(0,-3);
             }}
   \put(78,69.5){\TikZ{[scale=.95]
        \path[use as bounding box](-3,-2.1)--(2,2.1);
		\foreach\y in{-2,...,1}
		 \foreach\x in{-2,...,1} \fill[gray](\x,\y)circle(.4mm);
		 \path(-1.4,-1.45)node
		 {\includegraphics[width=10mm]{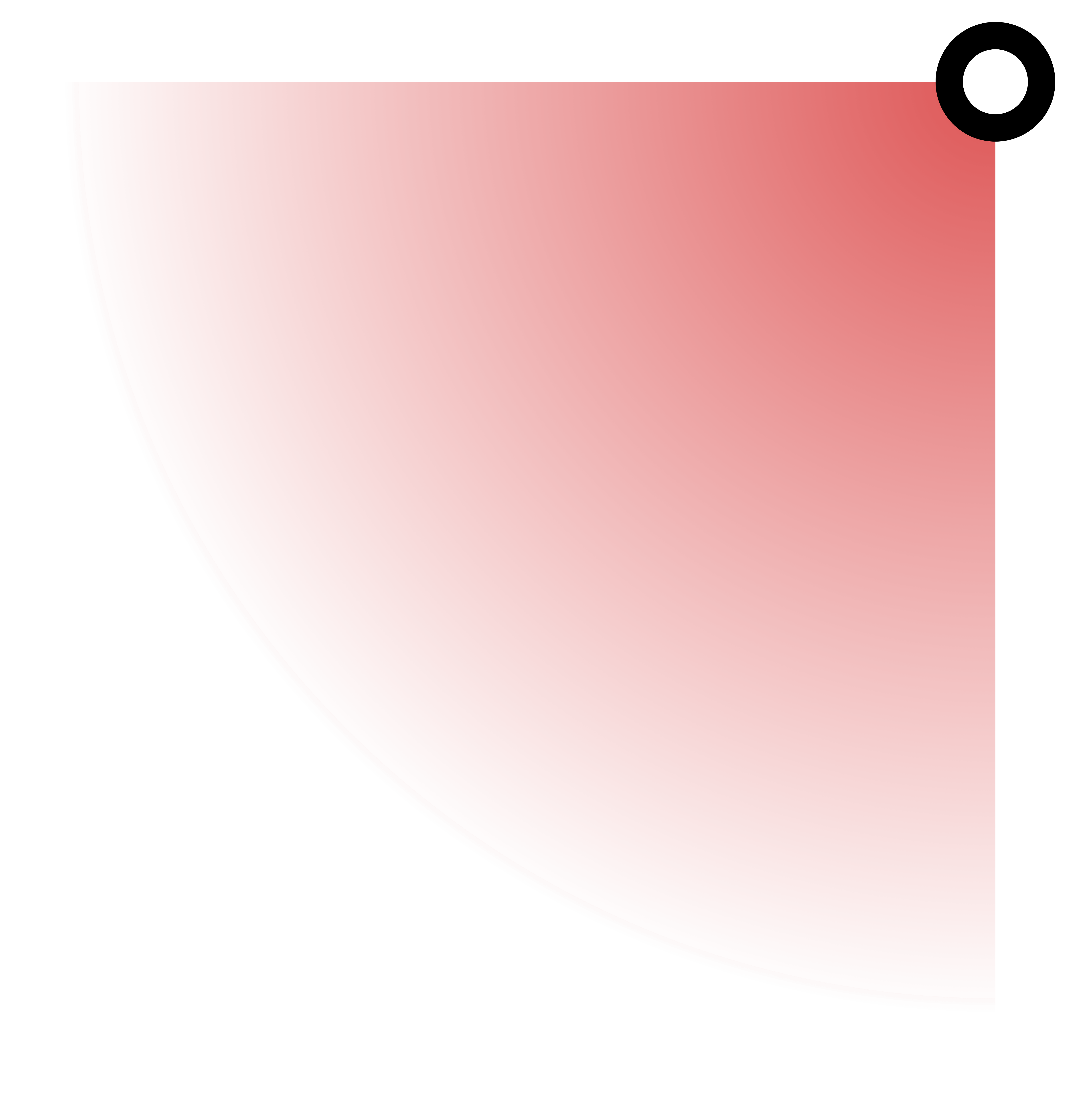}};
		\fill[yellow, opacity=.9](1,1)--(-2,-1)--(-1,-1)--(-1,-2);
		 \draw[thick,-stealth](0,0)--(1,1);
		 \draw(1,1)node[right]{\footnotesize$\n_1$};
		 \draw[blue, thick, midarrow=stealth](1,1)--(-2,-1);
		 \draw[thick,-stealth](0,0)--(-2,-1);
		 \draw(-2,-1)node[below]{\footnotesize$\n_2$};
		 \draw[blue, thick, midarrow=stealth](-2,-1)--(-1,-1);
		 \draw[blue, thick, midarrow=stealth](-1,-1)--(-1,-2);
		 \draw[thick,-stealth](0,0)--(-1,-2);
		 \draw(-1,-2)node[left]{\footnotesize$\n_5$};
		 \draw[blue, thick, midarrow=stealth](-1,-2)--(1,1);
		\filldraw[draw=Rouge, fill=red](-1,-1)circle(.5mm);
		 \draw[thick,-stealth](0,0)--(-1,-1);
		 \draw(-.95,-1)node[below left=-2pt]{\footnotesize$\n_*$};
		\filldraw[fill=white,thick](0,0)circle(.4mm);
		 \path(.85,-1.5)node{$\pDs{\FF[2]3}\<\lat\pFn{\FF[2]3}$};
		\draw[ultra thick, dashed,-Stealth](.15,-2.1)
			to[out=-65,in=65] node[right]{surgery}++(0,-2.5);
        \draw[red, opacity=.33, ultra thick, densely dotted,
			Stealth-Stealth](-4.1,.7)
			to[out=15,in=-135]++(2.6,-2.2);
		\draw[ultra thick, <<-](-.15,-1.9)
			to[out=-100,in=100] node[left=-3pt]
			{$\mM{\;\Ht{\n'_*},\cdots\\[1pt]
			      \Ht{\,\n_4\,},\cdots\\[1pt]
			      \Ht{\,\n_3\,},\cdots}$}++(0,-2.5);
             }}
   \put(0,69){\TikZ{[scale=.75, 
                     every node/.style={inner sep=1pt, outer sep=0,
                     }]
     \path[use as bounding box](-3,-3)--(2,2);
      \node[rounded corners=4pt, inner sep=2pt, draw=black, fill=white](1)
           at(-3,2) {\scriptsize$-1$};
	  \node[rounded corners=4pt, inner sep=2pt, draw=black, fill=white](6)
            at(-2,1) {\tiny$-2$};
	  \node[rounded corners=4pt, inner sep=2pt, draw=black, fill=white](7)
            at(-1,0) {\tiny$-2$};
	  \node[rounded corners=4pt, inner sep=2pt, draw=black, fill=white](8)
            at(0,-1) {\tiny$-2$};
	  \node[rounded corners=4pt, inner sep=2pt, draw=black, fill=white](9)
            at(1,-2) {\tiny$-2$};
	  \node[rounded corners=4pt, inner sep=2pt, draw=black, fill=white](2)
           at(2,-3) {\scriptsize$-1$};
	  \node[rounded corners=4pt, inner sep=2pt, draw=black, fill=white](10)
            at(1,-1) {\tiny$-2$};
	  \node[circle, draw=black, fill=white](3) at(0,1)  {\scriptsize$1$};
	  \node[circle, draw=black, fill=white](5) at(1,0)  {\scriptsize$1$};
	  \node[rounded corners=4pt, inner sep=2pt, draw=black, fill=white](11)
            at(-1,1) {\tiny$-2$};
		\draw[blue, thick, midarrow=stealth](1)--node[below left=1pt]
			{\scriptsize$1$}(6);
		\draw[blue, thick, midarrow=stealth](6)--node[below left=1pt]
			{\scriptsize$1$}(7);
		\draw[blue, thick, midarrow=stealth](7)--node[below left=1pt]
			{\scriptsize$1$}(8);
		\draw[blue, thick, midarrow=stealth](8)--node[below left=1pt]
			{\scriptsize$1$}(9);
		\draw[blue, thick, midarrow=stealth](9)--node[below left=1pt]
			{\scriptsize$1$}(2);
		\draw[blue, thick, midarrow=stealth](2)--node[above right=1pt]
			{\scriptsize$1$}(10);
		\draw[blue, thick, midarrow=stealth](10)--node[left=1pt]
			{\scriptsize$1$}(3);
		\draw[red, thick, midarrow=stealth](3)--node[above right=0pt]
			{\scriptsize$-1$}(5);
		\draw[blue, thick, midarrow=stealth](5)--node[below=1pt]
			{\scriptsize$1$}(11);
		\draw[blue, thick, midarrow=stealth](11)--node[above=1pt]
			{\scriptsize$1$}(1);
     }}
   \put(130,77){\TikZ{[scale=.75, every node/.style=
			{inner sep=1pt, outer sep=0}]
		\path[use as bounding box](-2,-2)--(2,2);
		\node[circle, draw=black, fill=white](1)
           at(1,1) {\scriptsize$3$};
		\node[circle, draw=black, fill=white](2)
           at(-2,-1) {\scriptsize$0$};
		\node[red, rounded corners=4pt, inner sep=2pt, draw=red, 
		   fill=white](*)
           at(-1,-1) {\scriptsize$-3$};
		\node[circle, draw=black, fill=white](5) at(-1,-2){\scriptsize$0$};
		\draw[blue, thick, midarrow=stealth](1)--node[below=2pt]
			{\scriptsize$1$}(2);
		\draw[blue, thick, midarrow=stealth](2)--node[below=1pt]
			{\scriptsize$1$}(*);
		\draw[blue, thick, midarrow=stealth](*)--node[left=1pt]
			{\scriptsize$1$}(5);
		\draw[blue, thick, midarrow=stealth](5)--node[above left=0pt]
			{\scriptsize$1$}(1);
     }}
   \put(30,0){\TikZ{[scale=.95]
        \path[use as bounding box](-3,-3.1)--(2,2.1);
		\foreach\x in{1,...,2} \fill[gray](\x,-3)circle(.4mm);
		\foreach\x in{0,...,2} \fill[gray](\x,-2)circle(.4mm);
		\foreach\x in{-1,...,2} \fill[gray](\x,-1)circle(.4mm);
		\foreach\x in{-2,...,2} \fill[gray](\x,0)circle(.4mm);
		\foreach\x in{-3,...,2}{\fill[gray](\x,1)circle(.4mm);
								 \fill[gray](\x,2)circle(.4mm);};
		 \path(.5,.5)node
		 {\includegraphics[width=12mm]{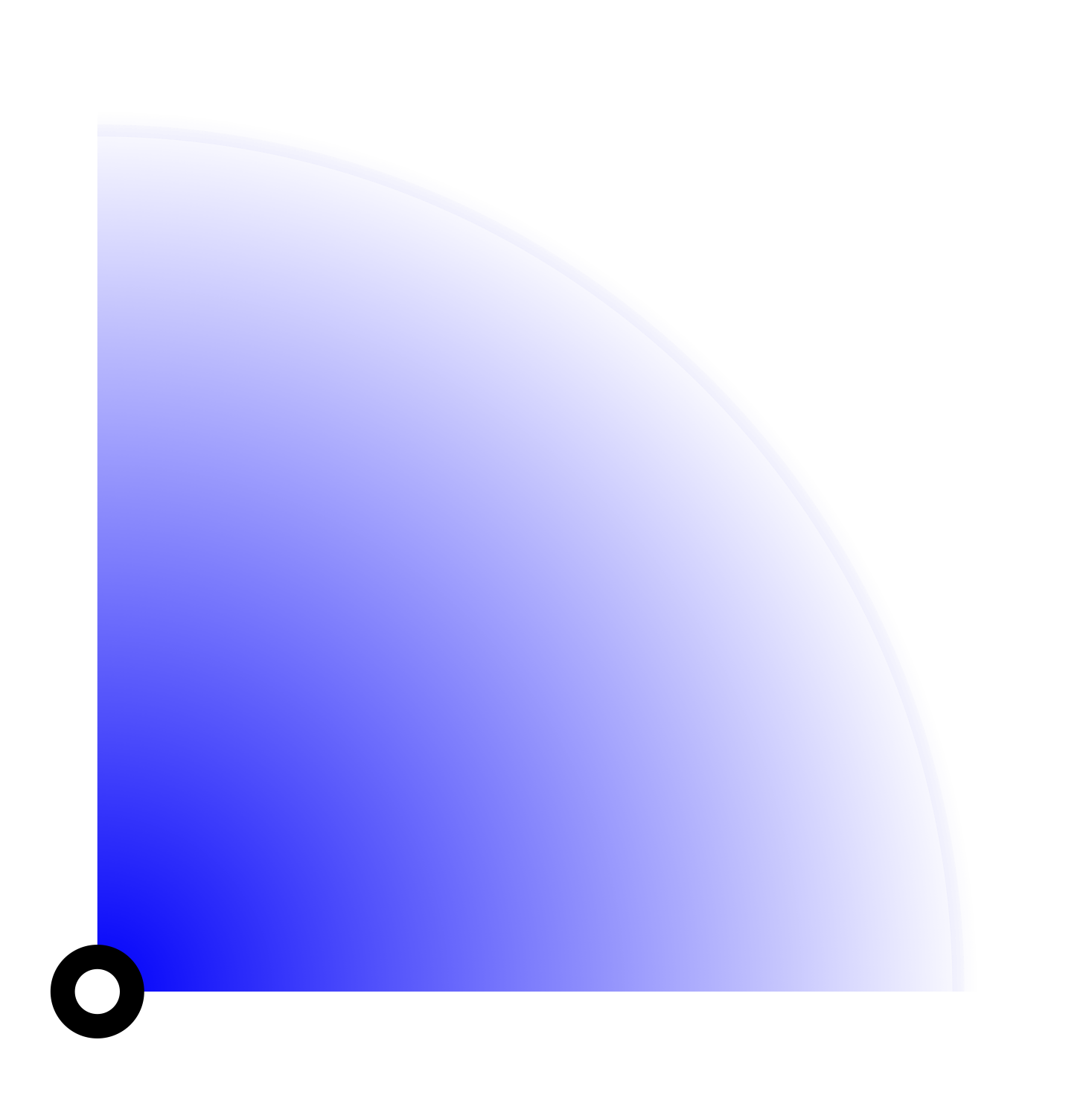}};
		\fill[Turque!25, opacity=.9](0,0)--(0,1)--(-1,-1);
		 \draw[Purple, very thick](0,1)--(-.6,-.2);
		 \draw[Purple, very thick, midarrow=stealth](-.6,-.2)--(-1,-1);
		 \path[blue](-1.35,-.5)node[right]{\footnotesize$\n_3^\wtd$};
		\fill[Turque!67, opacity=.33](0,0)--(0,1)--(2,-3);
		\fill[Turque!25, opacity=.9](0,0)--(2,-3)--(-3,2);
		 \draw[blue, thick, midarrow=stealth](2,-3)
				to node[above right=-1mm]{\footnotesize$\n_2^\wtd$}(0,1);
		 \draw[thick, -stealth](0,0)--(0,1);
		 \draw[thick,-stealth](0,0)--(2,-3);
		 \draw(0,1)node[above right=-1mm]{\footnotesize$\n_{23}^\wtd$};
		 \draw(-1,-1)node[left]{\footnotesize$\n_{34}^\wtd$};
		\draw[blue, thick, midarrow=stealth](-3,2)--(2,-3);
		 \path[blue](-1.1,.1)node[below left=-1mm]{\footnotesize$\n_1^\wtd$};
		 \draw(2,-3)node[right]{\footnotesize$\n_{12}^\wtd$};
		 \draw(1,0)node[above right=-1mm]{\footnotesize$\n_{45}^\wtd$};
		\foreach\z in{0,...,3} \draw[densely dotted](0,0)--(1-\z,\z-2);
		 \draw[densely dotted](0,0)--(1,-1);
		\fill[Turque!25,opacity=.9](0,0)--(-3,2)--(1,0)--(-1,-1);
		 \draw[thick,-stealth](0,0)--(1,0);
		 \draw[blue, thick, midarrow=stealth](1,0)
				to node[above]{\footnotesize$\n_5^\wtd$}(-3,2);
		 \draw[thick,-stealth](0,0)--(-3,2);
		 \draw(-3,2)node[below=1mm]{\footnotesize$\n_{51}^\wtd$};
		 \draw[Purple, very thick, midarrow=stealth](-1,-1)--(1,0);
		 \path[blue](.1,-.65)node{\footnotesize$\n_4^\wtd$};
		 \draw[densely dotted](-1,1)--(0,0);
		 \draw[thick, densely dotted, -stealth](0,0)--(-1,-1);
		 \filldraw[fill=white,thick](0,0)circle(.4mm);
		 \path(-1,-1.5)node{$\pDN{\,\Ht{F}_3^{\sss(2)}}\<\lat
							   \pFn{{\tW\Ht{F}_3^{\sss(2)}}}$};
		 \draw[ultra thick, densely dotted,<->](.8,.7)
                    to[out=30,in=150] node[above]{transpolar}++(3,0);
             }}
   \put(78,9.5){\TikZ{[scale=.95]
        \path[use as bounding box](-3,-2.1)--(2,2.1);
		\foreach\y in{-2,...,2}
		 \foreach\x in{-2,...,2} \fill[gray](\x,\y)circle(.4mm);
		\fill[yellow, opacity=.9](1,1)--(-2,-1)--(-1,-1)--(-1,-2);
		 \draw[thick,-stealth](0,0)--(1,1);
		 \draw(1,1)node[right]{\footnotesize$\n_1$};
		 \draw[blue, thick](1,1)--(-1.1,-.4);
		 \draw[blue, thick, midarrow=stealth](-1.1,-.4)--(-2,-1);
		 \draw[thick,-stealth](0,0)--(-2,-1);
		 \draw(-2,-1)node[below]{\footnotesize$\n_2$};
		 \draw[blue, thick, midarrow=stealth](-2,-1)--(-1,-1);
		 \draw[blue, ultra thick, midarrow=stealth](-1,-1)--(-1,-2);
		 \draw[thick,-stealth](0,0)--(-1,-2);
		 \draw(-1,-2)node[left]{\footnotesize$\n_5$};
		 \draw[blue, thick, midarrow=stealth](-1,-2)--(1,1);
		\fill[yellow, opacity=.9](-1,-1)--(2,-1)--(-1,2);
		 \draw[Purple, very thick, midarrow=stealth](-1,-1)--(2,-1);
		 \draw[Purple, ultra thick, midarrow=stealth](-1,2)--(-1,-1);
		 \draw[thick,-stealth](0,0)--(2,-1);
		 \draw(2,-1)node[right]{\footnotesize$\n_3$};
		 \draw[thick,-stealth](0,0)--(-1,2);
		 \draw(-1,2)node[left]{\footnotesize$\n_4$};
		 \draw[purple, densely dotted, very thick, -stealth](0,0)--(-1,-1);
		 \draw[blue, thick](-1,-1)circle(.55mm);
		 \draw[blue, thick](-1,-1)circle(1.1mm);
		 \draw[blue](-1.1,-1)node[left, rotate=45]
			{\footnotesize$\n'_*,\,\n_*$};
		 \path[purple](-.7,-2)node[rotate=45]
			{\footnotesize$\vec\n_*$ is a branch-cut};
		 \path[purple](-.25,1.5)node[right]{$\n_*\<\in\n_{23}$,~
			$\n'_*\<\in\n_{45}$};
		 \draw[Purple, very thick, midarrow=stealth](2,-1)--(-1,2);
		 \foreach\z in{0,1} {\draw[thick, densely dotted](0,0)--(\z,-1);
			\draw[thick, densely dotted](0,0)--(1-\z,\z);
			\draw[thick, densely dotted](0,0)--(-1,\z);};
		\draw[blue, opacity=.33, ultra thick, densely dotted, 
			Stealth-Stealth](-5.8,-.7)
			to[out=-15,in=210]++(6.23,1.25);
		\filldraw[fill=white,thick](0,0)circle(.4mm);
		\path(1.35,-1.5)node{$\pDs{\,\Ht{F}_3^{\sss(2)}}\<\lat
                               \pFn{\Ht{F}_3^{\sss(2)}}$};
             }}
   \put(0,9){\TikZ{[scale=.75,
                    every node/.style={inner sep=1pt, outer sep=0}]
		\path[use as bounding box](-3,-3)--(2,2);
	  \node[rounded corners=4pt, inner sep=2pt, draw=black, fill=white](1)
           at(-3,2) {\scriptsize$-1$};
	  \node[rounded corners=4pt, inner sep=2pt, draw=black, fill=white](6)
            at(-2,1) {\tiny$-2$};
	  \node[rounded corners=4pt, inner sep=2pt, draw=black, fill=white](7)
            at(-1,0) {\tiny$-2$};
	  \node[rounded corners=4pt, inner sep=2pt, draw=black, fill=white](8)
            at(0,-1) {\tiny$-2$};
	  \node[rounded corners=4pt, inner sep=2pt, draw=black, fill=white](9)
            at(1,-2) {\tiny$-2$};
	  \node[rounded corners=4pt, inner sep=2pt, draw=black, fill=white](2)
           at(2,-3) {\scriptsize$-1$};
	  \node[circle, draw=black, fill=white](3) at(0,1)  {\scriptsize$2$};
	  \node[rounded corners=4pt, inner sep=2pt, draw=black, fill=white](10)
            at(1,-1) {\tiny$-2$};
	  \node[circle, draw=black, fill=white](4) at(-1,-1){\scriptsize$1$};
	  \node[circle, draw=black, fill=white](5) at(1,0)  {\scriptsize$2$};
	  \node[rounded corners=4pt, inner sep=2pt, draw=black, fill=white](11)
            at(-1,1) {\tiny$-2$};
		\draw[blue, thick, midarrow=stealth](1)--node[below left=1pt]
			{\scriptsize$1$}(6);
		\draw[blue, thick, midarrow=stealth](6)--node[below left=1pt]
			{\scriptsize$1$}(7);
		\draw[blue, thick, midarrow=stealth](7)--node[below left=1pt]
			{\scriptsize$1$}(8);
		\draw[blue, thick, midarrow=stealth](8)--node[below left=1pt]
			{\scriptsize$1$}(9);
		\draw[blue, thick, midarrow=stealth](9)--node[below left=1pt]
			{\scriptsize$1$}(2);
		\draw[blue, thick, midarrow=stealth](2)--node[above right=1pt]
			{\scriptsize$1$}(10);
		\draw[blue, thick, midarrow=stealth](10)--node[left=1pt]
			{\scriptsize$1$}(3);
		\draw[Purple, very thick, midarrow=stealth](3)--node[right=1pt]
			{\scriptsize$1$}(4);
		\draw[Purple, very thick, midarrow=stealth](4)--node[above=1pt]
			{\scriptsize$1$}(5);
		\draw[blue, thick, midarrow=stealth](5)--node[below=1pt]
			{\scriptsize$1$}(11);
		\draw[blue, thick, midarrow=stealth](11)--node[above=1pt]
			{\scriptsize$1$}(1);
     }}
   \put(130,17){\TikZ{[scale=.75, every node/.style=
			{inner sep=1pt, outer sep=0}]
		\path[use as bounding box](-2,-2)--(2,2);
	  \node[circle, draw=black, fill=white](1)
           at(1,1) {\scriptsize$3$};
	  \node[circle, draw=black, fill=white](2)
           at(-2,-1) {\scriptsize$0$};
	  \node[rounded corners=4pt, inner sep=2pt, draw=black, double, fill=white](*)
           at(-1,-1) {\tiny$-2$};
	  \node[rounded corners=4pt, inner sep=2pt, draw=black, fill=white](6)
           at(0,-1) {\tiny$-2$};
	  \node[rounded corners=4pt, inner sep=2pt, draw=black, fill=white](7)
           at(1,-1) {\tiny$-2$};
	  \node[rounded corners=4pt, inner sep=2pt, draw=black, fill=white](3)
           at(2,-1)  {\scriptsize$-1$};
	  \node[rounded corners=4pt, inner sep=2pt, draw=black, fill=white](8)
           at(1,0) {\tiny$-2$};
	  \node[rounded corners=4pt, inner sep=2pt, draw=black, fill=white](9)
           at(0,1) {\tiny$-2$};
	  \node[rounded corners=4pt, inner sep=2pt, draw=black, fill=white](4)
           at(-1,2)  {\scriptsize$-1$};
	  \node[rounded corners=4pt, inner sep=2pt, draw=black, fill=white](10)
           at(-1,1) {\tiny$-2$};
	  \node[rounded corners=4pt, inner sep=2pt, draw=black, fill=white](11)
           at(-1,0) {\tiny$-2$};
	  \node[circle, draw=black, fill=white](5) at(-1,-2){\scriptsize$0$};
		\draw[blue, thick, midarrow=stealth](1)--node[below=2pt]
			{\scriptsize$1$}(2);
		\draw[blue, thick, midarrow=stealth](2)--node[below=1pt]
			{\scriptsize$1$}(*);
		\draw[Purple, thick, midarrow=stealth](*)--node[above=1pt]
			{\scriptsize$1$}(6);
		\draw[Purple, very thick, midarrow=stealth](6)--node[below=1pt]
			{\scriptsize$1$}(7);
		\draw[Purple, very thick, midarrow=stealth](7)--node[below=1pt]
			{\scriptsize$1$}(3);
		\draw[Purple, very thick, midarrow=stealth](3)
		    --node[above right=1pt]{\scriptsize$1$}(8);
		\draw[Purple, very thick, midarrow=stealth](8)
		    --node[below left=1pt]{\scriptsize$1$}(9);
		\draw[Purple, very thick, midarrow=stealth](9)
		    --node[above right=1pt]{\scriptsize$1$}(4);
		\draw[Purple, very thick, midarrow=stealth](4)--node[left=1pt]
			{\scriptsize$1$}(10);
		\draw[Purple, very thick, midarrow=stealth](10)--node[left=1pt]
			{\scriptsize$1$}(11);
		\draw[Purple, very thick, midarrow=stealth](11)--node[right=1pt]
			{\scriptsize$1$}(*);
		\draw[blue, thick, midarrow=stealth](*)--node[left=1pt]
			{\scriptsize$1$}(5);
		\draw[blue, thick, midarrow=stealth](5)--node[above left=0pt]
			{\scriptsize$1$}(1);
     }}
   \put(10,3)
	 {$\C3{C_2(\tW\Ht{F}_3^{\sss(2)})}\<=11
	    \<=C_1^{~2}(\Ht{F}_3^{\sss(2)})$}
   \put(110,3)
	 {$C_1^{~2}(\tW \Ht{F}_3^{\sss(2)})\<=13
	   \<=\C3{C_2(\Ht{F}_3^{\sss(2)})}$}
  \end{picture}
 \end{center}
 \caption{Surgical replacement of the negative (CW) 
 $\n_*^\wtd\<\in\pDN{\FF[2]3}$ edge (top, mid-left, red) with a positive (CCW) union $\n_3^\wtd\<{\uplus_{\n_{34}^\wtd}}\n_4^\wtd$ (bottom, mid-left), which subdivides the (reflex, $270^\circ$, CCW) {\em\/complement\/} of the CW 2-cone over $\n_*^\wtd$ by inserting $\n_{34}^\wtd$. Conversely, $\tW \Ht{F}_3^{\sss(2)}\onto\MF[2]3$ is a blowdown/contraction of a sort, its arrow labeled by the cone being removed, e.g., $\Ht{\n_{34}^\wtd}$.
The transpolar of this surgery is shown at right, with nine cones collapsed. }
 \label{f:2F3R}
\end{figure}

We close this section with suggestions for another interpretation of ``surgery,'' which seems natural in that the cut-and-paste operations with the cones in (multi)fans are literally followed within a maximally stratified $\rT$-decomposition of the (unitary) torus manifolds.
To this end, reconsider the example $(\FF[2]3,\MF[2]3)$ from~\eqref{e:2FmDDs} and Figure~\ref{f:*2FmDD}, shown again in the top half of Figure~\ref{f:2F3R} upon the $\GL(2;\ZZ)$ transformation $\bM{x\\y}\<\to\bM{-1&-1\\~~2&~~1\\}\bM{x\\y}$ to save some space.
The doubling of $\n_*\<\in\pFn{\,\Ht{F}_3^{\sss(2)}}$ (Figure~\ref{f:2F3R}, bottom mid-right) is transpolar to the double-covering of the 1st quadrant (outward fading shade in Figure~\ref{f:2F3R}, bottom mid-left) by parts of
 $\sfA{\n_{12}^\wtd,\n_{23}^\wtd}$ and $\sfA{\n_{45}^\wtd,\n_{51}^\wtd}$, which are in different (Riemann sheet like) layers of $\pFn{{\tW \Ht{F}_3^{\sss(2)}}}$.
In turn, the (CCW-oriented, large, right-triangular) ``sail'' in $\pFn{{\Ht{F}_3^{\sss(2)}}}$ (bottom mid-right) is transpolar to the dart-shaped portion
 $\sfA{\n_{23}^\wtd,\n_{34}^\wtd,\n_{45}^\wtd}
   \<\subset\pFn{{\tW \Ht{F}_3^{\sss(2)}}}$ (bottom mid-left).

The upward solid arrows indicate exceptional set collapsing:
In the left-hand half of Figure~\ref{f:2F3R}, $\pFn{\MF[2]3}$ may be regarded as the result of removing the 1-cone $\n_{34}^\wtd$ from 
$\pFn{\tW\Ht{F}_3^{\sss(2)}}$, i.e., by collapsing the corresponding exceptional set ($\rT$-characteristic submanifold) in the almost complex 
$\Ht{F}_3^{\sss(2)}$. In the right-hand half, $\pFn{\FF[2]3}$ is similarly obtained from $\pFn{\Ht{F}_3^{\sss(2)}}$ by removing the nine 1-cones,
$\n'_*,\cdots,\n_4,\cdots,\n_3,\cdots$, in the top-layer triangular region. This corresponds to collapsing nine exceptional sets in the almost complex $\Ht{F}_3^{\sss(2)}$ to obtain the algebraic variety $\FF[2]3$. Although encoded similarly to the standard blowdown/blowup, the so-suggested
$\rT$-equivariant surgery operations must be more general than those in Ref.\cite{Jang:2023aa}. We now turn to a different possible interpretation, following the downward dashed arrows.

\paragraph{Collapsing Submanifolds:}
The cone-degree data are also straightforward to read off directly from Figure~\ref{f:2F3R} and are shown in Table~\ref{t:2AFmlist}.
\begin{table}[htb]
$$
 \begin{array}{r|r@{~}l@{}c@{}r@{~}l|lcr@{~}r@{~}l}
\dim& \MC2c{\pFn{\Ht{F}_3^{\sss(2)}}\<\smt\pDs{\,\Ht{F}_3^{\sss(2)}}} &\MC1c{\fif\wtd}
    & \MC2{c|}{\pDN{\Ht{F}_3^{\sss(2)}}\<\lat\pFN{\Ht{F}_3^{\sss(2)}}} &\dim &
    & \MC3c{\sum d(\n_*)d(\n_*^\wtd)}\\[-1pt] \toprule \nGlu{-2pt}
  0 & d(0)&=1                  && d(0^\wtd)&=11
     & 3 &\to &11\\[-1pt] \midrule \nGlu{-2pt}
  1 & d(\n_1)&=1               && d(\n_1^\wtd)&=5
     & 2 &\to &5
      &\MR3*[2pt]{$\left.\rule{0pt}{8mm}\right\}\,11$}
       &\MR3*[2pt]{$=C_1^{~2}(\Ht{F}_3^{\sss(2)})$} \\[-1pt]
    &d(\n_2)\<=d(\n_5) &=1 &&d(\n_2^\wtd)\<=d(\n_4^\wtd)&=2 &&\to &2{\times}2 \\[-1pt]
    &d(\n_3)\<=d(\n_4) &=1 &&d(\n_3^\wtd)\<=d(\n_4^\wtd)&=1 &&\to &2{\times}1 \\[-1pt]
 \midrule \nGlu{-2pt}
  2 &d(\n_{12})\<=d(\n_{51})&=1&&d(\n_{12}^\wtd)\<=d(\n_{51}^\wtd)&=1&1&\to&2{\times}1      
      & \MR3*[2pt]{$\left.\rule{0pt}{8mm}\right\}\,13$} 
       &\MR3*[2pt]{$=C_2(\Ht{F}_3^{\sss(2)})$}\\[-1pt]
    &d(\n_{23})\<=d(\n_{45})&=4&&d(\n_{23}^\wtd)\<=d(\n_{45}^\wtd)&=1&&\to&2{\times}4 \\[-1pt]
    & d(\n_{34})&=3            && d(\n_{34}^\wtd)&=1   &   &\to & 3 \\[-1pt]
 \midrule \nGlu{-2pt}
  3 & d(\n_{1\cdots5})&=13 && d(\n_{1\cdots5}^\wtd)&=1 & 0 &\to &13\\[-1pt]
\bottomrule \nGlu{1mm}
 \end{array}
$$
\caption{The cones and their degrees, spanned by the transpolar pair
         $(\pDs{\Ht{F}_3^{\sss(2)}},\pDN{\Ht{F}_3^{\sss(2)}})$}
\label{t:2AFmlist}
\end{table}
Here, $C_1^{~2}(\Ht{F}_3^{\sss(2)}){+}C_2(\Ht{F}_3^{\sss(2)})\<=2\,{\cdot}12$, indicating that $\Td(\Ht{F}_3^{\sss(2)})\<=\c^h(\Ht{F}_3^{\sss(2)})\<=2$\cite{rP+RV-12, rM-MFans, Masuda:2000aa, rHM-MFs, Masuda:2006aa}, in complete agreement with the intersection number computations indicated at far left and far right in Figure~\ref{f:2F3R}.
These double-winding multitopes (bottom of Figure~\ref{f:2F3R})
span uniformly oriented smoothly subdivided multifans, and so
correspond to almost complex torus manifolds, $\tW \Ht{F}_3^{\sss(2)}$ (left) and its transpolar, $\Ht{F}_3^{\sss(2)}$ (right).
By collapsing the indicated $\rT$-characteristic submanifolds, this transpolar pair of almost complex torus manifolds, produces:
$(\Ht{F}_3^{\sss(2)},\tW \Ht{F}_3^{\sss(2)})\<\onto(\FF[2]3,\MF[2]3)$.

\paragraph{Conjoining Surgery:}
Consider decomposing 
 $\pFn{\Ht{F}_3^{\sss(2)}}
   \<=\n_{12}\<\uplus\n_{23}\<\uplus\n_{34}\<\uplus\n_{45}\<\uplus\n_{51}$ along the ``branch-cut,'' $\n_*$, indicating the cross-over of the two layers of the multifan depictions in Figure~\ref{f:2F3Rs}.
\begin{figure}[htbp]
$$
 \vC{\begin{picture}(150,40)(0,0)
   \put(0,10){\TikZ{[scale=1]\path[use as bounding box](-3,-1)--(2,2);
              \foreach\y in{-2,...,2}
               \foreach\x in{-2,...,2} \fill[gray](\x,\y)circle(.4mm);
              \fill[yellow, opacity=.9](1,1)--(-2,-1)--(-1,-1)--(-1,-2);
               \draw[thick,-stealth](0,0)--(1,1);
               \draw(1,1)node[right]{\footnotesize$\n_1$};
               \draw[blue, thick](1,1)--(-1.1,-.4);
                \draw[blue, thick, midarrow=stealth](-1.1,-.4)--(-2,-1);
               \draw[thick,-stealth](0,0)--(-2,-1);
               \draw(-2,-1)node[left]{\footnotesize$\n_2$};
               \draw[blue, thick, midarrow=stealth](-2,-1)--(-1,-1);
               \draw[blue, ultra thick, midarrow=stealth](-1,-1)--(-1,-2);
               \draw[thick,-stealth](0,0)--(-1,-2);
               \draw(-1,-2)node[left]{\footnotesize$\n_5$};
               \draw[blue, thick, midarrow=stealth](-1,-2)--(1,1);
              \fill[yellow, opacity=.9](-1,-1)--(2,-1)--(-1,2);
               \draw[blue, thick, midarrow=stealth](-1,-1)--(2,-1);
               \draw[blue, ultra thick, midarrow=stealth](-1,2)--(-1,-1);
               \draw[thick,-stealth](0,0)--(2,-1);
               \draw(2,-1)node[right=1mm]{\footnotesize$\n_3$};
               \draw[thick,-stealth](0,0)--(-1,2);
               \draw(-1,2)node[left]{\footnotesize$\n_4$};
              \draw[purple, densely dotted, very thick, -stealth]
                  (0,0)--(-1,-1);
               \draw[purple, thick](-1,-1)circle(.55mm);
               \draw[purple, thick](-1,-1)circle(1.1mm);
               \draw[purple](-1.1,-1)node[left, rotate=45]
                                     {\footnotesize$\n'_*,~\n_*$};
                \path[purple](-.25,1.5)node[right]{$\n_*\<\in\n_{23}$,~
                                                   $\n'_*\<\in\n_{45}$};
               \draw[blue, thick, midarrow=stealth](2,-1)--(-1,2);
               \foreach\z in{0,1} {\draw[thick, densely dotted]
                   (0,0)--(\z,-1);
               \draw[thick, densely dotted](0,0)--(1-\z,\z);
               \draw[thick, densely dotted](0,0)--(-1,\z);};
              \filldraw[fill=white,thick](0,0)circle(.4mm);
              \path(1.2,-1.6)node{$\pDs{\,\Ht{F}_3^{\sss(2)}}\<\lat
                                    \pFn{\Ht{F}_3^{\sss(2)}}$};
             }}
   \put(60,10){\TikZ{[scale=1]\path[use as bounding box](-3,-1)--(1,2);
              \path(-2.9,0)node{\Huge$=$};
              \foreach\y in{-2,...,1}
               \foreach\x in{-2,...,1} \fill[gray](\x,\y)circle(.4mm);
                \fill[white](-1,-1)circle(.45mm);
              \fill[yellow, opacity=.9](0,0)--(-1.02,-.97)--(-2,-1)
                  --(1,1)--(-1,-2)--(-.97,-1.03);
               \draw[thick,-stealth](0,0)--(1,1);
               \draw(1,1)node[right]{\footnotesize$\n_1$};
               \draw[blue, thick, midarrow=stealth](1,1)--(-2,-1);
               \draw[thick,-stealth](0,0)--(-2,-1);
               \draw(-2,-1)node[left]{\footnotesize$\n_2$};
               \draw[blue, thick, midarrow=stealth](-2,-1)--(-1.03,-.97);
               \draw[blue, ultra thick, midarrow=stealth]
                  (-.97,-1.03)--(-1,-2);
               \draw[thick,-stealth](0,0)--(-1,-2);
               \draw(-1,-2)node[left]{\footnotesize$\n_5$};
               \draw[blue, thick, midarrow=stealth](-1,-2)--(1,1);
              \draw[Purple, densely dashed, very thick, -stealth]
                  (0,0)--(-1.03,-.97);
               \path[Purple](-1.15,-.85)node{\footnotesize$\n_*$};
              \draw[teal, densely dashed, very thick, -stealth]
                  (0,0)--(-.97,-1.03);
               \path[teal](-.75,-1.1)node{\footnotesize$\n'_*$};
              \filldraw[fill=white,thick](0,0)circle(.4mm);
              \path(.8,-1.5)node{\Large$\pFn{\FF[2]3\<\ssm\IP^1_*}$};
             }}
   \put(118,10){\TikZ{[scale=1]\path[use as bounding box](-1.5,-1)--(2,2);
              \path(-2,0)node{\Large``$\biguplus_{2\IP^1_*}\!$''\;};
              \foreach\y in{-1,...,2}
               \foreach\x in{-1,...,2} \fill[gray](\x,\y)circle(.4mm);
                \fill[white](-1,-1)circle(.45mm);
              \fill[yellow, opacity=.9](0,0)--(-.97,-1.02)--(2,-1)
                  --(-1,2)--(-1.03,-.97);
               \draw[blue, thick, midarrow=stealth](-.97,-1.03)--(2,-1);
               \draw[blue, thick, midarrow=stealth](2,-1)--(-1,2);
               \draw[blue, ultra thick, midarrow=stealth]
                  (-1,2)--(-1.02,-.98);
               \draw[thick,-stealth](0,0)--(2,-1);
               \draw(2,-1)node[right=1mm]{\footnotesize$\n_3$};
               \draw[thick,-stealth](0,0)--(-1,2);
               \draw(-1,2)node[left]{\footnotesize$\n_4$};
              \draw[teal, densely dashed, very thick, -stealth]
                  (0,0)--(-1.03,-.97);
               \path[teal](-1.25,-.85)node{\footnotesize$\n'_*$};
              \draw[Purple, densely dashed, very thick, -stealth]
                  (0,0)--(-.97,-1.03);
               \path[Purple](-.85,-1.25)node{\footnotesize$\n_*$};
               \foreach\z in{0,1} {\draw[thick, densely dotted]
                  (0,0)--(\z,-1);
                \draw[thick, densely dotted](0,0)--(1-\z,\z);
                \draw[thick, densely dotted](0,0)--(-1,\z);};
              \filldraw[fill=white,thick](0,0)circle(.4mm);
              \path(.5,-1.5)node{\Large$\pFn{\tW\IP^2\<\ssm\IP^1_*}$};
             }}
  \end{picture}}
$$
 \caption{\baselineskip=13pt Details of the surgery depicted at right in Figure~\ref{f:2F3R}: The two plain fans (middle and right) glue along two separate copies of $\n_*$ (a $\IP^1$ each) to form the double-winding fan of the almost complex $\Ht{F}_3^{\sss(2)}$ (left)}
 \label{f:2F3Rs}
\end{figure}
Both sub-fans, in the middle and on the right, are incomplete, since
 $\n_*\<\subset\n_{23}$ and $\n'_*\<\subset\n_{45}$ lie in different layers of the original multifan on the left in Figure~\ref{f:2F3Rs}.
Their gluing (along $\n_*$ and $\n'_*$, in each respective layer) then (re-)introduces two $\IP^1$s, as indicated by the subscript of the adjoining:
 $\Ht{F}_3^{\sss(2)} = (\FF[2]3)\<{\boxplus_{2\IP^1_*}}(\tW[-2mu]\IP^2)$
may seem {\em\/somewhat akin to\/} a connected sum of the two components along the two different-layer images of $\IP^1_*\mapsfrom\n_*$.\footnote{However, this ``conjoining'' surgery seems to be novel, as the centers of the right-hand side fans in Figure~\ref{f:2F3Rs} and~\ref{f:2F3++} are being identified. This corresponds to identifying the $(\IC^*)^n$-like ``dense open subsets'' (i.e., almost everything) in $\FF[2]3$ and $\tW[-2mu]\IP^2$, which is thus {\em\/not at all\/} the standard ``connected sum'' of manifolds, nor yet the $\#$-operation on polytopes in Ref.\cite{Buchstaber:2001aa}. Since $\c(\IC^*)^n\,=\c(S^1)^n\<=0$, this dense open subset does not enter the indicated Euler number computations.}
This agrees with the simple Chern/Euler number computation ($\Ht{F}_3^{\sss(2)}$ is almost complex\cite[Thm.\;1]{rM-MFans}, so $C_1^{~2}\<=\c$\,),
\begin{equation}
  13\5[2pt]{\sss\text{left}}{\6{\sss\text{Fig.}\,\ref{f:2F3R}}=}
  \c(\Ht{F}_3^{\sss(2)})
   \isBy{\sss\text{Fig.}\,\ref{f:2F3Rs}}\big(\c(\FF[2]3){-}\c(\IP^1_*)\big)
   +\c(2\IP^1_*) +\big(\c(\tW[-2mu]\IP^2){-}\c(\IP^1_*)\big)
   =(4{-}2)+(9{-}2) +2(2).
\end{equation}
While the uniform CCW-orientation of multifans such as in Figure~\ref{f:2F3Rs} implies an almost complex structure\cite[Thm.\;1]{rM-MFans}, its winding number, $(w\<=2)\<=\Td(\Ht{F}_3^{\sss(2)})$, 
implies it is not a toric variety, and a more precise identification of corresponding torus manifolds such as
 $\Ht{F}_3^{\sss(2)}$ requires further systematic study.
The transpolar multitope $\pDN{\Ht{F}_3^{\sss(2)}}$, at bottom left in Figure~\ref{f:2F3R}, may be similarly decomposed
 $\tW \Ht{F}^{\sss(2)}_3 =
   \big(\MF[2]3\big)\<{\boxplus_{\IP^1_{23},\,\IP^1_{45}}}\big(\IP^2\big)$, as shown in Figure~\ref{f:2F3++},
\begin{figure}[htbp]
$$
 \vC{\begin{picture}(150,54)(0,-4)
   \put(0,0){\TikZ{[scale=.95]\path[use as bounding box](-3,-3)--(2,2);
              \foreach\y in{-3,...,2}
               \foreach\x in{-3,...,2} \fill[gray](\x,\y)circle(.4mm);
              \fill[Turque!25, opacity=.9](0,0)--(0,1)--(-1,-1);
               \draw[blue, thick](0,1)--(-.6,-.2);
               \draw[blue, thick, midarrow=stealth](-.6,-.2)--(-1,-1);
                \path[blue](-.75,-.6)node[left]{\footnotesize$\n_3^\wtd$};
              \fill[Turque!25, opacity=.9](0,0)--(0,1)--(2,-3)--(-3,2);
               \draw[blue, thick, midarrow=stealth](2,-3)
                     to node[above right=-1mm]
                     {\footnotesize$\n_2^\wtd$}(0,1);
               \draw[Magenta, thick, -stealth](0,0)--(0,1);
               \draw[thick,-stealth](0,0)--(2,-3);
                \draw(0,1)node[above right=-1mm]
                    {\footnotesize$\n_{23}^\wtd$};
               \draw(-1,-1)node[below left=-2pt]
                    {\footnotesize$\n_{34}^\wtd$};
              \draw[blue, thick, midarrow=stealth](-3,2)--(2,-3);
               \path[blue](0,-1)node[below left=-1mm]
                    {\footnotesize$\n_1^\wtd$};
               \draw(2,-3)node[left]{\footnotesize$\n_{12}^\wtd$};
               \draw(1,0)node[above right=-1mm]
                    {\footnotesize$\n_{45}^\wtd$};
              \draw[thick,-stealth](0,0)--(-3,2);
               \draw(-3,2)node[below=1mm]{\footnotesize$\n_{51}^\wtd$};
              \foreach\z in{0,...,3}
               \draw[densely dotted](0,0)--(1-\z,\z-2);
               \draw[densely dotted](0,0)--(1,-1);
              \fill[Turque!25,opacity=.9](0,0)--(-3,2)--(1,0)--(-1,-1);
               \draw[Magenta, thick,-stealth](0,0)--(1,0);
               \draw[blue, thick, midarrow=stealth](1,0)
                     to node[above]{\footnotesize$\n_5^\wtd$}(-3,2);
               \draw[blue, thick, midarrow=stealth](-1,-1)--(1,0);
                \path[blue](.1,-.65)node{\footnotesize$\n_4^\wtd$};
               \draw[densely dotted](-1,1)--(0,0);
               \draw[thick, densely dotted, -stealth](0,0)--(-1,-1);
              \filldraw[fill=white,thick](0,0)circle(.4mm);
              \path(1,-2.5)node[left]
                    {$\pDN{\,\Ht{F}_3^{\sss(2)}}
                       \pDs{\tW \Ht{F}_3^{\sss(2)}}
                        \<\lat\pFn{{\tW \Ht{F}_3^{\sss(2)}}}$};
              \path(-1.75,1.75)node[right]
                    {\footnotesize uniform double cover};
             }}
   \put(60,0){\TikZ{[scale=.95]\path[use as bounding box](-3,-3)--(2,2);
              \path(-3.67,0)node{\Huge$=$};
              \foreach\y in{-3,...,2}
               \foreach\x in{-3,...,2} \fill[gray](\x,\y)circle(.4mm);
              \fill[Turque!25, opacity=.9](0,0)--(0,1)--(2,-3)--(-3,2);
               \draw[blue, thick, midarrow=stealth](2,-3)
                     to node[above right=-1mm]
                    {\footnotesize$\n_2^\wtd$}(0,1);
               \draw[Magenta, densely dotted, thick, -stealth](0,0)--(0,1);
               \draw[thick, -stealth](0,0)--(2,-3);
                \draw(0,1)node[above right=-1mm]
                    {\footnotesize$\n_{23}^\wtd$};
              \fill[red,opacity=.6](0,0)--(0,1)--(1,0);
                \draw[Rouge, very thick, midarrow=stealth](0,1)
                     to node[above=-1pt, rotate=-45]
                    {\footnotesize$~~~\bS{\Tw\n_3^\wtd}$}(1,0);
              \draw[blue, thick, midarrow=stealth](-3,2)--(2,-3);
               \path[blue](0,-1)node[below left=-1mm]
                    {\footnotesize$\n_1^\wtd$};
              \draw(2,-3)node[right]{\footnotesize$\n_{12}^\wtd$};
               \path(1,0)node[above right=-1mm]
                    {\footnotesize$\n_{45}^\wtd$};
              \foreach\z in{0,...,3}
               \draw[densely dotted](0,0)--(1-\z,\z-2);
               \draw[densely dotted](-1,1)--(0,0)--(1,-1);
              \fill[Turque!25,opacity=.9](0,0)--(-3,2)--(1,0);
               \draw[Magenta, densely dotted, thick,-stealth](0,0)--(1,0);
               \draw[blue, thick, midarrow=stealth](1,0)
                     to node[above]{\footnotesize$\n_4^\wtd$}(-3,2);
              \draw[thick,-stealth](0,0)--(-3,2);
               \path(-3,2)node[below=2pt]{\footnotesize$\n_{51}^\wtd$};
               \draw[densely dotted](-1,1)--(0,0);
              \filldraw[fill=white,thick](0,0)circle(.4mm);
              \path(1,-2.5)node[left]{$\pDN{\FF[2]3}\<=\pDs{\MF[2]3}
                                             \<\lat\pFn{\MF[2]3}$};
              \path(-1.75,1.75)node[right]
                    {\footnotesize effectively simple cover};
             }}
   \put(128,19){\TikZ{[scale=.95]\path[use as bounding box](-1,-1)--(1,1);
              \path(-2,0)node{\Large``$\biguplus_{\C7{2\IP^1}}\!$''\;};
              \foreach\y in{-1,...,1}
               \foreach\x in{-1,...,1} \fill[gray](\x,\y)circle(.4mm);
              \fill[Turque!25, opacity=.9](0,1)--(-1,-1)--(1,0);
               \draw[blue, thick, midarrow=stealth](0,1)
                     to node[above left=-3pt]
                     {\footnotesize$\n_3^\wtd$} (-1,-1);
               \draw[Magenta, densely dotted, thick, -stealth](0,0)--(0,1);
                \draw(0,1)node[above right=-1mm]
                    {\footnotesize$\n_{23}^\wtd$};
               \draw(1,0)node[above right=-1mm]
                    {\footnotesize$\n_{45}^\wtd$};
               \draw(-1,-1)node[below left=-2pt]
                    {\footnotesize$\n_{34}^\wtd$};
               \draw[Magenta, densely dotted, thick,-stealth](0,0)--(1,0);
               \draw[blue, thick, midarrow=stealth](-1,-1)
                     to node[below right=-3pt]
                    {\footnotesize$\n_4^\wtd$}(1,0);
               \draw[blue, very thick, midarrow=stealth](1,0)
                     to node[above=-2pt, rotate=-45]
                    {\footnotesize$\bS{-\Tw\n_3^\wtd}$}(0,1);
               \draw[thick, -stealth](0,0)--(-1,-1);
              \filldraw[fill=white,thick](0,0)circle(.4mm);
              \path(0,-2)node{$(\pDN{\tW\IP^2}\<=\pDs{\IP^2}
                                 \<\lat\pFn{\IP^2})$};
              \path(0,1.75)node{\footnotesize simple cover};
             }}
  \end{picture}}
$$
 \caption{\baselineskip=13pt Details of the surgery depicted at left in Figure~\ref{f:2F3R}: relating (by means of the spanned multifans) the pre-complex $\MF[2]3$ (middle) to the almost complex $\Ht{F}_3^{\sss(2)}$ (left) and the standard $\IP^2$ (right)}
 \label{f:2F3++}
\end{figure}
which agrees with the corresponding simple Chern/Euler number calculation
($\tW \Ht{F}_3^{\sss(2)}$ is almost complex\cite[Thm.\;1]{rM-MFans}, so $C_2\<=\c$ again) 
\begin{equation}
  11\5[2pt]{\sss\text{right}}{\6{\sss\text{Fig.}\,\ref{f:2F3R}}=}
  \c(\tW \Ht{F}_3^{\sss(2)})
  \isBy{\sss\rm Fig.\,\ref{f:2F3++}}
    \big(C_2(\MF[2]3) -\c(\IP^1_*)\big)
    +\big(\c(\IP^2) -\c(\IP^1_*)\big) +\c(2\IP^1)
  =(8{-}2)+(3{-}2) +2(2).
\end{equation}
In this surgery, the CW 2-cone over $\Tw\n_3^\wtd\<\in\pDN{\FF[2]3}$ effectively cancels the CCW 2-cone over 
${-}\Tw\n_3^\wtd\<\in\pDN{\tW\IP^3}$, and is replaced by the composite, 
$\n_3^\wtd\<\uplus\n_4^\wtd$, that is effectively a subdivision of a
{\em\/reflex\/} (and so concave) cone,
$\measuredangle(\n_{23}^\wtd,\n_{45}^\wtd)_{\sss\rm CCW}$.
This suggest interpreting $\tW \Ht{F}_3^{\sss(2)}$ (Figure~\ref{f:2F3++}, far left) as something akin to a blowup of $\MF[2]3$ at a point in the chart $U_{\color{Rouge}\Tw\n_3^\wtd}$ (Figure~\ref{f:2F3++}, middle), as mentioned above.
These computations indicate that $C_2(\MF[2]3)$ behaves exactly like the standard Euler characteristic, although $\MF[2]3$ is only precomplex.

\section{Conclusions and Outlook}
\label{s:Coda}
Throughout this article, we have examined and tested various computational characteristics associated with candidate pairs of embedding spaces, where the transposition mirror construction\cite{rBH, rBH-LGO+EG, rCOK, Krawitz:2009aa, rMK-PhD, rFJR-07b, Ebeling:2012Mir, rLB-MirrBH, rACG-BHK, rA+P-MM, rF+K-BHK} of Calabi--Yau models is generalized in pairs of simple anticanonical hypersurfaces. The overall framework presented in Corollary~\ref{CC:MMM} uses the triple $(Z_f,X,\cKs{X})$, which contains a 
Calabi--Yau hypersurface, $Z_f\<=\{f(x)\<=0\}\<\subset X$, 
the zero locus of $f(x)\<\in\G(\cKs{X})$, a suitable section of the anticanonical bundle, $\cKs{X}$. 
The mirror model is then part of the analogous triple
$(Z_{\tT f},\tX,\cKs{\tX})$. When $X$ is a (semi) Fano (complex-algebraic) toric variety encoded by a (MPCP-subdivided) reflexive polytope, $\pD$, then $\tX$ is the (complex-algebraic) toric variety analogously encoded by the (MPCP-subdivided) polar polytope
$\pD^\circ$\cite{rBaty01}.

Following the work of\cite{rBH-gB, Berglund:2022dgb}, we present computational support for the Conjecture~\ref{C:gCYh} that this framework extends to the much more general VEX multitopes: smooth, star-triangulable but perhaps flip-folded or otherwise multilayered multihedral objects that span multifans\cite{rM-MFans, Masuda:2000aa, rHM-MFs, Masuda:2006aa, rHM-EG+MF, rH-EG+MFs2, Nishimura:2006vs}. They stem from compact unitary manifolds with ``half-dimensional'' torus-action, $\rT$, identified with the complement of the $\U(1)^r$-gauge symmetry in the associated GLSM\cite{rPhases,rMP0} and so germane to string theory; see Remark~\ref{r:GLSM}. 
Generalizing the standard polar duality~\eqref{e:StdP} to act as an involution among VEX multitopes, the transpolar Definition~\ref{D:tP},
relates the corresponding toric embedding spaces, $X\fif\wtd\tX$ (specified in the GLSM by omitting the superpotential), in the transposition mirror framework in Figure~\ref{f:gMM}. That is, the mirror symmetry between the anticanonical hypersurfaces (zero loci of GLSM superpotentials), $Z_f\<\subset X$ and $Z_{\tT f}\<\subset\tX$,
{\em\/lifts\/} to the transpolar relations $X\fif\wtd\tX$ and 
$\cKs{X}\fif\wtd\cKs{\tX}$.
Note that this also relates the non-compact Calabi--Yau manifolds, 
$(X\<\ssm Z_f)\fif\wtd(\tX\<\ssm Z_{\tT f})$\cite{rTY1, rTY2}, and that the total spaces of $(\cKs{X},\cKs{\tX})$ also form a transpolar pair of non-compact Calabi--Yau manifolds\cite{Hubsch:2025teh, Hubsch:2025sph}: is there mirror symmetry on every ``floor'' in Figure~\ref{f:gMM}?

 By explicit computation in several infinite sequences and many other (2 and 3-dimensional) examples of VEX multitopes, we find that full agreement with the framework in Figure~\ref{f:gMM}. In particular, 
 \SS\:\ref{s:CCTH} verifies all Chern characteristics --- and even detailed contributions to Chern classes, computed from the VEX multitopes, including various Todd--Hirzebruch identities and a few other applications of the HRR theorem.
 \SS\:\ref{s:iN} verifies all divisor class intersection numbers and their relation with Chern characteristics.
 \SS\:\ref{s:VEX} finds agreement with Hibi's Lemma~\ref{L:Hibi}, which extends the standard results about the Poincar\'e polynomial, including the Dehn--Sommerville relations, typically related to Poincar\'e duality. However, the so-called unimodality relations (typically related to the Hard Lefschetz theorem) are satisfied by some, but not all VEX multitopes.
 Finally, \SS\:\ref{s:surg} finds that the various multitope features lend themselves to further understanding of the corresponding unitary torus manifolds/orbifolds. In particular, we find that the a flip-folded region in a multitope, $\pDs{X}$, corresponds to a localized obstruction to a global complex structure over $X$, rendering it {\em\/precomplex\/}; see Conjecture~\ref{C:obstruct} and \SS\:\ref{s:pCpx}. Since flip-folded regions in $\pDs{X}$ are transpolar to concave regions in 
$\pDN{X}\<=\pDs{\tX}$ and mirror symmetry relates (pre)symplectic to (pre)complex structure, it would follow that concave regions in a multitope, $\pDs{X}$, are obstructions for a global symplectic structure over $X$.

Our Claims, Conjectures and multiple Remarks are here supported mostly by numerous concrete computations --- and the fact that no counter-example has been found, except for the explicitly stated lack of unimodality of the ``$h$-vector'' for the $\EE{m}$-sequence of examples, so its relation to Betti numbers fails. We should like to hope that this provides ample motivation for a more rigorous examination of these claims and conjectures, their possible (dis)proof --- or correction, and thereby a more rigorous foundation for the vast breadth of application of the transposition mirror framework in Figure~\ref{f:gMM} among VEX multitopes.

\begin{figure}[htb]
 \begin{center}
  \begin{picture}(160,80)
   \put(0,0){\TikZ{\path[use as bounding box](0,0);
             \draw[Rouge, thick, densely dotted, <->]
             (2.9,3.3)to[out=-90,in=45]++(.5,-1.33);
             \path(3.9,1.9)node[Rouge, right, rotate=-48]
             {\footnotesize$d(\q)\<=2$};
              \draw[Rouge, thick, -stealth](3.97,1.82)to
              [out=135,in=30]++(-.4,.05);
             }}
   \put(0,0){\includegraphics[height=35mm]{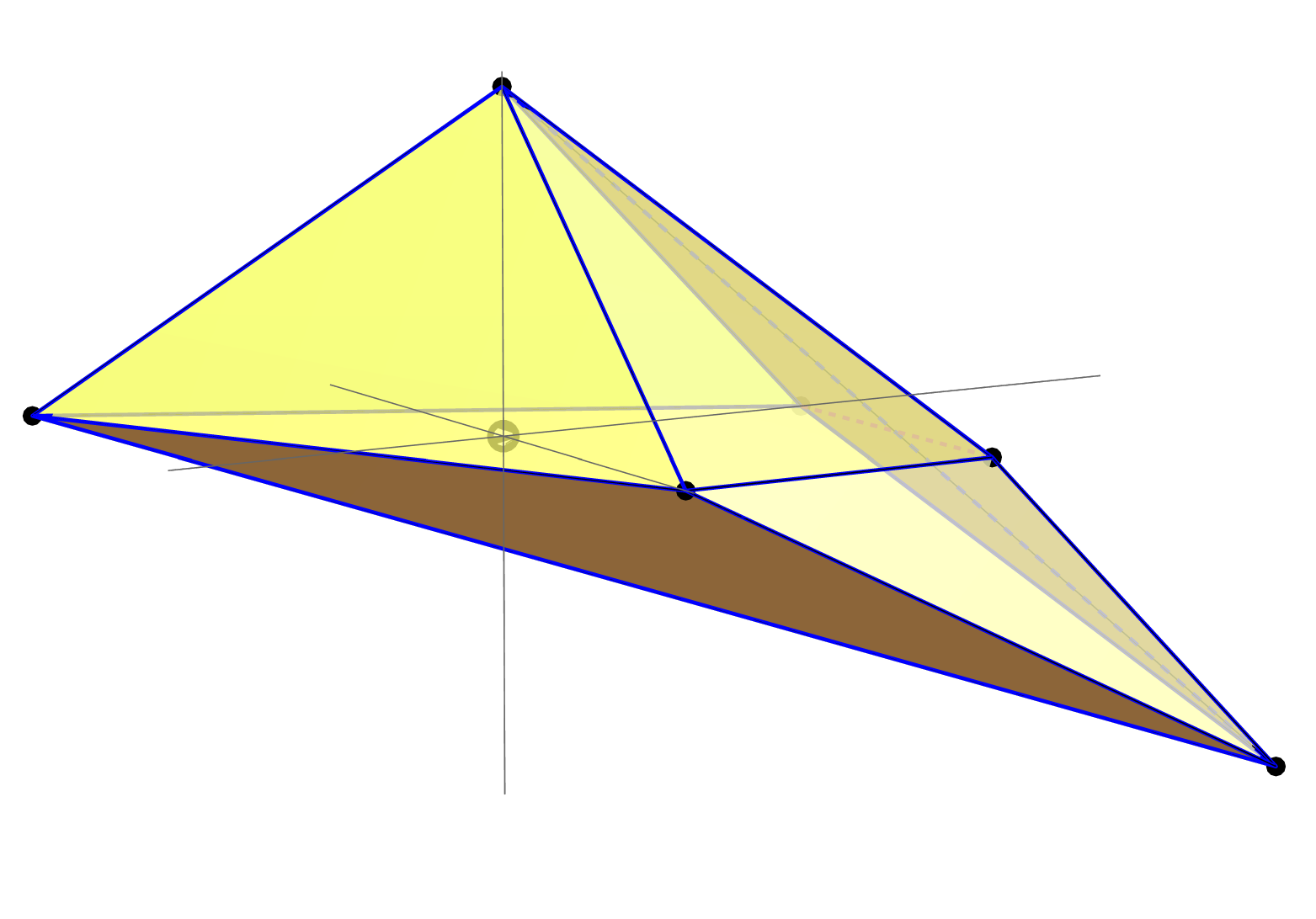}}
    \put(0,16.5){\footnotesize$\n_1$}
    \put(22,17.5){\footnotesize$\n_2$}
    \put(29,17.5){\footnotesize\color{Rouge!50}$\n_3$}
    \put(36,15){\footnotesize$\n_6$}
    \put(18,32.5){\footnotesize$\n_4$}
    \put(48,7){\footnotesize$\n_5$}
   \put(20,30){\includegraphics[height=50mm]{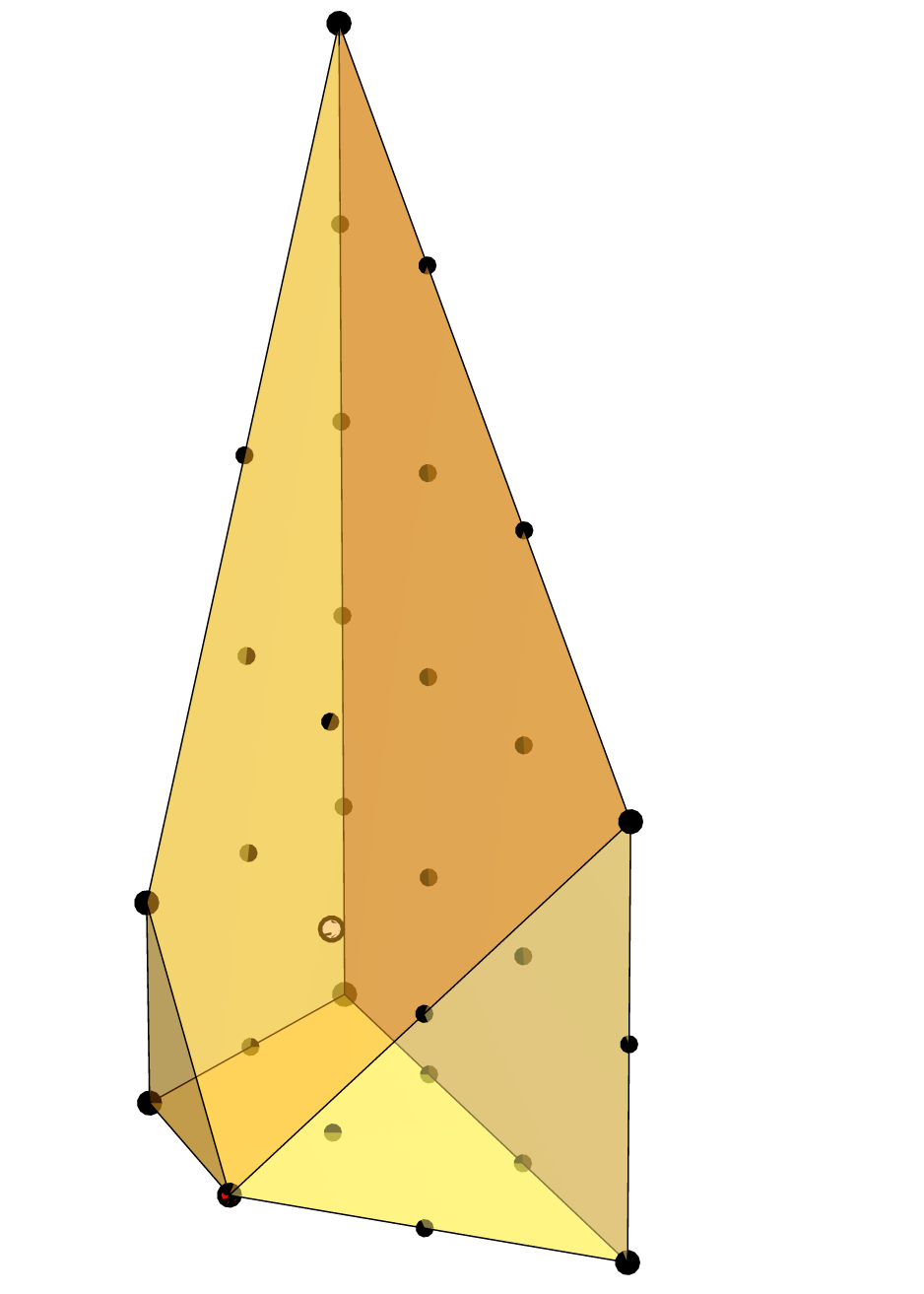}}
    \rput(25.5,36){\scriptsize$\m_4$}
    \rput(25.5,46.5){\scriptsize$\m_5$}
    \rput(28,33){\scriptsize$\m_3$}
   \put(55,0){\includegraphics[height=35mm]{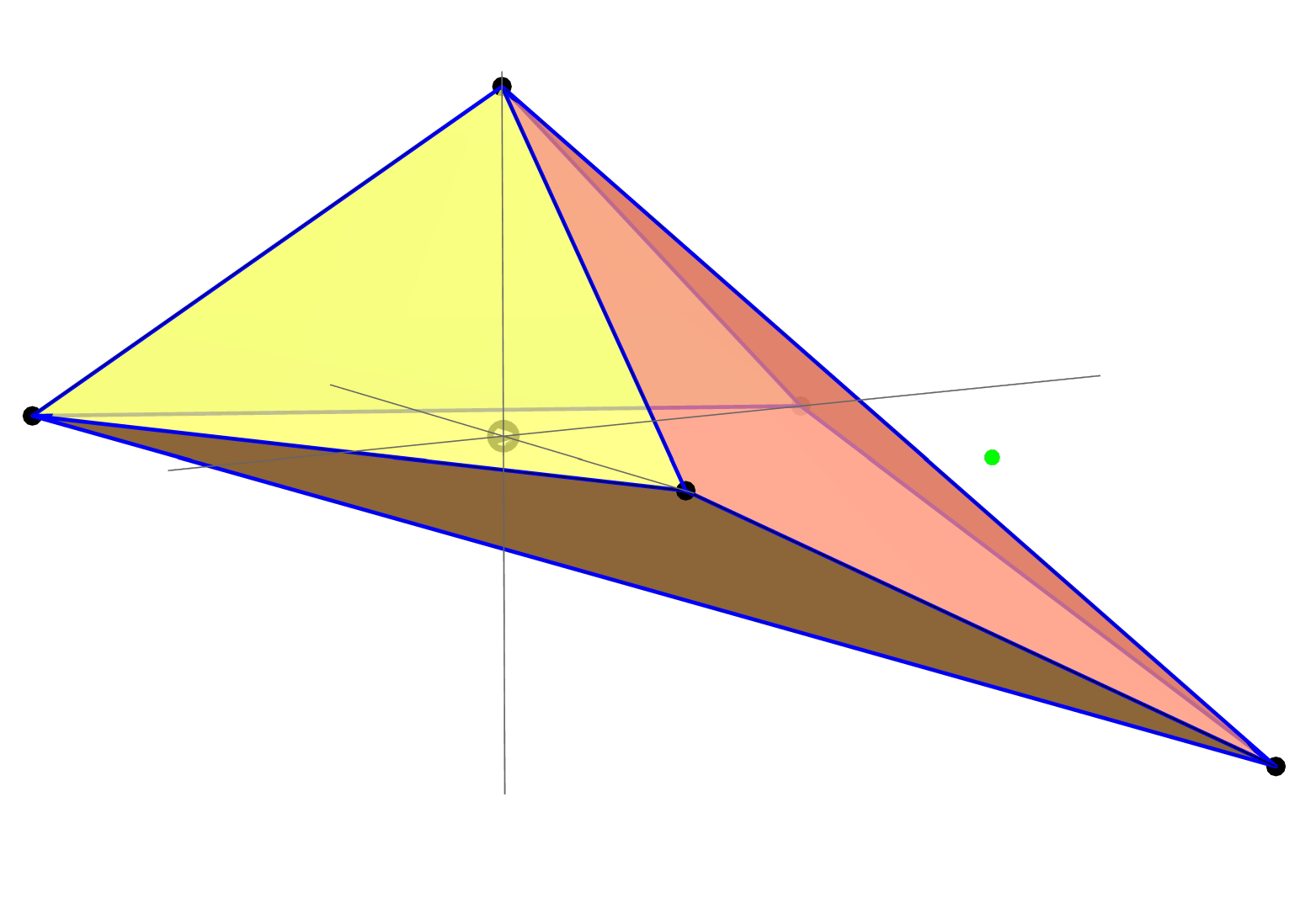}}
    \put(55,16.5){\footnotesize$\n_1$}
    \put(77,17.5){\footnotesize$\n_2$}
    \put(84,17.5){\footnotesize\color{Rouge!50!gray}$\n_3$}
    \put(94,16.5){\footnotesize$\C2{\n_6}$}
     \put(91.95,16.85){\footnotesize$\C2{\star}$}
    \put(73,32.5){\footnotesize$\n_4$}
    \put(103,7){\footnotesize$\n_5$}
    \rput(78.5,39){\scriptsize\C7{(fractional)}}
    \rput(78.5,35.5){\footnotesize$\C7{\n_{254}^\wtd\<=\m_*}$}
   \put(75,30){\includegraphics[height=50mm]{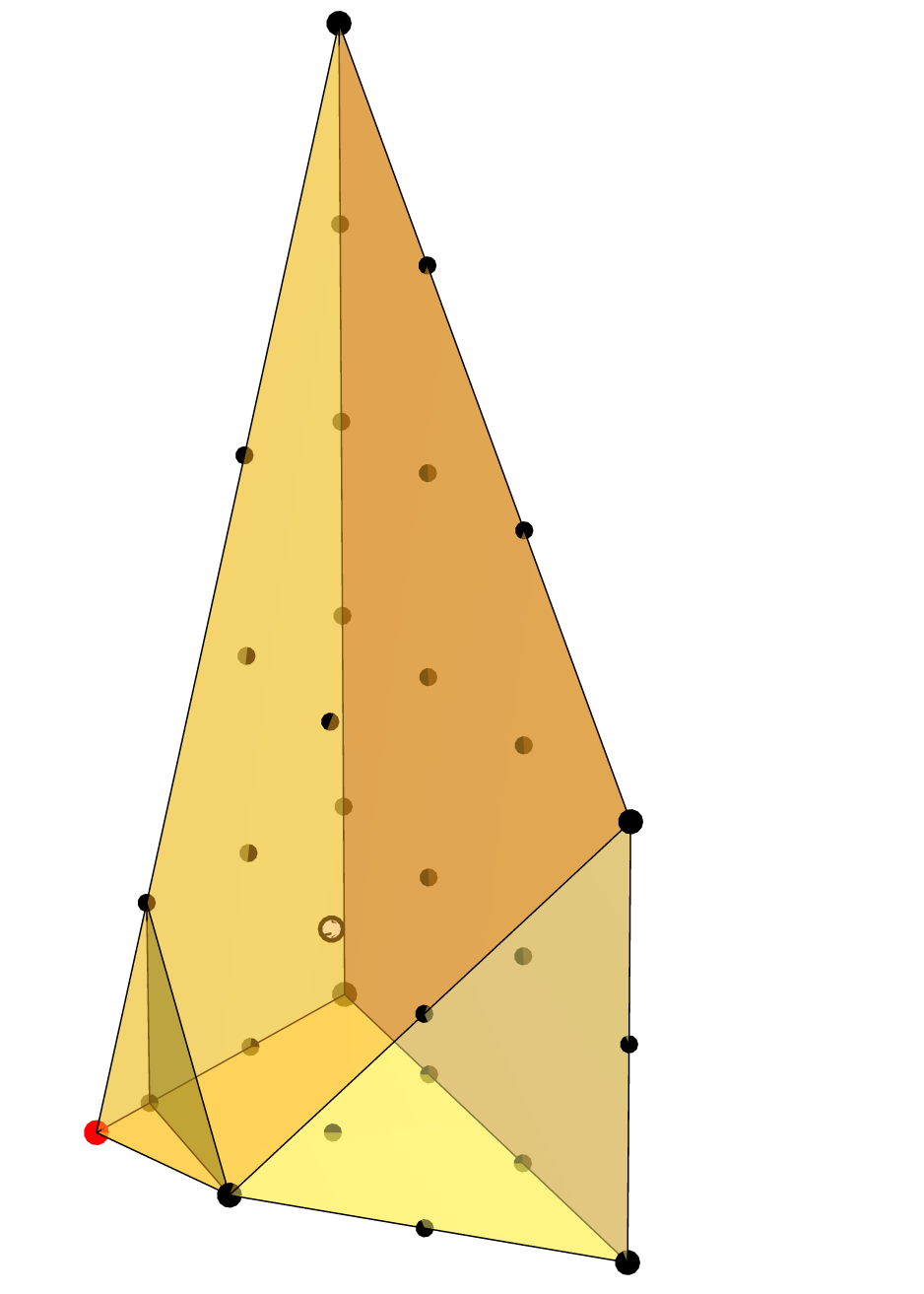}}
    \rput(80.5,46.5){\scriptsize$\m_5$}
    \rput(85,32){\scriptsize$\m_3$}
   \put(110,0){\includegraphics[height=35mm]{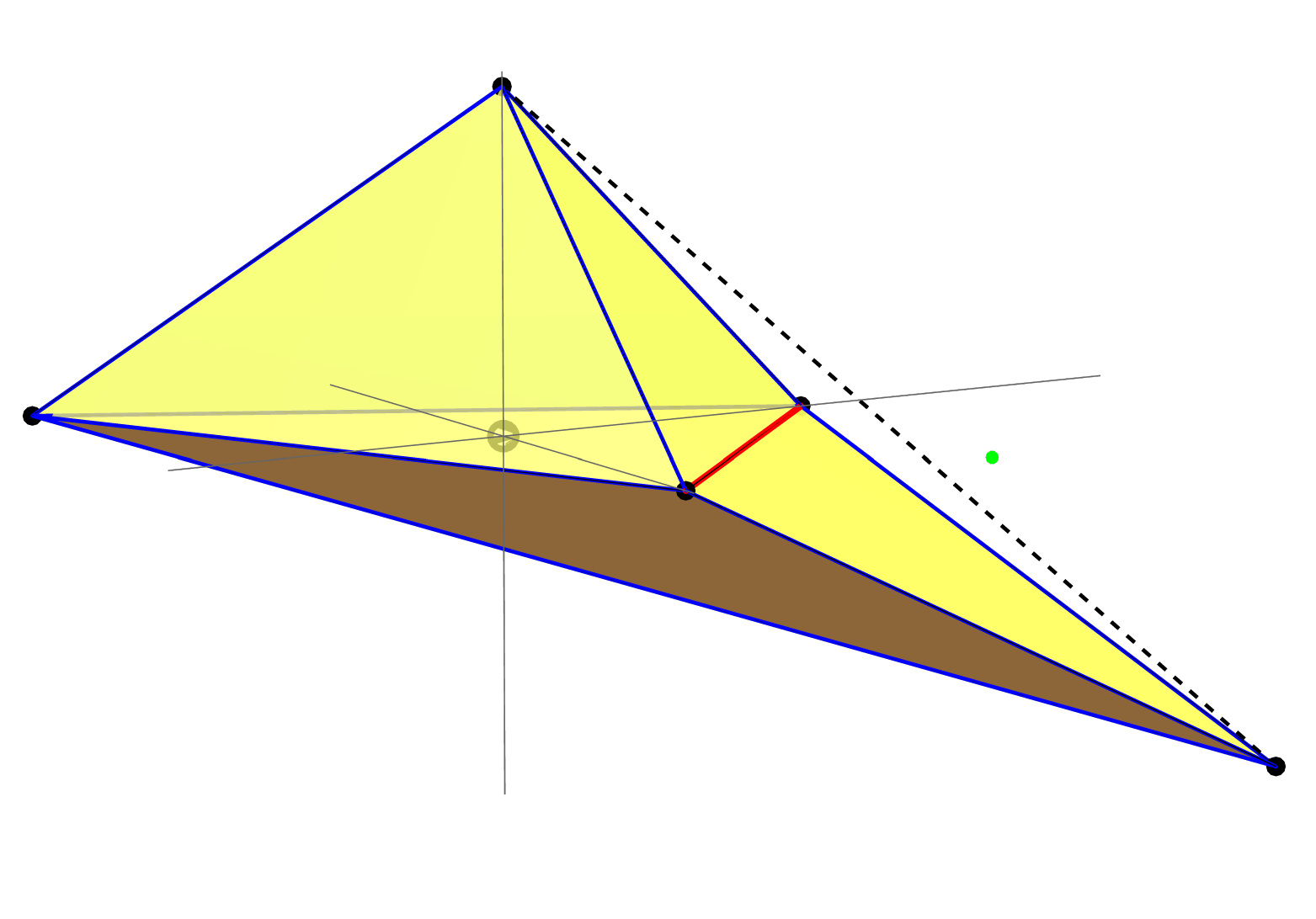}}
    \put(110,16.5){\footnotesize$\n_1$}
    \put(132,17.5){\footnotesize$\n_2$}
    \put(140,17.5){\footnotesize$\n_3$}
     \put(146.95,16.85){\footnotesize$\C2{\star}$}
    \put(128,32.5){\footnotesize$\n_4$}
    \put(158,7){\footnotesize$\n_5$}
   \put(135,30){\includegraphics[height=50mm]{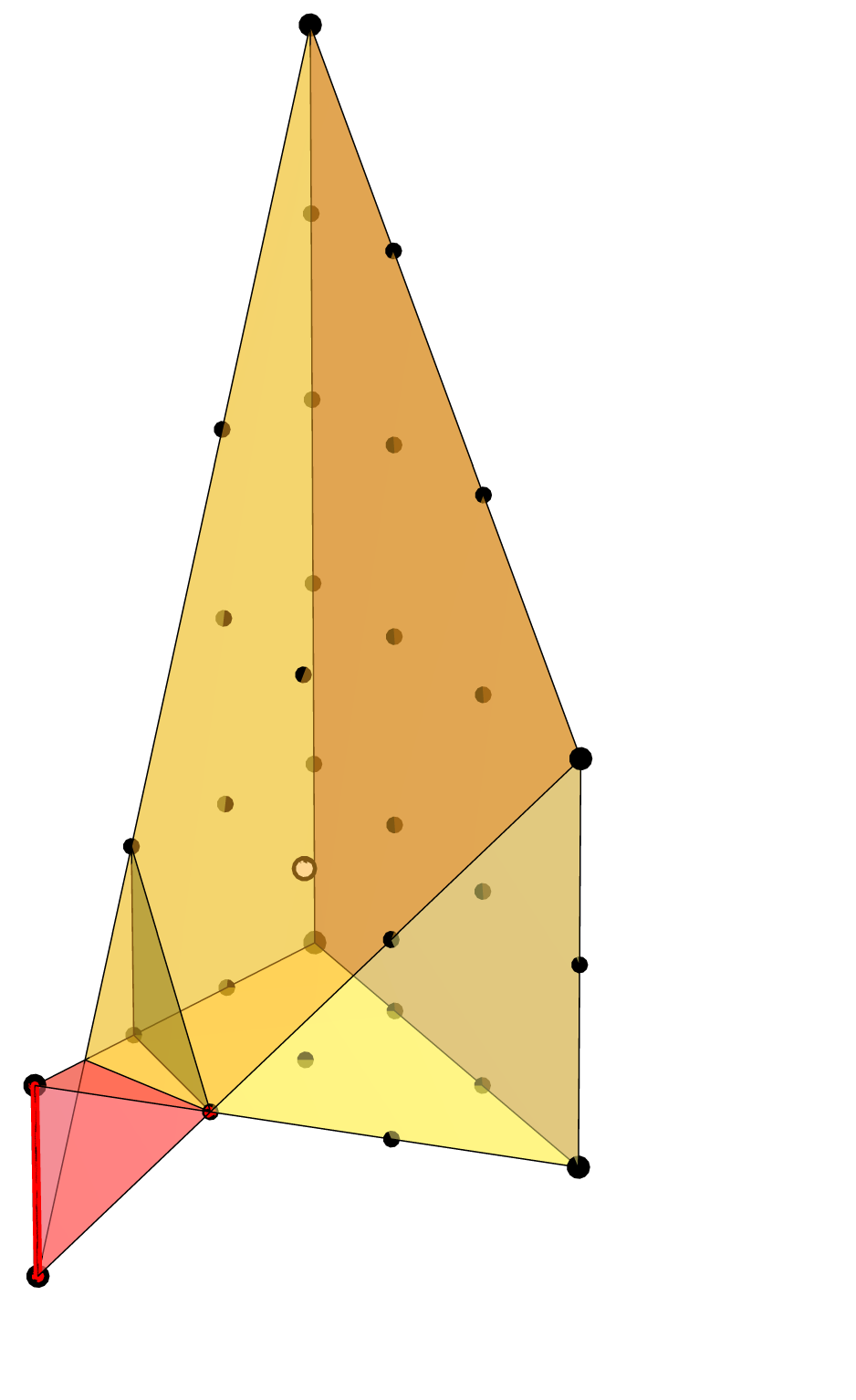}}
    \rput(138,43){\scriptsize$\C7{\m_*}$}
    \rput(136,40){\scriptsize$\C7{\m_1}$}
    \put(137.5,33.5){\scriptsize$\C7{\m_2}$}
    \put(141.5,38){\scriptsize$\C7{\m_3}$}
   \put(0,0){\TikZ{\path[use as bounding box](0,0);
             \path(.8,3)node[rotate=40]{\Large$\pDs{\Ht{\fB}}$};
             \path(2.9,7)node[left]{\Large$(\pDs{\Ht{\fB}})^\circ\<=\pDN{\Ht{\fB}}$};
             \draw[ultra thick, densely dotted, <->](.7,3.4)to[out=110,in=-135]
                 node[above, rotate=55]{polar}++(1.5,3.3);
             \path(3.3,2.85)node{\color{Rouge}\scriptsize$\m_3\<=\n_{3465}^\wtd$};
             \path(.6,.7)node[right]{singular, convex};
             \path(5.5,.2)node[left]{\footnotesize(has two smooth/regular triangulations)};
             \path(6.2,2.8)node[rotate=40]{\Large$\pDs{\fB}$};
             \path(8.5,7)node[left]{\Large$(\pDs{\fB})^\circ\<=\pDN{\fB}$};
             \draw[ultra thick, densely dotted, <->](6.1,3.2)to[out=110,in=-135]
                 node[above, rotate=55]{polar}++(1.6,3.5);
             \draw[Rouge, thick, densely dotted, <->](7.85,3.5)to[out=-90,in=75]++(.3,-1.4);
             \draw[Green, thick, densely dotted, <->](8.3,3.8)to[out=0,in=75]++(1,-1.95);
             \path(8.4,2.54)node[Rouge, right, rotate=-42]{\footnotesize``tent''};
             \path(6.3,1.4)node[Rouge, right, rotate=-15]{$d(\q)\<=2$};
              \draw[Rouge, thick, -stealth](7.7,1.0)to[out=-15,in=-90]++(1,.65);
             \path(6,.7)node[right]{singular, convex};
             \path(5.6,.2)node[right]{\footnotesize(no smooth/regular triangulation)};
             \draw[blue, ultra thick, densely dashed, -stealth]
                 (5.9,2.3)--node[above]{blowup}++(-2.3,0);
             \draw[Magenta, ultra thick, densely dashed, -stealth]
                 (9.1,2.3)--node[above]{VEXing}++(2.3,0);
             \path(11.8,3)node[rotate=40]{\Large$\pDs{\FF[3]{1,2}}$};
             \path(14.3,7)node[left]{\Large$(\pDs{\MF[3]{1,2}})\<=\pDN{\FF[3]{1,2}}$};
             \draw[ultra thick, densely dotted, <->](11.6,3.3)to[out=110,in=-135]
                 node[above, rotate=55]{\em transpolar}++(1.6,3.3);
             \draw[Rouge, thick, densely dotted, <->](13.55,3.7)to[out=-180,in=120]++(.3,-1.82);
             \path(14.5,3.05)node{\scriptsize${\color{Rouge}\m_{12}}\<=\n_{23}^\wtd$};
             \path(13.4,3.05)node[teal, right, rotate=-43]{\footnotesize``hammock''};
             \draw[Green, thick, densely dotted, <->](14.2,4.4)to[out=0,in=45]++(.625,-2.59);
             \path(11.3,1.5)node[right, rotate=-15]{$d(\q)\<=1,~\text{all}~\q$};
             \path(11,.2)node[right]{smooth/regular, non-convex};
             }}
  \end{picture}
 \end{center}
 \caption{Middle: the polar of the (red-shaded) $d(\q)\<=2$ singular facet is fractional.
     Left: its blowup results in a polar pair of reflexive lattice polytopes.
     Right: the VEX (reduced) re-triangulation $\pDs{\fB}\<\to\pDs{\FF[3]{1,2}}$ is non-convex and its transpolar, the Newton multitope, has a flip-folded (red-shaded) extension}
 \label{f:3pairs}
\end{figure}

\paragraph{VEXing:} Let us conclude with one more motivation for including flip-folded and otherwise multi-layered oriented multitopes (and the toric spaces to which they correspond) by considering the three (trans)polar pairs of multitopes depicted in Figure~\ref{f:3pairs}, which generalizes earlier work on this topic\cite{rBKK-tvMirr}.
\subparagraph{Middle:} Start with the toric variety encoded by the star-triangulable convex spanning polytope $\pDs{\fB}$, mid-bottom in Figure~\ref{f:3pairs}, with vertices
\begin{equation}
  \pDs{\fB}= \Conv\!\big(
  \n_1\<=({-}1,{-}1,0),~
  \n_2\<=(1,0,0),~
  \n_3\<=(0,1,0),~
  \n_4\<=(0,0,1),~
  \n_5\<=(1,2,-1)\big).
\end{equation}
It spans the (outward, positively oriented) fan
\begin{equation}
   \pFn{\fB}=\{\n_{124},\n_{143},\n_{135}\,\n_{152},\n_{345},\n_{254}\},
 \label{e:SfB}
\end{equation}
of central 3-cones over the facets,  $\n_{ijk}\<=\sfa\q_{ijk}$, so we use them interchangeably.
Since $d(\n_{254})\<=2$, this cone encodes a $\IC^3/\ZZ_2$-like singular chart $U_{254}\<\subset\fB$, and its polar image in the Newton polytope $\pDN{\fB}\<=(\pDs{\fB})^\circ$ is the fractional (non-lattice) point,
$\q_{254}^\wtd\<=\m_*\<=(-1,-\frac12,-1)$, which cannot encode a regular anticanonical monomial on $\fB$; $\m_*\<\mapsto \sqrt{x_1\!^5x_3}$ in terms of the standard Cox variables, $x_i\<\mapsfrom\n_i$.
$(\pDs{\fB},\pDN{\fB})$ is a polar pair of polytopes which however are not  reflexive: $\q_{254}\<\in\pDs{\fB}$ is at lattice distance~2\cite{rBaty01} from the origin and its polar image, $\q_{254}^\wtd\<=\m_*$, is not a lattice point.

\subparagraph{Left:} By removing the (magenta-colored, lower-left) fractional vertex $\m_*$ from
 $\pDN{\fB}$, we truncate $\pDN{\fB}\<\to\pDN{\Ht{\fB}}$ to the nearest lattice plane at distance~1\cite{rBaty01} from the origin, as shown at left in Figure~\ref{f:3pairs}. The polar of this is the convex spanning polytope $\pDs{\Ht{\fB}}$, depicted at bottom-left, where the degree-2 facet
 $\q_{254}\<\in\vd\pDs{\fB}$ has been subdivided by introducing the new vertex, $\n_6\<=(1,1,0)$. As $\n_6$ lies outside $\pDs{\fB}$ but within the cone being subdivided, $\n_{254}$, this encodes a standard blowup, $\Ht{\fB}\<=\Bl[\fB]$. The convex polytope $\pDs{\Ht{\fB}}$ now has a {\em\/quadrangular\/} degree-2 facet, $\q_{3465}$ (behind, top-back), which however has two distinct smooth triangulating star-subdivisions, and the corresponding two distinct fans:
\begin{alignat}9
 \pFn{\Ht{\fB}'}
  &=\big\{ \n_{124},\n_{143},\n_{135}\,\n_{152},
            \2{\n_{345}}, \n_{264}, \n_{256}, \2{\n_{465}} \big\}\,&\supset
  &\vC{\TikZ{\path[use as bounding box](-.25,-.25)--(1,.25);
              \draw(0,0)--++(.4,-.2)--++(.6,.2)--++(-.5,.25)--cycle;
              \draw(0,0)--++(.5,-.4)--++(.5,.4); \draw(.4,-.2)--(.5,-.4);
              \draw[Blue, thick](0,0)--(1,0);
              \fill(0,0)circle(.4mm); \fill[Sage](.4,-.2)circle(.4mm);
              \fill(1,0)circle(.4mm); \fill(.5,.25)circle(.4mm); \fill(.5,-.4)circle(.4mm);
              \path(0,0)node[left=-2pt]{\tiny$4$};
              \path(1,0)node[right=-2pt]{\tiny$5$};
              \path(.5,.25)node[left=0pt]{\tiny$3$};
              \path(.4,-.2)node[Sage, above=-2pt]{\tiny$6$};
              \path(.5,-.4)node[right=-1pt]{\tiny$2$};
            }} \label{e:SfBb}\\[3mm]
 \pFn{\Ht{\fB}''}
  &=\big\{ \n_{124},\n_{143},\n_{135}\,\n_{152},
            \2{\n_{346}}, \n_{264}, \n_{256}, \2{\n_{365}}\big\}\,&\supset
  &\vC{\TikZ{\path[use as bounding box](-1.25,-.25)--(1,.25);
              \draw(0,0)--++(.4,-.2)--++(.6,.2)--++(-.5,.25)--cycle;
              \draw(0,0)--++(.5,-.4)--++(.5,.4); \draw(.4,-.2)--(.5,-.4);
              \draw[Green, thick](.4,-.2)--(.5,.25);
              \fill(0,0)circle(.4mm); \fill[Sage](.4,-.2)circle(.4mm);
              \fill(1,0)circle(.4mm); \fill(.5,.25)circle(.4mm); \fill(.5,-.4)circle(.4mm);
              \path(0,0)node[left=-2pt]{\tiny$4$};
              \path(1,0)node[right=-2pt]{\tiny$5$};
              \path(.5,.25)node[left=0pt]{\tiny$3$};
              \path(.4,-.2)node[Sage, above=5pt, left=-3pt]{\tiny$6$};
              \path(.5,-.4)node[right=-1pt]{\tiny$2$};
            }} \label{e:SfBb'}
\end{alignat}
The underlined cones distinguish the triangulations; the transformation
 $\Ht{\fB}'\iff\Ht{\fB}''$ is known as a flop;
$(\pDs{\Ht\fB},\pDN{\Ht\fB})$ is a polar pair of reflexive polytopes for both triangulations.

\subparagraph{Right:} Instead, we may re-triangulate the convex polytope $\pDs{\fB}$ by re-subdividing
\begin{equation}
   \underbrace{\vtl_{0254}\<\uplus\vtl_{0345}}_{\subset\,\pDs{\fB},\quad
    \vC{\TikZ{[scale=.75]\path[use as bounding box](0,-.25)--(1,.25);
              \fill[red!33](0,0)--++(.5,-.4)--++(.5,.4)--cycle;
              \draw(0,0)--++(.5,-.4)--++(.5,.4)--++(-.5,.25)--cycle;
              \draw[Blue, thick](0,0)--(1,0);
              \fill(0,0)circle(.4mm); \fill(.5,-.4)circle(.4mm);
              \fill(1,0)circle(.4mm); \fill(.5,.25)circle(.4mm);
              \path(0,0)node[left=-2pt]{\tiny$4$};
              \path(1,0)node[right=-2pt]{\tiny$5$};
              \path(.5,.25)node[left=0pt]{\tiny$3$};
              \path(.5,-.4)node[right=-1pt]{\tiny$2$};
            }}} ~\to~
    \underbrace{\vtl_{0234}\<\uplus\vtl_{0325}}_{\subset\,\pDs{\FF[3]{1,2}},\quad
    \vC{\TikZ{[scale=.75]\path[use as bounding box](0,-.25)--(1,.25);
              \draw(0,0)--++(.5,-.4)--++(.5,.4)--++(-.5,.25)--cycle;
              \draw[Red, thick](.5,.25)--++(0,-.65);
              \fill(0,0)circle(.4mm); \fill(.5,-.4)circle(.4mm);
              \fill(1,0)circle(.4mm); \fill(.5,.25)circle(.4mm);
              \path(0,0)node[left=-2pt]{\tiny$4$};
              \path(1,0)node[right=-2pt]{\tiny$5$};
              \path(.5,.25)node[left=0pt]{\tiny$3$};
              \path(.5,-.4)node[right=0pt]{\tiny$2$};
            }}}
    ~\uplus~ \underbrace{\vtl_{2345}}_{\text{outer simplex}},
 \label{e:S3Fb}
\end{equation}
where the last, ``outer'' 3-simplex, $\vtl_{2345}$, does not contain the origin and so does not define a central cone. Its removal results in the corresponding star-triangulating smooth fan:
\begin{equation}
   \pFn{\FF[3]{1,2}}=\{\n_{124},\n_{143},\n_{135}\,\n_{152},\n_{234},\n_{325}\}.
 \label{e:S3F12}
\end{equation}
Removing this ``outer'' 3-simplex, replaces the ``tent-like'' (convex!) region in $\pDs{\fB}$ by a ``hammock-like'' (non-convex!) region in $\pDs{\FF[3]{1,2}}$. This VEX re-triangulation $\pDs{\fB}\<\lat\pFn{\fB}\<\to\pFn{\FF[3]{1,2}}\<\smt\pDs{\FF[3]{1,2}}$ (with ``outer'' simplices removed) may be called {\em\bfseries\/VEX{ing}.} The resulting polytope, $\pDs{\FF[3]{1,2}}$ depicted at bottom-right in Figure~\ref{f:3pairs}, is smooth, VEX, but not convex: its edge $\n_{23}$ is concave, and its image in the transpolar (Newton) multitope is the downward pointed ``extension'' 
$\m_{12}\<\in\pDN{\FF[3]{1,2}}$; see Figure~\ref{f:3pairs}, upper-right.

\subparagraph{To summarize:}
On one hand, the standard blowup (Figure~\ref{f:3pairs}, left) extends the spanning polytope, refines the star-triangulating fan, and reduces the transpolar Newton polytope
\begin{alignat}9
  \text{blowup:}&\quad
  \pDs{\fB}&&<
  \big( \pDs{\fB}\<\uplus \vtl_{254\C26} \big)&&=\pDs{\Ht{\fB}}
   \quad&&\fif{\circ}&\quad
  \pDN{\Ht{\fB}}&=\big( \pDN{\fB}\<\ssm\vtl_{\C7*345} \big)~
  &<\pDN{\fB}.
\iText{In turn, VEX{ing} (Figure~\ref{f:3pairs}, right) reduces the spanning polytope, re-triangulates the star-triangulating fan, and extends the transpolar Newton polytope}
  \text{VEX{ing}:}&\quad
  \pDs{\fB}&&>
  \big( \pDs{\fB}\<\ssm \vtl_{2345} \big)&&=\pDs{\FF[3]{1,2}}
   \quad&&\fif{\wtd}&\quad
  \pDN{\FF[3]{1,2}}&=\big( \pDN{\fB}\<\uplus\C1{\,\vtl_{123*}} \big)~
  &>\pDN{\fB}.
\end{alignat}
The blowup depicted on the left in Figure~\ref{f:3pairs} results in a polar pair of reflexive polytopes, $(\pDs{\Ht{\fB}},\pDN{\Ht{\fB}})$, which encode the transpolar pairs of (smooth, complex-algebraic) toric varieties, $(\Ht{\fB}',\tW\Ht{\fB})$ and $(\Ht{\fB}'',\tW\Ht{\fB})$,
and can be used to construct multiple mirror pairs of K3 surfaces, constructed as anticanonical hypersurfaces as specified in Figure~\ref{f:gMM} and Remark~\ref{r:MMM}.
 Analogously, the VEX{ing} depicted on the right in Figure~\ref{f:3pairs} results in a transpolar pair of VEX multitopes, $(\pDs{\FF[3]{1,2}},\pDN{\FF[3]{1,2}})$, which correspond to the pair,
  $(\FF[3]{1,2},\MF[3]{1,2})$, of a (smooth, complex-algebraic, albeit non-Fano) toric variety, $\FF[3]{1,2}$, and its transpolar (smooth, precomplex) torus manifold, $\MF[3]{1,2}$;
these encode a web of multiple mirror pairs of K3 surfaces, constructed as anticanonical hypersurfaces as specified in Figure~\ref{f:gMM} and Remark~\ref{r:MMM}.

The comparison of~\eqref{e:SfBb} or~\eqref{e:SfBb'} with~\eqref{e:S3Fb} exhibits that VEX{ing} amounts to {\em\/re-triangulating the fan\/} spanned by the convex $\pDs{\fB}$ into the fan spanned by the non-convex $\pDs{\FF[3]{1,2}}$.

The constructions depicted in Figure~\ref{f:3pairs} have a noteworthy variation: Moving the vertex $\n_5\<\to(2,2,{-}1)$ causes $d(\n_{254})\<=2\<=d(\n_{345})$ but also places the subdividing vertex $\n_6$ within the edge $\q_{45}$. This allows subdividing both (now 
 $\deg\<=2$) facets, $\q_{254}$ and $\q_{345}$, without introducing new vertices, and so provides this stretched version of the convex polytope $\pDs{\fB}$ with a smooth triangulating subdivision that encodes a smooth, Fano toric variety. VEX{ing} as above now produces the non-convex $\pDs{\FF[3]{2,2}}$.

\paragraph{Acknowledgments:}
 We should like to thank C.F.~Doran, A.~Gholampour, K.~Iga, A.~Malmendier, H.~Schenck, J.~Rosenberg and especially Y.~Cui for helpful discussions on these topics, and M.~Masuda for his generous help with torus manifolds and the rich and diverse literature on this topic.
 PB\ would like to thank the CERN Theory Group for their hospitality over the past several years. His work is supported in part by the Department of Energy grant DE-SC0020220.
 TH\ is grateful to the Department of Mathematics, University of Maryland, College Park MD, 
 and the Physics Department of the Faculty of Natural Sciences of the University of Novi Sad, Serbia, for the recurring hospitality and resources.

\appendix
\section{Sub-Simplex Reduction Transpose-Mirror Examples}
\label{s:MM}
We further illustrate the web of constructions afforded by the mirror framework Corollary~\ref{CC:MMM} and Remark~\ref{r:MMM}; it is tempting to expect this framework to incorporate many if not all the ``multiple mirrors,'' as well as the ``coincidences'' reported in the literature, such as in\cite{Belakovskiy:2020Coi,Belavin:2023zmo}.

\subsection{A Network of Transpose-Mirrors}
\label{s:2MM}
Supporting Corollary~\ref{CC:MMM} and extending the analysis in\cite{Berglund:2022dgb},
consider two pairwise sub-simplex reductions of the transpolar pair of VEX multitopes $(\pDs{\FF[3]3},\pDN{\FF[3]3})$:
\begin{figure}[htbp]
 \begin{center}
  \begin{picture}(160,90)(0,2)
   \put(-5,55){\includegraphics[width=30mm]{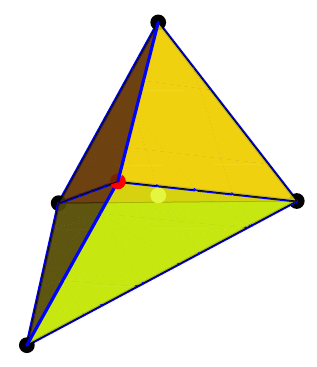}}
    \put(-5,80){\Large$\pDs{\FF[3]3}$}
    \put(7,70){\C1{$\n_1$}}
    \put(22,67){$\n_2$}
    \put(-5,70){$\n_3$}
    \put(12,87){$\n_4$}
    \put(-1,55){$\n_5$}
    \put(25,73){\huge$\leadsto$}
   \put(35,56){\includegraphics[width=25mm]{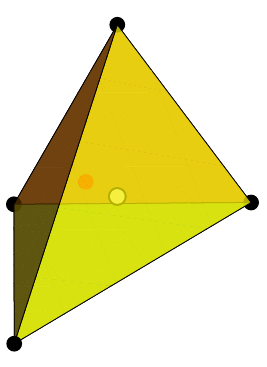}}
    \put(28,85){$\Red\limits_{1|\,\text{---}}\pDs{\FF[3]3}$}
    \put(50,85){$=\Conv(\pDs{\FF[3]3})$}
    \put(55,80){$=\pDs{\IP^3_{(3:1:1:1)}}$}
    \put(45,60){$\C1{x_1\to1}$}
   \put(93,22){\includegraphics[width=20mm]{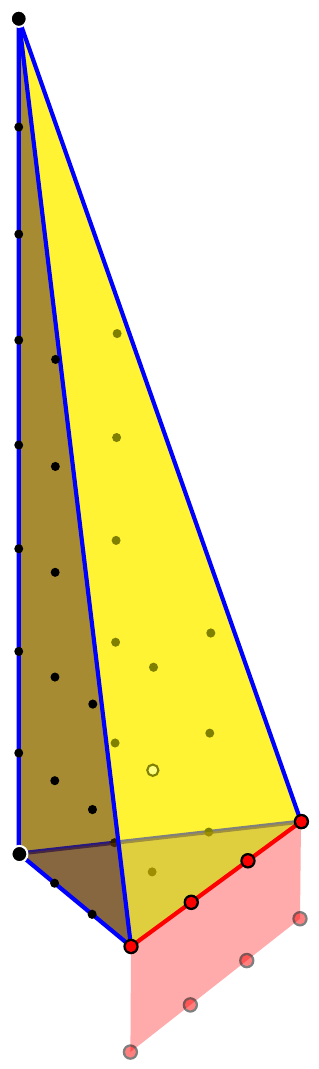}}
    \put(98,15){$\Red\limits_{-|35}\pDs{\FF[3]3}$}
    \put(98,8){$\C1{y_3,y_5\to1}$}
   \put(114,22){\includegraphics[width=20mm]{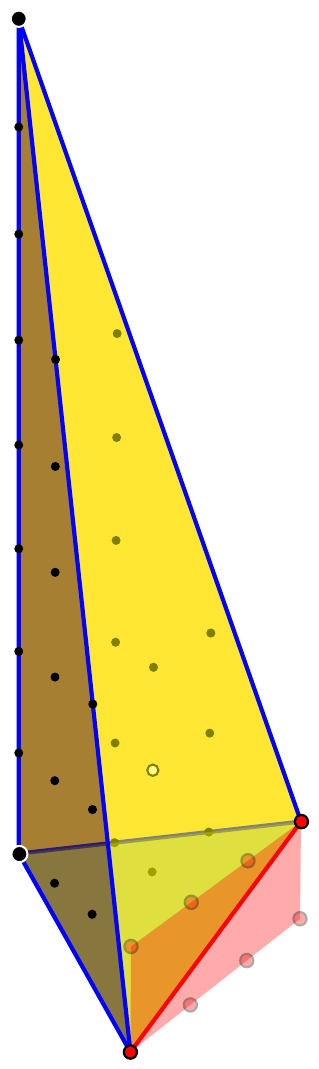}}
    \put(119,15){$\Red\limits_{-|45}\pDs{\FF[3]3}$}
    \put(119,8){$\C1{y_4,y_5\to1}$}
    \put(130,53){\huge\reflectbox{$\leadsto$}}
   \put(140,1){\includegraphics[width=26mm]{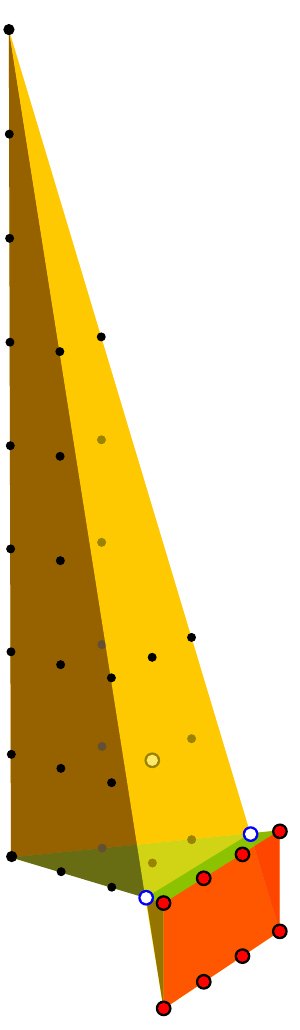}}
    \put(145,80){\Large$\pDN{\FF[3]3}$}
    \put(142,86){\small$\m_1$}
    \put(140,13){\small$\m_2$}
    \put(149.5,2){\small\C1{$\m_3$}}
    \put(155,10){\small{\color{red!60!black}$\m_4$}}
    \put(163,7){\small\C1{$\m_5$}}
    \put(163,20){\small\C1{$\m_6$}}
   \put(-4,48){$\begin{array}[t]{@{}r@{\,}l@{\,}l@{}}
            \MC3l{\Red_{1|3,5}(\pDs{\FF[3]3},\pDN{\FF[3]3}):~
            \{f_1(x)\<=0\} \fif{\sss\rm mm} \{\tT f_1(y)\}/\ZZ_3}
             \\[1mm] \toprule
            f_1(x)
            &= a_1 x_4{}^8 +a_2 x_5{}^8 +a_4\frac{x_2{}^3}{x_5} 
              +a_6\frac{x_3{}^3}{x_5}
            &\subset \IP^3_{(3:3:1:1)}[8] \\[1mm]
            \tT f_1(y)
            &= b_2 y_4{}^3 +b_3 y_6{}^3 +b_4 y_2{}^8 
              +b_5\frac{y_1{}^8}{y_4\,y_6}
            &\subset \IP^3_{(3:5:8:8)}[24]
                \end{array}$}
   \put(-4,20){$\begin{array}[t]{@{}r@{\,}l@{\,}l@{}}
            \MC3l{\Red_{1|4,5}(\pDs{\FF[3]3},\pDN{\FF[3]3}):~
            \{f_2(x)\<=0\}/\ZZ_4 \fif{\sss\rm mm} \{\tT f_2(y)\}/\ZZ_3}
             \\[1mm] \toprule
            f_2(x)
            &= a_1 x_4{}^8 +a_2 x_5{}^8 +a_3\frac{x_2{}^3}{x_4} 
              +a_6\frac{x_3{}^3}{x_5}
            &\subset \IP^3_{(3:3:1:1)}[8] \\[1mm]
            \tT f_2(y)
            &= b_2 y_3{}^3 +b_3 y_6{}^3 +b_4\frac{y_1{}^8}{y_3} 
              +b_5\frac{y_2{}^8}{y_6}
            &\subset \IP^3_{(1:1:2:2)}[6]
                \end{array}$}
 \end{picture}
 \end{center}
 \caption{The spanning polytope, $\pDs{\FF[3]m}$ (far left), and the (extended) Newton multitope,
 $\pDN{\FF[3]m}$ (far right); their sub-simplex reductions and corresponding pairs of transpose polynomials}
 \label{f:3F3MM}
\end{figure}
The spanning multitope (top far left) admits a single 0-enclosing sub-simplex reduction, 
$\pDs{\FF[3]3}\<\to\pDs{\IP^2_{(3:3:1:1)}}$, while the (extended) Newton multitope (far right) admits many such reductions. Two of these are indicated in Figure~\ref{f:3F3MM}, mid-right; their combination with the unique 0-enclosing sub-simplex reduction of $\pDs{}$ then leads to the two transposition pairs in Figure~\ref{f:3F3MM}.
 The symmetry groups are easily found from the rows (vs.\ columns) of the inverse of the exponents' matrix\cite{Krawitz:2009aa, rMK-PhD} and are as follows:
\begin{alignat}9
 \Red\limits_{1|3,5}:&\quad
   \AR{@{}c@{~~}c@{~~}c@{~~}c@{}}
            {\frac13&\frac23&0&0\\[1pt]
             \frac1{24}&\frac1{24}&\frac18&0\\[2pt]\hline\rule{0pt}{2.25ex}
             \frac38&\frac38&\frac18&\frac18\\}
   \MM{~\\\bigg\{\begin{array}{@{}r@{\,=\,}l}
          G&\ZZ_3\,{\times}\,\ZZ_{24},\\[1mm] \hline\nGlu{1mm}
          Q&\ZZ_8,
         \end{array}}
 &&\quad\text{vs.}\quad
 &&\AR{@{}c@{~~}c@{~~}c@{~~}c@{}}
            {0&0&\frac13&\frac23\\[1pt]
             \frac18&0&0&0\\[2pt]\hline\rule{0pt}{2.25ex}
             \frac5{24}&\frac3{24}&\frac13&\frac13\\}
  \MM{~\\\bigg\{\begin{array}{@{}r@{\,=\,}l}
          G^\wtd&\ZZ_3\,{\times}\,\ZZ_8,\\[1mm] \hline\nGlu{1mm}
          Q^\wtd&\ZZ_{24}.
         \end{array}}
 \label{e:Z3.24.8}
\iText[-2pt]{Quotienting the transpose (on the right-hand side) by the indicated $\ZZ_3$ action then produces $\Tw{G}^\wtd\<=\ZZ_8\<=Q$ and $\Tw{Q}^\wtd\<=\ZZ_{24}\<\times\ZZ_3\<=G$. One could also quotient the ``original'' (on the left-hand side) by the indicated $\ZZ_3$-action, leading to $\Tw{G}^\wtd\<=\ZZ_3\<\times\ZZ_8\<=Q$ and $\Tw{Q}^\wtd\<=\ZZ_{24}\<=G$ instead.
In turn, the other sub-simplex reduction has the following symmetries:}
 \Red\limits_{1|4,5}:&\quad
   \AR{@{}c@{~~}c@{~~}c@{~~}c@{}}
            {\frac13&\frac13&0&0\\[1pt]
             \frac1{24}&\frac{23}{24}&\frac18&\frac78\\[2pt]\hline\rule{0pt}{2.25ex}
             \frac38&\frac38&\frac18&\frac18\\}
   \MM{~\\\bigg\{\begin{array}{@{}r@{\,=\,}l}
          G&\ZZ_3\,{\times}\,\ZZ_{24},\\[1mm] \hline\nGlu{1mm}
          Q&\ZZ_8,
         \end{array}}
 &&\quad\text{vs.}\quad
 &&\AR{@{}c@{~~}c@{~~}c@{~~}c@{}}
            {\frac1{24}&0&\frac13&0\\[1pt]
             0&\frac1{24}&0&\frac13\\[2pt]\hline\rule{0pt}{2.25ex}
             \frac16&\frac16&\frac13&\frac13\\}
   \MM{~\\\bigg\{\begin{array}{@{}r@{\,=\,}l}
          G^\wtd&(\ZZ_{24})^2/\ZZ_6,\\[1mm] \hline\nGlu{1mm}
          Q^\wtd&\ZZ_6.
         \end{array}}
 \label{e:Z4.24.6}
\end{alignat}
Here we quotient the ``original'' (on the left-hand side) by $\ZZ_4\<\subset\ZZ_{24}$, leading to
 $\Tw{G}\<=\ZZ_3\<\times\ZZ_6$ and $\Tw{Q}\<=\ZZ_8\<\times\ZZ_4$.
The geometric symmetry of the transpose (on the right-hand side) is
\begin{alignat}9
  (\ZZ_{24})^2\<\supset
  (g^\wtd_1{\times}g^\wtd_2)/Q^\wtd
  &=\exp\!\big\{2\p i\big[\,\big(\frc1{24},0,\frc13,0\big){\oplus}\big(0,\frc1{24},0,\frc13\big)
          ~~\text{mod}\:\big(\frc16,\frc16,\frc13,\frc13\big)\,\big]\big\}, \label{e:2Z24}\\
  \big(\frc16,\frc16,\frc13,\frc13\big)
  &=4\big(\frc1{24},0,\frc13,0\big){+}4\big(0,\frc1{24},0,\frc13\big). \label{e:Z6proj}
\end{alignat}
This has an effective $\ZZ_3$ subgroup action generated by the equivalence classes
\begin{equation}
  \big[\,4\big(\frc1{24},0,\frc13,0\big){-}4\big(0,\frc1{24},0,\frc13\big)
          ~~\text{mod}\:2\big(\frc16,\frc16,\frc13,\frc13\big)\,\big]
  =\big[\,\big(\frc16,\frc56,\frc13,\frc23\big)
          ~~\text{mod}\:\big(\frc12,\frc12,0,0\big)\,\big].
 \label{e:Z3eff}
\end{equation}
Quotienting by this $\ZZ_3$ produces $\Tw{Q}^\wtd\<=\ZZ_6\<\times\ZZ_3\<=\Tw{G}$, and leaves
 $\Tw{G}^\wtd\<=(\ZZ_{24})^2/(\ZZ_6{\times}\ZZ_3)\<\approx\ZZ_8\<\times\ZZ_4\<=\Tw{Q}$.

However, notice that $\{f_1(x){=}0\}$ and $\{f_2(x){=}0\}$ are simple (complex-structure) deformations of each other, whereby we obtain a {\em\/fragment\/} of the so-generated web of related pairs of mirror models:
\begin{equation}
  \begin{array}{c@{~}c@{~}c@{~}c@{~}c@{~}c@{~}c}
    \{\tT f_1(y){=}0\}/\ZZ_3 &\fif{\sss\rm mm} & \{f_1(x){=}0\}
     & \fIf[2pt]{\sss\rm cpx.\,str.}{\sss\rm~deform~} &\{f_2(x){=}0\}
     &\C7{\fif{\sss\rm mm?}} & \C7{\{\tT f_2(y){=}0\}/?}\\*[-3pt]
    \big\uA\crlap{/\ZZ_3} && \big\dA\crlap{/\ZZ_3} && \big\dA\crlap{/\ZZ_4}
     && \C7{\big\uA\crlap{/?}}\\*[0pt]
    \{\tT f_1(y){=}0\} &\fif{\sss\rm mm} & \{f_1(x){=}0\}/\ZZ_3
     &&\{f_2(x){=}0\}/\ZZ_4 &\fif{\sss\rm mm} & \{\tT f_2(y){=}0\}/\ZZ_3
  \end{array}
 \label{e:f1f2}
\end{equation}
Given the identification of finite group actions on the right-hand side of~\eqref{e:Z3eff}, finding the ``missing mirror'' in the top far-right corner of the diagram~\eqref{e:f1f2} would require enhancing $Q^\wtd\<=\ZZ_6\to\ZZ_{24}{\times}\ZZ_3$. Since the $\ZZ_6$ from the $\IP^2_{(2:2:1:1)}$-projectivization is generated by the sum~\eqref{e:Z6proj}, such a $\ZZ_{24}$-enhancement can only be generated by the sum of the two $\ZZ_{24}$ generators in~\eqref{e:2Z24}:
\begin{equation}
  \zZ{24}{\big(\frc1{24},0,\frc13,0\big){+}\big(0,\frc1{24},0,\frc13\big)}
  =\zZ{24}{\frc1{24},\frc1{24},\frc13,\frc13},
\end{equation}
which however is not traceless: $\frc1{24}{+}\frc1{24}{+}\frc13{+}\frc13=\frc34\not\in\ZZ$, i.e., does not satisfy the so-called Calabi--Yau condition and so could not be used as a quantum symmetry.

In remedy, we observe that the specification~\eqref{e:Z4.24.6} may be replaced with
\begin{equation}
   \Red\limits_{1|4,5}:\quad
   \AR{@{}c@{~~}c@{~~}c@{~~}c@{}}
            {\frac13&\frac13&0&0\\[1pt]
             \frac1{24}&\frac{23}{24}&\frac18&\frac78\\[2pt]\hline\rule{0pt}{2.25ex}
             \frac38&\frac38&\frac18&\frac18\\}
   \MM{~\\\bigg\{\begin{array}{@{}r@{\,=\,}l}
          G&\ZZ_3\,{\times}\,\ZZ_{24},\\[1mm] \hline\nGlu{1mm}
          Q&\ZZ_8,
         \end{array}}
   \quad\text{vs.}\quad
   \AR{@{}c@{~~}c@{~~}c@{~~}c@{}}
            {\frac1{24}&0&\frac13&0\\[1pt]
             0&\frac1{24}&0&\frac13\\[2pt]\hline\rule{0pt}{2.25ex}
             \frac16&\frac56&\frac13&\frac23\\}
   \MM{~\\\bigg\{\begin{array}{@{}r@{\,=\,}l}
          G^\wtd&(\ZZ'_{24}{\times}\ZZ_{24})/\ZZ_6',\\[1mm] \hline\nGlu{1mm}
          Q^\wtd&\ZZ_6',
         \end{array}}
 \label{e:Z4.24.6'}
\end{equation}
where we use the twisted projectivization/quantum (sub)group
\begin{equation}
 \big(\ZZ_6'{:}\,\frc16,\frc56,\frc13,\frc23\big)
  =\big(\ZZ'_6{:}\,4\big(\frc1{24},0,\frc13,0\big){-}4\big(0,\frc1{24},0,\frc13\big)\big)
  \subset\big(\ZZ'_{24}{:}\,\big(\frc1{24},0,\frc13,0\big){-}\big(0,\frc1{24},0,\frc13\big)\big). \label{e:Z6'}
\end{equation}
The complementary $\ZZ_{24}$ needs to be generated by any linear combination in
 $\big(\frc1{24},0,\frc13,0\big){\oplus}\big(0,\frc1{24},0,\frc13\big)$
such that all its integral powers differ from those of the generators of $\ZZ'_{24}$, i.e., that all elements of the so-generated $\ZZ_{24}$ differ from those of $\ZZ'_{24}$; this excludes
 $\big(\ZZ''_{24}{:}\,\frc1{24},\frc1{24},\frc13,\frc13\big)
  \<=\big(\ZZ''_{24}{:}\,\big(\frc1{24},0,\frc13,0\big){+}\big(0,\frc1{24},0,\frc13\big)\big)$, since
 $\big(\ZZ''_2{:}\,\frc{12}{24},\frc{12}{24},\frc{12}3,\frc{12}3\big)\<\subset\ZZ'_{24}$.
We thus choose
 $\big(\ZZ^{\sss(1)}_{24}{:}\,\frc1{24},0,\frc13,0\big)$ and note that it has the traceless subgroup,
 $\big(\ZZ^{\sss(1)}_3{:}\,\frc{8}{24},0,\frc{8}3,0\big)\<=
   \big(\ZZ^{\sss(1)}_3{:}\,\frc13,0,\frc23,0\big)$,
which is also independent of $Q^\wtd\<=\ZZ'_6$ in~\eqref{e:Z4.24.6'}. This allows identifying the geometric group on the right-hand (transposed) side of~\eqref{e:Z4.24.6'} as
\begin{equation}
  G^\wtd=(\ZZ'_{24}{\times}\ZZ_{24})/\ZZ_6'
  =(\ZZ'_{24}/\ZZ_6'\<\approx\ZZ'_4) \times \ZZ^{\sss(1)}_{24}.
\end{equation}
These definitions permit completing a \eqref{e:f1f2}-like multiple mirror web:
\begin{equation}
  \begin{array}{c@{~}c@{~}c@{~}c@{~}c@{~}c@{~}c}
    \{\tT f_1(y){=}0\}/\ZZ_3 &\fif{\sss\rm mm} & \{f_1(x){=}0\}
     & \fIf[2pt]{\sss\rm cpx.\,str.}{\sss\rm~deform~} &\{f_2(x){=}0\}
     &\fif{\sss\rm mm} & \{\tT f_2(y){=}0\}'/[(\ZZ'_{24}/\ZZ'_6){\times}\ZZ^{\sss(1)}_3]
  \\*[-1mm]
    \Big\uA\crlap{/\ZZ_3} && \Big\dA\crlap{/\ZZ_3} && \Big\dA\crlap{/\ZZ_4}
     && \Big\uA\crlap{/(\ZZ'_{24}/\ZZ_6')\approx\ZZ'_4}\\*
    \{\tT f_1(y){=}0\} &\fif{\sss\rm mm} & \{f_1(x){=}0\}/\ZZ_3
     &&\{f_2(x){=}0\}/\ZZ_4 &\fif{\sss\rm mm} & \{\tT f_2(y){=}0\}'/\ZZ^{\sss(1)}_3
  \end{array}
 \label{e:f1f2'}
\end{equation}
where appropriate desingularizing blowups are understood for each quotient, where the primes indicate the use of the twisted quantum group, twisting the Ref.\cite{Krawitz:2009aa, rMK-PhD} specifications~\eqref{e:Z4.24.6}--\eqref{e:f1f2}, but in agreement with the less prescriptive original algorithm\cite{rBH}.

The twisted specifications~\eqref{e:Z4.24.6'}--\eqref{e:f1f2'} are possible precisely because
\begin{alignat}9
  \tT f_2(y)
   &=\Big(b_2\,y_3\!^3 +b_4\,\frac{y_1\!^8}{y_3}\Big)
     +\Big(b_3\,y_6\!^3 +b_5\,\frac{y_2\!^8}{y_6}\Big)
\iText[-1mm]{decomposes as a sum of two separate summands, coupled only by the overall $\IP^3_{(1{:}1{:}2{:}2)}$-projectivization. The fact that the ``original,''}
  f_2(x)
   &=\Big(a_1\,x_4\!^8 +a_3\,\frac{x_2\!^3}{x_4}\Big)
     +\Big(a_2\,x_5\!^8 +a_6\,\frac{x_3\!^3}{x_5}\Big),
\end{alignat}
correspondingly also decomposes as a sum of two separate parts permits similarly twisting also the $(G,Q)$-specifications, which allows additional quotienting options that we however do not explore herein.

\subsection{0-Enclosing Sub-Simplex Reductions}
\label{s:0E}
The ``0-enclosing'' simplices, $\Red\pDN{X}$, of Remark~\ref{r:MMM} and footnote~\ref{fn:encl} on p.\;\pageref{fn:encl}, formed by appropriate subsets of the lattice points $\pDN{X}\<\cap M$ are not conceptually new, merely adapted for VEX multitopes:
 Such reductions in fact appeared in the original proposal of transpose-mirror construction\cite{rBH}, adopting the language and diagrams from Arnold's original classification\cite{rAGZV-Sing1} and Kreuzer and Skarke's analysis\cite{Kreuzer:1992bi, rKreSka98}. The simple case of $\IP^2[3]$ has five types of suitable (``invertible''\cite{Krawitz:2009aa, rMK-PhD}) transverse polynomials: 
\begin{equation}
 \mkern-12mu
  \begin{array}{@{}c@{\quad}c@{\quad}c@{\quad}c@{\quad}c@{}}
  \vC{\TikZ{[scale=.9]\path[use as bounding box](-1,-1.3)--(2,2.2);
        \filldraw[fill=yellow!50, draw=orange, thick, line join=round]
            (1,0)--(0,1)--(-1,-1)--cycle;
        \path(1,.5)node[right, orange!75!black]{$\pDs{\IP^2}$};
         \draw[thick, densely dotted, -stealth, orange!75!black]
            (1.1,.5)to[out=180,in=45]++(-.75,-.33);
        \filldraw[fill=Turque, draw=blue, thick, line join=round, opacity=.33]
            (-1,2)--(-1,-1)--(2,-1)--cycle;
        \draw[Green!75!black, line width=3pt, line join=round, opacity=.5]
            (-1,2)--(-1,-1)--(2,-1)--cycle;
        \foreach\x in{-1,...,2}\foreach\y in{-1,...,2}\fill[gray](\x,\y)circle(.4mm);
        \path(-1,2)node[right]{\scriptsize$1$};
        \path(-1,-1)node[above right]{\scriptsize$2$};
        \path(2,-1)node[above]{\scriptsize$3$};
        \path(0,1.5)node[right, Cobalt]{$\pDN{\IP^2}$};
         \draw[thick, densely dotted, -stealth, Cobalt]
            (.1,1.5)to[out=180,in=75]++(-.75,-.75);
        \filldraw[fill=white, thick](0,0)circle(.5mm);
            }}
 &
  \vC{\TikZ{[scale=.9]\path[use as bounding box](-1,-1.3)--(2,2.2);
        \filldraw[fill=yellow!50, draw=orange, thick, line join=round]
            (1,0)--(0,1)--(-1,-1)--cycle;
        \filldraw[fill=Turque, draw=blue, thick, line join=round, opacity=.15]
            (-1,2)--(-1,-1)--(2,-1)--cycle;
        \draw[Green!75!black, line width=3pt, line join=round, opacity=.5]
            (-1,1)--(-1,-1)--(2,-1)--cycle;
        \foreach\x in{-1,...,2}\foreach\y in{-1,...,2}\fill[gray](\x,\y)circle(.4mm);
        \path(-1,2)node[right]{\scriptsize$1$};
        \path(-1,-1)node[above right]{\scriptsize$2$};
        \path(2,-1)node[above]{\scriptsize$3$};
        \path(0,1.5)node[right, Green!75!black]{$\Red\pDN{\IP^2}$};
         \draw[very thick, densely dotted, -stealth, Green!75!black]
            (.5,1.3)to++(-.25,-1.1);
        \filldraw[fill=white, thick](0,0)circle(.5mm);
            }}
 &
  \vC{\TikZ{[scale=.9]\path[use as bounding box](-1,-1.3)--(2,2.2);
        \filldraw[fill=yellow!50, draw=orange, thick, line join=round]
            (1,0)--(0,1)--(-1,-1)--cycle;
        \filldraw[fill=Turque, draw=blue, thick, line join=round, opacity=.15]
            (-1,2)--(-1,-1)--(2,-1)--cycle;
        \draw[Green!75!black, line width=3pt, line join=round, opacity=.5]
            (-1,1)--(-1,0)--(2,-1)--cycle;
        \foreach\x in{-1,...,2}\foreach\y in{-1,...,2}\fill[gray](\x,\y)circle(.4mm);
        \path(-1,2)node[right]{\scriptsize$1$};
        \path(-1,-1)node[above right]{\scriptsize$2$};
        \path(2,-1)node[above]{\scriptsize$3$};
        \path(0,1.5)node[right, Green!75!black]{$\Red'\!\!\pDN{\IP^2}$};
         \draw[very thick, densely dotted, -stealth, Green!75!black]
            (.5,1.3)to++(-.25,-1.1);
        \filldraw[fill=white, thick](0,0)circle(.5mm);
            }}
 &
  \vC{\TikZ{[scale=.9]\path[use as bounding box](-1,-1.3)--(2,2.2);
        \filldraw[fill=yellow!50, draw=orange, thick, line join=round]
            (1,0)--(0,1)--(-1,-1)--cycle;
        \filldraw[fill=Turque, draw=blue, thick, line join=round, opacity=.15]
            (-1,2)--(-1,-1)--(2,-1)--cycle;
        \draw[Green!75!black, line width=3pt, line join=round, opacity=.5]
            (-1,1)--(0,-1)--(2,-1)--cycle;
        \foreach\x in{-1,...,2}\foreach\y in{-1,...,2}\fill[gray](\x,\y)circle(.4mm);
        \path(-1,2)node[right]{\scriptsize$1$};
        \path(-1,-1)node[above right]{\scriptsize$2$};
        \path(2,-1)node[above]{\scriptsize$3$};
        \path(0,1.5)node[right, Green!75!black]{$\Red''\!\!\pDN{\IP^2}$};
         \draw[very thick, densely dotted, -stealth, Green!75!black]
            (.5,1.3)to++(-.25,-1.1);
        \filldraw[fill=white, thick](0,0)circle(.5mm);
            }}
 &
  \vC{\TikZ{[scale=.9]\path[use as bounding box](-1,-1.3)--(2,2.2);
        \filldraw[fill=yellow!50, draw=orange, thick, line join=round]
            (1,0)--(0,1)--(-1,-1)--cycle;
        \filldraw[fill=Turque, draw=blue, thick, line join=round, opacity=.15]
            (-1,2)--(-1,-1)--(2,-1)--cycle;
        \draw[Green!75!black, line width=3pt, line join=round, opacity=.5]
            (-1,1)--(0,-1)--(1,0)--cycle;
        \foreach\x in{-1,...,2}\foreach\y in{-1,...,2}\fill[gray](\x,\y)circle(.4mm);
        \path(-1,2)node[right]{\scriptsize$1$};
        \path(-1,-1)node[above right]{\scriptsize$2$};
        \path(2,-1)node[above]{\scriptsize$3$};
        \path(0,1.5)node[right, Green!75!black]{$\Red'''\!\!\pDN{\IP^2}$};
         \draw[very thick, densely dotted, -stealth, Green!75!black]
            (.5,1.3)to++(-.25,-.9);
        \filldraw[fill=white, thick](0,0)circle(.5mm);
            }}\\*[2mm]
    x_1\!^3 {+}x_2\!^3 {+}x_3\!^3
  & x_1\!^2x_2 {+}x_2\!^3 {+}x_3\!^3
  & x_1\!^2x_2 {+}x_1x_2\!^2 {+}x_3\!^3
  & x_1\!^2x_2 {+}x_2\!^2x_3 {+}x_3\!^3
  & x_1\!^2x_2 {+}x_2\!^2x_3 {+}x_3\!^2x_1\\*[0mm]
   \vC{\TikZ{\path[use as bounding box](0,-.5)--(2,.75);
               \foreach\x in{0,...,2}{\fill(\x,0)circle(.9mm);
                                      \draw[very thick](\x,.2)circle(2mm);}
               \foreach\x in{1,...,3}\path(\x-1,-.3)node{\scriptsize\x};
            }}
  &\vC{\TikZ{\path[use as bounding box](0,-.5)--(2,.75);
               \foreach\x in{0,...,2}\fill(\x,0)circle(.9mm);
               \draw[very thick, midarrow=stealth](0,0)--(1,0);
               \draw[very thick](1,.2)circle(2mm);
               \draw[very thick](2,.2)circle(2mm);
               \foreach\x in{1,...,3}\path(\x-1,-.3)node{\scriptsize\x};
            }}
  &\vC{\TikZ{\path[use as bounding box](0,-.5)--(2,.75);
               \foreach\x in{0,...,2}\fill(\x,0)circle(.9mm);
               \draw[very thick, midarrow=stealth](0,0)--(1,0);
               \draw[very thick, midarrow=stealth]
                   (1,0)to[out=90,in=0]++(-.3,.3)--++(-.4,0)to[out=180,in=90]++(-.3,-.3);
               \draw[very thick](2,.2)circle(2mm);
               \foreach\x in{1,...,3}\path(\x-1,-.3)node{\scriptsize\x};
            }}
  &\vC{\TikZ{\path[use as bounding box](0,-.5)--(2,.75);
               \foreach\x in{0,...,2}\fill(\x,0)circle(.9mm);
               \draw[very thick, midarrow=stealth](0,0)--(1,0);
               \draw[very thick, midarrow=stealth](1,0)--(2,0);
               \draw[very thick](2,.2)circle(2mm);
               \foreach\x in{1,...,3}\path(\x-1,-.3)node{\scriptsize\x};
            }}
  &\vC{\TikZ{\path[use as bounding box](0,-.5)--(2,.75);
               \foreach\x in{0,...,2}\fill(\x,0)circle(.9mm);
               \draw[very thick, midarrow=stealth](0,0)--(1,0);
               \draw[very thick, midarrow=stealth](1,0)--(2,0);
               \draw[very thick, midarrow=stealth]
                   (2,0)to[out=90,in=0]++(-.3,.3)--++(-1.4,0)to[out=180,in=90]++(-.3,-.3);
               \foreach\x in{1,...,3}\path(\x-1,-.3)node{\scriptsize\x};
            }}
  \end{array}
\end{equation}
The Newton triangle, $\pDs{\IP^2}$ (bigger, pale blue), is depicted over the spanning triangle,
$\pDN{\IP^2}$ (smaller, yellow); the thicker (green) triangle outlines $\Red\pDN{\IP^2}$,  the so-encoded defining polynomial underneath it, and its graphical depiction \`a~la\cite{rAGZV-Sing1,rBH} underneath that. Arnold's original classification\cite{rAGZV-Sing1} guarantees that every other sub-simplex reduction is either equivalent to one of these or not transverse.

\subsection{0-Exposing Sub-Simplex Reductions}
\label{s:0nE}
Supporting Remark~\ref{r:MMM} and footnote~\ref{fn:encl}, p.\,\pageref{fn:encl},
consider the pair of examples obtained by fixing a sub-simplex reduction of $\pDs{\FF[2]3}$, and reducing $\pDN{\FF[2]3}$ to two simplicial subsets,
$\Red'(\pDN{\FF[2]3})$ and $\Red''(\pDN{\FF[2]3})$, each of which containing $0\<\in M$ in one of its facets:
\begin{equation}
\setbox9\hbox{\tiny$\!\sss(3{:}1{:}1)$}
 \vC{\begin{picture}(160,60)
   \put(22,40){\TikZ{\path[use as bounding box](-3,-1)--(1,1);
              \foreach\x in{-3,...,1}\fill[gray!50](\x,-1)circle(.5mm);
              \foreach\x in{-2,...,1}\fill[gray!50](\x,0)circle(.5mm);
              \foreach\x in{-1,...,1}\fill[gray!50](\x,1)circle(.5mm);
             \filldraw[fill=red!20, draw=Magenta, line join=round, line width=2pt, opacity=.75]
                 (1.05,-.015)--(0,1.04)--({180+atan(1/3)+.2}:3.3)--cycle;
             \filldraw[fill=yellow!75, draw=blue, line join=round, thick]
                 (1,0)--(0,1)--(-1,0)--(-3,-1)--cycle;
             \draw[black, -stealth](0,0)--(1,0);
              \path[black](1,0)node[below]{\tiny$1$};
             \draw[black, -stealth](0,0)--(0,1);
              \path[black](0,1)node[right]{\tiny$2$};
             \draw[black, -stealth](0,0)--(-1,0);
              \path[red](-1,0)node[above left=-3pt]{\tiny$3$};
             \draw[black, -stealth](0,0)--(-3,-1);
              \path[black](-3,-1)node[above]{\tiny$4$};
             \filldraw[thick, fill=white, draw=black](0,0)circle(.5mm);
             \path[Magenta](-3,.85)node[right=-3mm]{\footnotesize$\Red(\pDs{\FF[2]3})=$};
             \path[Magenta](-3.1,.45)node[right=-3mm]{\footnotesize$\Conv(\pDs{\FF[2]3})$};
             \path[Magenta](-3.2,.1)node[right=-3mm]{\footnotesize$\<=\pDs{\IP^2_{\copy9}}$};
             \path[blue](0,-.6)node[right]{$\pDs{\FF[2]3}$};
            }}
   \put(75,10){\TikZ{\path[use as bounding box](-1,-1)--(1,4);
              \foreach\y in{-2,...,4}
               \foreach\x in{-1,...,1} \fill[gray!50](\x,\y)circle(.5mm);
             \fill[Turque, opacity=.75](0,0)--(-1,4)--(-1,-1)--(1,-1);
             \draw[blue, thick](-1,4)--(-1,-1)--(1,-1);
             \fill[red, opacity=.5](0,0)--(1,-1)--(1,-2);
             \draw[red, thick](1,-1)--(1,-2);
             \draw[black, -stealth](0,0)--(1,-1);
             \fill[Turque, opacity=.75](0,0)--(1,-2)--(-1,4);
             \draw[blue, thick](1,-2)--(-1,4);
             \fill[red!20, opacity=.5]
                 (-.98,3.92)--(-.98,-.98)--(.633,-.98)--cycle;
             \draw[draw=Magenta, line join=round, line width=2pt, opacity=.75]
                 (-1,3.98)--(-1,-1)--(.67,-1)--cycle;
             \draw[draw=Green, line join=round, ultra thick, loosely dashed]
                 (1,-1.02)--(-1.04,1)--(-1.04,-.03)--cycle;
              \path[Green!75!black](-1,.2)node[left=1mm]
                 {\rotatebox{90}{\scriptsize$\Red'(\pDN{\FF[2]3})$}};
             \draw[draw=Rouge, line join=round, ultra thick, densely dotted]
                 (1.02,-.98)--(-1,1.05)--(.02,1.05)--cycle;
              \path[Rouge](-.1,.6)node[right]
                 {\rotatebox{-60}{\scriptsize$\Red''(\pDN{\FF[2]3})$}};
             \draw[black, -stealth](0,0)--(-1,4);
             \draw[black, -stealth](0,0)--(-1,-1);
             \draw[black, -stealth](0,0)--(1,-2);
             \filldraw[thick, fill=white, draw=black](0,0)circle(.5mm);
             \path[Magenta!75!black](.3,-1.4)node[left]{\scriptsize$(\sfrac23,-1)$};
              \draw[Magenta!75!black, very thick, densely dotted, -stealth]
                  (.2,-1.4)to[out=0,in=-110](2/3,-1);
             \path[blue](-.1,1.8)node[right]{$\pDN{\FF[2]3}$};
             \path[Magenta](-.8,3.75)node[right]
                 {\scriptsize$\big(\Red(\pDN{\FF[2]3})\big){}^\circ$};
             \path[Magenta](-.6,3.3)node[right]{\footnotesize$\<=(\pDs{\IP^2_{\copy9}})^\circ$};
             \path[Magenta](-.4,2.85)node[right]{\footnotesize$\<=\pDN{\IP^2_{\copy9}}$};
             \draw[Magenta, densely dotted, thick, -stealth]
                 (.2,2.65)to[out=-90,in=30]++(-.55,-.6);
            }}
   \put(0,10){$\begin{array}{@{}r@{\,}l@{\,}l@{}}
                 \MC2l{\Red'(\pDs{\FF[2]3},\pDN{\FF[2]3})}\\[2pt] \toprule
                 f_1(x)&=x_2\!^2\,x_4\!^3 +x_2\,x_4\!^4 +\frac{x_1\!^2}{x_4}
                        &\in\IP^2_{(3:1:1)}\\[2mm]
                 \tT f_1(y)&=y_3\!^2 +y_1\!^2\,y_2 +\frac{y_1\!^3\,y_2\!^4}{y_3}
                        &\in\IP^2_{(1:0:1)}\\
               \end{array}$}
   \put(100,10){$\begin{array}{@{}r@{\,}l@{\,}l@{}}
                 \MC2l{\Red''(\pDs{\FF[2]3},\pDN{\FF[2]3})}\\[2pt] \toprule
                 f_2(x)&=x_2\!^2\,x_4\!^3 +x_1\,x_2\!^2 +\frac{x_1\!^2}{x_4}
                        &\in\IP^2_{(3:1:1)}\\[2mm]
                 \tT f_2(y)&=y_2\,y_3\!^2 +y_1\!^2\,y_2\!^2 +\frac{y_1\!^3}{y_3}
                        &\in\IP^2_{(1:0:1)}\\
               \end{array}$}
   \put(0,28){$\IE_1=\bM{ 0 & 2 & 3 \\[1pt] 0 & 1 & 4 \\[1pt] 2 & 0 & -1 }$, so
              $\IE_1\!^{-1}\equiv\bM{ \fRc{9}{10} & \fRc{4}{5}  & \fRc{4}{5} \\
                                      \fRc{1}{5}  & \fRc{2}{5}  & \fRc{2}{5} \\
                                      \fRc{1}{2}  & 0 & 0 }$}
 \put(100,28){$\IE_2=\bM{ 0 & 2 & 3 \\[1pt] 1 & 2 & 0 \\[1pt] 2 & 0 & -1 }$, so
              $\IE_2\!^{-1}\equiv\bM{ \fRc{1}{5}  & \fRc{9}{10} & \fRc{2}{5} \\
                                      \fRc{4}{5}  & \fRc{3}{5}  & \fRc{3}{5} \\
                                      \fRc{3}{5}  & \fRc{7}{10} & \fRc{1}{5} }$}
 \end{picture}}
\end{equation}
The thicker (magenta) outline highlights: the convex hull,
 $\Red(\pDs{\FF[2]3})\<=\Conv(\pDs{\FF[2]3})\<=\pDs{\IP^2_{(3{:}1{:}1)}}$ (left),
and its non-lattice polar (right),
 $\big(\Red(\pDs{\FF[2]3})\big){}^\circ\<=\pDN{\IP^2_{(3{:}1{:}1)}}\<=
  \Conv\big((-1,4),(-1,-1),(\fRc23,-1)\big)$. The latter polygon has no lattice point in the right-hand half-space and cannot span the lattice (multi)fan of {\em\/any\/} compact toric space, which prompts the extension to VEX multitopes\cite{rBH-gB}.

Although both $f_1(x)$ and $\tT f_1(y)$ are transverse, the degrees in the latter are consistent only with $q(y_2)\<=0$, i.e., 
$(y_1,y_2,y_3)\<\in\IP^2_{(1{:}0{:}1)}$: $y_2$ is an uncharged ``spectator'' (super)field in the corresponding GLSM with the superpotential $W\<=P\,\tT f_1(y)$, parametrizing a non-compact direction.
 The same holds for the other reduction, except now $f_2(x)$ also fails to be transverse.
 Finally, neither $\IE_1\!^{-1}$ nor $\IE_2\!^{-1}$ defines a collection of finite group actions fit for the hallmark exchange of ``geometric'' and ``quantum'' symmetries.
 Thus, neither $(f_1(x),\tT f_1(y))$ nor $(f_2(x),\tT f_2(y))$ can be used to construct transpose-mirrors \`a~la\cite{rBH, rBH-LGO+EG, Krawitz:2009aa, rMK-PhD, rFJR-07b, Ebeling:2012Mir}.
 
 While we are not aware of a rigorous proof that the sub-simplex reductions of both the spanning and the (extended) Newton multitope must enclose the lattice origin for the transpose-mirror construction\cite{rBH, rBH-LGO+EG, Krawitz:2009aa, rMK-PhD, rFJR-07b, Ebeling:2012Mir}, we have not found any counter-example to this, whence statement~(1) of Conjecture~\ref{C:gMM}.

\begingroup
\footnotesize\raggedright
\def\rasp{\leavevmode\raise.45ex\hbox{$\rhook$}}
\providecommand{\href}[2]{#2}\begingroup\raggedright\endgroup

\endgroup

\end{document}